\documentclass{aa}

\usepackage{graphicx}
\usepackage{txfonts}
\usepackage{natbib}
\bibpunct{(}{)}{;}{a}{}{,}

\begin{document}

\title{Optical multiband surface photometry of a sample of \\ Seyfert galaxies}

\subtitle{I. Large-scale morphology and local environment analysis of\\ matched Seyfert and inactive galaxy samples\thanks{Based on observations made with the 2-m telescope of the Institute of Astronomy and National Astronomical Observatory, Bulgarian Academy of Sciences.}\fnmsep\thanks{Based on observations made with ESO Telescopes at the La Silla or Paranal Observatories under programmes 69.D-0453(B), 079.B-0196(A), 081.B-0350(B), 71.B-0202(B), and 077.B-0356(B).}\fnmsep\thanks{Based on observations made with the NASA/ESA Hubble Space Telescope, obtained from the data archive at the Space Telescope Science Institute. STScI is operated by the Association of Universities for Research in Astronomy, Inc. under NASA contract NAS 5-26555.}}

\author{L.~Slavcheva-Mihova\inst{}
\and B.~Mihov\inst{}}

% \offprints{L.~Slavcheva-Mihova}

\institute{Institute of Astronomy and National Astronomical Observatory, Bulgarian Academy of Sciences, \\72 Tsarigradsko Chaussee Blvd., 1784 Sofia, Bulgaria \\
\email{lslav,bmihov@astro.bas.bg}}

\date{Received \ldots / Accepted \ldots}

\abstract
{Parallel analysis of the large-scale morphology and local environment of matched active and control galaxy samples plays an important role in studies of the fueling of active galactic nuclei. }
{
We carry out a detailed morphological characterization of a sample of 35 Seyfert galaxies and a matched sample of inactive galaxies in order to compare the evidence of non-axisymmetric perturbation of the potential and, in the second part of this paper, to be able to perform a multicomponent photometric decomposition of the Seyfert galaxies.}
{
We constructed contour maps, $BVR_{\rm \scriptstyle C}I_{\rm \scriptstyle C}$ profiles of the surface brightness, ellipticity, and position angle, as well as colour index profiles. We further used colour index images, residual images, and structure maps, which helped clarify the morphology of the galaxies. We studied the presence of close companions using literature data.}
{By straightening out the morphological status of some of the objects, we derived an improved morphological classification and built a solid basis for a further multicomponent decomposition of the Seyfert sample. We report hitherto undetected (to our knowledge) structural components in some Seyfert galaxies~-- a bar (Ark\,479), an oval/lens (Mrk\,595), rings (Ark\,120, Mrk\,376), a nuclear bar and ring (Mrk\,352), and nuclear dust lanes (Mrk\,590). 
We compared the large-scale morphology and local environment of the Seyfert sample to those of the control one and found that (1) the two samples show similar incidences of bars, rings, asymmetries, and close companions; (2) the Seyfert bars are generally weaker than the bars of the control galaxies; (3) the bulk of the two samples shows morphological evidence of non-axisymmetric perturbations of the potential or close companions; (4) the fueling of Seyfert nuclei is not directly related to the large-scale morphology and local environment of their host galaxies.

}
{}


\titlerunning{Optical multiband surface photometry of Seyfert galaxies. I}

\authorrunning{L.~Slavcheva-Mihova \& B.~Mihov}

\maketitle

\section{Introduction}

The generally accepted active galactic nucleus (AGN) model requires gas accretion onto a supermassive black hole (SMBH). 
There are a number of fueling mechanisms of different relative importance depending on mass accretion rates and spatial scales. 
Major mergers are most commonly involved to explain the high accretion rates of the most luminous quasars; the more accretion rate decreases, the higher the number of efficient mechanisms (e.g., dynamical friction, viscous torques).
At Seyfert (Sy) luminosities, bars, tidal interactions, and minor mergers become important \citep[see, e.g., the reviews of][]{M_04,J_06}.
Bars have long been considered an efficient mechanism for inward gas transport down to about 1\,kpc \citep{S_84,SFB_89,PST_95}; among the possibilities for further driving the gas within the gravitational influence of the central source are nested bars \citep{SFB_89} and central spiral dust lanes \citep{RM_99}. The relation between galaxy interactions and the onset of nuclear activity is founded upon the key studies of \citet{TT_72} and \citet{G_79}. Minor mergers could induce gas inflow to the nuclear regions \citep[e.g.,][]{HM_95}. 
Finding clear evidence of a minor merger is generally hard, since the sinking satellite detectability depends on the stage and geometry of the merger and on the parameters of the galaxies involved; e.g., the minor merger is hardly recognizable in its final stages \citep[][]{WMH_96}. Numerical simulations show that (minor) mergers, together with tidal interactions, could induce tails, bridges, shells, bars, and various types of disturbed spiral structure and asymmetries \citep[e.g.,][]{TT_72,HQ_89,MWH_95,HM_95}. Thus, asymmetries have often been associated with mergers \citep[e.g.,][]{CBJ_00,C1_03,DCL_07}.  

The question of statistical differences between Sy and inactive galaxies considering non-axisymmetric perturbations of the potential is somewhat controversial. It is a prevalent view that neither bars \citep[e.g.,][]{MR_97}, companions \citep[e.g.,][]{RYH2_98,S_01}, minor mergers \citep[e.g.,][]{C_00}, nested bars \citep{ES_02}, nor nuclear dust spiral arms \citep{MRM2_03} are specific signatures of Sy galaxies. Some studies, however, prompt an excess of bars \citep[e.g.,][]{LSK_02}, outer rings \citep{HM1_99}, companions \citep[e.g.,][]{RVB_95}, and, considering early type galaxies, circumnuclear features \citep[][]{XP_02} and dust \citep[e.g.,][]{SSF_07} in Sy galaxy samples. \citet{T_99} even suggested (minor) mergers as a unified formation mechanism of (low-luminosity) AGNs in the local Universe. 

It is now believed that the SMBH and its host galaxy have coevolved. A relation between SMBH mass and bulge luminosity was first established for inactive galaxies \citep{KR_95} and then extended to active galaxies \citep{L_98,W_99}. Similar relations link SMBH mass with bulge velocity dispersion and light concentration \citep[for a review see][]{FF_05}.
In particular, the accurate photometric separation of the bulge from the other galactic components is of utmost importance for studying the ``SMBH mass~-- bulge luminosity'' relation in depth. The bulge luminosity obtained by bulge-disk decomposition tends to be systematically lower \citep[][]{W_02} and leads to less scatter of the above relation \citep[e.g.,][]{MH_03,EGC_04}, than the luminosity estimate based on the empirical relation between the bulge-to-total luminosity ratio (B/T) and the Hubble stage \citep[$T$,][]{SV_86}.

Generally, the way to have a precise differentiation of the flux of the individual components is photometric decomposition, which, in its simplest form, uses analytical functions for the radial surface brightness (SB) profiles of bulge and disk. Typically the SB distribution of disks is satisfactorily fitted by an exponential function \citep{F_70}. 
The S\'ersic law \citep[or $r^{1/n}$,][]{S_68} has supplanted the $r^{1/4}$ law \citep[][]{dV_48} in the approximation of bulge SB distribution since the works of \citet{AS_94} and \citet{APB_95}.
Bulges of early-type spirals, however, appear more exponential than previously assumed \citep{BGD_03,LSB_05}.
Furthermore, lower values of B/T than in earlier studies have been reported \citep{LSB_05,LSB_06,LSB_07,WJK_09}.
The reason for the observed lower mean values of $n$ and B/T is most likely related to the multicomponent decomposition used \citep{LSB_05,LSB_07}:
the omission of bars (and ovals/lenses) leads to modifying bulge (mostly) and disk parameters and furthermore to B/T inflation \citep[see also][]{G_08,WJK_09}. 
Therefore, the way to have precise parameter estimates as a result of decomposition is to take all significant components in the galaxy under consideration into account, as well as to choose the right functional form for each of them. 

Detailed morphological characterization, i.e., disclosure of the features present, is important in the context of AGN fueling mechanisms, correlations among structural parameters, and galaxy morphological classification. The last one named should not be overlooked, keeping the correlation of a great deal of parameters with $T$ in mind. 
Two basic kinds of approaches can be discerned: detailed case-by-case research on relatively small galaxy samples, which can adequately reveal and model the components present but would be an arduous task for a great number of objects, and studies of large samples in an automated manner, which would lead to results of higher statistical weight but could hardly take all structures in the individual galaxies into account \citep[see also][]{G_08}.

Our study is of the first type. Its aim is to explore the morphological features related to AGN fueling mechanisms and the relations among the structural parameters, including the ``SMBH mass~-- bulge luminosity'' relation, involving detailed morphological characterization and SB decomposition of a sample of Sy galaxies. A parallel discussion of a matched sample of inactive galaxies is presented to study the eventual differences in the large-scale morphology and local environment of the Sy and inactive galaxies. The morphological characterization is based on scrutinizing various types of images, maps, residuals, and profiles. The results of a multicomponent SB decomposition will be presented in a companion paper.

The paper is organized as follows. Sample selection is presented in Sect.\,\ref{sample}. Observations and primary data reduction are outlined in Sect.\,\ref{obsred}. Surface photometry steps are followed through in Sect.\,\ref{surf}. Bar characterization is described in Sect.\,\ref{bars}. Section\,\ref{res} presents the surface photometry outputs. The local environment of the galaxies is commented on in Sect.\,\ref{loc_env}. A discussion follows in Sect.\,\ref{disc}. A summary of our results is given in Sect.\,\ref{concl}. A set of contour maps and profiles of the Sy galaxies is presented in Appendix\,\ref{profiles}. Individual Sy galaxies are discussed in Appendix\,\ref{indiv}. 

Throughout the paper the linear sizes and projected linear separations in kpc have been calculated using the cosmology-corrected scale given in NASA/IPAC Extragalactic Database \citep[NED; $H_{\rm 0}\,$=$\,73\,\rm km\,\rm s^{-1}\,Mpc^{-1}$,  $\Omega_{\,\rm matter}\,$=$\,0.27$,  $\Omega_{\,\rm vacuum}\,$=$\,0.73$,][]{SBD_07}.

\section{Sample selection}
\label{sample}

We selected Sy galaxies with reverberation-based black hole masses compiled by \citet{H_99} and updated by \citet{PFG_04}, as well as relatively poorly studied Sy galaxies regarding morphological characterization and multicomponent SB profile decomposition from \citet{VV_98},
on which we imposed the following constraints:
\begin{itemize}
 \item redshift $z\,$$<$\,0.1, so that the galaxies are close enough to provide spatial resolution adequate for a proper morphological characterization and multicomponent SB decomposition;
\item galaxy isophotal (at 25 $B$ mag arcsec$^{-2}$) diameters larger than 20\arcsec, i.e., well-resolved host galaxies;
\item inclination less than 70$\degr$ to avoid highly inclined galaxies, for which structures can be difficult to recognize;
\item suitable for observation at the Rozhen National Astronomical Observatory (NAO), Bulgaria.
\end{itemize}

The Sy sample consists of 35 galaxies. A control sample of inactive galaxies was selected from the Center for Astrophysics (CfA) Redshift Survey \citep{HDL_83,HGC_95} to compare their morphology and environment to those of the Sy sample galaxies. The inactive galaxies were matched on a one-to-one basis to the Sy galaxies in $T$, radial heliocentric velocity $V\rm _{r}$, absolute $B$-band magnitude $M ^{B}_{\rm abs}$, and ellipticity $\epsilon$.
For two of the Sy galaxies (III\,Zw\,2 and Mrk\,1513, among the most distant ones), we could not find any appropriate counterparts in the CfA Redshift Survey, so we selected their matched galaxies from the Sloan Digital Sky Survey \citep[SDSS,][]{YAA_00}.
 $M ^{B}\rm_{abs}$ of an inactive galaxy was matched to $M^{B}\rm_{abs}\,$+$0\fm5$ of an Sy galaxy and the median of these values is given below and plotted in Fig.\,\ref{M_AN_5ouraCfA_hbm}. The value of $0\fm5$ is a mean one, based on our preliminary decomposition results for the contribution of the AGNs to the total Sy galaxy magnitudes. 
The median values of the matched parameters of the Sy/control sample are $T$\,$=$\,$0/0$, $V\rm _{r}\,$=$\,8089$/$7934\,\rm km\,\rm s^{-1}$, $M ^{B}\rm_{abs}\,$=$\, -20\fm88$/$-21\fm03$, and $\epsilon\,$=$\,0.19/0.20$. Their distribution is shown in Figs.\,\ref{MT_AN_ourour_hbm}-\ref{E_AN_ourour_hbm}.

Basic information about the sample galaxies is given in Tables\,\ref{T_morph_AG} and \ref{T_morph_NG}. 
The morphological type is taken from the Third Reference Catalogue of Bright Galaxies \citep[RC3,][]{RC3}. If there is no classification in RC3 or it is doubtful, we take the morphological type given in NED; if none of the above types exists, we list the one in HyperLeda\footnote{http://leda.univ-lyon1.fr} \citep[][]{PPP_03} or SIMBAD.

\section{Observations and primary data reduction}
\label{obsred}

The Sy sample observations were performed at NAO with the 2-m Ritchey-Chr\'etien telescope equipped with $1024\,$$\times$$\,1024$ Photometrics AT200 CCD camera (CCD chip SITe~SI003AB with a square pixel size of $24\,\mu\rm m$ that corresponds to $0\farcs309$ on the sky) or with $1340\,$$\times$$\,1300$ Princeton Instruments VersArray:1300B CCD camera (CCD chip EEV CCD36-40 with a square pixel size of $20\,\mu\rm m$ that corresponds to $0\farcs258$ on the sky). Two galaxies were observed
employing a two-channel focal reducer \citep{JCB_00} and $512\,$$\times$$\,512$ Princeton Instruments VersArray:512B CCD camera (CCD chip EEV CCD77-00 with a square pixel size of $24\,\mu\rm m$, which corresponds to $0\farcs884$ on the sky). Standard Johnson-Cousins $BVR_{\rm \scriptstyle C}I_{\rm \scriptstyle C}$ filters were used. For a couple of objects we were not able to obtain images of good quality and used archival data. 
The observation log (including archival data) is presented in Table~\ref{T_obs}.

\begin{table*}
\caption{Characteristics of the Sy galaxy sample in order of increasing right ascension.}
\label{T_morph_AG}
\centering
\begin{tabular}{llllllccccc} 

\hline\hline             
\noalign{\smallskip}~~Galaxy & ~~~~~~~Other Names & ~~~~~~~~~~~$z^{\rm N}$ & Sy$^{\rm N}$ & Morph. Type$^{\rm RC3}$ & Morph. Type$^{\rm our}$   & B & R & A & C & Any  \\
\noalign{\smallskip} ~~~~~(1) & ~~~~~~~~~~~~~~~(2) & ~~~~~~~~~~(3) & (4) & ~~~~~~~~(5)  & (6) & (7) & (8) & (9) & (10)& (11)      \\
\noalign{\smallskip}
\hline
\noalign{\smallskip}

 Mrk\,335  & PG\,0003+199              & $0.025785\,(~~63)$  & 1.2 & S0/a$^{\rm N}$	      & SA0\,pec		       & $\circ$   & $\circ$   & $\bullet$ & $\circ$   & $\bullet$ \\
 III\,Zw\,2& Mrk\,1501,~PG\,0007+106   &  0.089338           & 1.2 & E$^{\rm Sim}$	      & SA0\,pec		       & $\circ$   & $\circ$   & $\bullet$ & $\bullet$ & $\bullet$ \\
 Mrk\,348  & NGC\,262,~UGC\,00499      & $0.015034\,(~~13)$  & 2   & SA(s)0/a:  	      & SA(s)a  		       & $\circ$   & $\bullet$ & $\bullet$ & $\bullet$ & $\bullet$ \\
 I\,Zw\,1  & Mrk\,1502,~PG\,0050+124   & $0.061142\,(~~67)$  & 1   & Sa$^{\rm N}$	      & SA(s)ab 		       & $\circ$   & $\circ$   & $\bullet$ & $\bullet$ & $\bullet$ \\
 Mrk\,352  & CGCG\,501--058            & $0.014864\,(~~20)$  & 1   & SA0		      & SA0			       & $\circ$   & $\circ$   & $\circ$   & $\circ$   & $\circ$   \\
 Mrk\,573  & UGC\,01214                & $0.017179\,(~~37)$  & 2   & (R)SAB(rs)0$^{+}$:       & (R)SAB(r)0		       & $\bullet$ & $\bullet$ & $\circ$   & $\bullet$ & $\bullet$ \\
 Mrk\,590  & NGC\,863,~UGC\,01727      & $0.026385\,(~~40)$  & 1.2 & SA(s)a:		      & SA(s)a  		       & $\circ$   & $\circ$   & $\circ$   & $\bullet$ & $\bullet$ \\
 Mrk\,595  & CGCG\,414--040            & $0.026982\,(~~80)$  & 1.5 & Sa$^{\rm N}$	      & SAB0/a  		       & $\bullet$ & $\circ$   & $\circ$   & $\circ$   & $\bullet$ \\
 3C\,120   & Mrk\,1506,~UGC\,03087     & $0.033010\,(~~30)$  & 1   & S0:		      & SA0\,pec		       & $\circ$   & $\circ$   & $\bullet$ & $\circ$   & $\bullet$ \\
 Ark\,120  & Mrk\,1095,~UGC\,03271     & $0.032713\,(~~57)$  & 1   & Sb\,pec$^{\rm N}$        & SA(r)0\,pec		       & $\circ$   & $\bullet$ & $\bullet$ & $\circ$   & $\bullet$ \\
 Mrk\,376  & IRAS\,07105+4547          & $0.055980\,(~~23)$  & 1.5 & S0:$^{\rm N}$	      & ($\rm R^{\prime}$)SAB(r)a      & $\bullet$ & $\bullet$ & $\circ$   & $\circ$   & $\bullet$ \\
 Mrk\,79   & UGC\,03973                & $0.022189\,(~~27)$  & 1.2 & SBb		      & SB(rs)b 		       & $\bullet$ & $\bullet$ & $\bullet$ & $\circ$   & $\bullet$ \\
 Mrk\,382  & CGCG\,207--005            & $0.033687\,(~~53)$  & 1   & SBc$^{\rm Sim}$	      & ($\rm R^{\prime}$)SAB(r)bc     & $\bullet$ & $\bullet$ & $\circ$   & $\circ$   & $\bullet$ \\
 NGC\,3227 & UGC\,05620                & $0.003859\,(~~10)$  & 1.5 & SAB(s)a\,pec	      & SAB(s)a\,pec		       & $\bullet$ & $\circ$   & $\bullet$ & $\bullet$ & $\bullet$ \\
 NGC\,3516 & UGC\,06153                & $0.008836\,(~~23)$  & 1.5 & (R)SB(s)0$^{0}$:	      & (R)SAB(r)0		       & $\bullet$ & $\bullet$ & $\circ$   & $\circ$   & $\bullet$ \\
 NGC\,4051 & UGC\,07030                & $0.002336\,(~~~~4)$ & 1.5 & SAB(rs)bc  	      & SAB(s)bc		       & $\bullet$ & $\circ$   & $\bullet$ & $\circ$   & $\bullet$ \\
 NGC\,4151 & UGC\,07166                & $0.003319\,(~~10)$  & 1.5 & ($\rm R^{\prime}$)SAB(rs)ab: & ($\rm R^{\prime}$)SB(rs)ab & $\bullet$ & $\bullet$ & $\circ$   & $\bullet$ & $\bullet$ \\
 Mrk\,766  & NGC\,4253,~UGC\,07344     & $0.012929\,(~~53)$  & 1.5 & ($\rm R^{\prime}$)SB(s)a:    & ($\rm R^{\prime}$)SAB(s)ab & $\bullet$ & $\bullet$ & $\bullet$ & $\circ$   & $\bullet$ \\
 Mrk\,771  & Ark\,374,~PG\,1229+204    & $0.063010\,(153)$   & 1   & Spiral$^{\rm N}$	      & ($\rm R^{\prime}$)SAB0/a\,pec  & $\bullet$ & $\bullet$ & $\bullet$ & $\circ$   & $\bullet$ \\
 NGC\,4593 & Mrk\,1330                 & $0.009000\,(127)$   & 1   & (R)SB(rs)b 	      & ($\rm R^{\prime}$)SAB(rs)b     & $\bullet$ & $\bullet$ & $\bullet$ & $\bullet$ & $\bullet$ \\
 Mrk\,279  & UGC\,08823,~PG\,1351+695  & $0.030451\,(~~83)$  & 1.5 & S0 		      & (R)SAB0\,pec		       & $\bullet$ & $\bullet$ & $\bullet$ & $\bullet$ & $\bullet$ \\
 NGC\,5548 & Mrk\,1509,~UGC\,09149     & $0.017175\,(~~23)$  & 1.5 & ($\rm R^{\prime}$)SA(s)0/a   & SA0/a\,pec  	       & $\circ$   & $\circ$   & $\bullet$ & $\bullet$ & $\bullet$ \\
 Ark\,479  & CGCG\,107--010            & $0.019664\,(133)$   & 2   & S0$^{\rm HL}$	      & SAB(s)ab 		       & $\bullet$ & $\circ$   & $\circ$   & $\bullet$ & $\bullet$ \\
 Mrk\,506  & CGCG\,170--020            & $0.043030\,(~~40)$  & 1.5 & SAB(r)a		      & (R)SA(r)0/a		       & $\circ$   & $\bullet$ & $\circ$   & $\bullet$ & $\bullet$ \\
 3C\,382   & CGCG\,173--014            & $0.057870\,(160)$   & 1   & \ldots		      & SA0\,pec 		       & $\circ$   & $\circ$   & $\bullet$ & $\circ$   & $\bullet$ \\
 3C\,390.3 & VII\,Zw\,838              &  0.056100           & 1   & S0:$^{\rm Sim}$	      & SA0			       & $\circ$   & $\circ$   & $\circ$   & $\bullet$ & $\bullet$ \\
 NGC\,6814 & MCG\,--02--50--001        & $0.005214\,(~~~~7)$ & 1.5 & SAB(rs)bc  	      & SAB(rs)bc		       & $\bullet$ & $\bullet$ & $\circ$   & $\circ$   & $\bullet$ \\
 Mrk\,509  & IRAS\,20414--1054	       & $0.034397\,(~~40)$  & 1.2 & \ldots		      & SA0			       & $\circ$   & $\circ$   & $\circ$   & $\circ$   & $\circ$   \\
 Mrk\,1513 & II\,Zw\,136,~PG\,2130+099 & $0.062977\,(100)$   & 1   & (R)Sa		      & ($\rm R^{\prime}$)SA(s)a       & $\circ$   & $\bullet$ & $\circ$   & $\circ$   & $\bullet$ \\
 Mrk\,304  & II\,Zw\,175,~PG\,2214+139 & $0.065762\,(~~27)$  & 1   & \ldots		      & SA0			       & $\circ$   & $\circ$   & $\circ$   & $\circ$   & $\circ$   \\
 Ark\,564  & UGC\,12163                & $0.024684\,(~~67)$  & 1.8 & SB 		      & ($\rm R^{\prime}$)SB(s)b       & $\bullet$ & $\bullet$ & $\circ$   & $\circ$   & $\bullet$ \\
 NGC\,7469 & Mrk\,1514,~UGC\,12332     & $0.016317\,(~~~~7)$ & 1.2 & ($\rm R^{\prime}$)SAB(rs)a   & ($\rm R^{\prime}$)SAB(rs)a & $\bullet$ & $\bullet$ & $\circ$   & $\bullet$ & $\bullet$ \\
 Mrk\,315  & II\,Zw\,187               & $0.038870\,(~~83)$  & 1.5 & E1\,pec?		      & SA(s)0/a\,pec		       & $\circ$   & $\circ$   & $\bullet$ & $\bullet$ & $\bullet$ \\
 NGC\,7603 & Mrk\,530,~UGC\,12493      & $0.029524\,(~~73)$  & 1.5 & SA(rs)b:\,pec	      & SA0\,pec		       & $\circ$   & $\circ$   & $\bullet$ & ?         & $\bullet$ \\
 Mrk\,541  & CGCG\,408--001            & $0.039427\,(~~40)$  & 1   & E/S0$^{\rm Sim}$	      & (R)SA(r)0     		       & $\circ$   & $\bullet$ & $\bullet$ & $\circ$   & $\bullet$ \\ 

\hline

\end{tabular}
\tablefoot{Columns\,7-11 reveal the presence of (7) bar, oval or lens; (8) inner and/or outer (pseudo-)ring; (9) asymmetry; (10) companion (? denotes an anomalous redshift system); (11) any of the previous features.
The superscripts N/HL/Sim stand for NED/HyperLeda/SIMBAD.} 
\end{table*}

\begin{table*}
 
\caption{Characteristics of the inactive galaxy sample.}
\label{T_morph_NG}
\centering
\begin{tabular}{@{}l@{\hspace{0.32cm}}l@{\hspace{0.2cm}}l@{\hspace{0.35cm}}l@{\hspace{0.2cm}}l@{\hspace{0.35cm}}l@{\hspace{0.1cm}}l@{\hspace{0.2cm}}c@{\hspace{0.18cm}}c@{\hspace{0.08cm}}c@{\hspace{0.02cm}}c@{\hspace{0.02cm}}c@{}} 

\hline\hline             
\noalign{\smallskip}Sy Galaxy & Inactive Galaxy & Source & Telescope & ~~~~~~~~~$z^{\rm N}$ & Morph. Type$^{\rm RC3}$ & Morph. Type$^{\rm our}$  & B & R & A & C & Any \\
\noalign{\smallskip} ~~~~~~(1) & ~~~~~~~~~(2) & ~~~~(3) & ~~~~(4) & ~~~~~~~~~(5) & ~~~~~~~~~~(6) & ~~~~~~~~(7) & (8) & (9) & (10) & (11) & (12)     \\
\noalign{\smallskip}
\hline
\noalign{\smallskip}

 Mrk\,335   & IC\,5017  		& ESO          & VLT-U4      & 0.025174\,(~\,87) & (R)SAB(rs)0$^{0}$ 		              & (R)SB(r)0			& $\bullet$ & $\bullet$ & $\circ$   & $\circ$	& $\bullet$ \\                                       
 III\,Zw\,2   & 2MASX\,J01505708+0014040	& SDSS         & 2.5-m       & 0.082226\,(102)   & S0/a$^{\rm HL}$			      & SA0				& $\circ$   & $\circ$	& $\circ$   & $\circ$	& $\circ$   \\                                    
 Mrk\,348   & NGC\,2144 		& DSS+E        & STs         & 0.015924	         & ($\rm R^{\prime}$)SA(rs)a:	              & ($\rm R^{\prime}$)SA(rs)a\,pec  & $\circ$   & $\bullet$ & $\bullet$ & $\bullet$ & $\bullet$ \\               
 I\,Zw\,1     & ESO\,155--\,G\,027	& DSS+E        & STs         & 0.062110\,(334)   & ($\rm R^{\prime}_1$?)SB(rs)ab$^{\rm N}$    & SB(r)b  			& $\bullet$ & $\bullet$ & $\bullet$ & $\bullet$ & $\bullet$ \\                 
 Mrk\,352   & 2MASX\,J04363658--0250350 & NED          & CFHT        & 0.015564\,(163)   & S0$^{\rm N}$			              & SA0				& $\circ$   & $\circ$	& $\circ$   & $\bullet$ & $\bullet$ \\                                   
 Mrk\,573   & ESO\,542--\,G\,015	& SDSS         & 2.5-m       & 0.018570\,(~\,90) & S0(r):$^{\rm Sim}$		              & SAB0				& $\bullet$ & $\circ$	& $\circ$   & $\circ$	& $\bullet$ \\                                      
 Mrk\,590   & NGC\,4186 		& SDSS         & 2.5-m       & 0.026292\,(~\,17) & SA(s)ab:  			              & SA(rs)a 			& $\circ$   & $\bullet$ & $\bullet$ & $\circ$	& $\bullet$ \\                                               
 Mrk\,595   & 2MASX\,J00342513--0735582 & NED          & CFHT        & 0.026218\,(150)   & SB0/a$^{\rm N}$			      & SAB0				& $\bullet$ & $\circ$	& $\circ$   & $\bullet$ & $\bullet$ \\                               
 3C\,120    & ESO\,202--\,G\,001	& DSS+E        & STs         & 0.033620\,(~\,87) & SAB(r)0$^{\rm 0}$:\,pec		      & SA0\,pec			& $\circ$   & $\circ$	& $\bullet$ & $\circ$	& $\bullet$ \\                              
 Ark\,120   & IC\,5065  		& ESO          & 3.6-m      & 0.032689\,(~\,47) & SB0:\,pec 			              & SAB0\,pec			& $\bullet$ & $\circ$	& $\bullet$ & $\bullet$ & $\bullet$ \\                                              
 Mrk\,376   & ESO\,545--\,G\,036	& ESO\tablefootmark{a}& Dutch       & 0.057166\,(~\,97) & ($\rm R^{\prime}$?)SA(s)a$^{\rm N}$        & ($\rm R^{\prime}$)SA(s)a\,pec	& $\circ$   & $\bullet$ & $\bullet$ & $\circ$	& $\bullet$ \\
 Mrk\,79    & ESO\,340--\,G\,036	& DSS+E        & STs         & 0.021722\,(~\,83) & SB(r)b				      & SB(r)b  			& $\bullet$ & $\bullet$ & $\bullet$ & $\circ$	& $\bullet$ \\                                           
 Mrk\,382   & ESO\,268--\,G\,032	& DSS+E        & STs         & 0.034657\,(~\,33) & SAB(s)bc$^{\rm N}$		              & SAB(s)bc 			& $\bullet$ & $\circ$	& $\circ$   & $\circ$	& $\bullet$ \\                                  
 NGC\,3227  & IC\, 5240     		& ESO          & NTT         & 0.005886\,(~\,24) & SB(r)a				      & SB(r)a  			& $\bullet$ & $\bullet$ & $\circ$   & $\circ$	& $\bullet$ \\                                              
 NGC\,3516  & ESO\,183--\,G\,030	& ESO          & 2.2-m      & 0.008966\,(107)   & SA0$^{-}$\,pec?			      & SA0				& $\circ$   & $\circ$	& $\circ$   & $\bullet$ & $\bullet$ \\                                       
 NGC\,4051  & IC\,1993     		& ESO          & 3.6-m      & 0.003602\,(~\,10) & ($\rm R^{\prime}$)SAB(rs)b	              & ($\rm R^{\prime}$)SA(s)bc	& $\circ$   & $\bullet$ & $\circ$   & $\circ$	& $\bullet$ \\               
 NGC\,4151  & NGC\,2775    		& SDSS         & 2.5-m       & 0.004516\,(~\,17) & SA(r)ab				      & SA(s)ab 			& $\circ$   & $\circ$	& $\circ$   & $\bullet$ & $\bullet$ \\                                              
 Mrk\,766   & UGC\,6520   		& SDSS         & 2.5-m       & 0.012255\,(133)   & SB?				              & ($\rm R^{\prime}$)SB(rs)ab	& $\bullet$ & $\bullet$ & $\circ$   & $\bullet$ & $\bullet$ \\                                 
 Mrk\,771   & ESO\,349--\,G\,011	& DSS+E        & STs         & 0.063821\,(~\,93) & SB(r)a$^{\rm N}$  		              & SB(r)a  			& $\bullet$ & $\bullet$ & $\circ$   & $\circ$	& $\bullet$ \\                                 
 NGC\,4593  & NGC\,4902 		& NED          & CTIO 0.9-m & 0.008916\,(~\,17) & SB(r)b				      & SB(r)b  			& $\bullet$ & $\bullet$ & $\bullet$ & $\bullet$ & $\bullet$ \\                                                
 Mrk\,279   & ESO\,324--\,G\,003	& DSS+E        & STs         & 0.029073	         & (R)SA(r)0$^{+,\rm N}$		      & SA(r)0  			& $\circ$   & $\bullet$ & $\circ$   & $\circ$	& $\bullet$ \\                                   
 NGC\,5548  & NGC\,466 	        	& DSS+E        & STs         & 0.017552\,(~\,87) & SA(rs)0$^{+}$:			      & SA(s)0/a			& $\circ$   & $\circ$	& $\bullet$ & $\circ$	& $\bullet$ \\                                      
 Ark\,479   & ESO\,297--\,G\,027	& ESO          & NTT         & 0.021221\,(~\,33) & SA(rs)b:  			              & SA(rs)ab			& $\circ$   & $\bullet$ & $\circ$   & $\circ$	& $\bullet$ \\                                        
 Mrk\,506   & ESO\,510--\,G\,048		& DSS+E        & STs         & 0.044991\,(103)   & SA(s)0/a:\,pec			      & SA(s)0/a\,pec			& $\circ$   & $\circ$	& $\bullet$ & $\bullet$ & $\bullet$ \\                                 
 3C\,382    & ESO\,292--\,G\,022	& DSS+E        & STs         & 0.056119\,(107)   & SA0$^{-}$\,pec:			      & SA0				& $\circ$   & $\circ$	& $\circ$   & $\circ$	& $\circ$   \\                                         
 3C\,390.3  & ESO\,249--\,G\,009	& DSS+E        & STs         & 0.054534\,(~\,81) & (R)SB0$^+:^{\rm N}$		              & (R)SAB0 			& $\bullet$ & $\bullet$ & $\circ$   & $\circ$	& $\bullet$ \\                                
 NGC\,6814  & NGC\,7421    		& ING          & JKT, WHT    & 0.005979\,(~\,29) & SB(rs)bc  			              & SB(rs)bc\,pec			& $\bullet$ & $\bullet$ & $\bullet$ & $\bullet$ & $\bullet$ \\                                     
 Mrk\,509   & ESO\,147--\,G\,013	& DSS+E        & STs         & 0.035485\,(~\,93) & S0$^{0}$:\,pec			      & SA0\,pec			& $\circ$   & $\circ$	& $\bullet$ & $\circ$	& $\bullet$ \\                                      
 Mrk\,1513  & 2MASX\,J14595983+2046121	& SDSS         & 2.5-m       & 0.061600\,(200)   & \ldots				      & SA(s)a  			& $\circ$   & $\circ$	& $\bullet$ & $\circ$	& $\bullet$ \\                                       
 Mrk\,304   & ESO\,292--\,G\,007	& DSS+E        & STs         & 0.068381\,(100)   & S0$^{\rm N}$			              & SA0				& $\circ$   & $\circ$	& $\circ$   & $\bullet$ & $\bullet$ \\                                          
 Ark\,564   & ESO\,552--\,G\,053	& DSS+E        & STs         & 0.024147\,(~\,90) & SB(r)b				      & ($\rm R^{\prime}$)SB(r)b	& $\bullet$ & $\bullet$ & $\bullet$ & $\circ$	& $\bullet$ \\                             
 NGC\,7469  & NGC\,897 		        & DSS\tablefootmark{b}+E& STs         & 0.015868\,(~\,53) & SA(rs)a				      & SA(rs)a 			& $\circ$   & $\bullet$ & $\circ$   & $\bullet$ & $\bullet$ \\                                        
 Mrk\,315   & ESO\,423--\,G\,016	& DSS+E        & STs         & 0.039204\,(~\,73) & (R)SB(s)0/a			              & (R)SAB(s)0/a			& $\bullet$ & $\bullet$ & $\circ$   & $\bullet$ & $\bullet$ \\                                 
 NGC\,7603  & ESO\,113--\,G\,050	& DSS+E        & STs         & 0.028873\,(~\,90) & S0$^{-}$\,pec			      & SA0				& $\circ$   & $\circ$	& $\bullet$ & $\circ$	& $\bullet$ \\                                           
 Mrk\,541   & UGC\,9532\,NED04 	        & SDSS         & 2.5-m       & 0.041889\,(150)   & S0$^{\rm Sim}$			      & (R)SAB0 			& $\bullet$ & $\bullet$ & $\circ$   & $\bullet$ & $\bullet$ \\                           
 
\hline

\end{tabular}
\tablefoot{(3) DSS+E designates DSS\,I,\,II, and digitized ESO-Uppsala Survey; (4) VLT-U4~-- Very Large Telescope Unit 4; STs~-- the Schmidt telescopes used to produce the Palomar Observatory Sky Survey\,I,\,II, and the ESO-Uppsala Survey; CFHT~-- Canada France Hawaii Telescope; NTT~-- New Technology Telescope; CTIO~-- Cerro Tololo Inter-American Observatory; JKT~-- Jacobus Kapteyn Telescope; WHT~-- William Herschel Telescope; (5) The redshift of 2MASX\,J14595983+2046121 was taken from SDSS; 
Cols.\,8-12 are the same as Cols. 7-11 of Table\,\ref{T_morph_AG}. \\
\tablefoottext{a}{We inspected the $R$ image of \citet{BBH_99}.}
\tablefoottext{b}{DSS\,II optical data not available for this object.}}
 
 \end{table*}

Multiple frames were acquired for all objects of interest. Standard fields were observed two
or three times each night at different airmass values not exceeding $X\,$=$\,2$. The galaxy fields were
observed in the same airmass range. Zero-exposure frames were taken regularly during the observing runs, and
$I_{\rm \scriptstyle C}$ frames of relatively blank night sky regions were taken for fringe
frame composition. Flat field frames were taken in the morning and/or evening twilight. Both flat fields
and $I_{\rm \scriptstyle C}$ blank frames were offset from one exposure to the next, so
that stars could be filtered out. 

The primary reduction of the data, as well as most of the surface photometry, was performed by means of packages within the ESO-MIDAS\footnote{The European Southern Observatory Munich Image Data Analysis System.} environment. The mean bias level was estimated using the virtual pre-scan bias section. 
The median of the zero-exposure frames was used to remove the residual bias pattern. Dark current correction was not needed. 
The median of the normalized flat field frames in each passband was used for flat fielding. Median combined blank $I_{\rm \scriptstyle C}$ frames were used for fringe correction; defringing increases the signal-to-noise ratio (S/N) in the outer galaxy regions and allows better estimation of sky background. Cosmic ray events and bad pixels were replaced by a local median value ($\rm \scriptstyle FILTER/COSMIC$ command). All images of a particular object were aligned to match the highest S/N frame, generally the $R_{\rm \scriptstyle C}$ one, and combined to obtain single $BVR_{\rm \scriptstyle C}I_{\rm \scriptstyle C}$ images. Primary reduction refers both to galaxy and standard field frames. 

To construct the point spread function (PSF) of the combined frames, we picked up a number of bright, isolated, and unsaturated stars, fitted a 2D Moffat profile \citep{M_69} to them, and weight-averaged the full widths at half maximum (FWHMs) and power-law indices, $\beta$, in each passband. The minimal FWHM (over all passbands) and the corresponding $\beta$ are listed in Table~\ref{T_obs}. If $\beta\,$$\rightarrow$$\,\infty$, then Moffat profile$\,\rightarrow\,$Gaussian \citep{TAC_01}, and if $\beta=1$, then Moffat profile$\,\equiv\,$Lorentzian.

We used archival data for the control sample. For about half of the galaxies CCD data from the SDSS, the European Southern Observatory (ESO), NED, and the Isaac Newton Group of Telescopes (ING) image archives were used. SDSS uses a dedicated 2.5-m telescope and a large-format CCD camera \citep{GCR_98,GSM_06} at the Apache Point Observatory in New Mexico to obtain images in five broad bands \citep[$ugriz$,][]{FIG_96}. The imaging data were processed by a photometric pipeline \citep{LGI_01,SLB_02}. The primary reduction of the data from the ING and ESO archives was performed as described above. The data taken through NED were reduced by the corresponding authors.
Digitized Palomar Observatory Sky Survey (DSS)\,I,\,II, and digitized ESO-Uppsala Survey \citep{LV_89} data were used for the rest of the inactive galaxies.
Data sources of the inactive galaxies are listed in Table\,\ref{T_morph_NG}. The supervening data reduction described below concerns both galaxy samples, while calibration is applied only to the Sy galaxies.

\section{Surface photometry}
\label{surf}

\subsection{Adaptive filtering}
\label{adapt}

To increase the S/N of the galaxy images we applied the adaptive filter technique \citep{LRC_93}, implemented in the Astrophysical Institute of Potsdam (AIP) package. The task uses H-transform to calculate the local S/N at each point of the image and determines the size of the impulse response of the filter at this point. Thus, adaptive filter extensively smooths sky background, to a lesser extent galaxy outskirts, and does not treat the highest resolution features. 

Adaptive filter has many advantages over the other most commonly used filters in surface photometry of galaxies \citep{SMP_05}; furthermore, it introduces no systematic errors \citep{CHL_88,TT_01}.

\subsection{Contaminating feature cleaning}
\label{clean}

Aside from intrinsic structures, like dust lanes, star formation regions, etc., isophotal shape can also be modified in a systematic fashion by contaminating features: contaminating objects (foreground stars or projected/companion  galaxies) or other features (diffraction spikes from bright stars, scattered light, meteor or satellite trails, etc.). The closer the contaminating feature to the galaxy core, the more carefully we treated it.

We used the following techniques to clean out contaminating features: interpolation, PSF subtraction, symmetric replacement, deblending, and annular cleaning.

{\em Interpolation}. This is the technique used most often. The algorithm in use (within the AIP package) generates masks covering the contaminating features and iteratively fills the background inside the masks by interpolating the background from the regions surrounding them. For each galaxy we generated one set of masks for all passbands (generally using the $R_{\rm \scriptstyle C}$ image), ensuring that the analysis is restricted to the same portions of the images.

The cleaned frames were visually inspected to remove any remaining sources of contamination in all passbands. Some of the defects~-- such as objects appearing with different brightnesses in the individual passbands (e.g., too ``blue''/``red''), meteor or satellite trails, diffraction spikes, bleeding along columns, scattered light, etc.~-- are unique for each passband/frame and demand individual approach. For instance, owing to a bright star in the field of Ark\,564, ghost images of the telescope main mirror and bleeding along columns were produced. These defects were corrected with the help of a set of individual masks.

{\em PSF subtraction.} 
We applied this technique in cases of stellar object contamination, especially when superposed on galaxy regions with large gradients. The contaminating objects were cleaned out by aligning, scaling, and subtracting the frame PSF. 

{\em Symmetric replacement.} We replaced the contaminated region with the region that is symmetric with respect to the galaxy centre and free of contaminating features in case the contaminated region does not cover too large a part of the galaxy area and is far enough from the galaxy centre. The galaxy in the two regions should be symmetric with respect to the galaxy centre. 

{\em Deblending.} We iteratively fitted ellipses to the contaminating object and the galaxy, starting with the brighter object, and subtracted the fitted models from the original frame until convergence was achieved. 

{\em Annular cleaning.} The contaminated region was covered with elliptical annuli of fixed width, centred on the galaxy nucleus. In each annulus the pixel values, deviating by more than a predefined threshold above the mode, were replaced by the mode \citep[for details see][]{MVB_97}.
\\

When we wanted to keep a single feature in the images and not take it into account in the ellipse fitting, we used the option of $\rm \scriptstyle FIT/ELL3$ command to exclude the sector containing the given feature from the fit (e.g., the extended feature in Mrk\,335).

We treated contaminating features when it was obvious that they did not belong to the galaxy. 
Some faint galaxy regions/projected objects can be cleaned/left by mistake, but their exclusion/inclusion will make little difference to the parameters derived.

\begin{figure*}[htbp]
   \centering
\vspace{0.1cm}
\begin{minipage}[t]{5.6cm}   
\includegraphics[width=5.6cm]{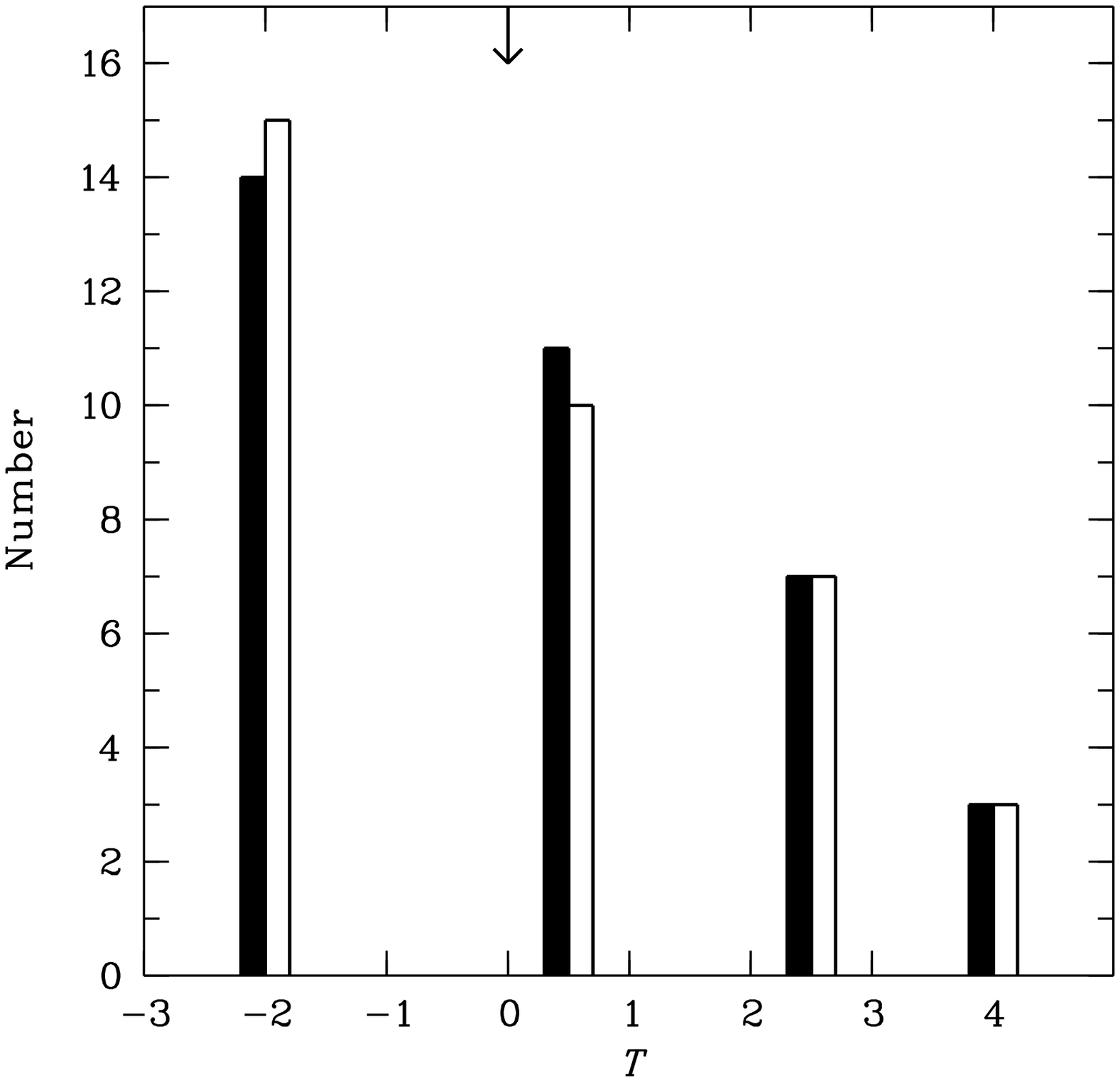}
      \caption{Distribution of $T$ assigned by this study for the Sy (black columns) and control (empty columns) sample. The bins shown correspond to $T$\,=\,--2, $T$\,=\,0,\,1, $T$\,=\,2,\,3, and $T$\,=\,4.  The arrow designates the median value of the two samples.}
\label{MT_AN_ourour_hbm}
\end{minipage}
\hspace{0.6cm}
\begin{minipage}[t]{5.6cm}   
\includegraphics[width=5.6cm]{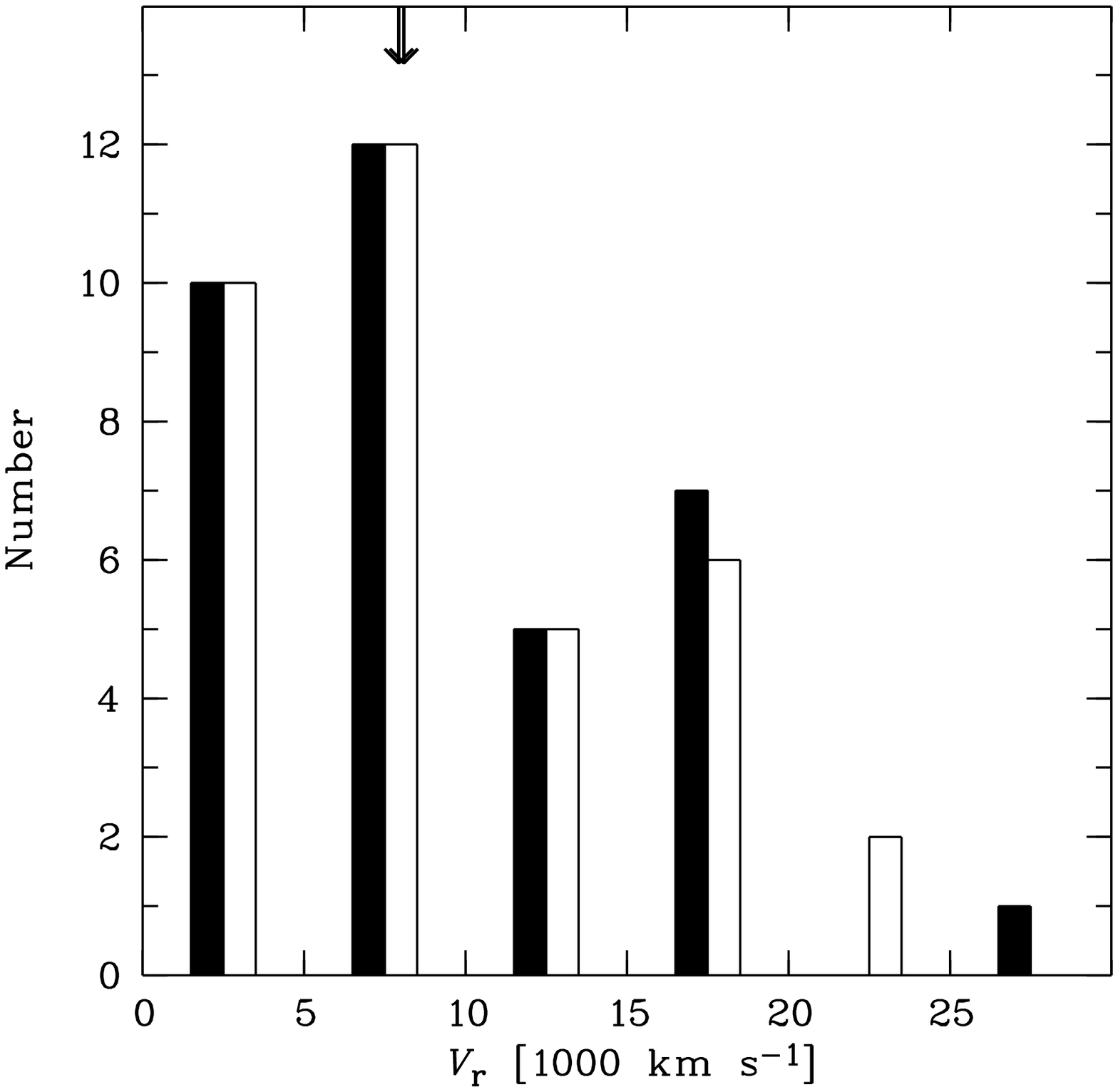}
      \caption{Distribution of $V\rm _{r}$ for the Sy sample (black columns, taken from NED) and the control one (empty columns, taken from the CfA Survey). The bin size is  $5000\,\rm km\,\rm s^{-1}$. The difference between the median values of the two samples is $\approx$$\,150\,\rm km\,\rm s^{-1}$, thus, the arrows, designating these values, appear blended.}
\label{V_AN_NEDCfA_hbm}
\end{minipage}
 \hspace{0.6cm}
 \begin{minipage}[t]{5.6cm}   
 \includegraphics[width=5.6cm]{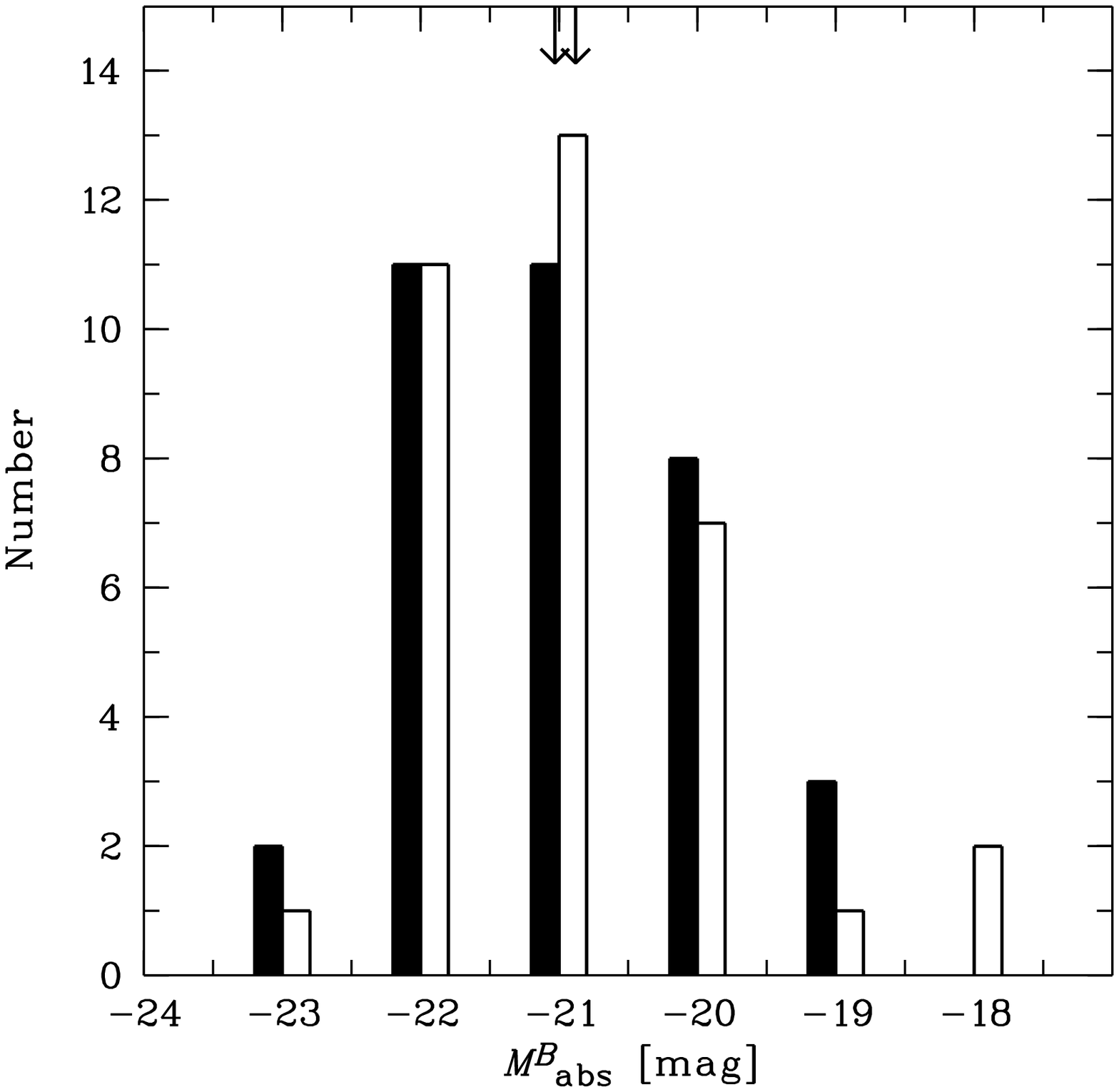}
       \caption{Distribution of $M ^{B}_{\rm abs}$ for the Sy sample (black columns, estimated on the basis of Slavcheva-Mihova \& Mihov, in prep.) and the control one (empty columns, estimated on the basis of the CfA Survey). The bin size is $1\rm ^{m}$. The left and right arrows designate the median values of the control and Sy (plus $0\fm5$, see text) sample, respectively.}
 \label{M_AN_5ouraCfA_hbm}
 \end{minipage}
\end{figure*}

\begin{figure}[htbp]
   \centering
\vspace{0.1cm}
\begin{minipage}[t]{5.6cm}   
\includegraphics[width=5.6cm]{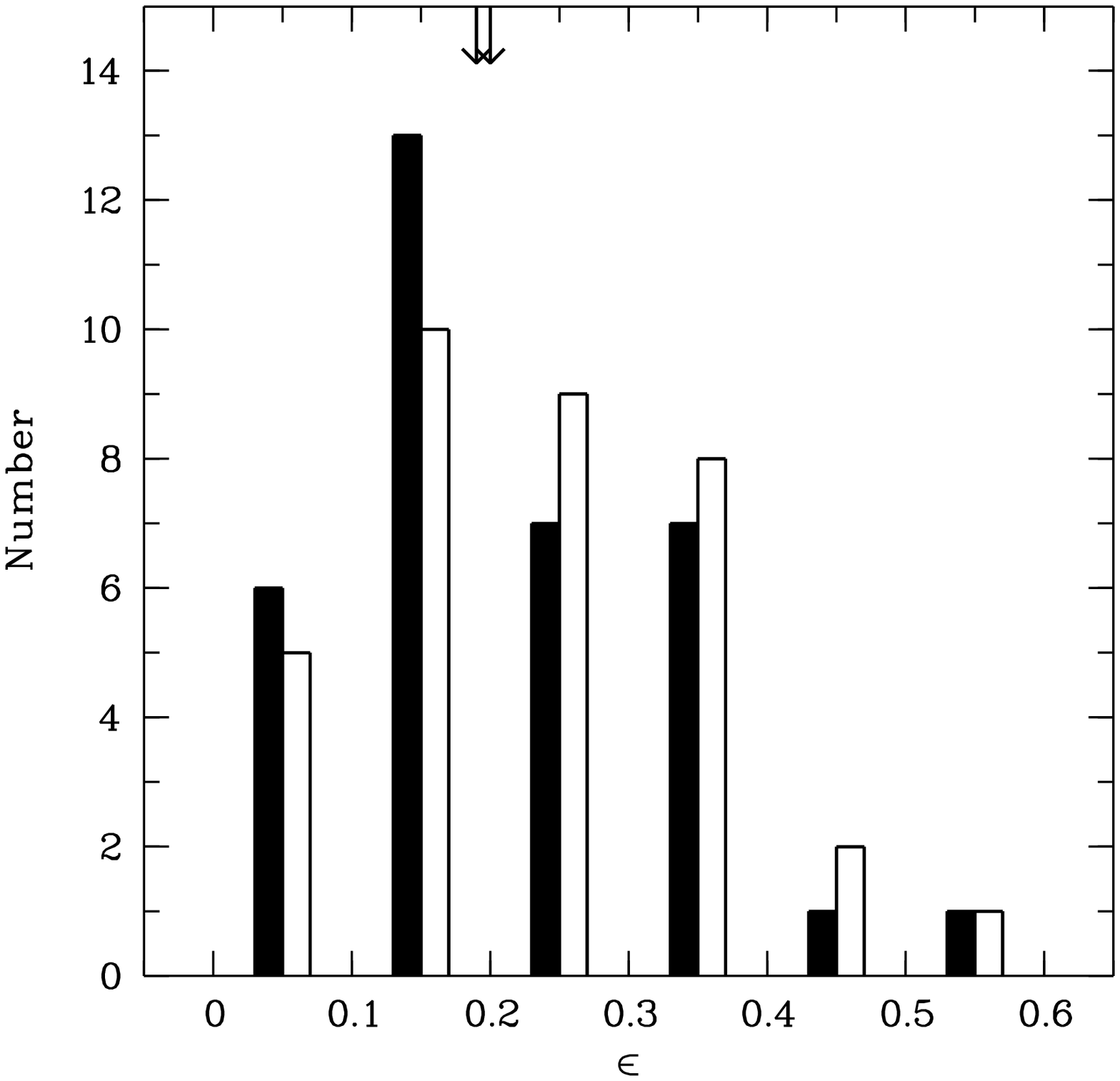}
      \caption{Distribution of $\epsilon$, estimated by Slavcheva-Mihova \& Mihov (in prep.), of the Sy (black columns) and control (empty columns) sample. The bin size is 0.1. The left and right arrows designate the median values of the Sy and control sample, respectively.}
\label{E_AN_ourour_hbm}
 \end{minipage}
\vspace{0.6cm}

\begin{minipage}[t]{5.6cm}   
 \includegraphics[width=5.6cm]{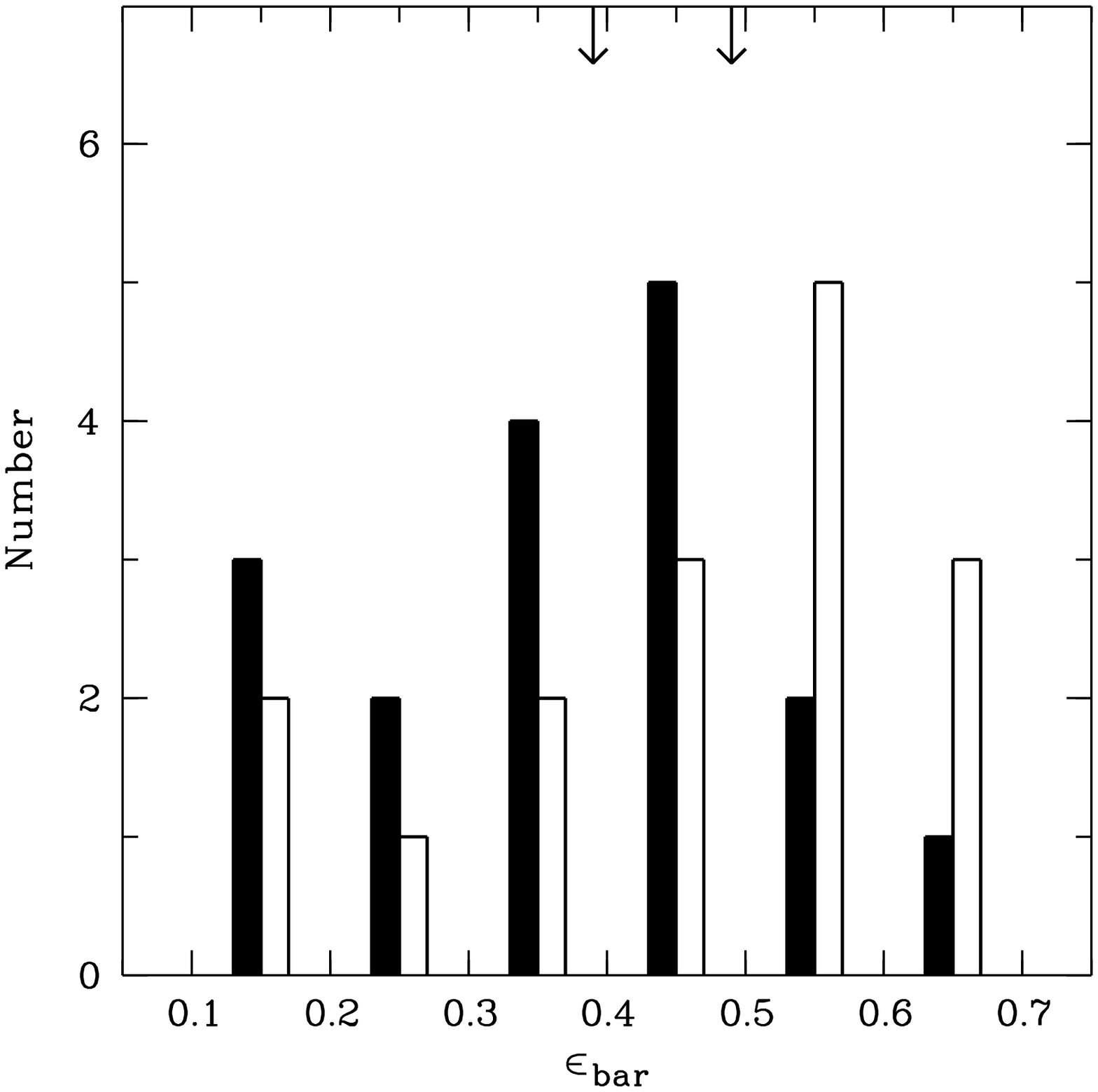}
       \caption{Distribution of the deprojected bar ellipticities $\epsilon_{\rm bar}$, estimated by Slavcheva-Mihova \& Mihov (in prep.), of the Sy (black columns) and control (empty columns) sample. The bin size is 0.1. The left and right arrows designate the median values of the Sy and control sample, respectively.}
 \label{Ebar_AN_2our_hbm}
 \end{minipage}
\end{figure}

\begin{table*}
\caption{Observation log of the Sy galaxies.}
\label{T_obs}
\centering
\begin{tabular}{lllllllr}
\hline\hline
\noalign{\smallskip}
~~Galaxy & ~~Civil Date & ~\,Filters & ~~~FWHM & ~~~~~~~$\beta$ & Calibration & CCD & Remarks~~~~~~~~ \\
\noalign{\smallskip}
       & [yyyy mm dd] &         & ~~~[arcsec] &         &     &       \\
\noalign{\smallskip}
~~~~~(1) & ~~~~~~~(2) & ~~~~(3) & ~~~~~~~(4) & ~~~~~~(5) & ~~~~~~(6) & ~~(7) & (8)~~~~~~~~~~~~~ \\
\noalign{\smallskip}
\hline
\noalign{\smallskip}

Mrk\,335           & 1998 08 22 & $BVR_{\rm \scriptstyle C}I_{\rm \scriptstyle C}$     & $1.17 \pm 0.06$ & $2.95 \pm 0.12$ & NGC\,7790        & AT           & \\
                   & 2007 08 20 & $~~\,VR_{\rm \scriptstyle C}I_{\rm \scriptstyle C}$& $1.14 \pm 0.03$ & $3.67 \pm 0.17$ & SS               & VA           & \\
III\,Zw\,2         & 1997 09 09 & $BVR_{\rm \scriptstyle C}I_{\rm \scriptstyle C}$   & $1.32 \pm 0.03$ & $3.66 \pm 0.35$ & SS               & AT$2\times$  & \\
Mrk\,348           & 1997 09 07 & $BVR_{\rm \scriptstyle C}I_{\rm \scriptstyle C}$   & $0.92 \pm 0.05$ & $3.24 \pm 0.46$ & NGC\,7790        & AT           & \\
I\,Zw\,1           & 1998 08 20 & $BVR_{\rm \scriptstyle C}I_{\rm \scriptstyle C}$     & $1.50 \pm 0.03$ & $2.75 \pm 0.12$ & NGC\,7790        & AT$2\times$  & \\
Mrk\,352\tablefootmark{a} & 2007 08 21 & $~~~~~\,R_{\rm \scriptstyle C}$                    & $0.86 \pm 0.06$ & $4.14 \pm 0.89$ & SS               & VA           & \\
                   & 2008 01 30 & $BVR_{\rm \scriptstyle C}I_{\rm \scriptstyle C}$   & $1.45 \pm 0.03$ & $2.97 \pm 0.06$ & SS               & VA           & \\
Mrk\,573           & 1997 09 07 & $BVR_{\rm \scriptstyle C}I_{\rm \scriptstyle C}$   & $1.60 \pm 0.08$ & $~~~~~~\,\ldots$& NGC\,7790        & AT           & \\
Mrk\,590\tablefootmark{a} & 1997 09 06 & $BVR_{\rm \scriptstyle C}I_{\rm \scriptstyle C}$   & $1.61 \pm 0.04$ & $2.44 \pm 0.07$ & SS               & AT$2\times$  & \\
Mrk\,595           & 1997 09 09 & $BVR_{\rm \scriptstyle C}I_{\rm \scriptstyle C}$   & $1.47 \pm 0.02$ & $3.78 \pm 0.17$ & SS               & AT$2\times$  & \\
3C\,120            & 1997 09 09 & $BVR_{\rm \scriptstyle C}I_{\rm \scriptstyle C}$   & $1.54 \pm 0.03$ & $3.71 \pm 0.14$ & SS               & AT$2\times$  & \\
                   & 2008 02 01 & $BVR_{\rm \scriptstyle C}I_{\rm \scriptstyle C}$   & $1.61 \pm 0.03$ & $3.47 \pm 0.09$ & SS               & VA           & \\
Ark\,120           & 1994 09 29 & $B~~\,R_{\rm \scriptstyle C}$                      & $0.95 \pm 0.02$ & $2.99 \pm 0.18$ & SS               & TEK4         & JKT,~$0\farcs33\,\rm px^{-1}$ \\
                   & 1991 12 08 & $~~\,V$                                            & $0.93 \pm 0.07$ & $3.87 \pm 0.31$ & SS               & SAIC1        & CFHT,~$0\farcs20\,\rm px^{-1}$ \\
Mrk\,376           & 2008 02 03 & $BVR_{\rm \scriptstyle C}I_{\rm \scriptstyle C}$   & $0.80 \pm 0.02$ & $3.84 \pm 0.12$ & SS               & VA           & \\
Mrk\,79            & 1999 02 16 & $BVR_{\rm \scriptstyle C}I_{\rm \scriptstyle C}$   & $3.57 \pm 0.29$ & $8.15 \pm 1.48$ & NGC\,7790, M\,92 & AT$2\times$  & \\
                   & 2008 02 01 & $BVR_{\rm \scriptstyle C}I_{\rm \scriptstyle C}$   & $1.24 \pm 0.04$ & $3.76 \pm 0.17$ & SS               & VA           & \\
Mrk\,382           & 1998 02 27 & $BVR_{\rm \scriptstyle C}I_{\rm \scriptstyle C}$   & $2.57 \pm 0.15$ & $5.39 \pm 0.99$ & SS               & AT$2\times$  & \\
                   & 2008 02 02 & $BVR_{\rm \scriptstyle C}I_{\rm \scriptstyle C}$   & $1.74 \pm 0.04$ & $3.50 \pm 0.16$ & SS               & VA           & \\
NGC\,3227          & 1999 04 17 & $BVR_{\rm \scriptstyle C}I_{\rm \scriptstyle C}$     & $1.86 \pm 0.08$ & $5.59 \pm 0.57$ & M\,92            & AT           & \\
NGC\,3516          & 2008 01 08 & $~~\,VR_{\rm \scriptstyle C}I_{\rm \scriptstyle C}$& $1.82 \pm 0.05$ & $3.54 \pm 0.31$ & SS               & FVA          & \\
NGC\,4051\tablefootmark{b}& 1995 05 06 & $B$                                                & $1.43 \pm 0.07$ & $3.43 \pm 0.16$ & ZP              & TEK4         & JKT,~$0\farcs241\,\rm px^{-1}$ \\
                   & 2000 03 30 & $~~~~~\,R_{\rm \scriptstyle C}$                    & $1.78 \pm 0.03$ & $3.09 \pm 0.30$ & SS               & SITe2        & JKT,~$0\farcs241\,\rm px^{-1}$ \\
                   & 2001 04 09 & $~~~~~~~~~~I_{\rm \scriptstyle C}$                 & $1.41 \pm 0.04$ & $4.42 \pm 0.43$ & ZP              & SITe2        & JKT,~$0\farcs241\,\rm px^{-1}$ \\
NGC\,4151          & 1999 03 10 & $BVR_{\rm \scriptstyle C}I_{\rm \scriptstyle C}$   & $1.06 \pm 0.03$ & $2.98 \pm 0.10$ & M\,67            & AT           & \\
                   & 1999 04 19 & $BVR_{\rm \scriptstyle C}I_{\rm \scriptstyle C}$     & $2.55 \pm 0.11$ & $2.68 \pm 0.47$ & M\,92            & AT$2\times$  & \\
Mrk\,766           & 1999 02 15 & $BVR_{\rm \scriptstyle C}I_{\rm \scriptstyle C}$   & $3.16 \pm 0.04$ & $4.23 \pm 0.25$ & NGC\,7790, M\,92 & AT$2\times$  & \\
Mrk\,771\tablefootmark{a,\,c}& 1990 06 23 & $~~\,V~~~~\,I_{\rm \scriptstyle C}$                & $0.56 \pm 0.01$ & $2.32 \pm 0.01$ & M\,92            & SAIC1        & CFHT,~$0\farcs13\,\rm px^{-1}$ \\
NGC\,4593          & 2008 01 08 & $~~\,VR_{\rm \scriptstyle C}I_{\rm \scriptstyle C}$& $2.56 \pm 0.08$ & $3.87 \pm 0.48$ & SS               & FVA          & \\
Mrk\,279\tablefootmark{a} & 2008 02 02 & $BVR_{\rm \scriptstyle C}I_{\rm \scriptstyle C}$   & $1.16 \pm 0.03$ & $3.01 \pm 0.05$ & SS               & VA           & \\
NGC\,5548          & 1999 04 19 & $BVR_{\rm \scriptstyle C}I_{\rm \scriptstyle C}$     & $2.44 \pm 0.26$ & $3.05 \pm 0.54$ & M\,92            & AT$2\times$  & \\
Ark\,479           & 2007 07 19 & $~~\,VR_{\rm \scriptstyle C}I_{\rm \scriptstyle C}$& $1.19 \pm 0.06$ & $7.05 \pm 0.48$ & SS               & VA           & \\
Mrk\,506           & 1997 06 01 & $BVR_{\rm \scriptstyle C}I_{\rm \scriptstyle C}$   & $2.47 \pm 0.17$ & $5.30 \pm 1.03$ & M\,92            & AT$2\times$  & \\
                   & 1998 07 18 & $BVR_{\rm \scriptstyle C}I_{\rm \scriptstyle C}$     & $1.59 \pm 0.03$ & $3.22 \pm 0.12$ & M\,92            & AT$2\times$  & \\
                   & 2007 06 17 & $BVR_{\rm \scriptstyle C}$                         & $1.56 \pm 0.03$ & $3.29 \pm 0.05$ & SS               & VA           & \\ 
3C\,382            & 1998 08 23 & $BVR_{\rm \scriptstyle C}I_{\rm \scriptstyle C}$     & $1.24 \pm 0.03$ & $3.49 \pm 0.36$ & NGC\,7790        & AT$2\times$  & \\
3C\,390.3          & 1998 08 20 & $BVR_{\rm \scriptstyle C}I_{\rm \scriptstyle C}$     & $1.71 \pm 0.06$ & $3.43 \pm 0.09$ & NGC\,7790        & AT$2\times$  & \\
NGC\,6814          & 1997 07 06 & $BVR_{\rm \scriptstyle C}I_{\rm \scriptstyle C}$   & $3.21 \pm 0.10$ & $3.01 \pm 0.33$ & SS               & AT$2\times$  & \\
                   & 1997 07 10 & $BVR_{\rm \scriptstyle C}I_{\rm \scriptstyle C}$   & $1.90 \pm 0.07$ & $3.83 \pm 0.41$ & M\,92            & AT$2\times$  & \\
                   & 1997 09 07 & $BVR_{\rm \scriptstyle C}I_{\rm \scriptstyle C}$   & $1.70 \pm 0.05$ & $~~~~~~\,\ldots$& NGC\,7790        & AT$2\times$  & \\
                   & 1998 07 18 & $BVR_{\rm \scriptstyle C}I_{\rm \scriptstyle C}$     & $1.24 \pm 0.01$ & $2.33 \pm 0.03$ & M\,92            & AT           & \\
Mrk\,509           & 1997 07 10 & $BVR_{\rm \scriptstyle C}I_{\rm \scriptstyle C}$   & $1.70 \pm 0.03$ & $2.92 \pm 0.07$ & M\,92            & AT$2\times$  & \\
                   & 1997 09 08 & $BVR_{\rm \scriptstyle C}I_{\rm \scriptstyle C}$   & $1.58 \pm 0.08$ & $9.40 \pm 0.90$ & SS               & AT$2\times$  & \\
                   & 1998 07 20 & $BVR_{\rm \scriptstyle C}I_{\rm \scriptstyle C}$     & $2.13 \pm 0.06$ & $3.66 \pm 0.16$ & M\,92            & AT$2\times$  & \\
Mrk\,1513          & 2007 08 20 & $~~\,VR_{\rm \scriptstyle C}I_{\rm \scriptstyle C}$& $1.19 \pm 0.02$ & $4.20 \pm 0.13$ & SS               & VA           & \\
Mrk\,304           & 1998 07 19 & $BVR_{\rm \scriptstyle C}I_{\rm \scriptstyle C}$     & $2.07 \pm 0.07$ & $2.86 \pm 0.19$ & M\,92            & AT$2\times$  & \\
Ark\,564           & 1998 07 18 & $BVR_{\rm \scriptstyle C}I_{\rm \scriptstyle C}$     & $2.11 \pm 0.06$ & $3.56 \pm 0.20$ & M\,92            & AT$2\times$  & \\
                   & 1998 08 20 & $BVR_{\rm \scriptstyle C}I_{\rm \scriptstyle C}$     & $2.26 \pm 0.14$ & $~~~~~~\,\ldots$& NGC\,7790        & AT$2\times$  & \\
NGC\,7469          & 1997 09 06 & $BVR_{\rm \scriptstyle C}I_{\rm \scriptstyle C}$   & $1.39 \pm 0.04$ & $2.57 \pm 0.32$ & SS               & AT$2\times$  & \\
                   & 1998 07 19 & $BVR_{\rm \scriptstyle C}I_{\rm \scriptstyle C}$     & $1.91 \pm 0.09$ & $2.99 \pm 0.20$ & M\,92            & AT$2\times$  & \\
                   & 1998 08 23 & $BVR_{\rm \scriptstyle C}I_{\rm \scriptstyle C}$     & $0.92 \pm 0.02$ & $4.68 \pm 0.29$ & NGC\,7790        & AT           & \\
                   & 2003 07 28 & $BVR_{\rm \scriptstyle C}I_{\rm \scriptstyle C}$     & $0.99 \pm 0.04$ & $3.41 \pm 0.29$ & M\,92            & AT           & \\
Mrk\,315           & 2007 08 22 & $~~~~~\,R_{\rm \scriptstyle C}$                    & $1.18 \pm 0.02$ & $3.37 \pm 0.08$ & SS               & VA           & \\
NGC\,7603          & 2007 07 19 & $~~\,V~~~~\,I_{\rm \scriptstyle C}$                & $1.41 \pm 0.04$ & $3.40 \pm 0.17$ & SS               & VA           & \\
Mrk\,541           & 2007 07 19 & $~~\,VR_{\rm \scriptstyle C}I_{\rm \scriptstyle C}$& $1.20 \pm 0.04$ & $3.42 \pm 0.09$ & SS               & VA           & \\
\hline
\end{tabular}
\tablefoot{(5) $\beta$ of the Moffat PSF (see text; ellipsis dots denote Gaussian PSF was assumed); (6) SS~-- secondary standards used, ZP~-- zero point taken from the FITS file header; (7) AT~-- AT200, AT$2\times$~-- AT200 $2\!\times\!2$ binned, VA~-- VersArray:1300B, and FVA~-- VersArray:512B attached to the two-channel focal reducer; (8) Telescopes used to obtain the archival data, and the corresponding scale factors. \\
 \tablefoottext{a}{Archival Hubble Space Telescope (HST) data have also been used.}
 \tablefoottext{b}{See \citet{KSB_04} for observational details.}
 \tablefoottext{c}{See \citet{HN_92} for observational details.}}

\end{table*}

\subsection{Sky background estimation}
\label{background}

Proper sky background estimation is important for the correct photometry of any extended object, especially the faint outer galaxy regions.
We estimated the sky background on the smoothed frame cleaned from contaminating features, approximating the sky intensity distribution (measured in boxes evenly placed in blank regions) by a tilted plane, using the least-squares method ($\rm \scriptstyle FIT/FLATSKY$ command). The frame areas having intensities above some threshold (given the trial estimate of the sky background and its error based on histogram analysis within the AIP package) were masked prior to  background estimation in order to avoid influence by contaminating features or the galaxy itself and to make the procedure more objective. To estimate sky background properly in the presence of eventual gradient, we placed the boxes relatively close to the galaxy outskirts. The adopted error of the sky background, $\sigma_{\rm sky}$, equals the standard deviation about the fitted plane.

Properly estimated sky background would show up in an asymptotically flat growth curve, whereas sky background estimation errors would cause a continuous increase or a maximum followed by a continuous decrease in the growth curve. The shape of the growth curve can therefore be used to fine-tune the sky background by simply adding/subtracting a constant to/from the frame \citep{BST_84}. In general, our procedures worked well though an unambiguous estimate of the sky background is not always possible, e.g., in cases of a bright companion, an overcrowded field, or not enough field free of the galaxy. In cases of a non-asymptotically flat growth curve, we preferred to re-determine the sky background rather than to add/subtract a constant.

\subsection{Ellipse fitting to isophotes}
\label{ellipse}

We performed ellipse fitting to the isophotes of the galaxy images ($\rm \scriptstyle FIT/ELL3$ command within the $\rm \scriptstyle SURFPHOT$ package) following \citet{BM_87}. The algorithm starts from the galaxy centre spacing the isophote levels in a logarithmically equidistant manner, i.e., with a fixed SB step. The fitting algorithm works on gradients ${\rm d}I/{\rm d}a\,$$<$$\,0$, where $I$ is the isophote intensity and $a$ the semi-major axis. 
Ellipses were generally fitted down to $I\,$$\approx$$\,\sigma_{\rm sky}$. The errors of the fitted ellipse parameters were computed following \citet{R_95}. The error of an intensity level takes into account the error of the mean intensity along the sampled ellipse and the sky background error \citep{SH1_89}. We obtained a model-subtracted residual image by subtracting the model image, reconstructed using the derived ellipse parameters, from the original image.

As a result of ellipse fitting, we derived profiles of the SB, $\epsilon$, and position angle (PA) as a function of $a$ for each galaxy.
Radial colour index (CI) profiles were obtained by subtracting two individually determined radial SB profiles after they had been calibrated (see Sect.\,\ref{calib}) and smoothed to the worse FWHM.

 \begin{table*}
 \caption{Close companions of the Sy galaxy sample.}
 \label{T_comp_AG}
\centering
\begin{tabular}{llrrrrr} 
\noalign{\smallskip}
\hline\hline             
\noalign{\smallskip}~~Galaxy & ~~~~~~~~~~~~~~~~~~~Companion & \multicolumn{2}{c}{$\Delta D$} & $|\Delta V_{\rm r}|$~~~~~ & \multicolumn{2}{c}{Ref.}  \\
\cline{3-4}
\cline{6-7}
\noalign{\smallskip}     &   & ~~~~[arcmin]     & [kpc]       & [$\rm km\,\rm s^{-1}$]~ &              \\
 \noalign{\smallskip} ~~~~~~(1) & ~~~~~~~~~~~~~~~~~~~~~~~~~(2)        & (3)~~        &  (4)~~          & (5)~~~~~~    & (6) & (7)        \\
\noalign{\smallskip}
 \hline
\noalign{\smallskip}
  	    \vspace{0.16cm}			  						     
III\,Zw\,2         & III\,Zw\,2B		     &  0.572 &  54.38& $199\pm106~\,$   & 10 & 10 \\
     	    \vspace{0.16cm}			  					      
Mrk\,348           & 2MASX\,J00485285+3157309	     &  1.218 &  20.05& $321\pm\,~51~\,$ & 12 & 19   \\
     	    \vspace{0.16cm}
I\,Zw\,1           & J005334+124133\tablefootmark{a}	     &  0.260 &  17.38&	$170\pm110$\tablefootmark{b}  & 5 &   \\
     	    \vspace{0.16cm}
Mrk\,573           & APMUKS(BJ)\,B014126.68+020933.5 &  3.734 &  70.95&$\approx$\,100\tablefootmark{b}\,~~~~~~~~& 16 &      \\
     	    \vspace{0.16cm}
Mrk\,590           & SDSS\,J021429.36--004604.7	     &  1.053 &  31.17&	$234\pm\,~39~\,$ & 8    & 1 \\
NGC\,3227          & NGC\,3226  		     &  2.177 &  12.78& $181\pm\,~38~\,$ & 14  & 14  \\
                   & 2MASX\,J10232246+1954510	     &  3.515 &  20.63& $42\pm\,~37~\,$  & 8    & 4 \\
                   & 2MASX J10232162+2001380 	     &  9.961 &  58.47& $ 9\pm\,~38~\,$  & 8    & 4 \\
                   & SDSS J102434.71+200157.8	     & 18.122 & 106.38& $128\pm\,~42~\,$ & 8    & 4 \\
    	    \vspace{0.16cm}
                   & SDSS\,J102315.38+201040.5       & 19.109 & 112.17& $41\pm\,~40~\,$  & 8    & 4 \\
NGC\,4151          & SDSS\,J121021.06+391252.1       & 11.688 &  57.62& $75\pm\,~35~\,$  & 8    & 3 \\
     	    \vspace{0.16cm}
     	           & SDSS\,J120959.87+391147.7       & 14.053 &  69.28& $343\pm\,~43~\,$ & 8    & 3 \\
     	    \vspace{0.16cm}
NGC\,4593          & MCG\,--01--32--033	             &  3.809 &  45.67&	$149\pm\,~67~\,$ & 21 & 13 \\
     	    \vspace{0.16cm}
Mrk\,279           & MCG\,+12--13--024		     &  0.761 &  26.88&	$159\pm\,~25~\,$ & 20  & 7 \\
NGC\,5548          & SDSS\,J141824.74+250650.7	     &  5.867 & 122.74&	$206\pm\,~22~\,$ & 8    & 4 \\
     	    \vspace{0.16cm}
     	           & 2MASX\,J14173385+2506515	     &  5.959 & 124.66& $40\pm\,~59~\,$  & 8    & 4 \\
     	    \vspace{0.16cm}
Ark\,479           & SDSS\,J153550.82+143035.3	     &  0.622 &  14.60& $ 48\pm\,~50~\,$ & 9 & 2 \\
Mrk\,506           & CGCG\,170--019 		     &  0.763 &  37.20& $561\pm\,~58~\,$ & 8    & 9 \\
     	    \vspace{0.16cm}
                   & SDSS\,J172229.44+305231.6	     &  2.277 & 111.00& $505\pm\,~58~\,$ & 8    & 2 \\
     	    \vspace{0.16cm}
3C\,390.3          & PGC\,062330           	     &  1.802 & 112.68& $549\pm\,~57~\,$ & 11  & 18  \\
     	    \vspace{0.16cm}
NGC\,7469          & IC\,5283			     &  1.315 &  23.29& $87\pm\,~37~\,$  & 15  & 14  \\
     	    \vspace{0.16cm}
Mrk\,315           & [CAM2005]\,DWARF	     &  1.024 &  44.10& $193\pm\,~35~\,$ & 6 & 6 \\
NGC\,7603\tablefootmark{c}& NGC\,7603:[LG2002]\,3     &  0.620 &  20.27& $87476\pm388~\,$ & 8    & 17  \\
     	           & SDSS\,J231859.26+001405.4 &  0.860 &  28.12& $56018\pm461~\,$ & 8    & 17  \\
     	           & NGC\,7603B 		     &  0.985 &  32.21& $7535\pm\,~47~\,$& 8    & 1 \\
\hline

\end{tabular}

\tablefoot{(3) and (4) Projected linear separation taken from NED; 
(5)  Absolute radial velocity difference calculated using redshift sources specified in Cols.\,6 and 7; (6) and (7) References regarding the redshift sources of the primary and the companion, respectively, generally taken through NED. \\
\tablefoottext{a}{We list the J2000 coordinates taken from Aladin for the name of the anonymous companion.}
\tablefoottext{b}{The absolute radial velocity difference itself was taken from the source given in Col.\,6.}
\tablefoottext{c}{Anomalous redshift galaxy system, not taken into account in the companion statistics.}}
\tablebib{(1)~\citet{SDSS1}; (2) \citet{SDSS7}; (3) \citet{SDSS5}; (4) \citet{SDSS6}; 
(5) \citet{CS_01}; (6) \citet{CAM_05};  (7) \citet{GLM_87}; (8) \citet{RC3}; (9) \citet{FKG_99}; 
(10) \citet{HBB_84}; (11) \citet{HB_91}; (12) \citet{HVG_99}; 
(13) \citet{6dF}; (14) \citet{K_96a}; (15) \citet{K_96b}; (16) \citet{KLT_08}; (17) \citet{LG_04}; 
(18) \citet{PP_73}; (19) \citet{P_82}; (20) \citet{SH2_88}; (21) \citet{SHD_92}.}
 \end{table*}

\begin{table*}

\caption{Close companions of the inactive galaxy sample.}
 \label{T_comp_NG}
\centering
\begin{tabular}{llrrrrrl} 
\noalign{\smallskip}
\hline\hline             
\noalign{\smallskip}~~Galaxy & ~~~~~~~~~~~~~~~~~~~Companion & \multicolumn{2}{c}{$\Delta D$} & $|\Delta V_{\rm r}|$~~~~~ & \multicolumn{2}{c}{Ref.}  \\
\cline{3-4}
\cline{6-7}
\noalign{\smallskip}     &   & ~~~~[arcmin]     & [kpc]       & [$\rm km\,\rm s^{-1}$]~ &              \\
 \noalign{\smallskip} ~~~~~~(1) & ~~~~~~~~~~~~~~~~~~~~~~~~~(2)        & (3)~~        &  (4)~~          & (5)~~~~~~    & (6) & (7)         \\
\noalign{\smallskip}
\hline
\noalign{\smallskip}
 
   	    \vspace{0.16cm}			  						     
 NGC\,2144        	 & ESO\,016--\,G\,009               &  8.492 & 160.24 & $23\pm\,~65$  & 12 & 17 \\
 ESO\,155--\,G\,027	 & ESO\,155--\,IG\,028\,NED02	   &  1.996 & 137.30 & $244\pm133$   &  6 & 22 \\
                         & ESO\,155--\,IG\,028\,NED01 	   &  2.290 & 157.53 &   $0\pm133$   &  6 & 20 \\
                         & 2MASX\,J03252010--5232057	   &  3.590 & 246.96 & $413\pm133$   &  6 & 22 \\
   	    \vspace{0.16cm}			  						         
                         & APMUKS(BJ)\,B032410.26--524522.5 &  4.047 & 278.39 & $260\pm133$   &  6 & 22 \\
 2MASX\,J04363658--0250350& CGCG\,393--045 		   &  2.109 &  38.07 & $167\pm\,~50$ & 10 & 10 \\
   	    \vspace{0.16cm}			  						           
                         & CGCG\,393--044 		   &  4.539 &  81.93 &  $96\pm\,~48$ & 10 & 10 \\
   	    \vspace{0.16cm}			  						     
 2MASX\,J00342513--0735582& MCG\,--01--02--032		   &  3.621 & 105.44 & $545\pm\,~74$ & 10 & 10 \\
   	    \vspace{0.16cm}			  						     
 IC\,5065		 & 2MASX\,J20520956--2951513 	   &  5.248 & 192.65 &  $36\pm\,~27$ &  8 & 14 \\
   	    \vspace{0.16cm}			  						     
 ESO\,183--\,G\,030 	 & IC\,4797			   & 14.879 & 154.44 &  $10\pm\,~37$ &  8 &  8 \\
 NGC\,2775       	 & SDSS\,J091019.53+070141.2	   &  0.608 &	4.01 & $110\pm\,~74$ &  8 &  2 \\
                         & SDSS\,J091028.77+071117.9	   &  9.275 &  61.12 & $168\pm\,~31$ &  8 & 11 \\
   	    \vspace{0.16cm}			  						     
                         & NGC\,2777        		   & 11.469 &  75.58 & $134\pm\,~32$ &  8 &  8 \\
 UGC\,6520       	 & SDSS\,J113240.41+622735.1	   &  3.187 &  47.61 & $140\pm\,~50$ &  9 &  1 \\	     
   	    \vspace{0.16cm}			  						     
                         & CGCG\,314--031    		   &  4.914 &  73.42 &  $23\pm\,~67$ &  9 &  8 \\	     
   	    \vspace{0.16cm}			  						     
 NGC\,4902		 & NGC\,4887			   & 10.410 & 123.05 &  $14\pm\,~49$ & 23 &  7 \\
   	    \vspace{0.16cm}			  						     
 ESO\,510--\,G\,048	 & ESO\,510--\,G\,050		   &  3.058 & 158.80 &  $29\pm103$   & 13 & 15 \\     
   	    \vspace{0.16cm}			  						     
 NGC\,7421       	 & NGC\,7418        		   & 19.371 & 117.20 & $342\pm\,~27$ & 21 & 18 \\ 	    
 ESO\,292--\,G\,007	 & 2MASX\,J23392481--4603161	   &  4.653 & 347.49 & $139\pm275$   & 19 &  4 \\
                         & 2MASX\,J23393103--4553430	   &  4.986 & 372.35 & $144\pm186$   & 19 &  4 \\
   	    \vspace{0.16cm}			  						         
                         & 2MASX\,J23401100--4559372	   &  7.854 & 586.54 & $410\pm187$   & 19 &  4 \\
 NGC\,897		 & 2dFGRS S517Z282		   &  0.772 &  13.83 & $374\pm\,~91$ &  6 &  5 \\
   	    \vspace{0.16cm}			  						         
                         & ESO\,355--\,G\,010		   &  8.237 & 147.52 &  $71\pm\,~38$ &  6 & 24 \\
   	    \vspace{0.16cm}			  						          
 ESO\,423--\,G\,016 	 & APMUKS(BJ)\,B052441.44--314926.2 &  3.732 & 167.53 & $102\pm\,~67$ &  6 & 16 \\
 UGC\,9532\,NED04	 & UGC\,9532\,NED02     	   &  0.237 &  11.46 & $194\pm\,~59$ & 10 & 10 \\
                         & UGC\,9532\,NED05		   &  0.782 &  37.80 & $482\pm\,~60$ & 10 & 10 \\
                         & UGC\,9532\,NED06		   &  1.192 &  57.62 & $482\pm\,~61$ & 10 &  8 \\
                         & UGC\,9532\,NED01		   &  1.302 &  62.94 &  $50\pm\,~58$ & 10 & 10 \\
                 	 & SDSS\,J144748.23+190352.2       &  1.804 &  87.21 & $411\pm\,~55$ & 10 &  3 \\

\hline

\end{tabular}

\tablefoot{
The columns are the same as in Table\,\ref{T_comp_AG}.}
\tablebib{(1)~\citet{SDSS2}; (2) \citet{SDSS3}; (3) \citet{SDSS7}; (4) \citet{AVL_99}; (5) \citet{CPJ_03}; 
(6) \citet{CPD_91}; (7) \citet{CWP_98}; (8) \citet{RC3}; (9) \citet{FKG_99}; (10) \citet{HMH_92}; 
(11) \citet{HRS_98}; (12) \citet{HGC_95}; (13) \citet{6dF}; (14) \citet{JSC_05}; 
(15) \citet{KCR_03}; (16) \citet{KMH_98}; (17) \citet{KS_90}; (18) \citet{KSK_04}; 
(19) \citet{LV_89}; (20) \citet{LDM_83}; (21) \citet{MZW_04}; (22) \citet{RGC_02}; 
(23) \citet{THC_07}; (24) \citet{ZSF_97}.}

\end{table*}

\subsection{Photometric calibration}
\label{calib}

The nights of photometric quality were calibrated using multi-star standard fields established in stellar clusters. We used the clusters M\,92\footnote{Note that we have added 0.002 mag to the $V$ magnitudes and to the $B\,$--$\,V$ colour indices of the M\,92 standard stars listed in \citet{MKK_94} according to the addendum of \citet{SH_88}.} \citep{MKK_94}, NGC\,7790 \citep{OBH_92,PSD_01}, and M\,67 \citep{CI_91} for this purpose. 

We performed aperture stellar photometry of the standard fields by means of DAOPHOT~II package \citep{S_87,S_91}. We applied the growth curve method to get the total instrumental magnitudes of the standard stars; its very idea rests on S/N arguments \citep{S_90}. 

The transformation coefficients to the standard Johnson-Cousins system were determined following
\citet{HFR_81} approach. The equations read as
\begin{eqnarray}
b-B~ & = & c_{\scriptstyle B}^{(0)}+c_{\scriptstyle B}^{(1)}\,X+c_{\scriptstyle B}^{(2)}\,(B~-V) \nonumber \\
\upsilon-V~ & = & c_{\scriptstyle V}^{(0)}+c_{\scriptstyle V}^{(1)}\,X+c_{\scriptstyle V}^{(2)}\,(V~-R_{\rm \scriptstyle C}) \nonumber \\
r-R_{\rm \scriptstyle C} & = & c_{\scriptstyle R_{\rm \scriptscriptstyle C}}^{(0)}+c_{\scriptstyle R_{\rm \scriptscriptstyle C}}^{(1)}\,X+c_{\scriptstyle R_{\rm \scriptscriptstyle C}}^{(2)}\,(V~-R_{\rm \scriptstyle C}) \nonumber \\
i-I\,_{\rm \scriptstyle C} & = & c_{\scriptstyle I_{\rm \scriptscriptstyle C}}^{(0)}+c_{\scriptstyle I_{\rm \scriptscriptstyle C}}^{(1)}\,X+c_{\scriptstyle I_{\rm \scriptscriptstyle C}}^{(2)}\,(R_{\rm \scriptstyle C}-I_{\rm \scriptstyle C}), \nonumber
\end{eqnarray}
where the small and capital letters denote the total instrumental and standard magnitudes of the cluster standard stars, respectively; $c^{(0)}$ is the zero point magnitude, $c^{(1)}$ the extinction coefficient, and $c^{(2)}$ the colour coefficient. The transformation coefficient errors were incorporated in the final SB error.

The data in nights with some weather or technical problems were transformed to the standard system using secondary standards after \citet{BSD_00}, \citet{GKM_01}, \citet{DSM_05a,DSM_05b}, and \citet{MS_08}; unpublished results of ours were also used. The standard fields, used for calibration, are denoted in Table~\ref{T_obs}.

\subsection{Subsidiary images} 
\label{images}

To facilitate revealing the individual galaxy features, we constructed the following subsidiary images: CI images, model-subtracted residual images (see Sect.\,\ref{ellipse}), unsharp masked residual images (unsharp mask-subtracted and unsharp mask-divided ones), and structure maps. In a couple of instances, a fitted 2D analytical model was subtracted.

The CI images were constructed after the corresponding frames had been aligned and smoothed to an equal (the worse) FWHM. Besides this, we smoothed the images using a median filter of a variable size depending on the size of the feature of interest. By subtracting the smoothed image from the original one, we constructed an unsharp mask-subtracted residual image. Analogically, by dividing the original image by the filtered one, we acquired an unsharp mask-divided residual image \citep[][]{S_93,SYA_94,LSB_05}. The subtraction procedure works best in the galaxy outskirts but fails near the centres as the absolute contribution of shot noise gets large; on the contrary, the division approach is best suited to examining structures in the central regions, as the relative contribution of shot noise is small there \citep{L_85}. 

Finally, we constructed a structure map, $\cal S$
\citep{PM_02}:
\begin{displaymath}
\cal S=\left({O \over {O \otimes M}}\right) \otimes M^{\rm T},
\end{displaymath}
where $\cal O$ is the original image, $\cal M$ the Moffat PSF, $\cal M^{\rm T}$ the
transpose of the PSF, ${\cal M}^{\rm T}(x,y)={\cal M}(-x,-y)$, and $\otimes$ the convolution operator.

\section{Bar characterization} 
\label{bars}

We consider a galaxy barred if there is an ellipticity maximum greater than 0.16 with an amplitude of at least 0.08 over a region of PA, constant within $20\degr$, following \citet{AMC_09}.

The deprojected ellipticities of the bar-like structures of several of the galaxies were found to be less than 0.15, which is typical of ovals and lenses, but we cannot be more specific without kinematic data \citep[][see also Sellwood \& Wilkinson 1993]{KK_04}.
According to \citet{KK_04} barred and oval galaxies evolve similarly and are essentially equivalent regarding gas inflow\footnote{Most of the oval galaxies are classified SAB.}; furthermore, both ovals and lenses are non-axisymmetric enough to drive secular evolution \citep[see also][]{WJK_09}. Thus, bars, ovals, and lenses are functionally equivalent in the context of AGN fueling and, for the purpose of bar statistics, we shall refer to them as bars. Deprojected bar ellipticity can be used as a first-order approximation of bar strength \citep[e.g.,][]{M_95,BBK_04}. We classify a bar as strong if its deprojected ellipticity is greater than 0.45 after \citet{LSK_02}. In bar and ring statistics, we take only large-scale structures into account. We adopt 1\,kpc as an upper limit for nuclear/secondary bar semi-major axis following \citet{GFW_00} and 1.5\,kpc as an upper limit for nuclear ring semi-major axis after \citet{ES_02}. 

\section{Surface photometry outputs}
\label{res}

We carried out a detailed morphological characterization of a sample of Sy galaxies and a matched inactive sample. We scrutinized various images, residuals, maps, and profiles in a case-by-case approach in order to reveal galaxy structures that could be related to the fueling of Sy nuclei. As a result we list the morphological classification assigned by this study (the error of $T$ is $\pm1$) and remarks concerning the presence of bars, rings (including pseudo-rings), asymmetries, and companions in the last six columns of Tables\,\ref{T_morph_AG} and \ref{T_morph_NG} for the Sy and control sample, respectively. 
 
We paid special attention to the Sy sample (see Appendices\,\ref{profiles} and \ref{indiv}) in the framework of the precise estimate of the structural parameters based on SB decomposition that is ongoing. Thus, the rest of the section concerns the Sy sample. On the basis of the ellipse fits, we determined global, isophotal, and bar parameters, which will be presented in Slavcheva-Mihova \& Mihov (in prep.). 
Furthermore, we reveal or straighten out the presence of structures in a part of the galaxies and discuss the influence of some features on the structural parameters estimation through decomposition. 
Brief comments on most of the galaxies in this context are given in Appendix~\ref{indiv}.

We found the following structures not hitherto reported to our knowledge: 
\begin{itemize}
\item a bar in \object{Ark 479} and an oval/lens in \object{Mrk 595},
\item inner rings in \object{Ark 120} and \object{Mrk 376},
\item a filament (\object{3C 382}) and loop-like features (\object{NGC 7603}),
\item a nuclear bar surrounded by a ring in \object{Mrk 352} and nuclear dust lanes in \object{Mrk 590}.
\end{itemize}
Furthermore, we clarified the morphological status of some objects. 
We consider Mrk\,376 barred, \object{Mrk 279} and \object{NGC 7469} harbouring ovals/lenses, \object{Mrk 506} non-barred, and that \object{NGC 3516} has an inner ring. We discussed the bars suggested for \object{Mrk 573} and \object{NGC 3227}, as well as the nature of the proposed galaxy merging into 3C\,382. 

The profiles of not all the barred galaxies express clear bar signatures. 
The ellipticity maximum may be masked by the beginnings of spiral arms (e.g., NGC\,6814), other features at the bar edges (e.g., Mrk\,771), or more central features (e.g., NGC\,7469). The SB bump may be weak or even absent (e.g., Mrk\,771, NGC\,6814). 
The spiral arm beginnings, emerging from the bar edges, produce a wavelength-dependent SB bump and an ellipticity peak, imposed over the end of the bar SB bump and the ellipticity maximum, respectively, accompanied by blue CI dips and an almost constant/slightly changing PA (e.g., Mrk\,79, NGC\,4593); the ellipticity maximum may be wavelength-dependent, too (e.g., Ark\,564). 
This masking of the bar end could result in an overestimation of the bar length determined through decomposition.

Generally, partial fitting of the spiral structure by the model results in SB bumps at a continuously changing PA, often accompanied by ellipticity maxima (e.g., Mrk\,348, I\,Zw\,1). Rings commonly produce SB bumps (e.g., Mrk\,506, Mrk\,541).
There are other features that could also influence the profiles~-- features at bar ends (Mrk\,771, Mrk\,279), shells (NGC\,5548), tidal features (e.g., III\,Zw\,2), dust (e.g., NGC\,3227), underlying features (3C\,120, Mrk\,315), and [\ion{O}{iii}] emission (Mrk\,573, Mrk\,595, Mrk\,766). 
Thus, the particularized features can modify the SB distribution, thereby altering the structural parameters obtained as a result of decomposition. To get  trustworthy estimates of these parameters, the above features should be considered in decomposition, either fitting some of them \citep[e.g., the recent version of GALFIT,][]{PHI_10} or excluding the corresponding regions.

\section{Local environment} 
\label{loc_env}

We explored the local environment of both the Sy galaxy sample and the inactive one. We looked for close physical companions (further referred to as just companions) within (1) a projected linear separation of five galaxy diameters after \citet{S_01} and (2) an absolute radial velocity difference of $|\Delta V_{\rm r}|\,$=$\,600\,\rm km\,\rm s^{-1}$. The latter is the typical pairwise velocity dispersion of the galaxies in the combined CfA2+SSRS2\footnote{Second CfA and Southern Sky Redshift Surveys.} and is about twice as the pairwise velocity dispersion, not including clusters \citep[Marzke et~al. 1995; see also][]{DP_83}. We did not impose a brightness difference limit criterion as it would introduce a bias against dwarf galaxies, which are believed to play a significant role in minor merger processes \citep[e.g.,][]{CAM_05}.

The companions of the Sy and inactive galaxy sample are listed in Tables\,\ref{T_comp_AG} and  \ref{T_comp_NG}, respectively.  In the separation estimation we used the  $25\,B\,\rm mag\,arcsec^{-2}$ isophotal diameters (also taking their errors into account), corrected for Galactic extinction and inclination. We generally used the diameters given in HyperLeda. There are instances of a larger companion than its primary (e.g., 2MASX\,J04363658--0250350). Thus, applying the separation criterion, we inspected the environment of each primary in a large enough field and took the greater of the diameters for each candidate pair into account.

The radial velocity difference was calculated using the special relativistic convention \citep[e.g.,][]{K_96b}:
\begin{displaymath}
V_{\rm r}=\frac{{(1+z)^2-1}}{{(1+z)^2+1}}\,c.
\end{displaymath}	
We used redshift, corrected to the reference frame defined by the 3K microwave background radiation given in NED.

\section{Discussion}
\label{disc}
\subsection{Seyfert sample}

The small share of ellipticals among Sy galaxies has been known for a long time \citep{A_77,MMP_95}, though it has been suggested that bright radio-quiet quasars, like their radio-loud counterparts, tend to reside in elliptical galaxies, thus refuting the stated link between radio loudness and $T$ \citep[e.g.,][]{MKD_99}. In fact, a few of our Sy sample galaxies have been classified as elliptical, but we found all of them disk-dominated (see Table\,\ref{T_morph_AG}). The $\rm E\,$$\leftrightarrow$$\,\rm S0$ misclassification could frequently occur when only visual inspection is involved, especially of faint/distant galaxies \citep[][]{SGA_04}. Identifying the disk galaxies, misclassified as ellipticals, is substantial for the correct photometric decomposition, too \citep{EGC_04}. Besides this, bad seeing or low resolution could lead to blurring of a spiral structure (i.e., $\rm spiral\,$$\rightarrow$$\,\rm S0$ misclassification) and of bars/rings. The morphological type differences between this study and RC3 (or NED/HyperLeda/SIMBAD when there is no well-defined classification in RC3) can be followed in Table\,\ref{T_morph_AG}. The Hubble type of the sample ranges from S0 to Sbc with a median of S0/a (Fig.\,\ref{MT_AN_ourour_hbm}). As the bulk of our sample consists of Sy\,1 objects, the preference for early types may reflect that Sy\,1 galaxies tend to have smaller $T$ than Sy\,2 ones \citep{FK_89,HM1_99,KPC_06}. The trend of active galaxies being earlier types than inactive ones has already been noted by Terlevich et~al. \citep[1987; see also][]{HM1_99}.

\subsection{Bar fraction}
The bar fraction of galaxies is known to vary with wavelength and bar detection methods \citep{AMC_09,HJB_09}.
Furthermore, there is no consensus on the preponderance of bars in Sy galaxies.
 Studying the bar fraction, based on RC3 classification of Sy/inactive galaxies, \citet{HM1_99} found  67\%/69\% for the Extended 12 Micron Galaxy Sample and \citet{LSB_04} found  62\%/69\% for the Ohio State University Bright Galaxy Survey. Using ellipse fitting of matched samples in the near-infrared (NIR), \citet{MR_97} obtained similar bar fractions for the Sy (73\%) and the control (72\%) sample, while \citet{LSK_02} found an excess of bars in Sy galaxies (73\% to 50\%).  

The incidence of bars in our Sy and control galaxy sample is similar: $(49\,$$\pm$$\,8)\%$\footnote{The associated error is estimated from binomial distribution as $\sigma(f)=\sqrt{f(1-f)/N\,}$, where $f$ is the fraction of interest in a sample of size $N$.} and $(46\,$$\pm$$\,8)\%$, respectively.
These percentages may be lower than in other studies in the optical, owing to the large share of S0 galaxies that tend to show a lower incidence of bars, relative to later types \citep{HFS_97,KSP2_00,LSB_09,AMC_09}. 
Given our morphological mix and the distribution of bar fractions with Hubble type in \citet{HFS_97}, based on RC3 classification, and in \citet{AMC_09}, based on ellipse fitting, we estimated the expected bar fraction of our combined (Sy and control) sample to be 50\% for the former and  44\% for the latter authors, which is similar to our fractions. 

We present the distribution of deprojected bar ellipticities for both samples in Fig.\,\ref{Ebar_AN_2our_hbm}. Based on the deprojected bar ellipticity value of 0.45 as an objective (although not ideal) criterion for bar strength, we found that the frequency of weak bars in the Sy sample is higher than in the control one at about the 98\% confidence level. This result, however, is sensitive to the adopted limit as the Sy bars ellipticities are peaked around it; e.g., if we adopt 0.40 as a limit between strong and weak bars after \citet{MF_97}, the trend toward a deficiency of strong bars in Sy galaxies practically disappears. Either way, Sy bars (with median deprojected $\epsilon_{\rm bar}$ of 0.39) appear weaker than their inactive counterparts (with median deprojected $\epsilon_{\rm bar}$ of 0.49) at the 95\% confidence level\footnote{The significance of the difference was estimated using the one-tailed Student's $t$-test.}, which is consistent with \citet{SPK_00} and \citet{LSR_02,LSB_04}. This difference cannot be accounted for with the preference by early type galaxies for weak bars \citep[e.g.,][]{LSB_04,AMC_09} as our samples are matched in $T$. 

\subsection{Ring fraction}
The fractions of ringed Sy, $(49\,$$\pm$$\,8)\%$, and control, $(54\,$$\pm$$\,8)\%$, galaxies are the same within the errors. In particular, we found a similar incidence of inner rings in the Sy, $(34\,$$\pm$$\,8)\%$, and control, $(40\,$$\pm$$\,8)\%$, sample. 
The abundance of outer rings is formally the same within the errors for both samples. Still, outer rings occur about 1.5 times more often in the Sy sample, $(40\,$$\pm$$\,8)\%$, than in the inactive one, $(26\,$$\pm$$\,7)$\%.
Such a trend was suggested by \citet{SSS_80} and \citet{HM1_99}. The correlation in our results is less pronounced than in these papers, basically because their results are not based on matched samples (especially in $T$). 
Furthermore, the incidence of rings in our barred subsample of galaxies is higher than in the non-barred one at the 99.9\% and 97.6\% confidence levels for the Sy and control sample, respectively, which is expected, since rings have been considered the loci of concentration of gas or stars near the dynamical resonances \citep{S_81,C_08}.

\subsection{Local environment and asymmetries}

We found at least one close physical companion for $(44\,$$\pm$$\,9)\%$ of the Sy sample\footnote{NGC\,7603 was not taken into account owing to the anomalous redshift of the companion.} and $(43\,$$\pm$$\,8)\%$ of the control one. These are lower limits, when keeping in mind the radial velocity difference requirement and the underestimation of dwarf satellites, especially of more distant galaxies.
For instance, there is some evidence of candidate companions (meeting the separation criterion but with no redshift information) in both samples~-- tidal features (3C\,382, ESO\,202--\,G\,001, and ESO\,113--\,G\,050) or neighbours of comparable brightness to the primaries (Mrk\,376 and ESO\,324--\,G\,003). Besides this, there is an extended feature in Mrk\,335 that may actually be a companion seen through the galaxy disk (see Appendix\,\ref{m335}). Thus, considering the most obvious cases of candidate companions, both the Sy and control sample are again in a tantamount position.
Comparison with other results is hardly relevant as there are no universal criteria for defining a physical companion: 
the choice of a limit for the projected linear separation, radial velocity difference, and brightness difference between the primary galaxy and its companion is empirical, hence, arbitrary; moreover, the lack of redshift information has often been substituted by statistical suppositions of the share of projected objects. In fact, most research has aimed at relative studies of the environment of Sy vs. inactive galaxies. No consensus has been reached about the share of Sy galaxies with companions~-- the results can be grouped into three: those with an excess of Sy galaxies with companions relative to inactive galaxies, those with no difference between Sy and inactive galaxies, and those with an excess of Sy\,2 galaxies with companions compared both to Sy\,1 and inactive galaxies \citep[][and references therein]{S_04}.

Tidal interactions and minor mergers could produce various tidal features and disturbed structures.
We found, however, no correlation between asymmetries and the presence of companions for both samples.
One explanation lies in the delay between the onset of interaction and its optical manifestation in the host galaxy \citep[e.g.,][]{BSV_87}, and bulge prominence can further delay this \citep{HM_95}.  
Second, an ongoing merger would show up as an isolated asymmetric galaxy.  	
The fraction of asymmetric galaxies is the same within the errors for the Sy $(51\,$$\pm$$\,8)\%$ and control $(43\,$$\pm$$\,8)\%$ sample. Similar results were found by \citet[][]{VRV_00} and \citet[][]{C_00}. Furthermore, the fraction of asymmetric galaxies without companions is practically equal for both samples (between $(20\,$$\pm$$\,7)\%$ and $(26\,$$\pm$$\,8)\%$, depending on whether candidate companions are excluded from consideration or not).
Therefore, we could come to the corollary that minor mergers, at least not accompanied by companions, do not occur in the Sy sample more often than in the control one.

\subsection{General discussion}

It turns out that $(91\,$$\pm$$\,5)\%$ of the Sy and $(94\,$$\pm$$\,4)\%$ of the inactive galaxies have bars or/and rings, asymmetries, companions. Thus, the vast majority of galaxies in both samples show morphological evidence of non-axisymmetric perturbations of the potential or/and have close companions. 
The rest of the galaxies all show some signs of interaction: they either have a companion within about seven galaxy diameters (Mrk\,352, 2MASX\,J01505708+0014040, and ESO\,292--\,G\,022), have a candidate companion without redshift information \citep[Mrk\,509,][]{BDP_02,RMB_93}, or show \ion{H}{i} evidence of a past merger \citep[Mrk\,304,][]{LH_99}. Thus, unperturbed galaxies, both Sy and inactive, may turn out to be related to interaction.
Instances of fine structures indicative of past mergers in active galaxies that were previously classified as undisturbed have already been adduced \citep[][]{CBJ_07,BCJ_08}.

 Even if we consider only the morphological evidence of non-axisymmetric perturbations of the potential, its incidence is equal within the errors in the Sy, $(86\,$$\pm$$\,6)\%$, and control, $(83\,$$\pm$$\,6)\%$, sample. Similar results were found by \citet[][]{VRV_00}.

\subsection{Robustness of the results regarding the different data sources}

All Sy galaxies and about a half of the control ones were imaged with CCDs. DSS\,I,\,II, and digitized ESO-Uppsala Survey data were used for the rest of the control galaxies.
We examined to what extent the different data sources of the Sy and control galaxies may introduce systematic errors in the results.

For all galaxies having CCD data (53 total), we also processed the corresponding DSS data and independently estimated the Hubble type and the presence of structures and asymmetries. 
In the photographic data, we detected the same incidence of bars and asymmetries as in the CCD data\footnote{The photographic images are saturated for one galaxy classified as barred (NGC\,6814) and for one galaxy classified as asymmetric (Mrk\,315) in the CCD data, so we did not consider these cases.}. Of the detected rings in the CCD data, we could not trace two inner rings (Mrk\,376 and Mrk\,541) and one outer ring\footnote{The fraction of bars, inner rings, outer rings, and asymmetries in the CCD/photographic data is $(48\,$$\pm$$\,7)\%$/$(48\,$$\pm$$\,7)\%$, $(36\,$$\pm$$\,7)\%$/$(32\,$$\pm$$\,6)\%$, $(36\,$$\pm$$\,7)\%$/$(34\,$$\pm$$\,7)\%$, and $(44\,$$\pm$$\,7)\%$/$(44\,$$\pm$$\,7)\%$, respectively.} (Mrk\,506) in the corresponding photographic data. 
If we use these considerations to roughly correct for structures missed because of using photographic data, the expected number of inner rings increases with one, and the number of outer rings remains the same for the photographically imaged galaxies. Thus, regarding the control galaxies, this correction affects only the fraction of inner rings, and it gets $(43\,$$\pm$$\,8)\%$ (vs. $(34\,$$\pm$$\,8)\%$ for the Sy sample). This translates into a maximal\footnote{Extra detection of inner rings would affect the fraction of rings only for galaxies not having outer rings.} fraction of rings of $(57\,$$\pm$$\,8)\%$ for the control sample (vs. $(49\,$$\pm$$\,8)\%$ for the Sy sample).
The most that this correction could affect the final results is when the undetected inner ring is among the galaxies without morphological evidence of non-axisymmetric perturbations of the potential (and companions). Then the fraction of galaxies with bars or/and rings, asymmetries, companions gets $(91\,$$\pm$$\,5)\%$ vs. $(97\,$$\pm$$\,3)\%$ for the Sy vs. control sample. Considering just the morphological evidence of non-axisymmetric perturbations of the potential, the two samples show an equal incidence, $(86\,$$\pm$$\,6)\%$. 
As we can see, this correction does not significantly influence the final results, since all features affected by it occur with the same incidence within the errors in both samples. 
The abundances themselves are either equal in both samples or higher in the control one.
In the context of our study, looking for an eventual excess of features in the Sy sample, this result should mean that the incidence of the features of interest is not lower in the control sample than in the Sy one.

\subsection{Implication for the fueling of Sy nuclei}

Regarding the Sy and control sample, we found a similar incidence of bars, rings, asymmetries, and close companions, considered both on an individual basis and all together, with the Sy bars somewhat weaker. Thus, our results imply that the fueling of Sy nuclei is not directly related to large-scale mechanisms operating over the bulk of the gas. There are some hints, however, of a link between them. 

First, it is generally believed that the required fuel is a tiny fraction of the gas in the inner few 100\,pc, especially of spiral galaxies, and angular momentum reduction is the major challenge \citep[e.g.,][]{J_06}. For instance, typical molecular gas mass of $\approx\,$$10^8\,M_\odot$ was reported for the central regions of most NUGA\footnote{NUclei of GAlaxies project.} galaxies \citep[e.g.,][]{GCS04_05}. 
A part of this gas is expected to have resulted from secular evolution. 
A higher molecular gas concentration was found in the central kiloparsec of barred galaxies than in non-barred ones \citep[][see also Regan et~al. 2006]{SOI_99,SVR_05}; according to the first authors, more than half of the central gas was driven there by the bar. 
The gas in nuclear rings, the most evident tracers of recent gas inflow, can be brought under the influence of the SMBH by viscous torques in the scenario of \citet[][]{GCS04_05}.
Furthermore, higher central gas concentration has been associated with interactions and mergers \citep[e.g.,][]{GFN_00,SSH_07}.

Second, generally weaker bars in Sy than in inactive galaxies have been associated with larger amounts of cold gas in their host galaxies in the framework of central mass concentrations that could destroy $x_1$ bar orbits \citep[][]{SPK_00}. It has been shown, however, that bars are less fragile than previously thought, and the mass of the central concentration required to dissolve the bar must be very high \citep[e.g.,][]{SS_04,DMC_06,MJ_07}. Alternatively, the main destruction mechanism could be the transfer of angular momentum from the gas inflow to the bar \citep[e.g.,][]{BCS_05}, especially in the presence of radiative cooling \citep[e.g.,][]{DMC_06}. Thus, the weaker Sy bars may be related to the generally larger cold gas amounts reported in their disks \citep[e.g.,][see also Ho et~al. 2008]{HMM2_99} in the context of angular momentum transfer.

The relatively low accretion rates of Sy nuclei prompt a variety of small-scale processes able to drive the circumnuclear gas down to the very centre \citep[e.g.,][]{M_04}. This could be the main reason for a lack of a universal morphological pattern on these scales \citep[e.g.,][]{GCS_04}. 
Seyfert activity, however, has been associated with the presence of dust \citep[][]{SSF_07} and more disturbed gaseous kinematics \citep[][]{DME_07} in the circumnuclear regions.
In this regard, we started a study of the circumnuclear regions of a sample of Sy galaxies using HST archival images. The circumnuclear structures of Mrk\,352 and Mrk\,590 are the first results of this research.

\section{Summary}
\label{concl}

\begin{enumerate}
\item We presented a detailed morphological characterization of a sample of 35 Sy galaxies. We scrutinized various images, residuals, maps, and profiles in order to reveal galaxy structures that could be important for the fueling of Sy nuclei, as well as for the proper photometric decomposition, which is ongoing. The careful analysis of these data on an individual, case-by-case basis, has led to a more explicit morphological status of a part of the galaxies, resulting in improved morphological type accuracy, and to new structural components and features being unveiled:

\begin{itemize}
\item we revealed a bar in Ark\,479, an oval/lens in Mrk\,595, inner rings in Ark\,120 and Mrk\,376, and features of possible tidal origin in 3C\,382 and NGC\,7603 for the first time to our knowledge;
\item we discussed some structures of controversial/unclear morphology in Mrk\,573, Mrk\,376, NGC\,3227, NGC 3516, Mrk\,279, Mrk\,506, 3C\,382, and NGC\,7469.
\end{itemize}
\item We compared the large-scale morphology and local environment of the Sy sample and a control one, matched in $T$, $V\rm_{r}$, $M^{B}_{\rm abs}$, and $\epsilon$, with the following main results:
\begin{itemize}

\item we found similar fractions of bars in the Sy, $(49\,$$\pm$$\,8)\%$, and control, $(46\,$$\pm$$\,8)\%$, galaxy sample;
\item the Sy bars are weaker than the bars in the control sample with median deprojected bar ellipticity values of 0.39 vs. 0.49, respectively, at the 95\% confidence level;
\item the incidence of rings in the Sy and control sample is similar~-- $(49\,$$\pm$$\,8)\%$ and $(54\,$$\pm$$\,8)\%$, respectively;
\item practically equal parts of the Sy, $(44\,$$\pm$$\,9)\%$, and control, $(43\,$$\pm$$\,8)\%$, sample have at least one physical companion within a projected linear separation of five galaxy diameters and an absolute radial velocity difference of $|\Delta V_{\rm r}|\,$=$\,600\,\rm km\,\rm s^{-1}$;
\item there is no correlation between the presence of asymmetries and companions for both samples; minor mergers, at least without companions, do not occur in the Sy sample more often than in the control one;
\item the vast majority of both samples, $(91\,$$\pm$$\,5)\%$ of the Sy and $(94\,$$\pm$$\,4)\%$ of the control one, have bars, rings, asymmetries, or close companions; 
\item similar fractions of the Sy, $(86\,$$\pm$$\,6)\%$, and control, $(83\,$$\pm$$\,6)\%$, sample show morphological evidence of non-axisymmetric perturbations of the potential; 
\item the fueling of Sy nuclei does not appear directly related to the large-scale morphology and local environment of their host galaxies.	 
\end{itemize}
\end{enumerate}
In the framework of our results we have started a study of the circumnuclear regions of a sample of Sy galaxies using HST archival images. As first results of this research, we revealed a nuclear bar and ring in Mrk\,352 and nuclear dust lanes in Mrk\,590.

\begin{acknowledgements}
This research has made use of the NASA/IPAC Extragalactic Database (NED), which
is operated by the Jet Propulsion Laboratory, California Institute of Technology,
under contract with the National Aeronautics and Space Administration.

We acknowledge the use of the HyperLeda database (http://leda.univ-lyon1.fr).

This research has made use of the SIMBAD database, operated at the CDS, Strasbourg,
France.

The Digitized Sky Surveys were produced at the Space Telescope Science Institute under U.S. Government grant NAG W-2166. The images of these surveys are based on photographic data obtained using the Oschin Schmidt Telescope on Palomar Mountain and the UK Schmidt Telescope. The plates were processed into the present compressed digital form with the permission of these institutions.

Funding for the SDSS and SDSS-II was provided by the Alfred P. Sloan Foundation, the Participating Institutions, the National Science Foundation, the U.S. Department of Energy, the National Aeronautics and Space Administration, the Japanese Monbukagakusho, the Max Planck Society, and the Higher Education Funding Council for England. The SDSS Web Site is http://www.sdss.org/.

Some of the data presented in this paper were obtained from the Multimission Archive at the Space Telescope Science Institute (MAST). STScI is operated by the Association of Universities for Research in Astronomy, Inc., under NASA contract NAS5-26555. Support for MAST for non-HST data is provided by the NASA Office of Space Science via grant NNX09AF08G and by other grants and contracts.

This publication makes use of data products from the Two Micron All Sky Survey,
which is a joint project of the University of Massachusetts and the Infrared
Processing and Analysis Center/California Institute of Technology, funded by
the National Aeronautics and Space Administration and the National Science
Foundation.

This paper makes use of data obtained from the Isaac Newton Group Archive, which is maintained as part of the CASU Astronomical Data Centre at the Institute of Astronomy, Cambridge.

This research has made use of Aladin.

We acknowledge the use of ESO-MIDAS. 

We acknowledge the support by UNESCO-BRESCE for the regional collaboration.

The Two-Channel Focal Reducer was transferred to the Rozhen National Astronomical
Observatory under contract between the Institute of Astronomy and National Astronomical Observatory, Bulgarian Academy
of Sciences, and the Max Planck Institute for Solar System Research. 
\end{acknowledgements}

\appendix

\section{Contour maps and profiles of the Sy galaxies}
\label{profiles}
We present contour maps and profiles of the SB, CI, $\epsilon$, and PA of the Sy galaxies in Fig.\,\ref{contprof}.

\begin{figure*}[htbp]
\vspace{0.1cm}
   \centering
\includegraphics[width=5.6cm]{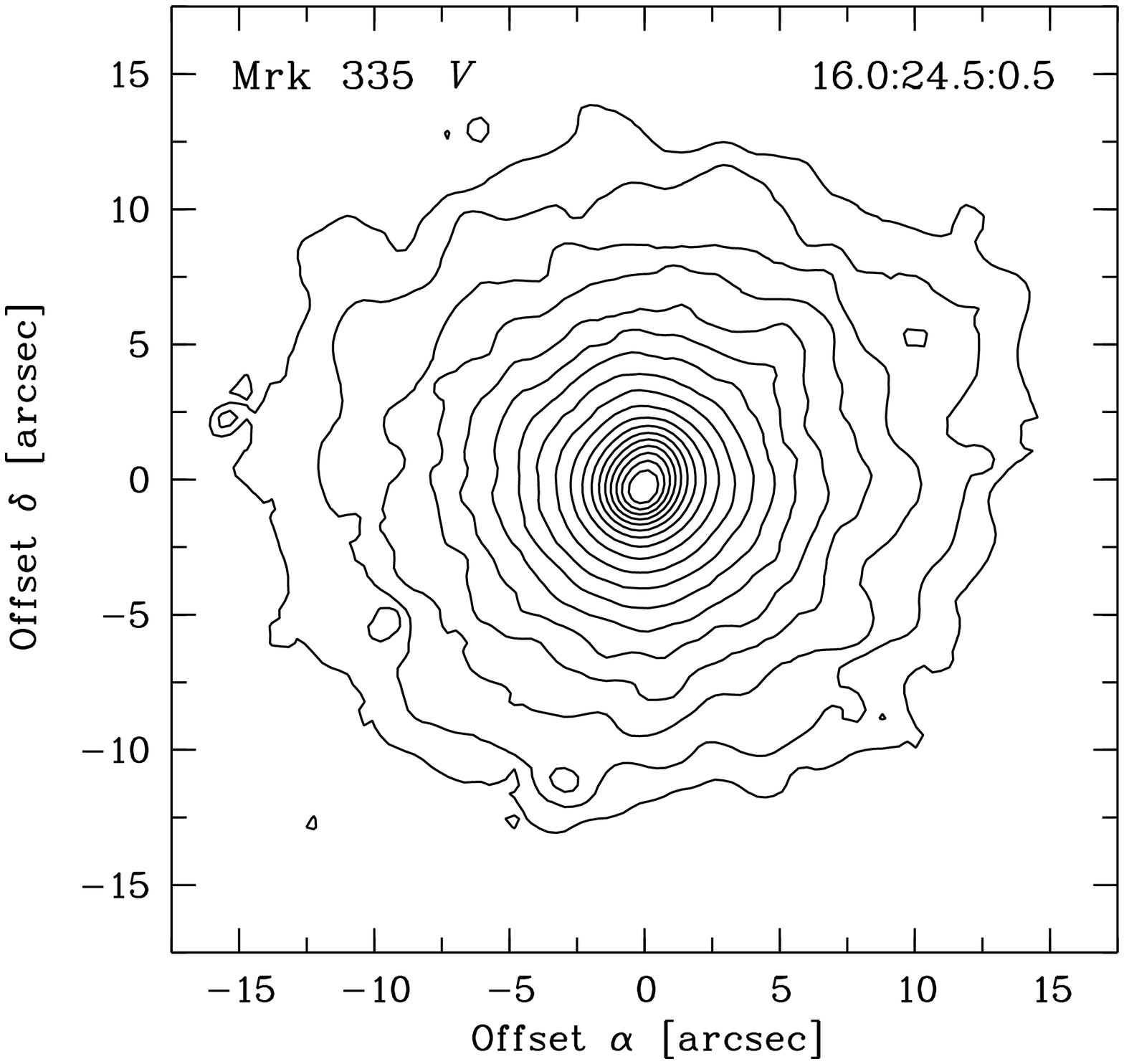}
\hspace{0.5cm}
\includegraphics[width=5.6cm]{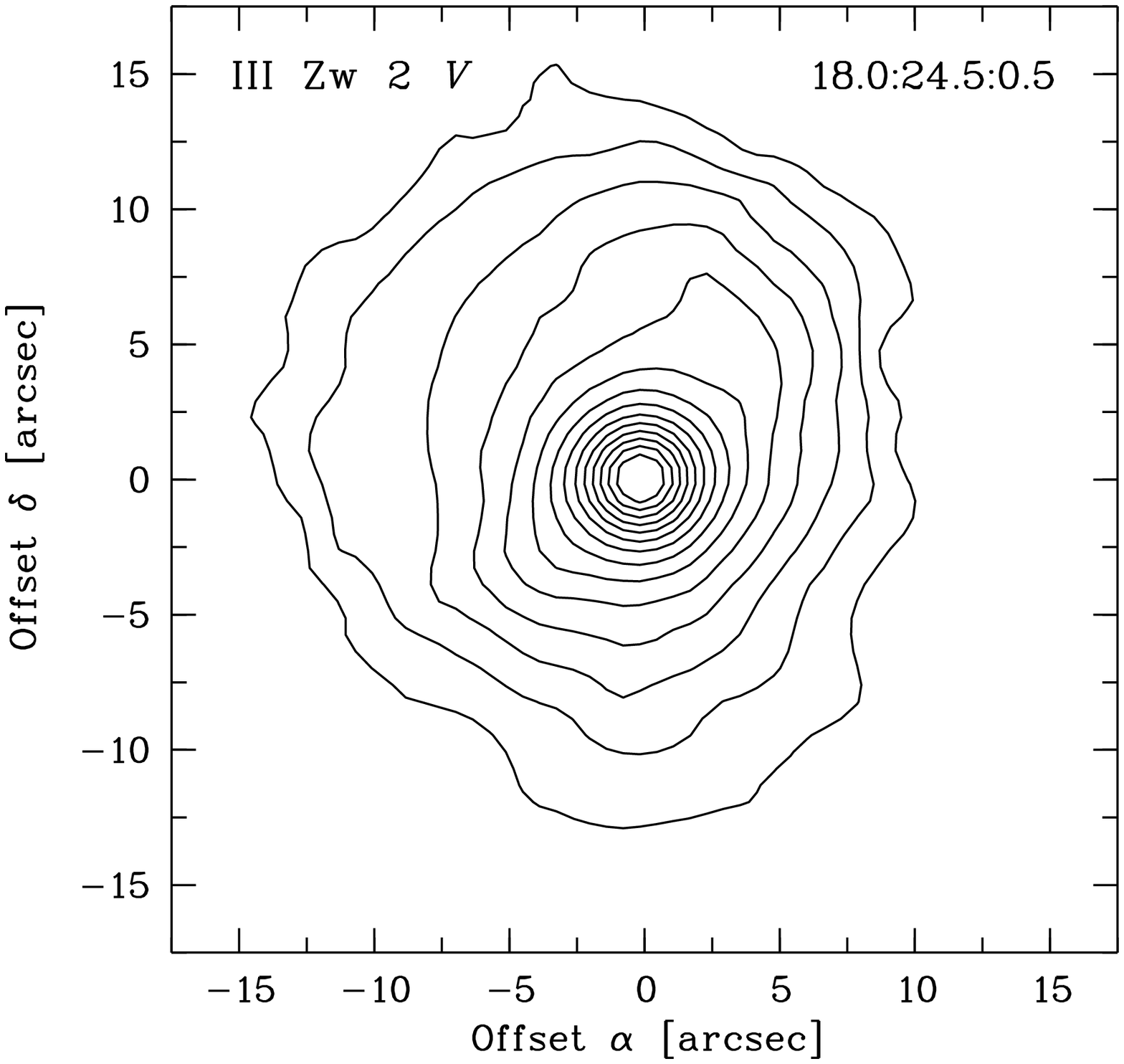}
\hspace{0.5cm}
\includegraphics[width=5.6cm]{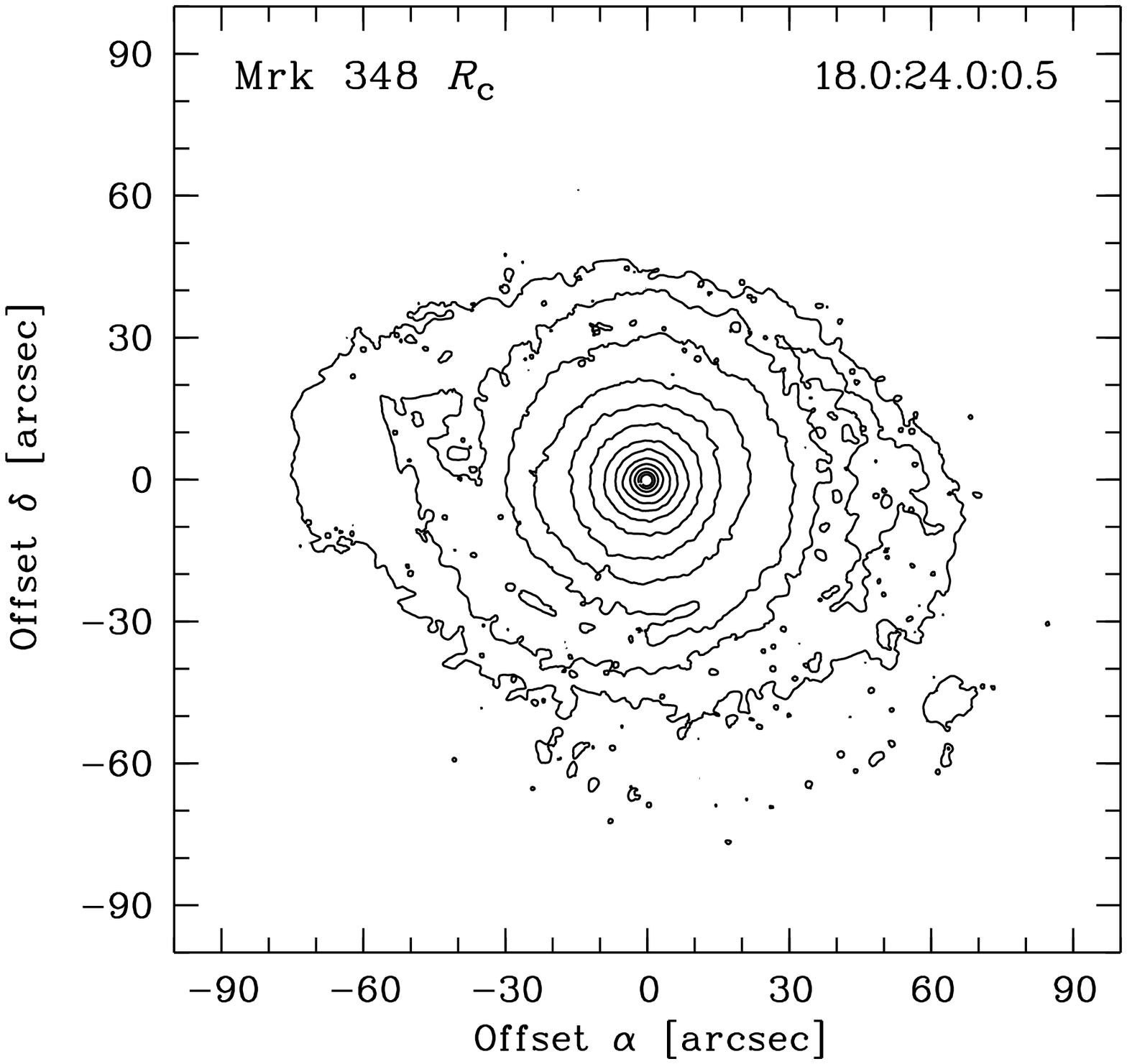}

\vspace{0.3cm}

\includegraphics[width=5.6cm]{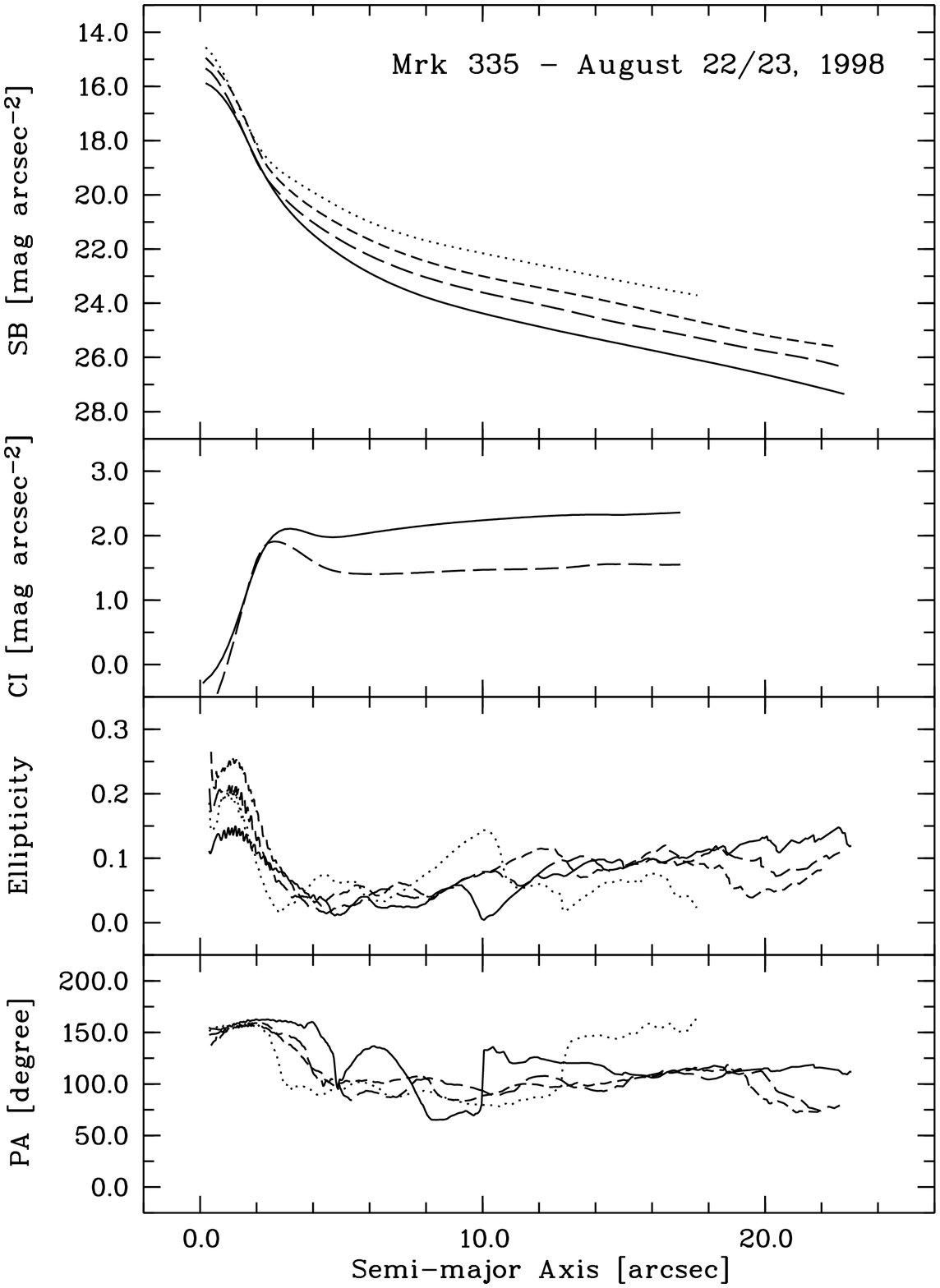}
\hspace{0.5cm}
\includegraphics[width=5.6cm]{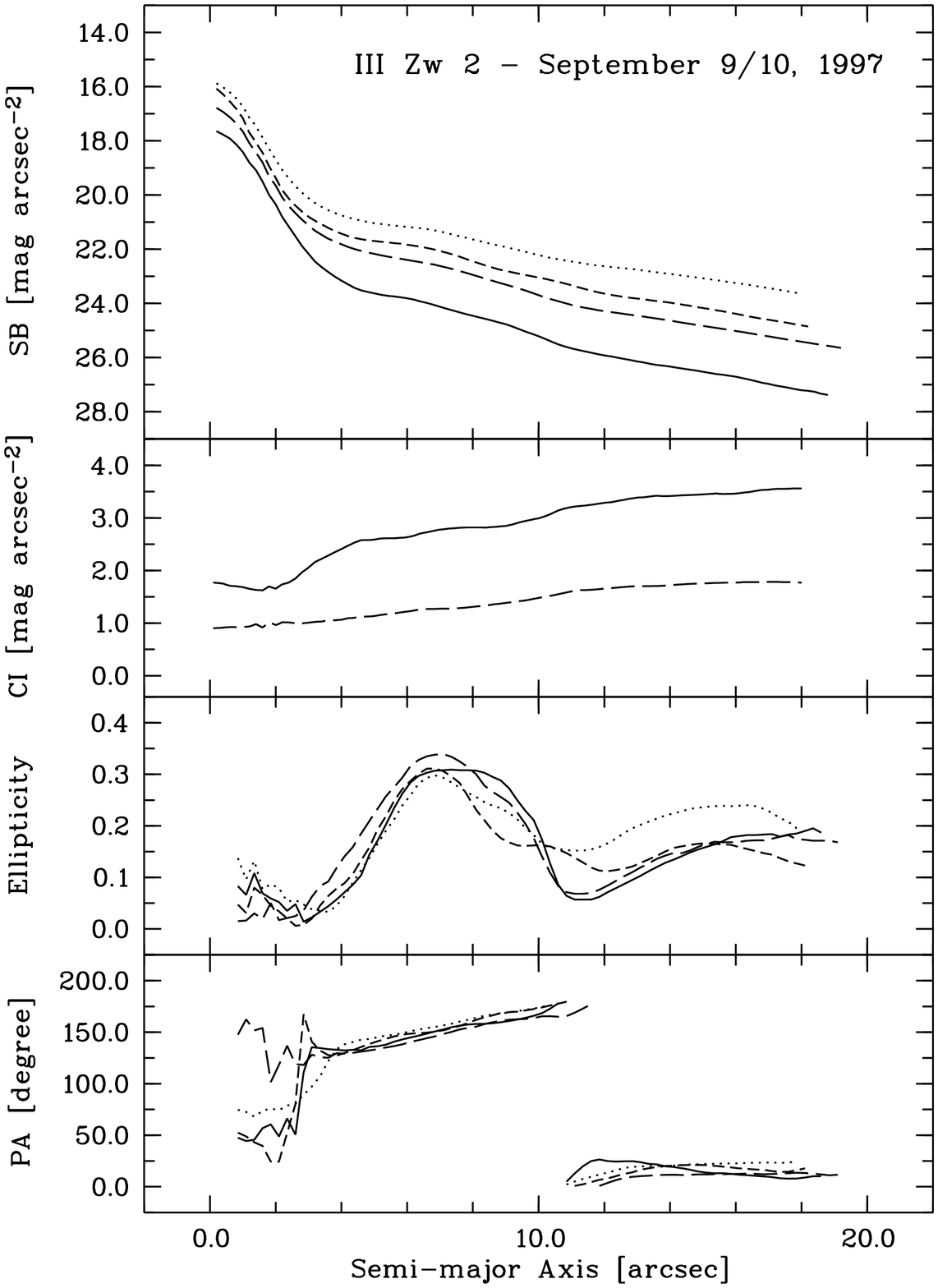}
\hspace{0.5cm}
\includegraphics[width=5.6cm]{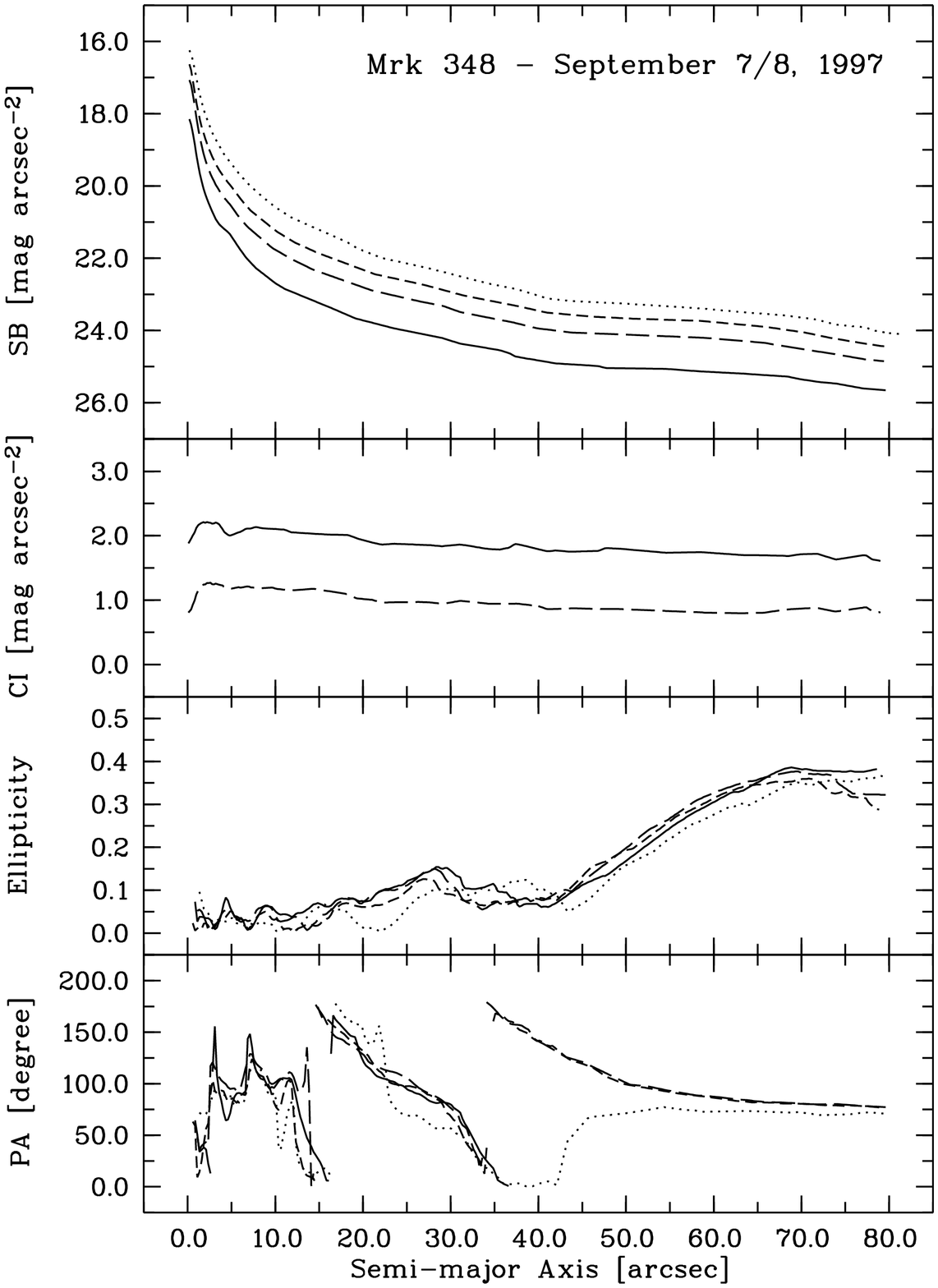}
\caption{Calibrated contour maps and profiles of the Sy galaxies, ordered by right ascension. Upper panels: contour maps. North is up and east to the left. The galaxy names and passbands are specified in the upper left; the numbers in the upper right denote the start SB, the end SB, and the SB step (fixed to 0.5) in units of $\rm mag\,\rm arcsec^{-2}$. Lower panels: profiles of SB, CI, $\epsilon$, and PA. The CI profiles shown are $B\,$--$\,I_{\rm \scriptstyle C}$ (solid) and  $V\,$--$\,I_{\rm \scriptstyle C}$ (dashed); for Ark\,120 $B\,$--$\,R_{\rm \scriptstyle C}$ (dash-dotted) is shown. For the rest of the profiles the solid, long-dashed, short-dashed, and dotted line is for the $B$-, $V$-, $R_{\rm \scriptstyle C}$-, and $I_{\rm \scriptstyle C}$-band, respectively. For Mrk\,352, Mrk\,771, and Mrk\,279 the HST profiles are also plotted (squares; their SB profiles are not calibrated).
%A sample of Fig. 1 appears here. The remainder appears on-line.
}
         \label{contprof}
   \end{figure*}
\setcounter{figure}{0}

\begin{figure*}[htbp]
\vspace{0.1cm}
   \centering
\includegraphics[width=5.6cm]{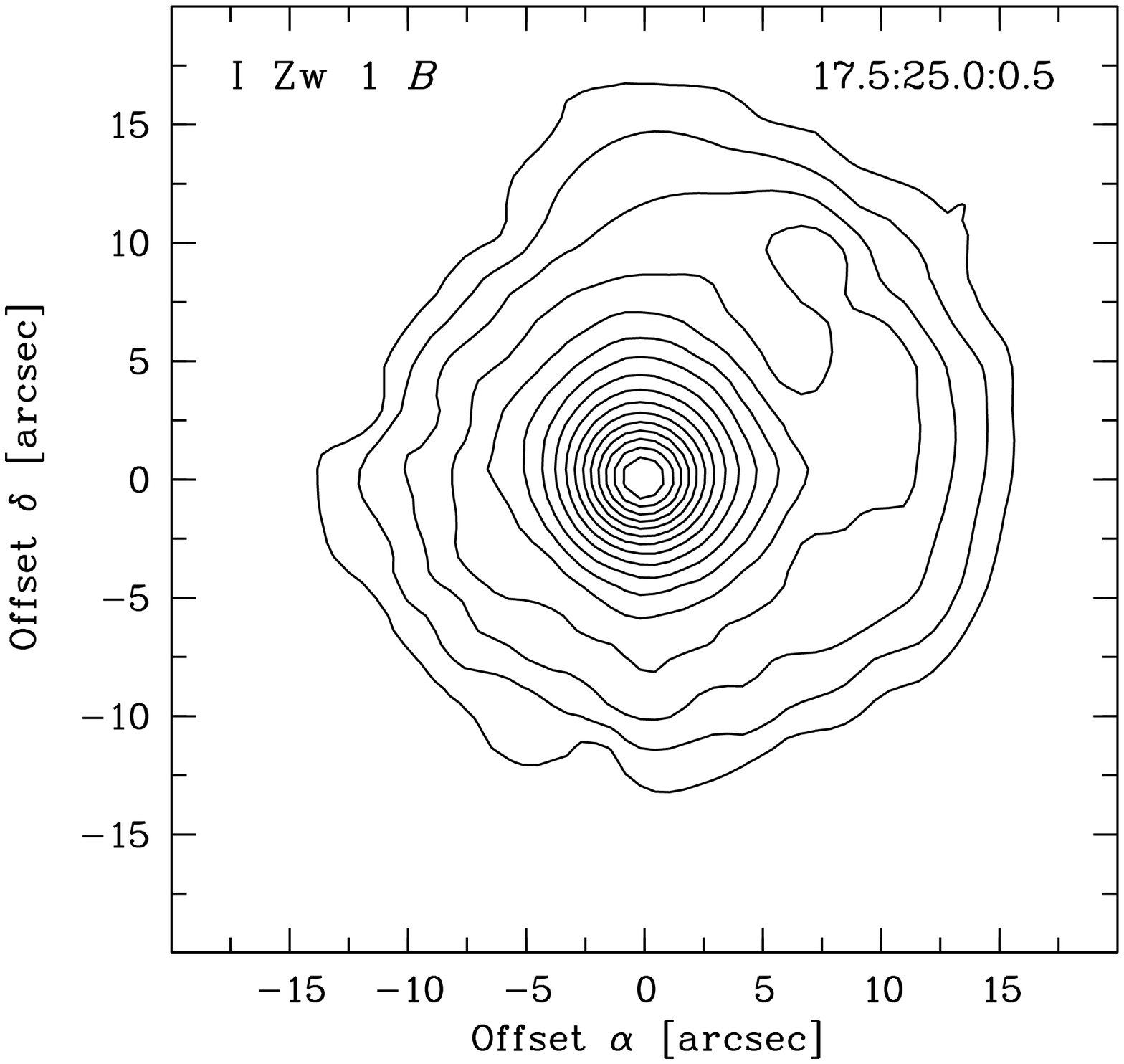}
\hspace{0.5cm}
\includegraphics[width=5.6cm]{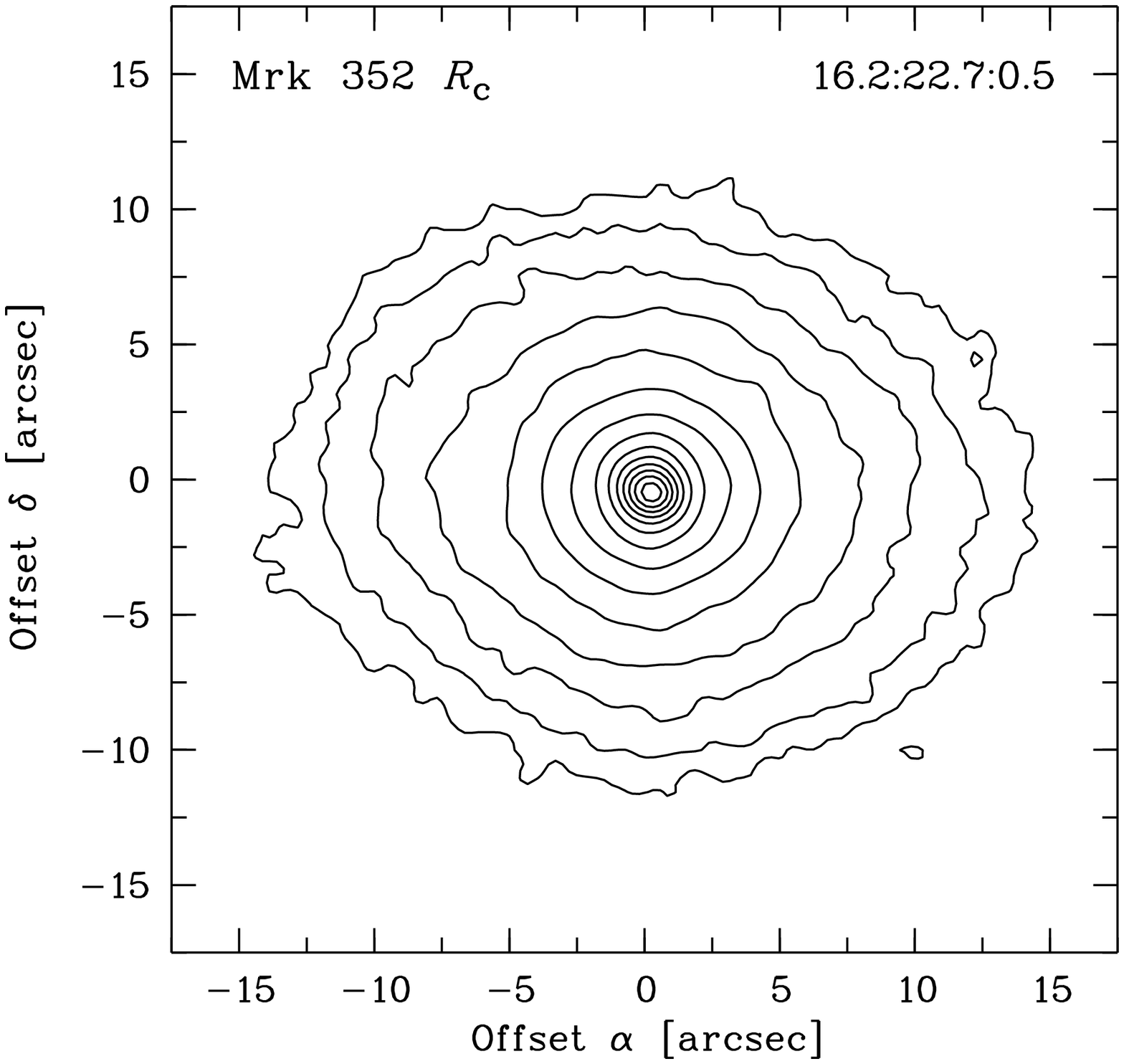}
\hspace{0.5cm}
\includegraphics[width=5.6cm]{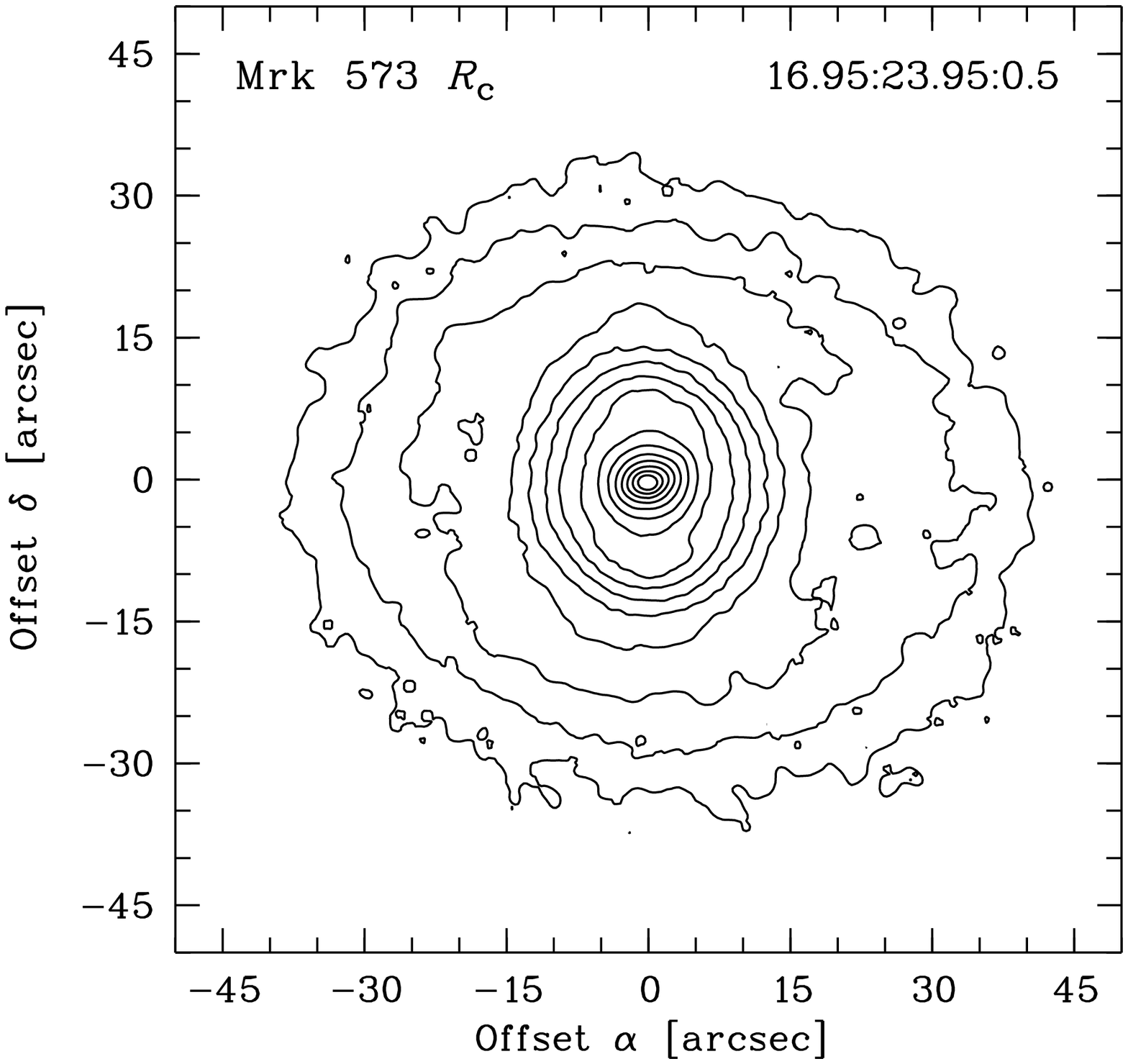}

\vspace{0.3cm}

\includegraphics[width=5.6cm]{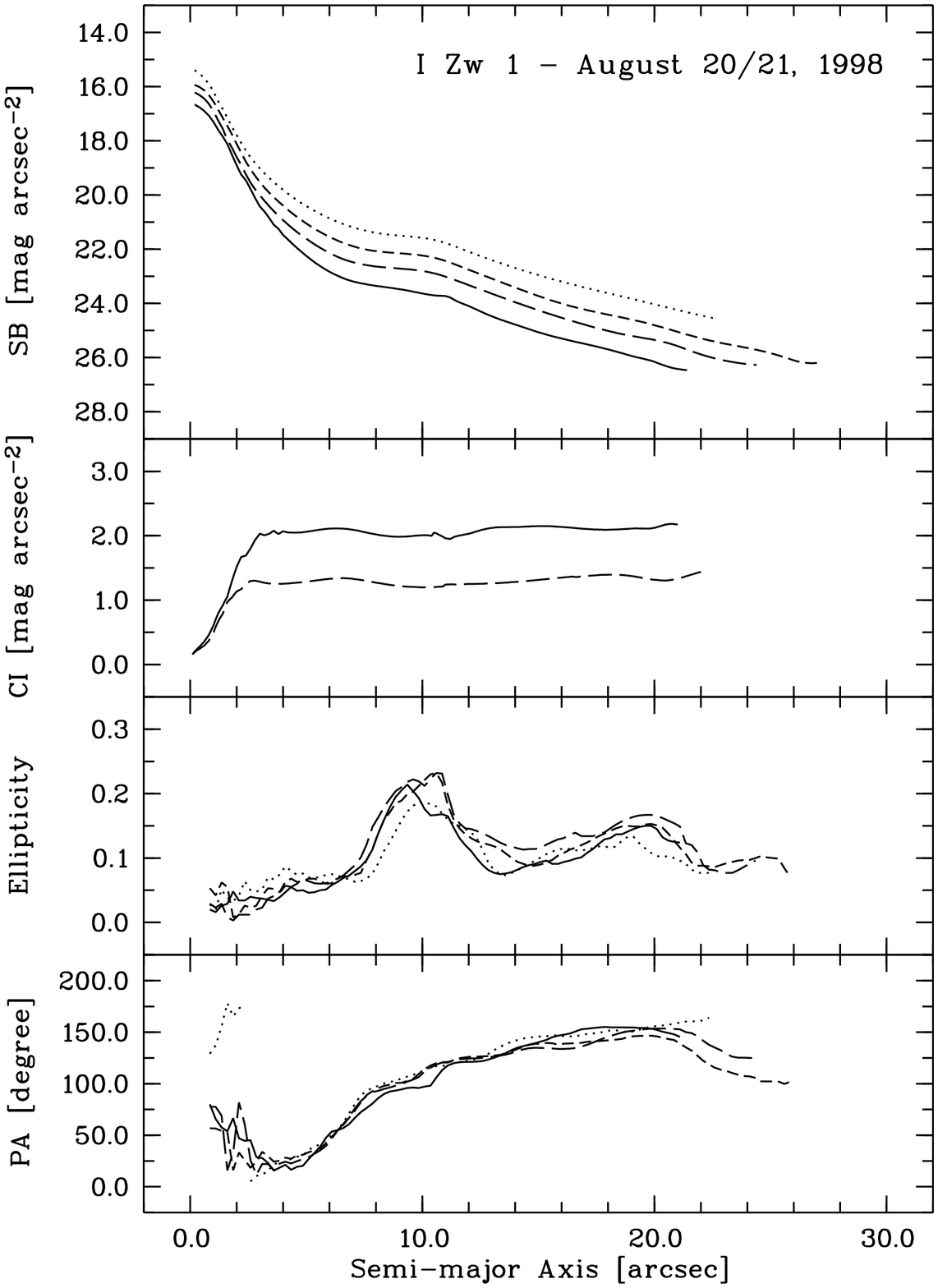}
\hspace{0.5cm}
\includegraphics[width=5.6cm]{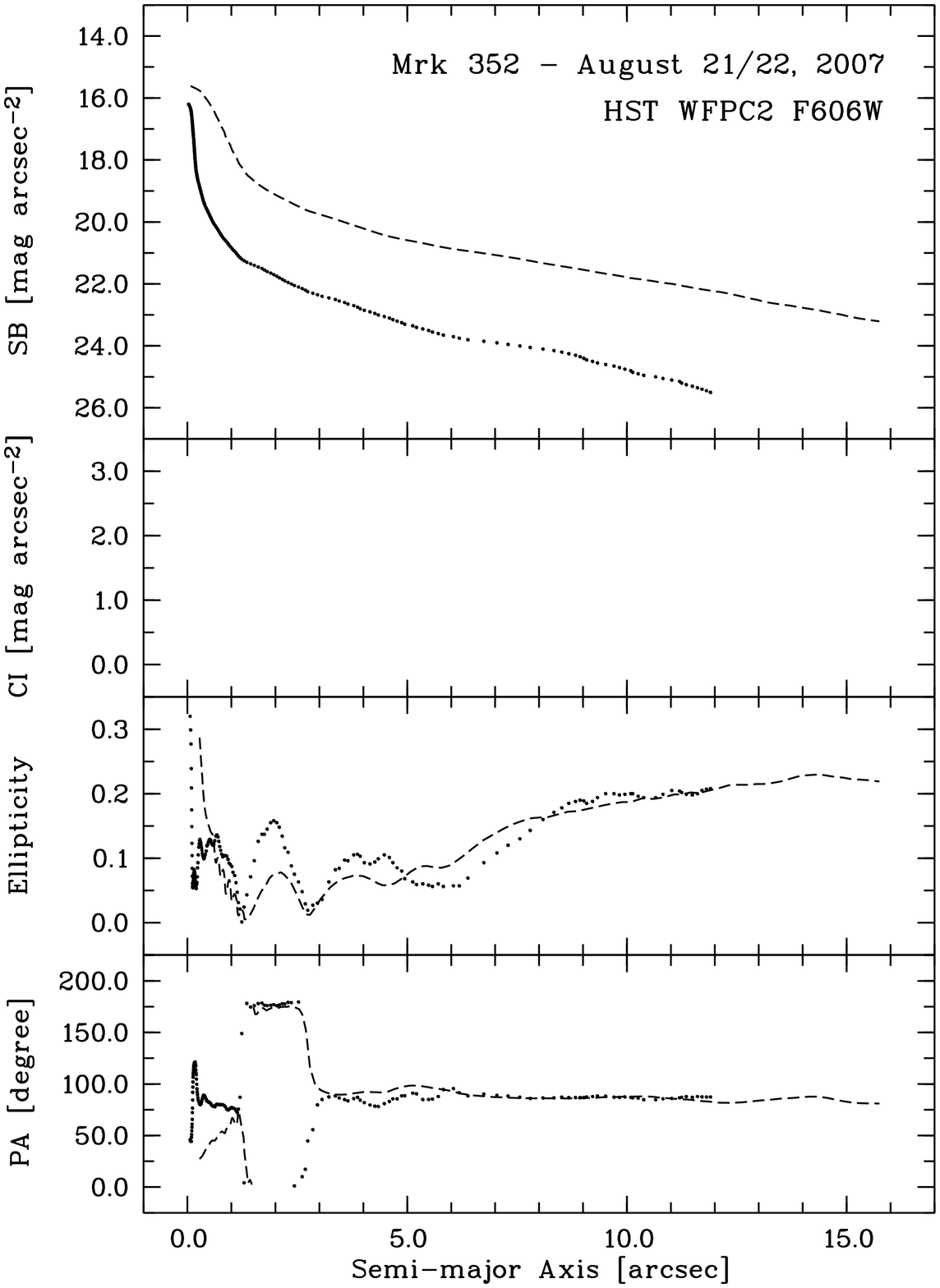}
\hspace{0.5cm}
\includegraphics[width=5.6cm]{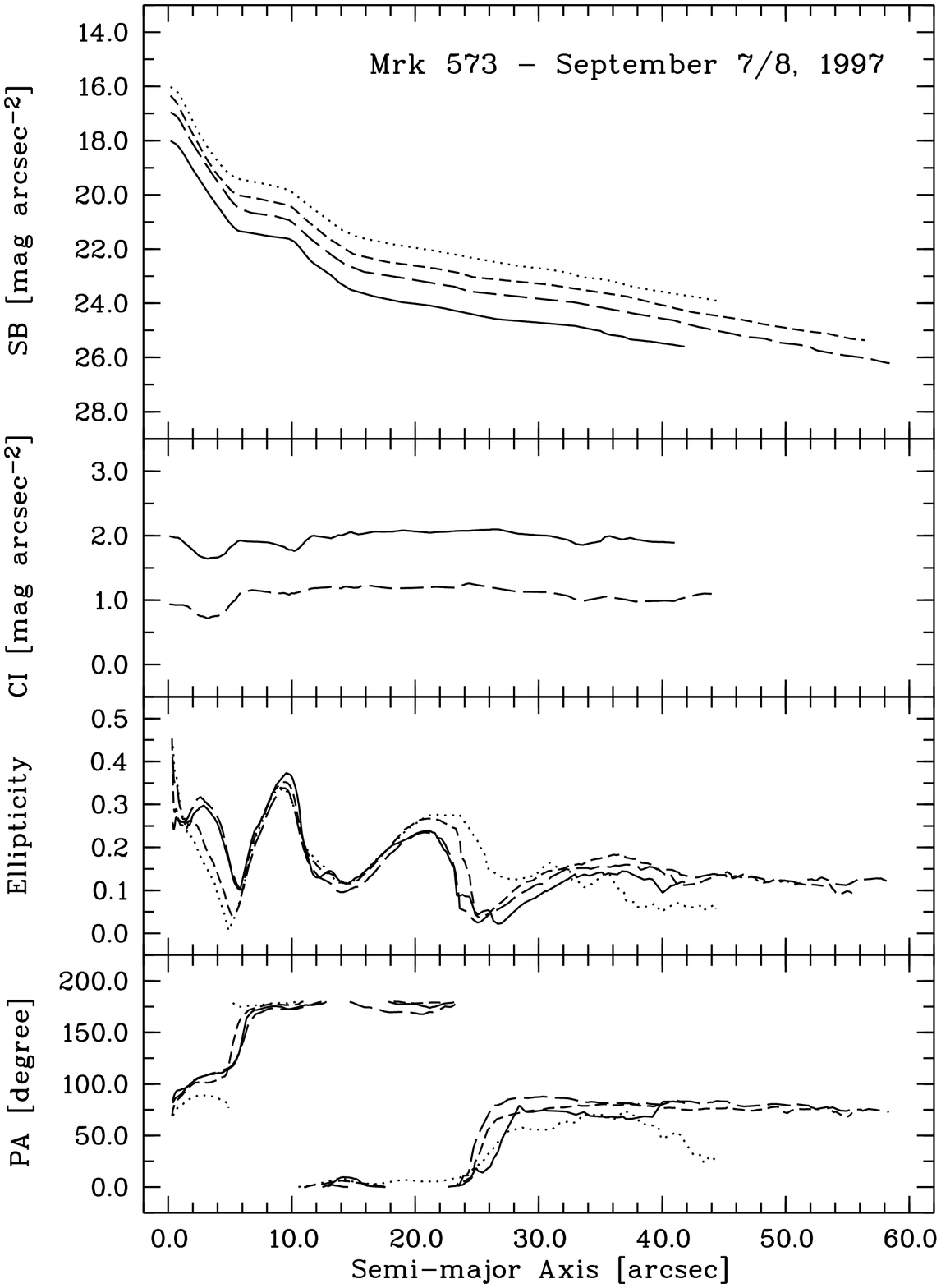}
      \caption{Continued.}
   \end{figure*}
\setcounter{figure}{0}

\begin{figure*}[htbp]
\vspace{0.1cm}
   \centering
\includegraphics[width=5.6cm]{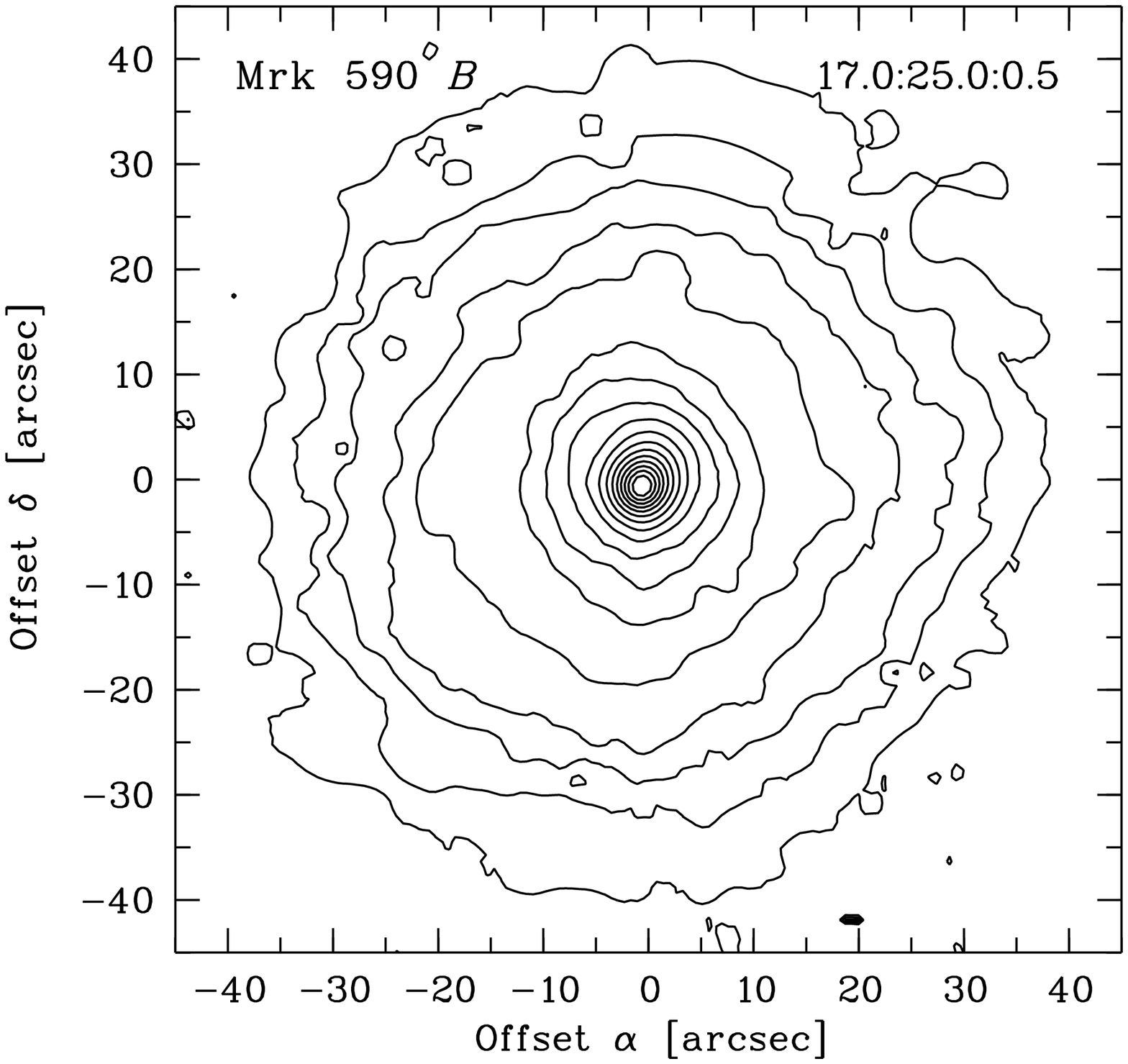}
\hspace{0.5cm}
\includegraphics[width=5.6cm]{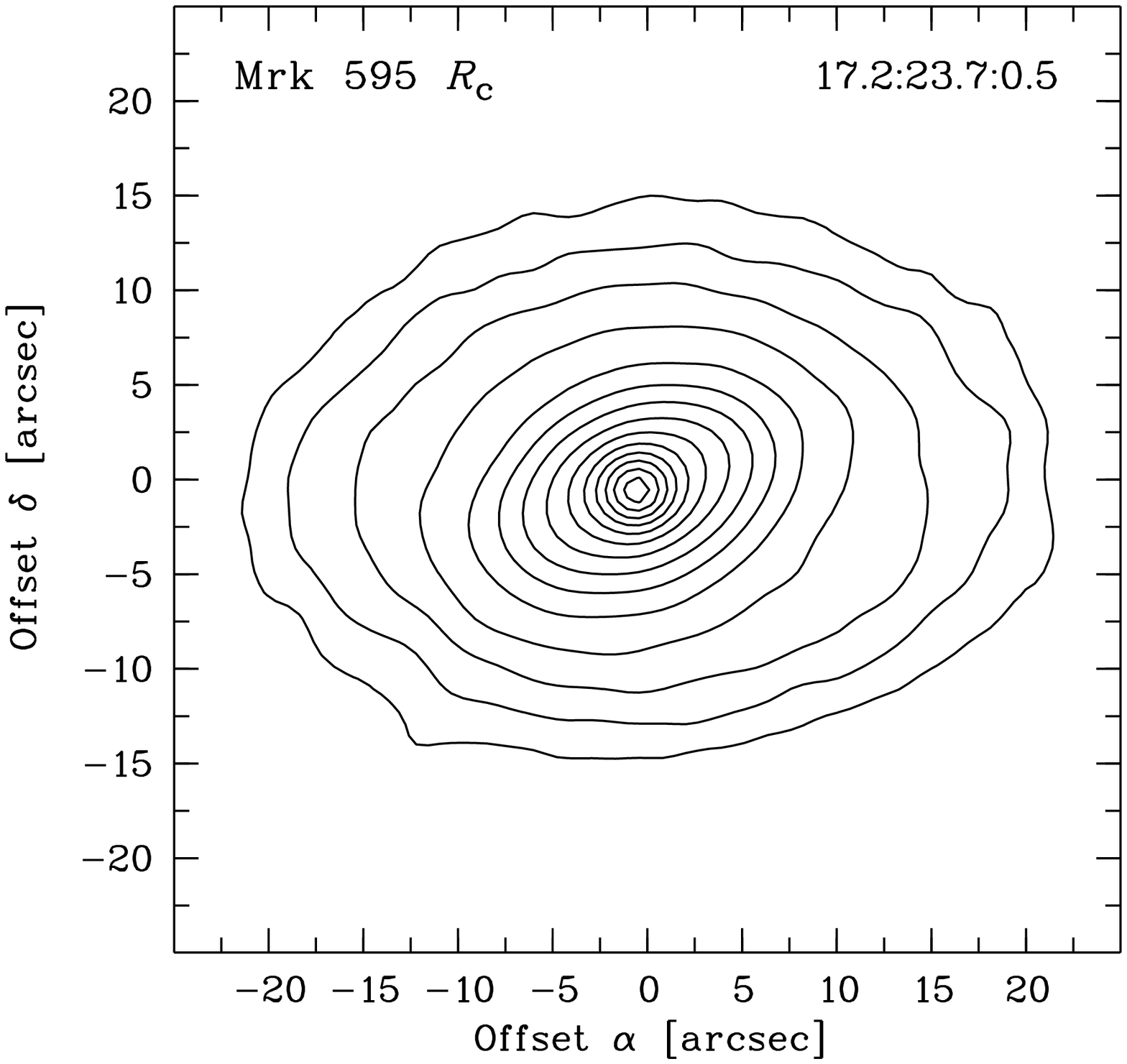}
 \hspace{0.5cm}
\includegraphics[width=5.6cm]{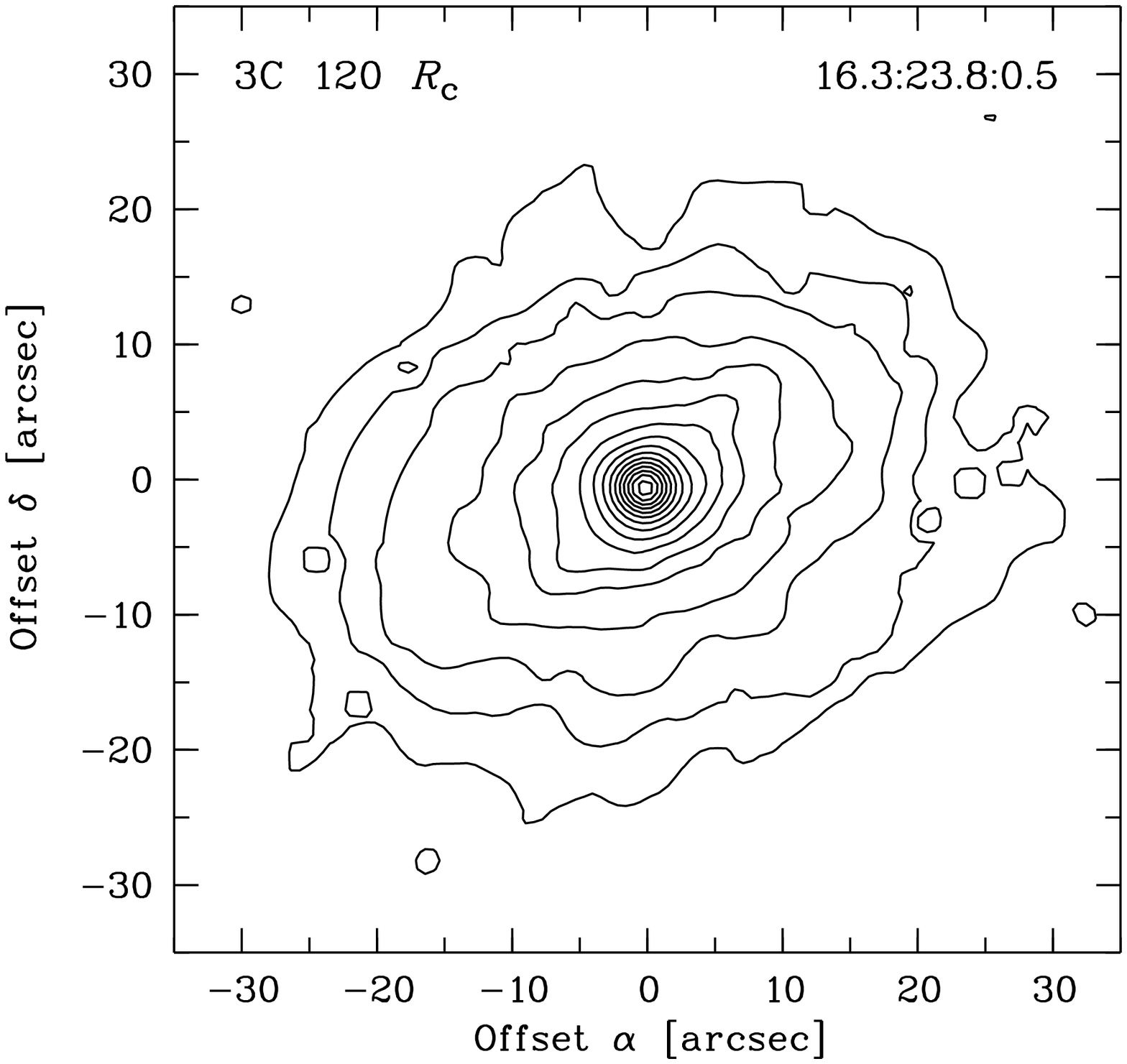}

\vspace{0.3cm}

\includegraphics[width=5.6cm]{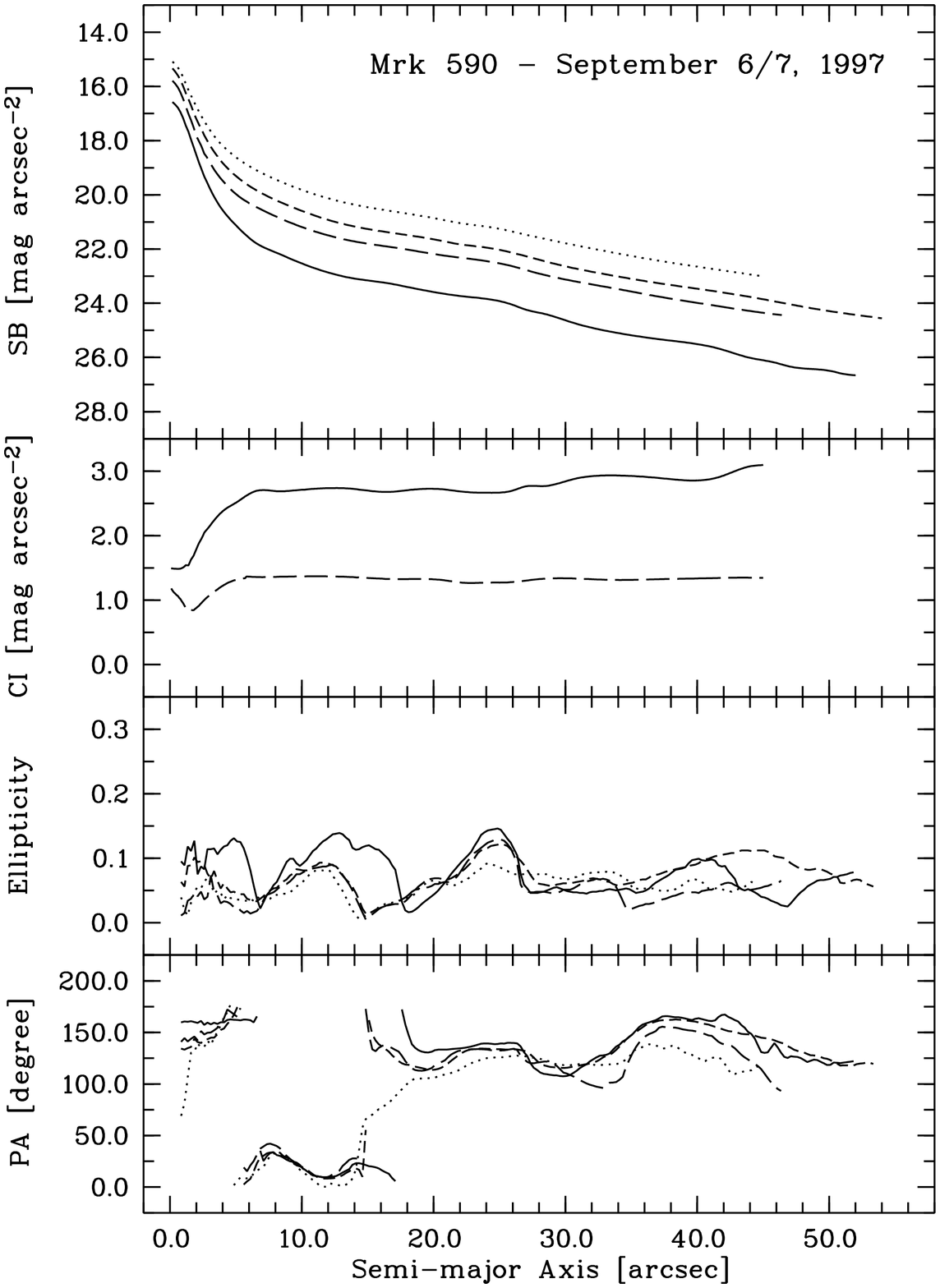}
\hspace{0.5cm}
\includegraphics[width=5.6cm]{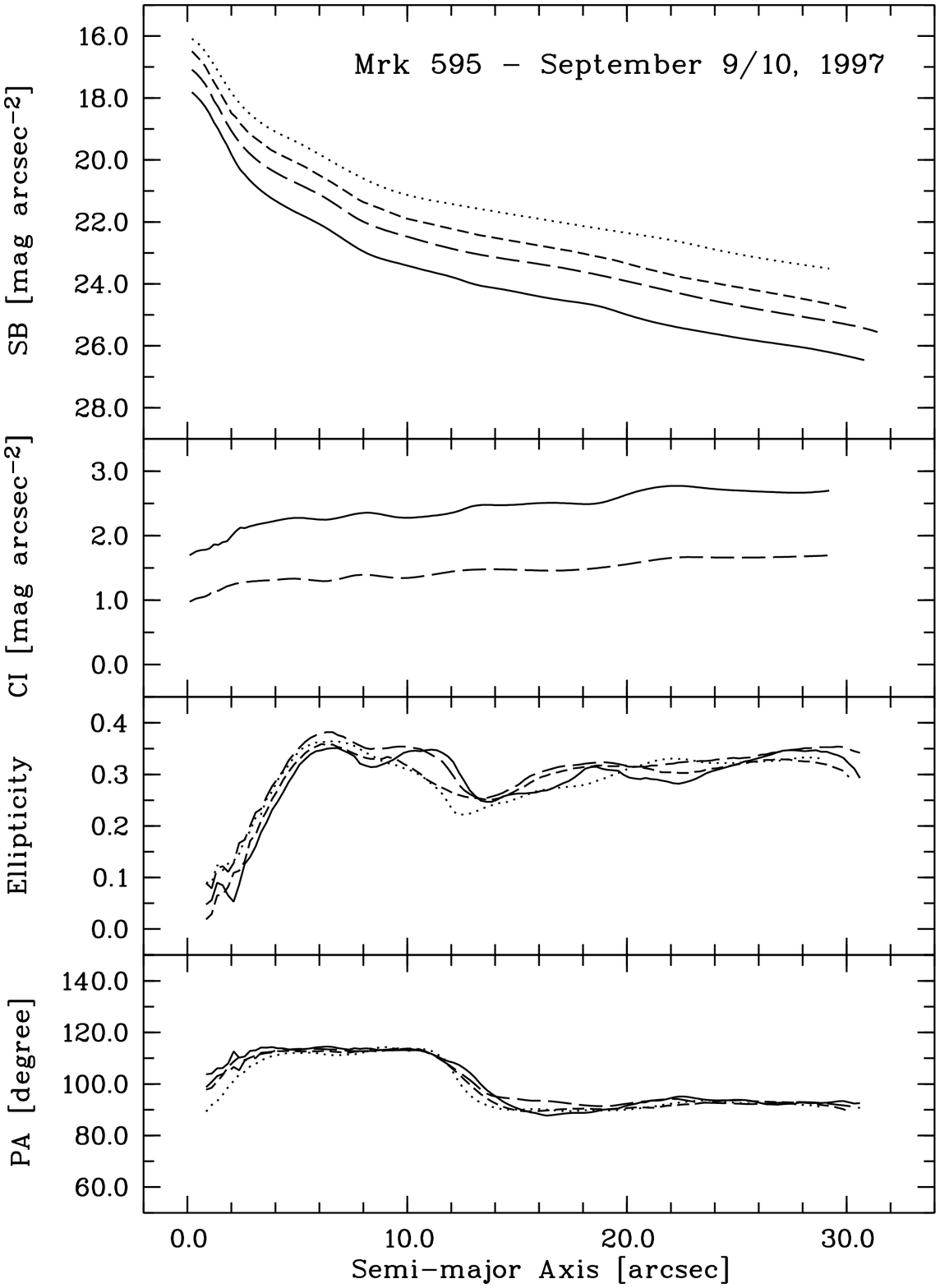}
 \hspace{0.5cm}
 \includegraphics[width=5.6cm]{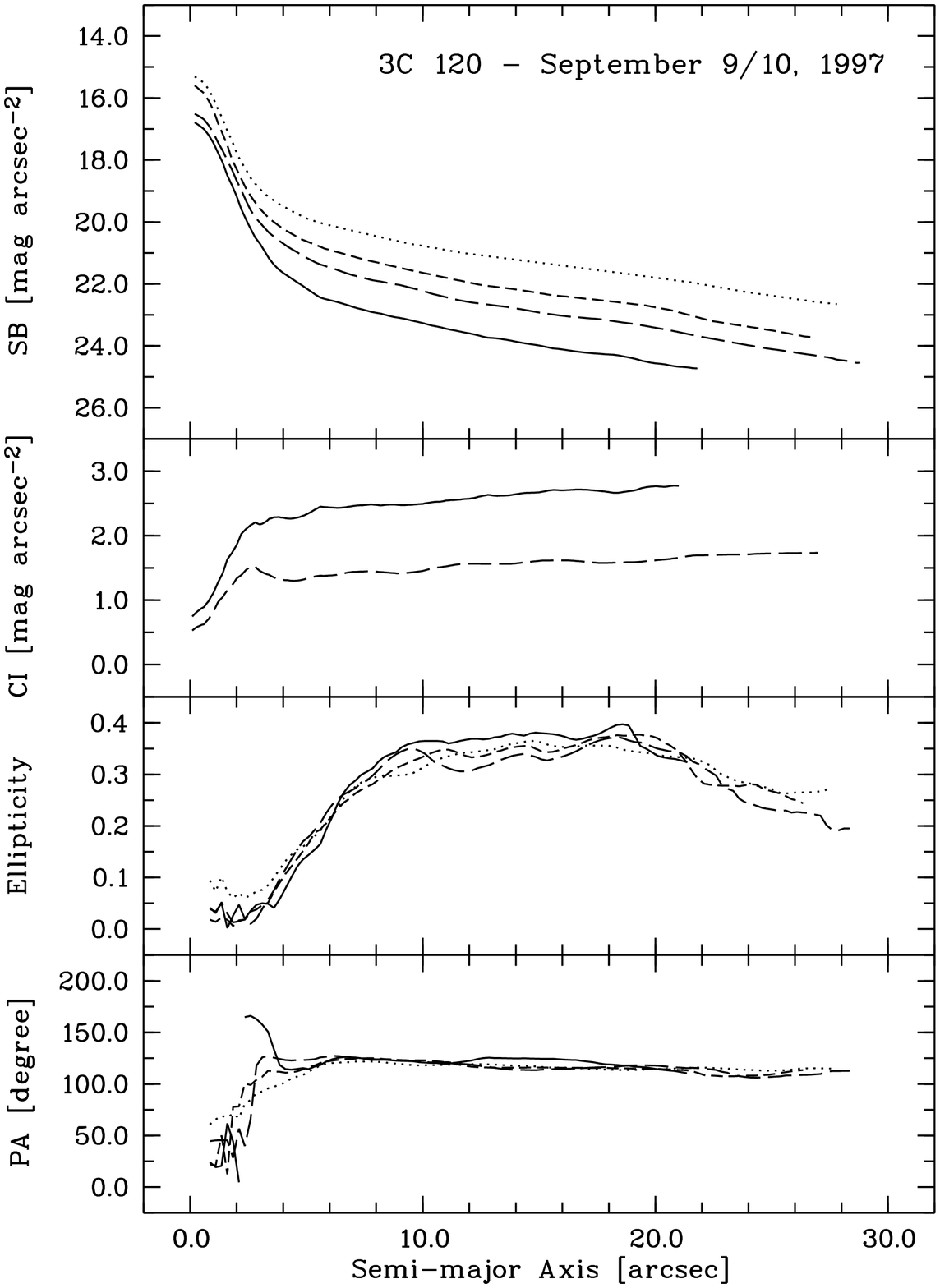}
    \caption{Continued.}
   \end{figure*}
\setcounter{figure}{0}
 
\begin{figure*}[htbp]
\vspace{0.1cm}
   \centering
\includegraphics[width=5.6cm]{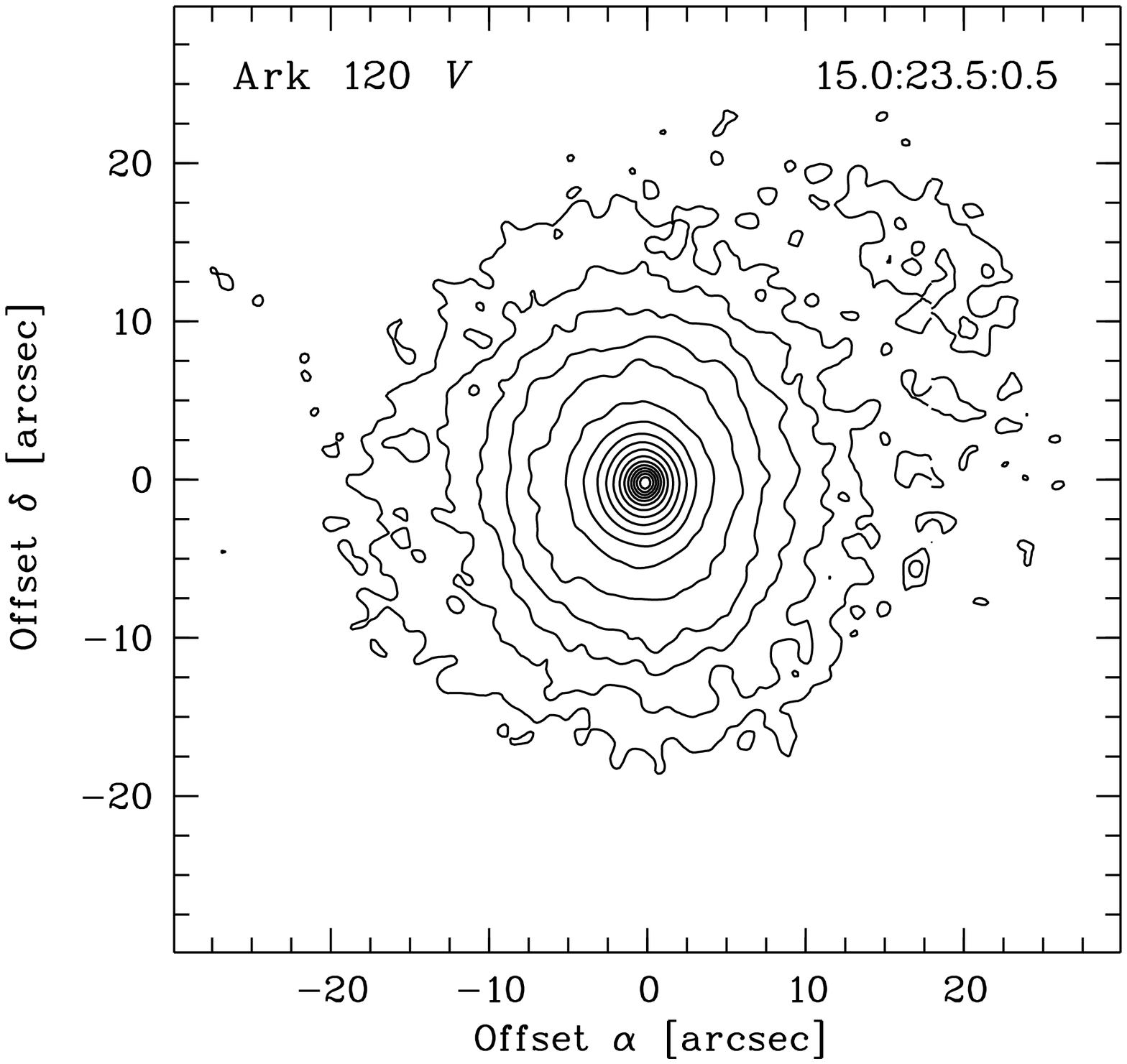}
\hspace{0.5cm}
\includegraphics[width=5.6cm]{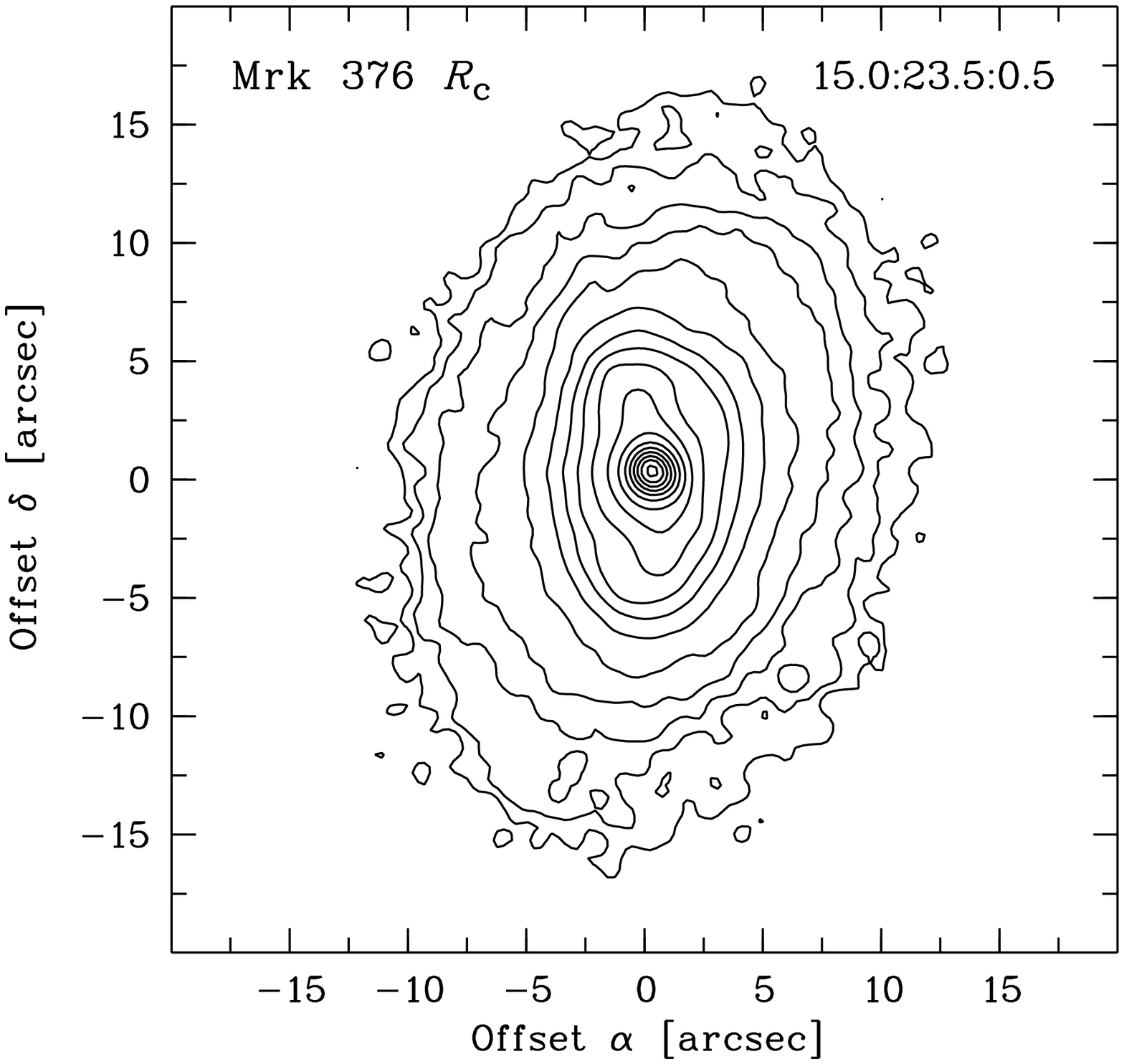}
\hspace{0.5cm}
\includegraphics[width=5.6cm]{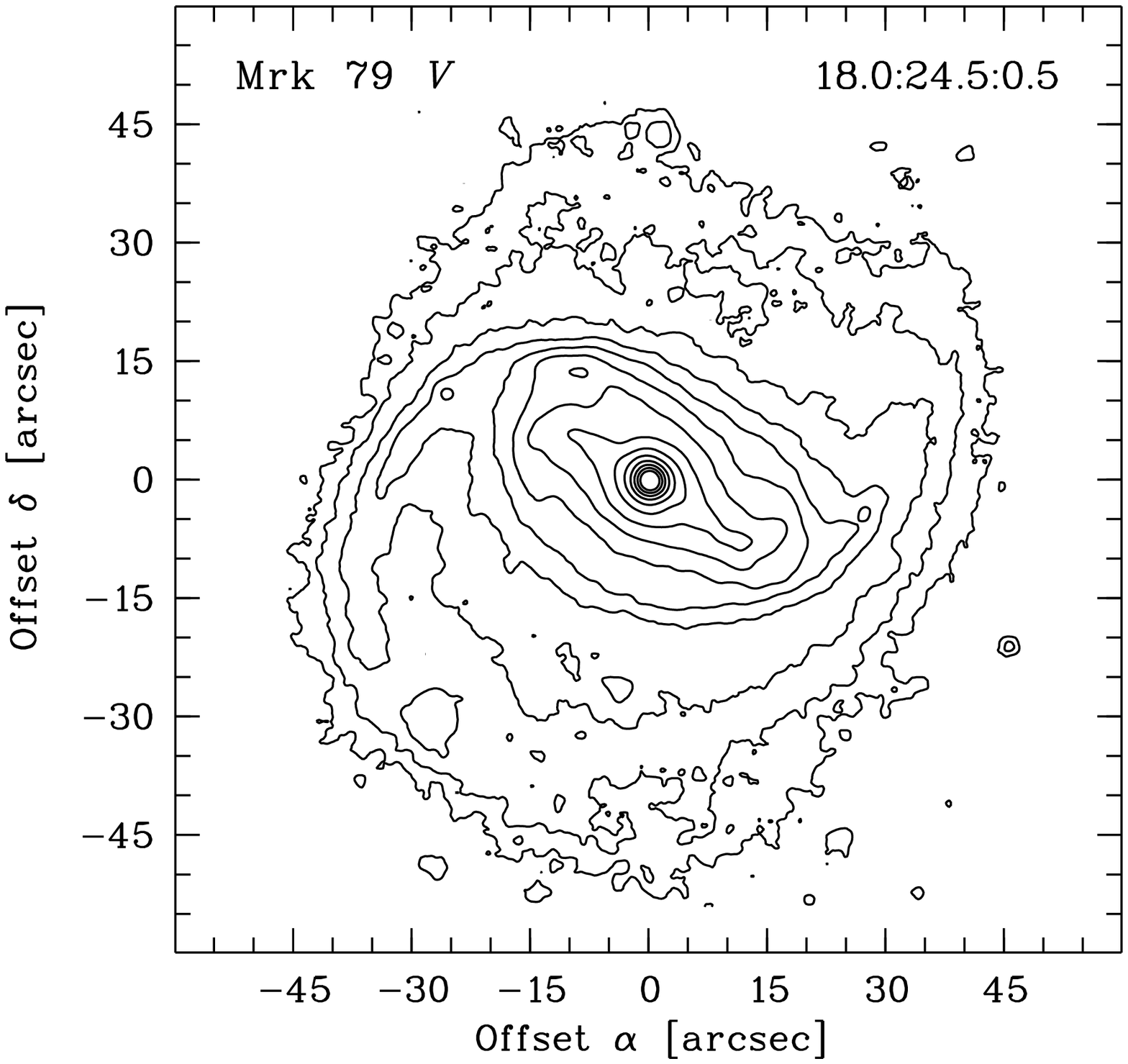}

\vspace{0.3cm}

\includegraphics[width=5.6cm]{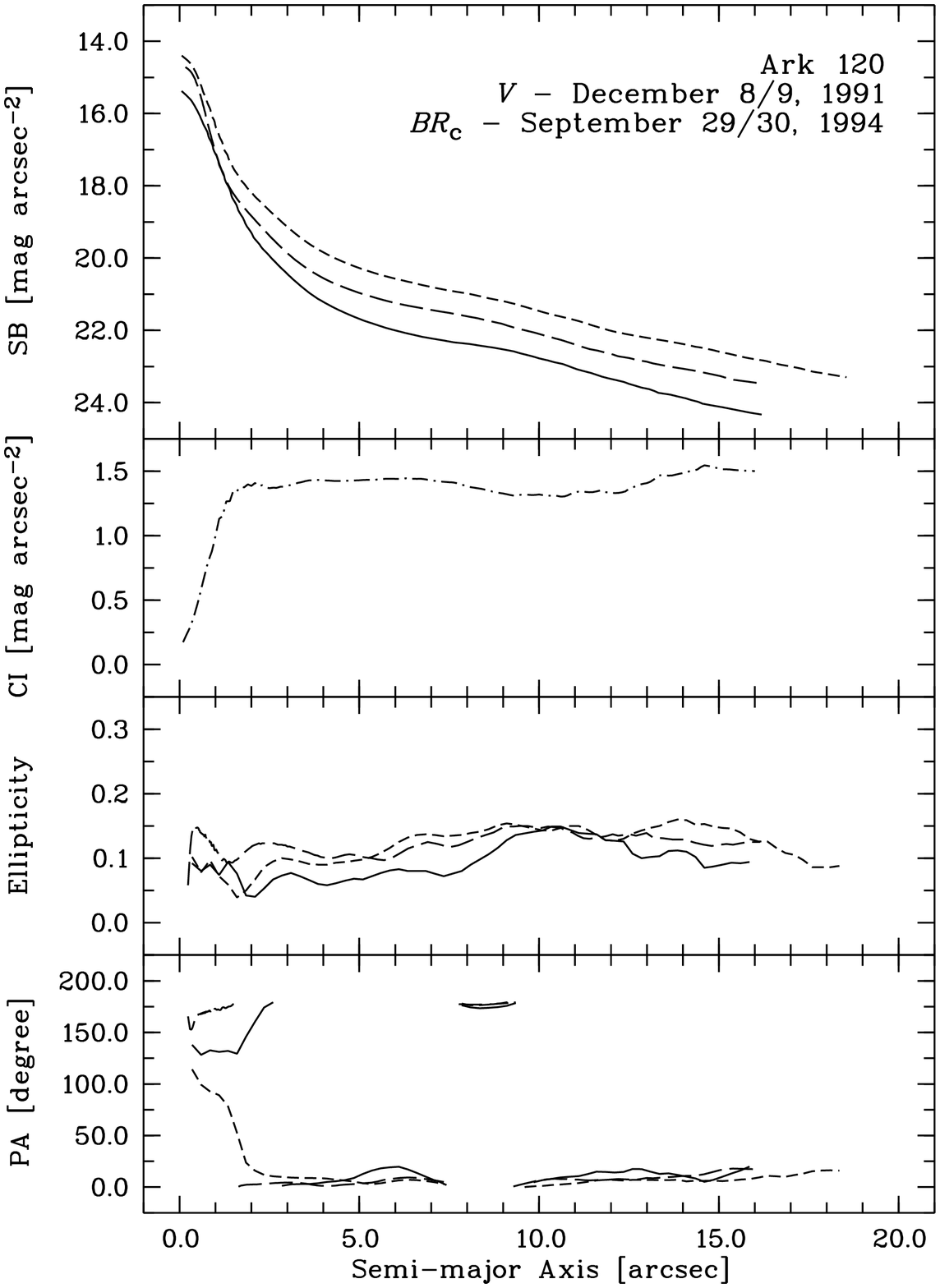}
\hspace{0.5cm}
\includegraphics[width=5.6cm]{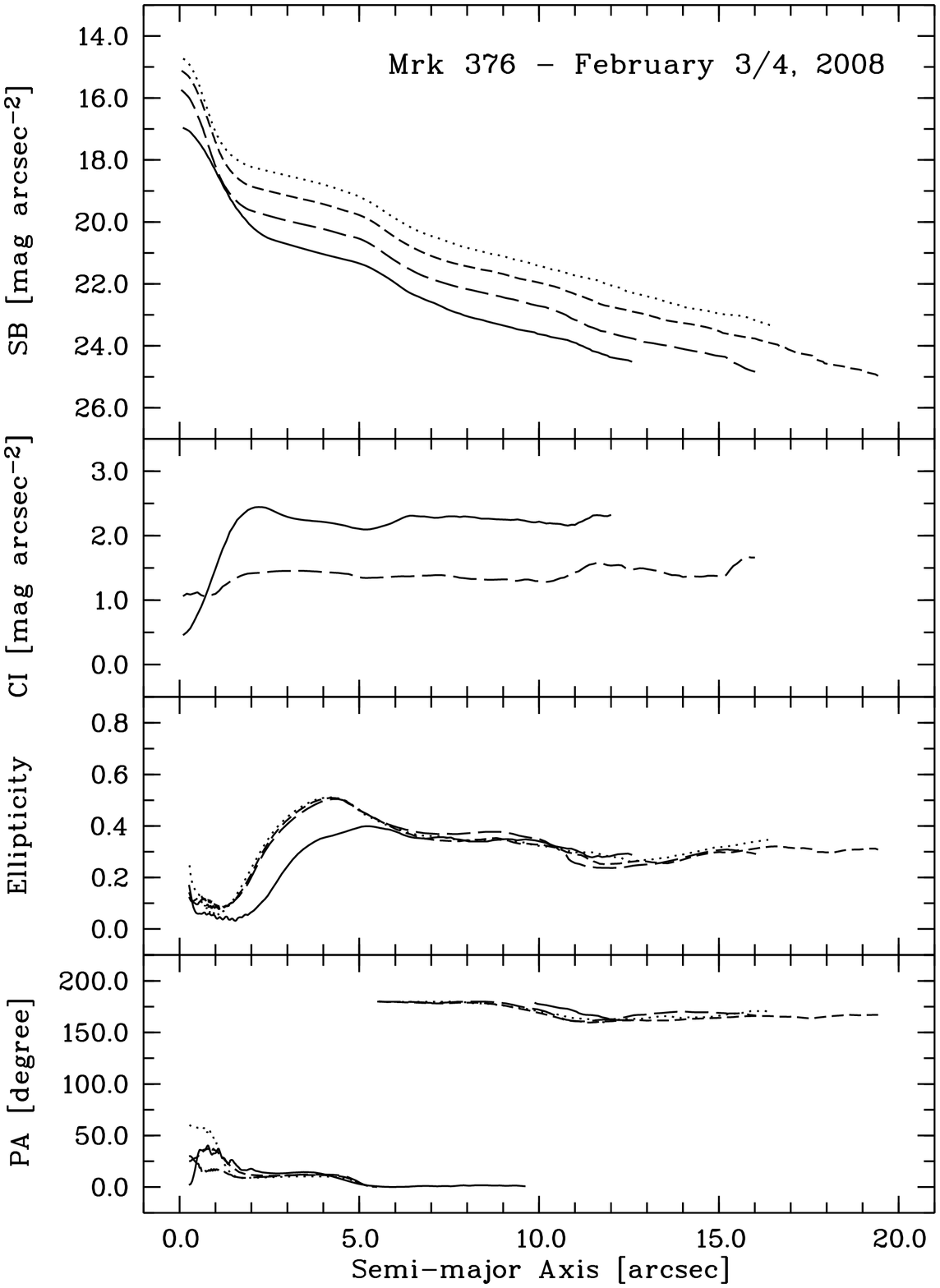}
\hspace{0.5cm}
\includegraphics[width=5.6cm]{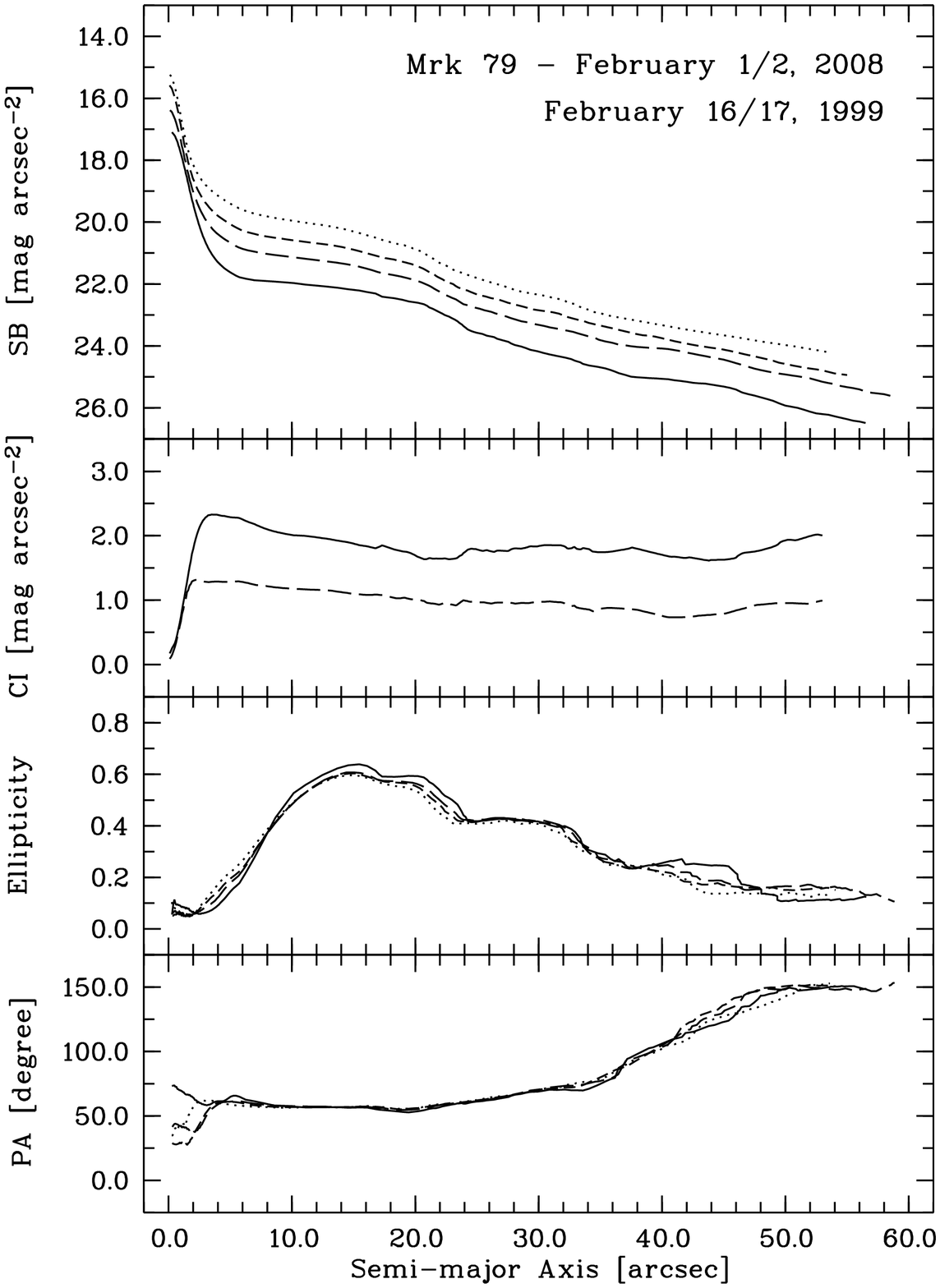}
      \caption{Continued.}
   \end{figure*}
\setcounter{figure}{0}
 
\begin{figure*}[htbp]
\vspace{0.1cm}
   \centering
\includegraphics[width=5.6cm]{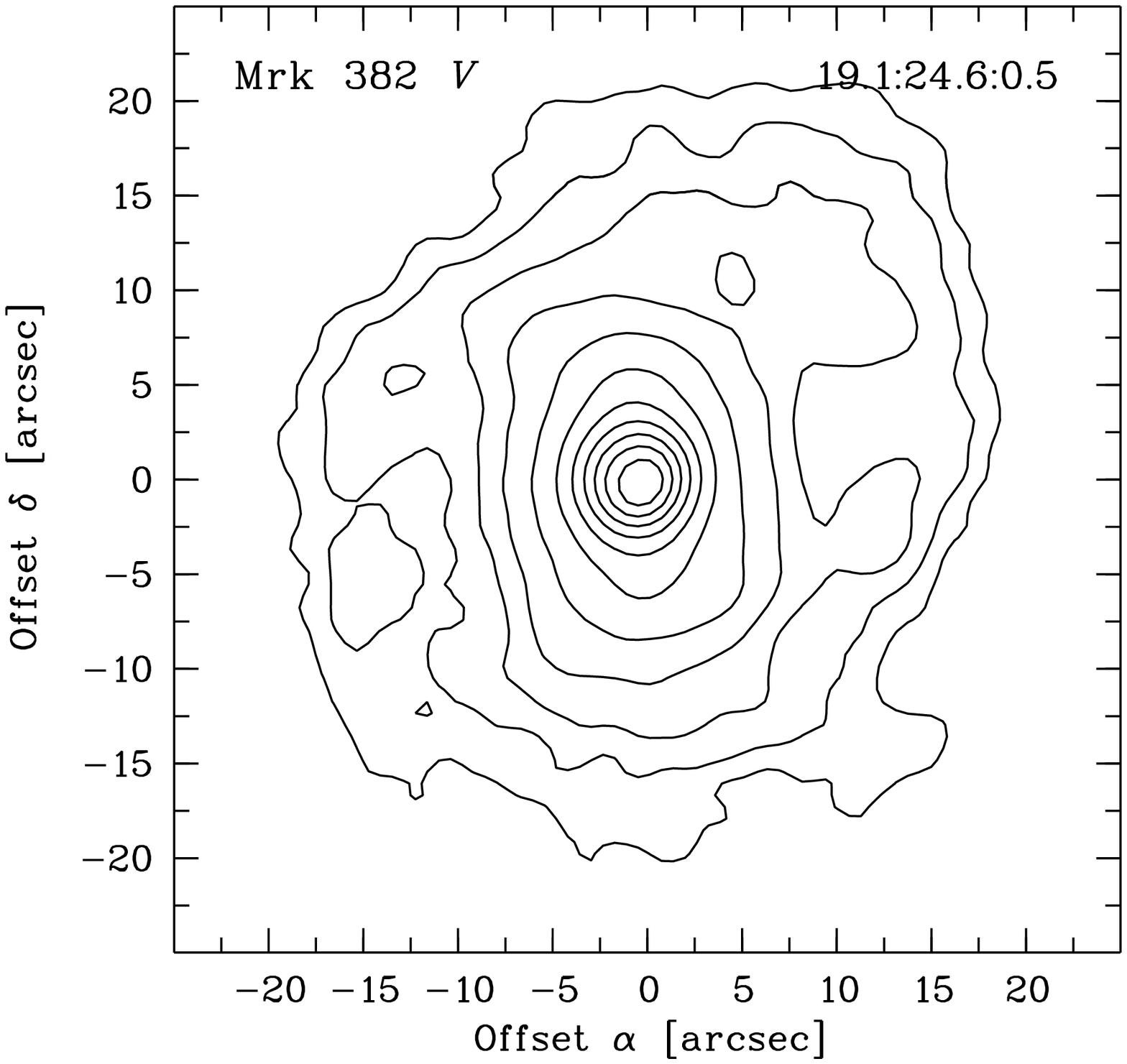}
\hspace{0.5cm}
\includegraphics[width=5.6cm]{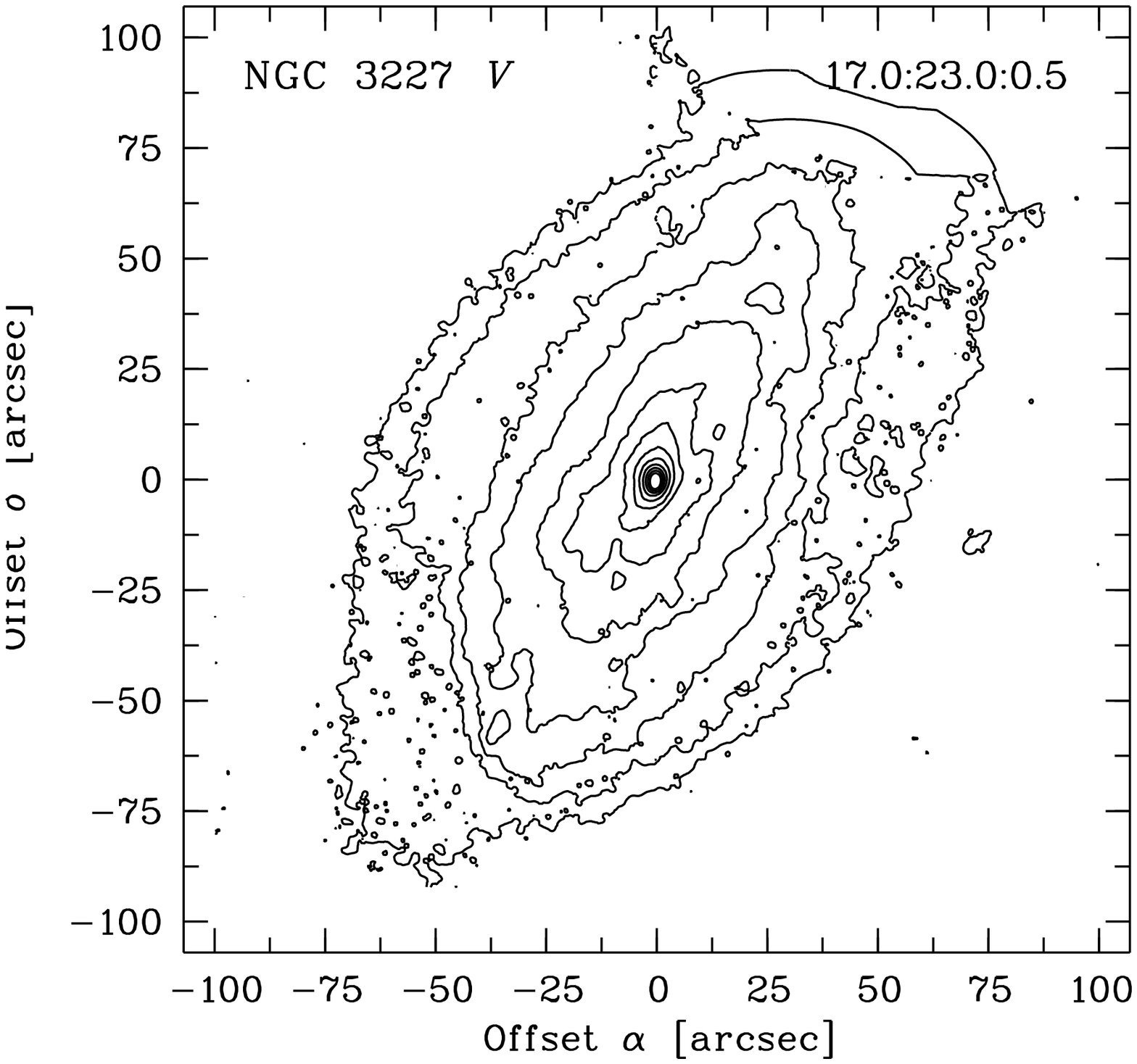}
\hspace{0.5cm}
\includegraphics[width=5.6cm]{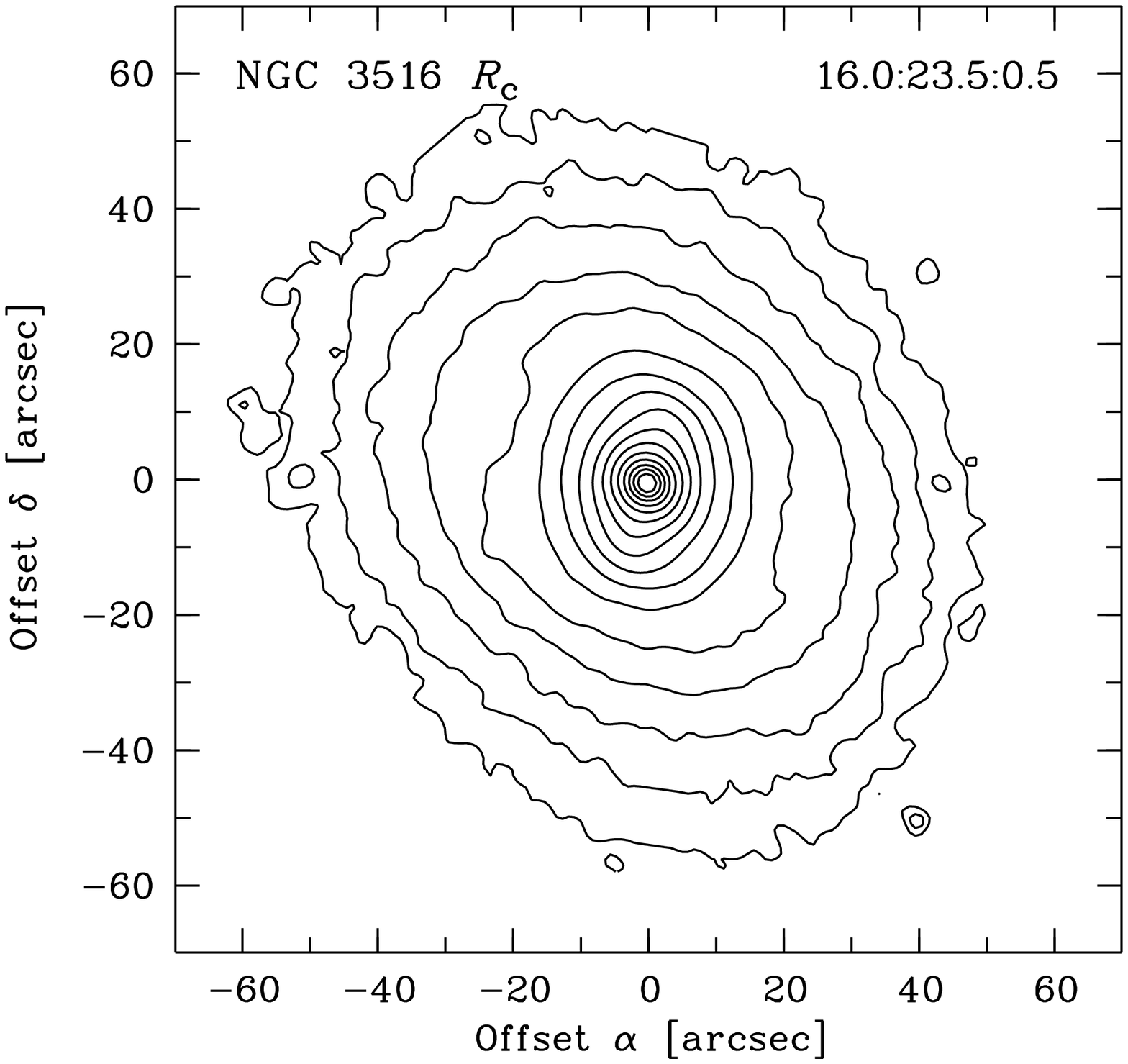}
\vspace{0.3cm}

\includegraphics[width=5.6cm]{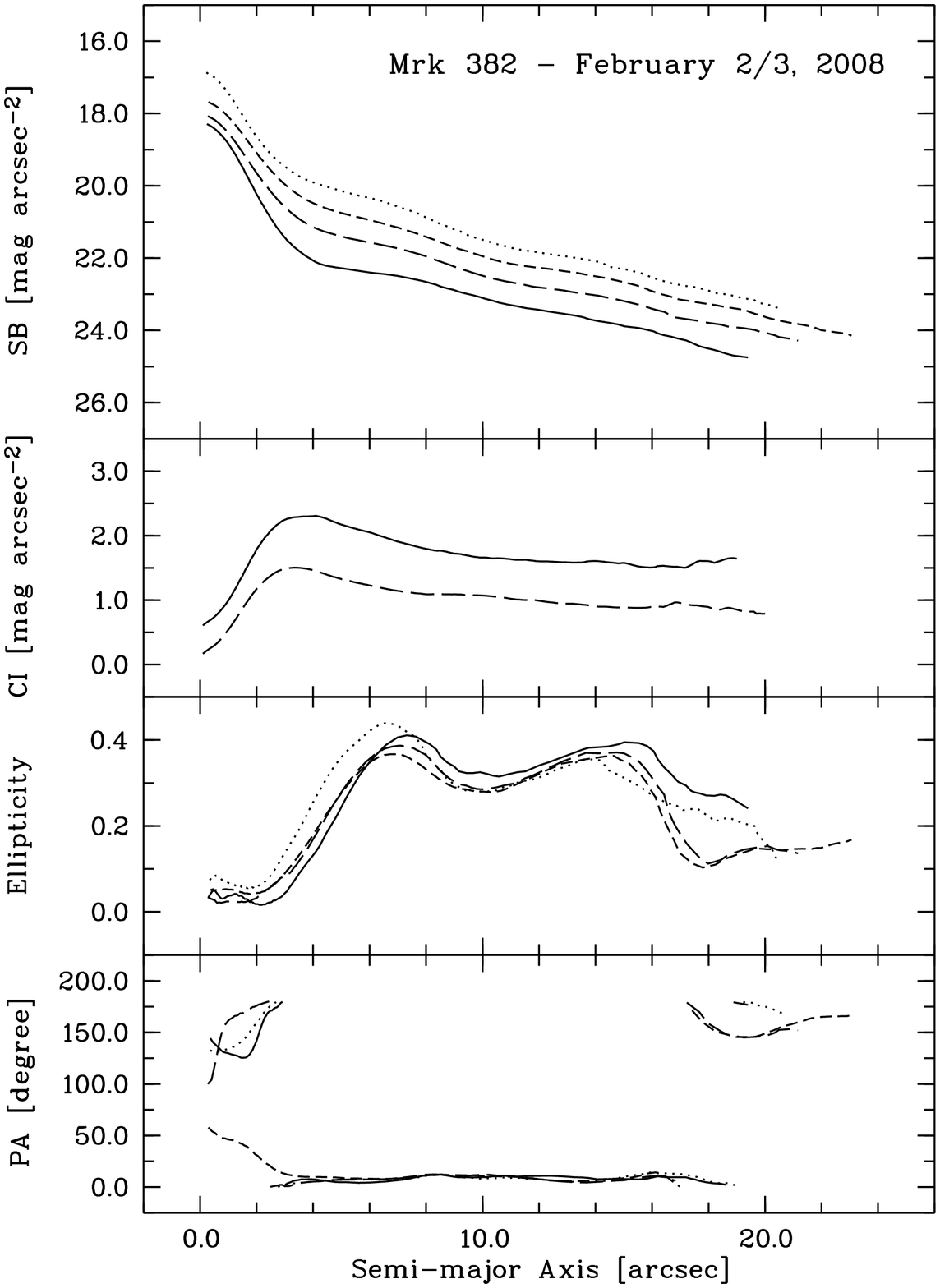}
\hspace{0.5cm}
\includegraphics[width=5.6cm]{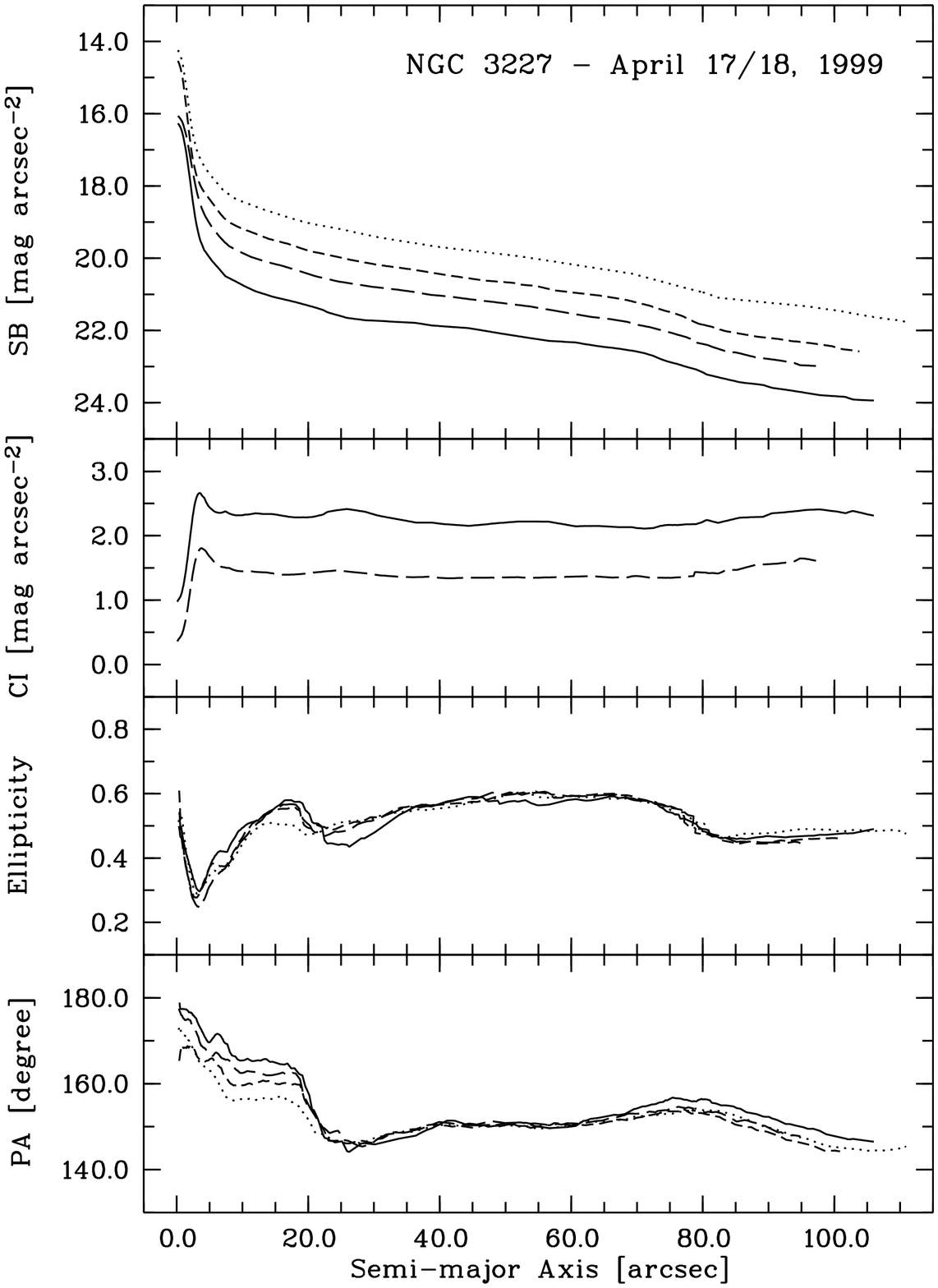}
\hspace{0.5cm}
\includegraphics[width=5.6cm]{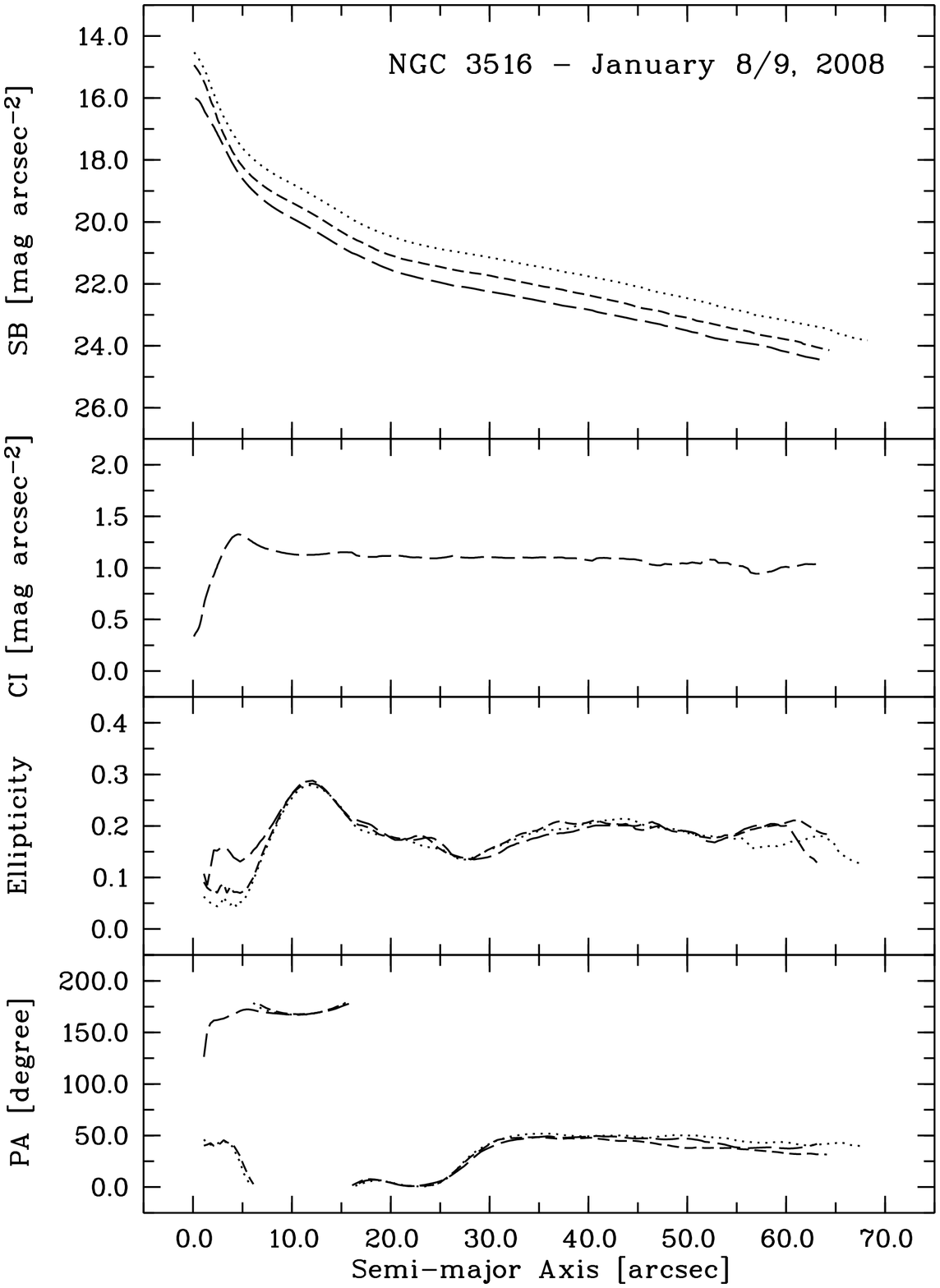}
      \caption{Continued.}
    \end{figure*}
\setcounter{figure}{0}
 
\begin{figure*}[htbp]
\vspace{0.1cm}
   \centering
\includegraphics[width=5.6cm]{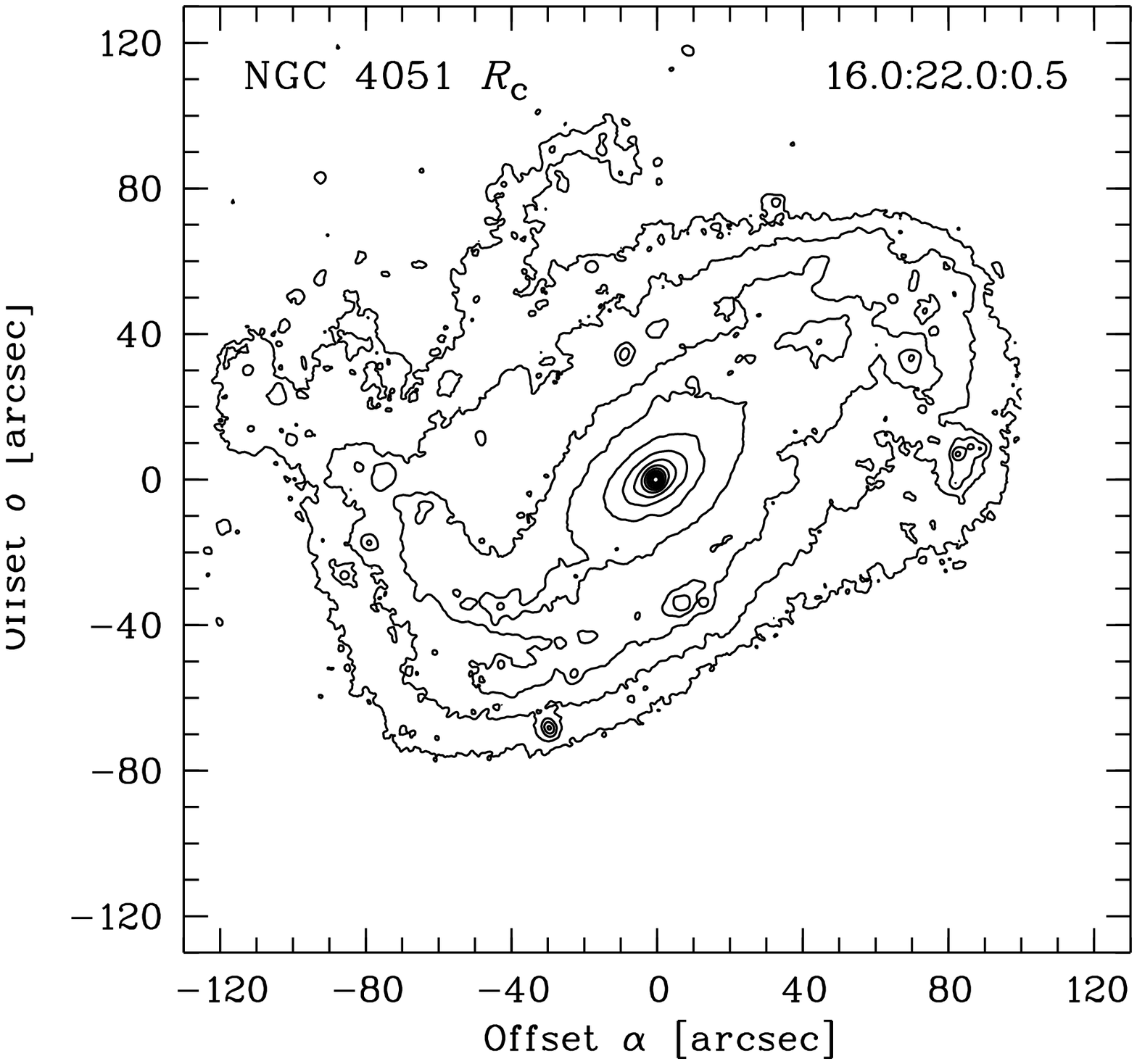}
\hspace{0.5cm}
\includegraphics[width=5.6cm]{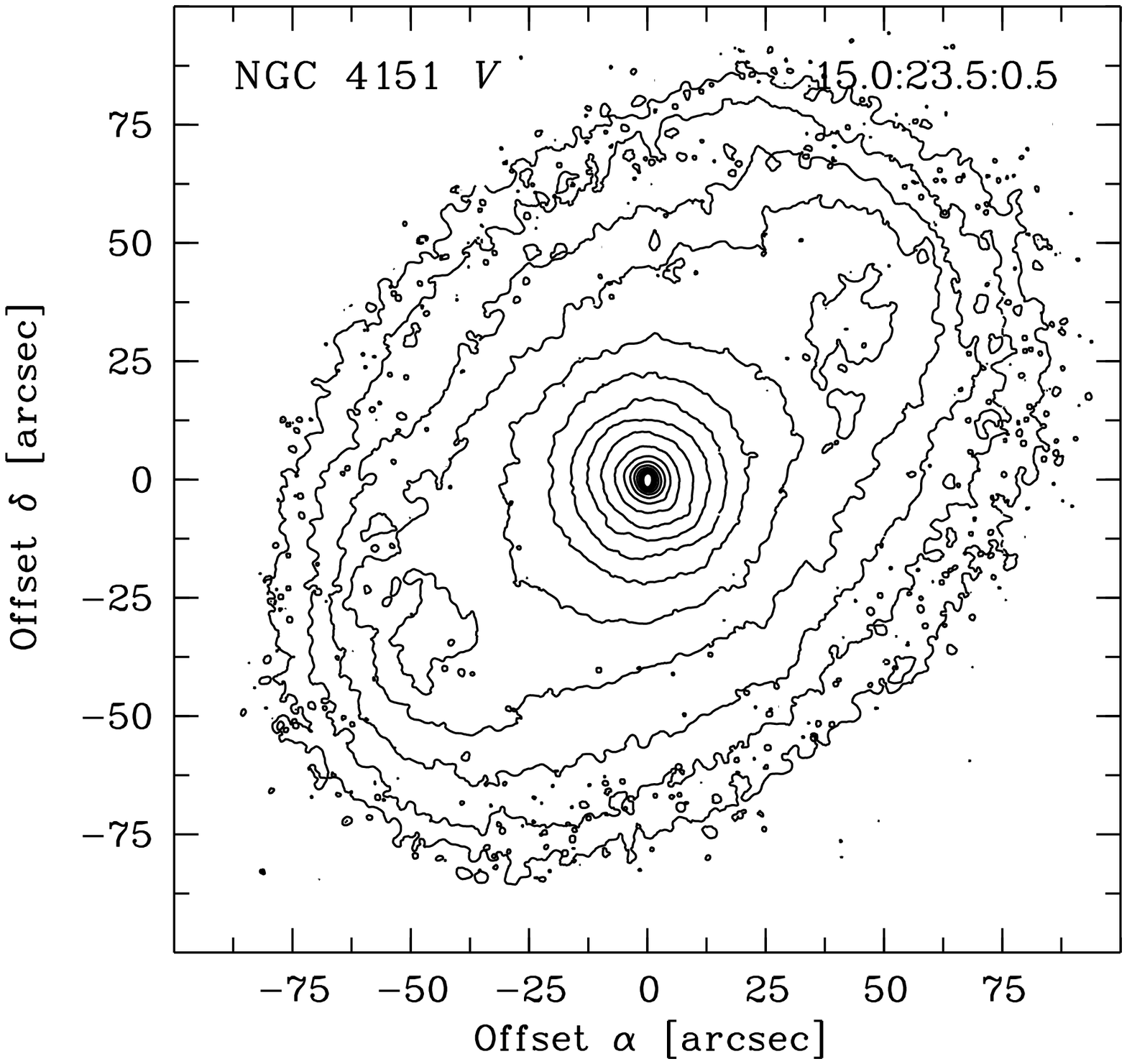}
\hspace{0.5cm}
\includegraphics[width=5.6cm]{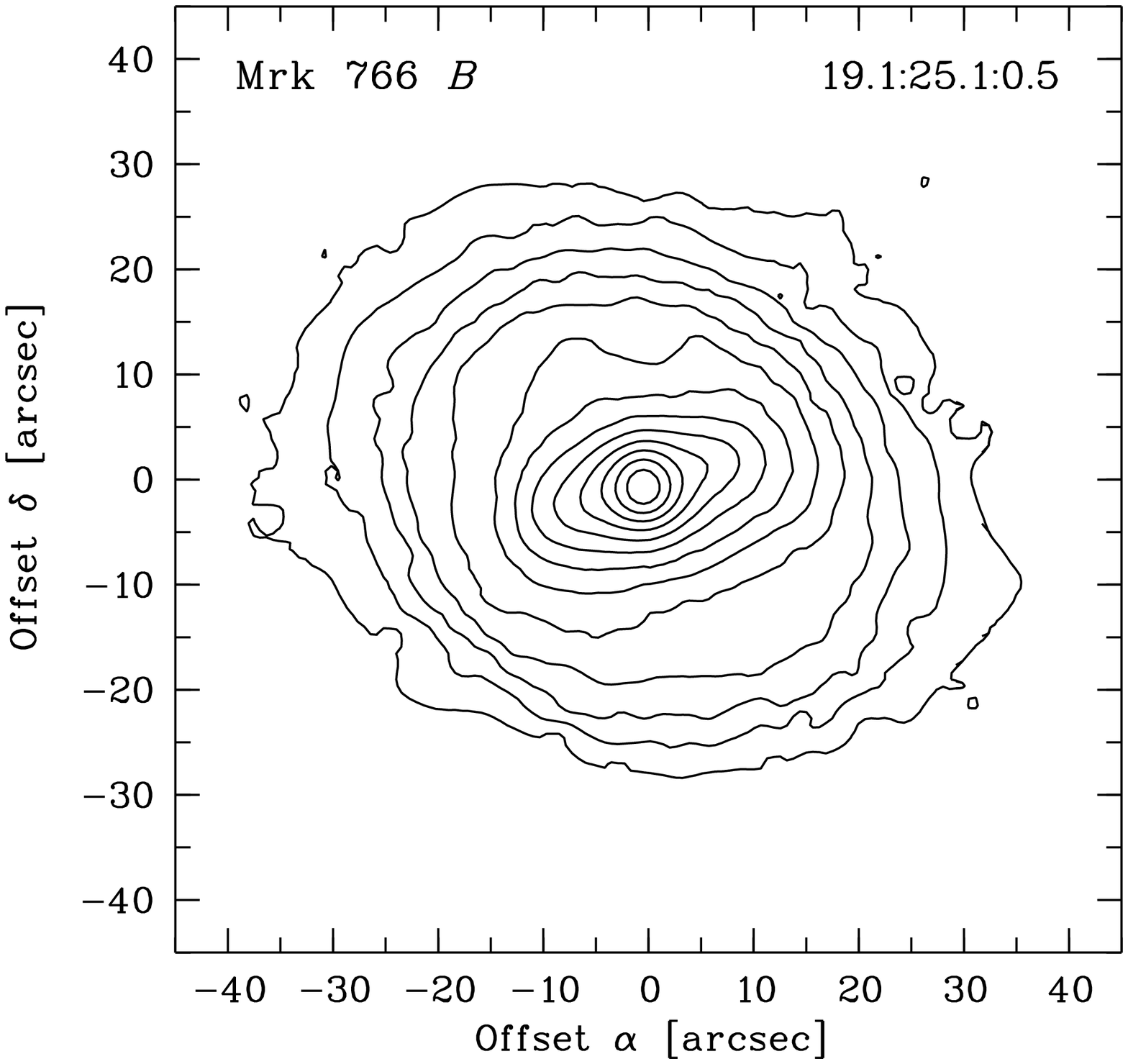}

\vspace{0.3cm}

\includegraphics[width=5.6cm]{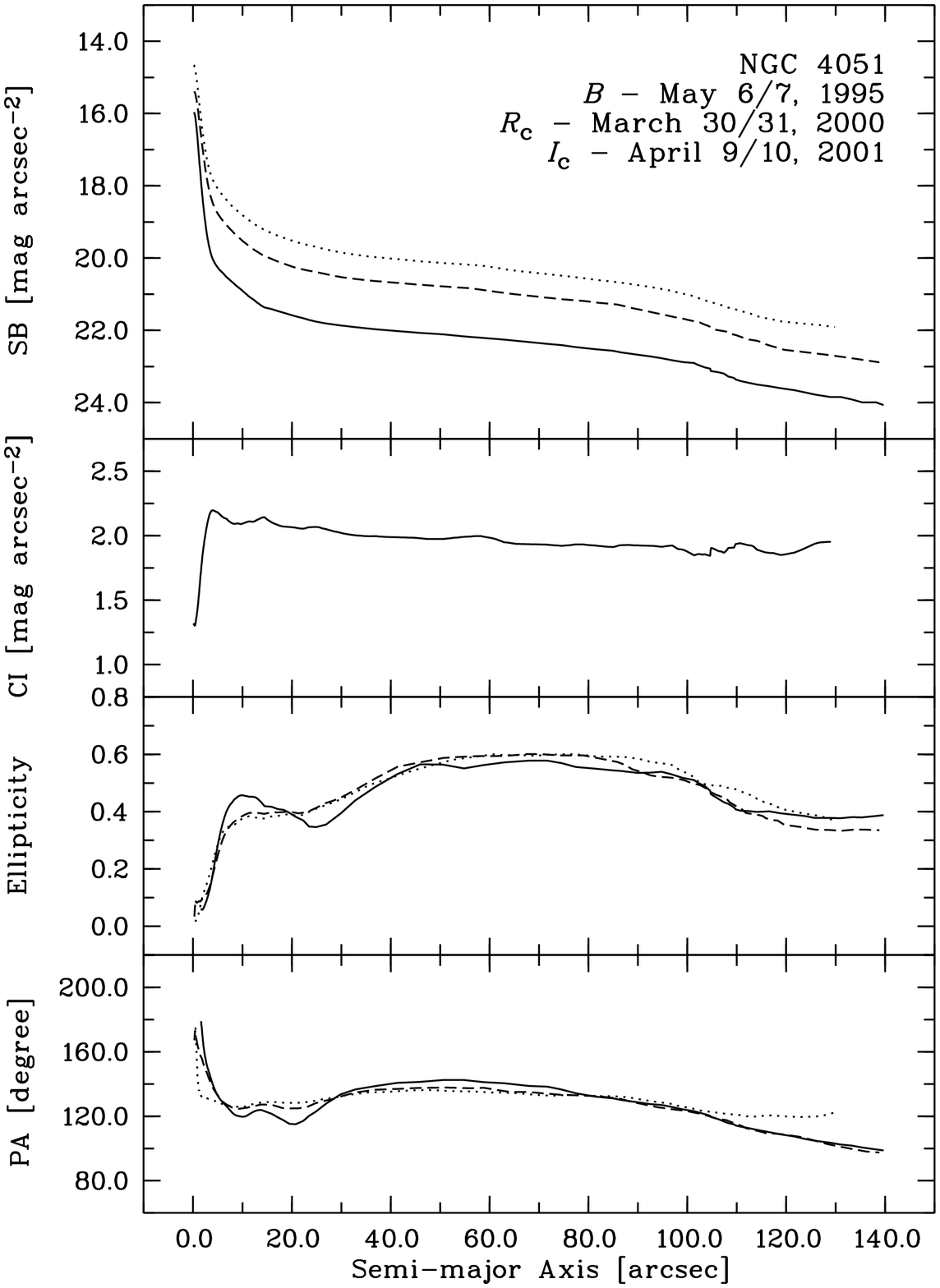}
\hspace{0.5cm}
\includegraphics[width=5.6cm]{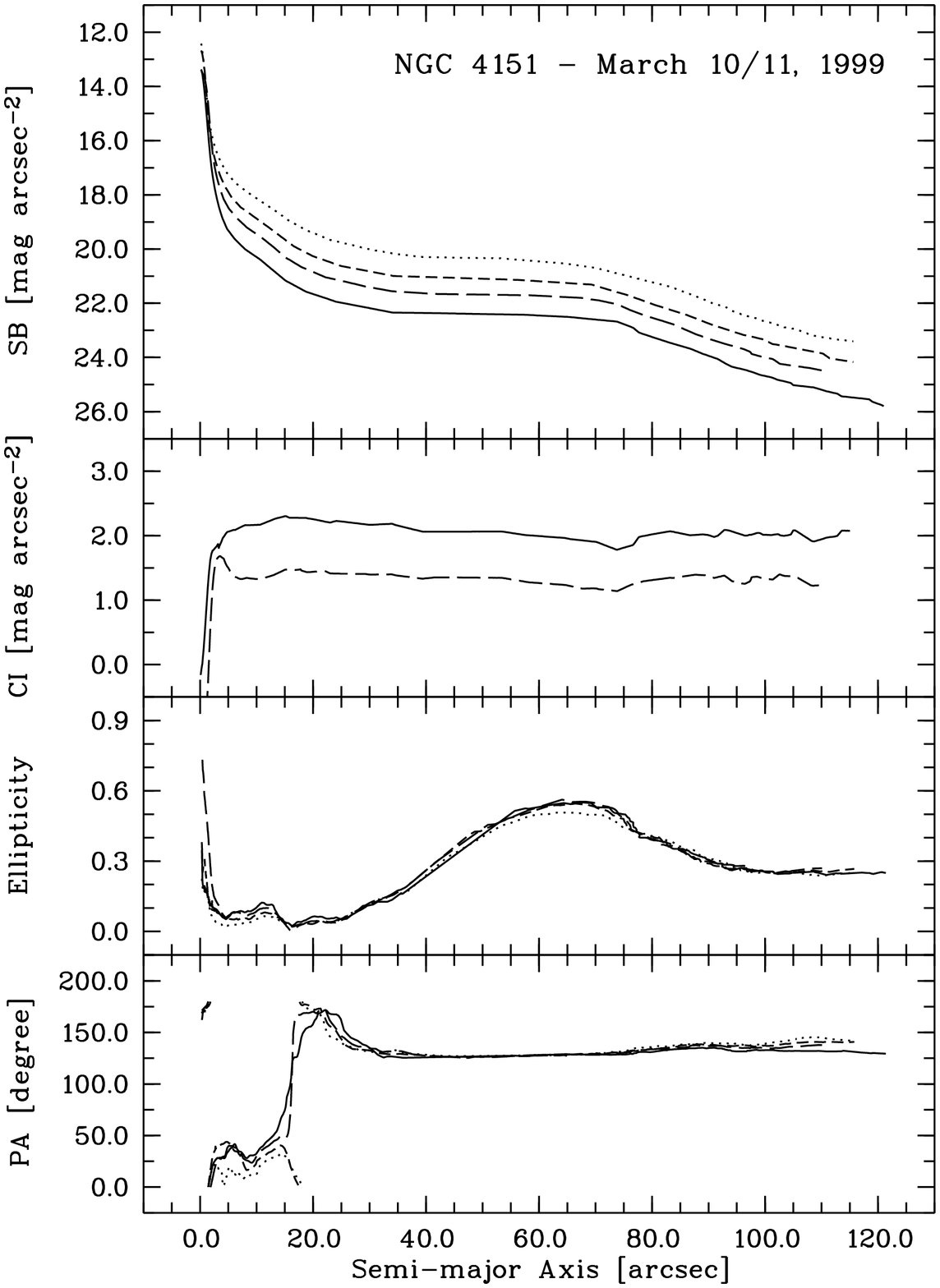}
\hspace{0.5cm}
\includegraphics[width=5.6cm]{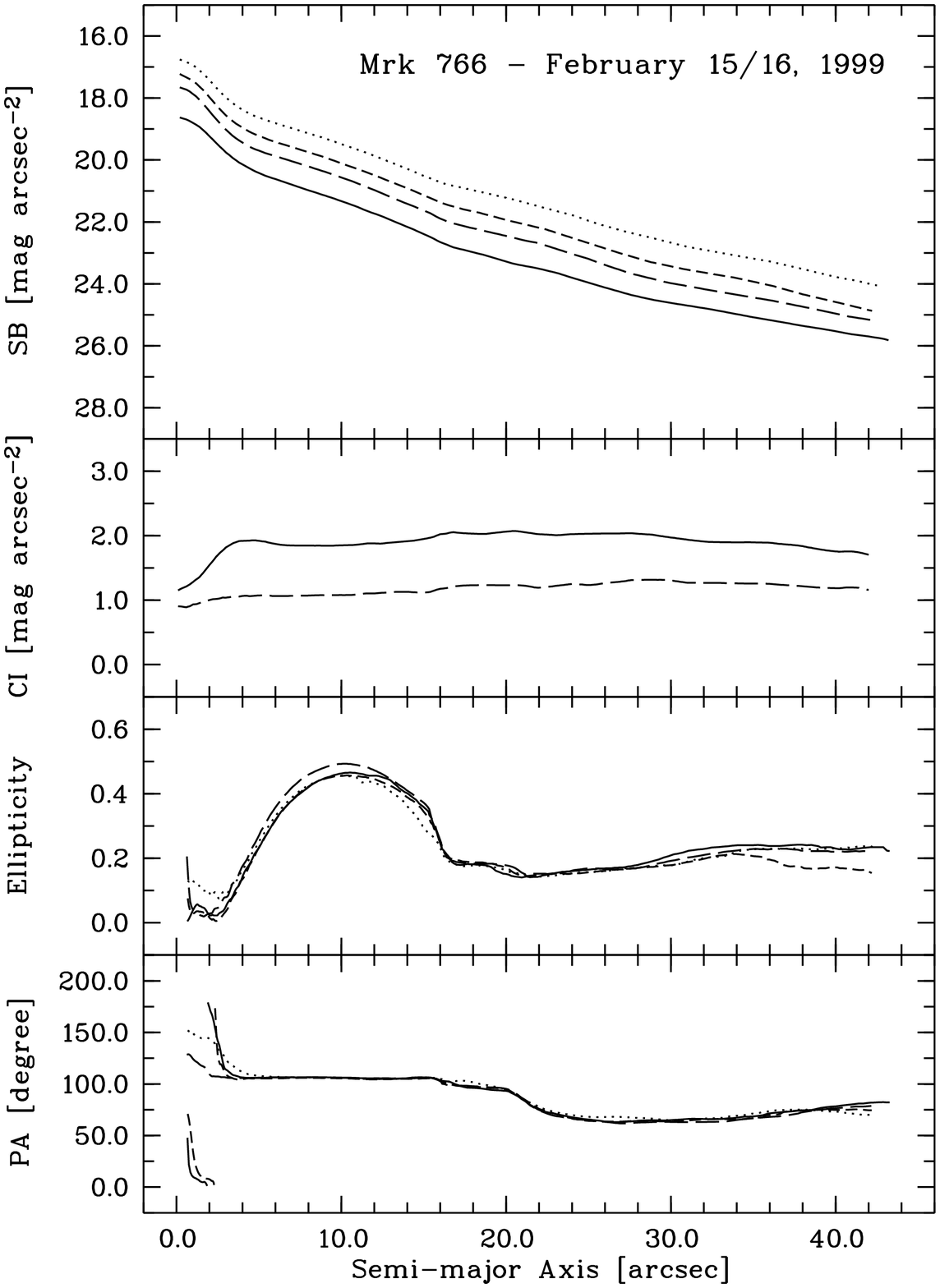}

      \caption{Continued.}
    \end{figure*}
\setcounter{figure}{0}
 
\begin{figure*}[htbp]
\vspace{0.1cm}
   \centering
\includegraphics[width=5.6cm]{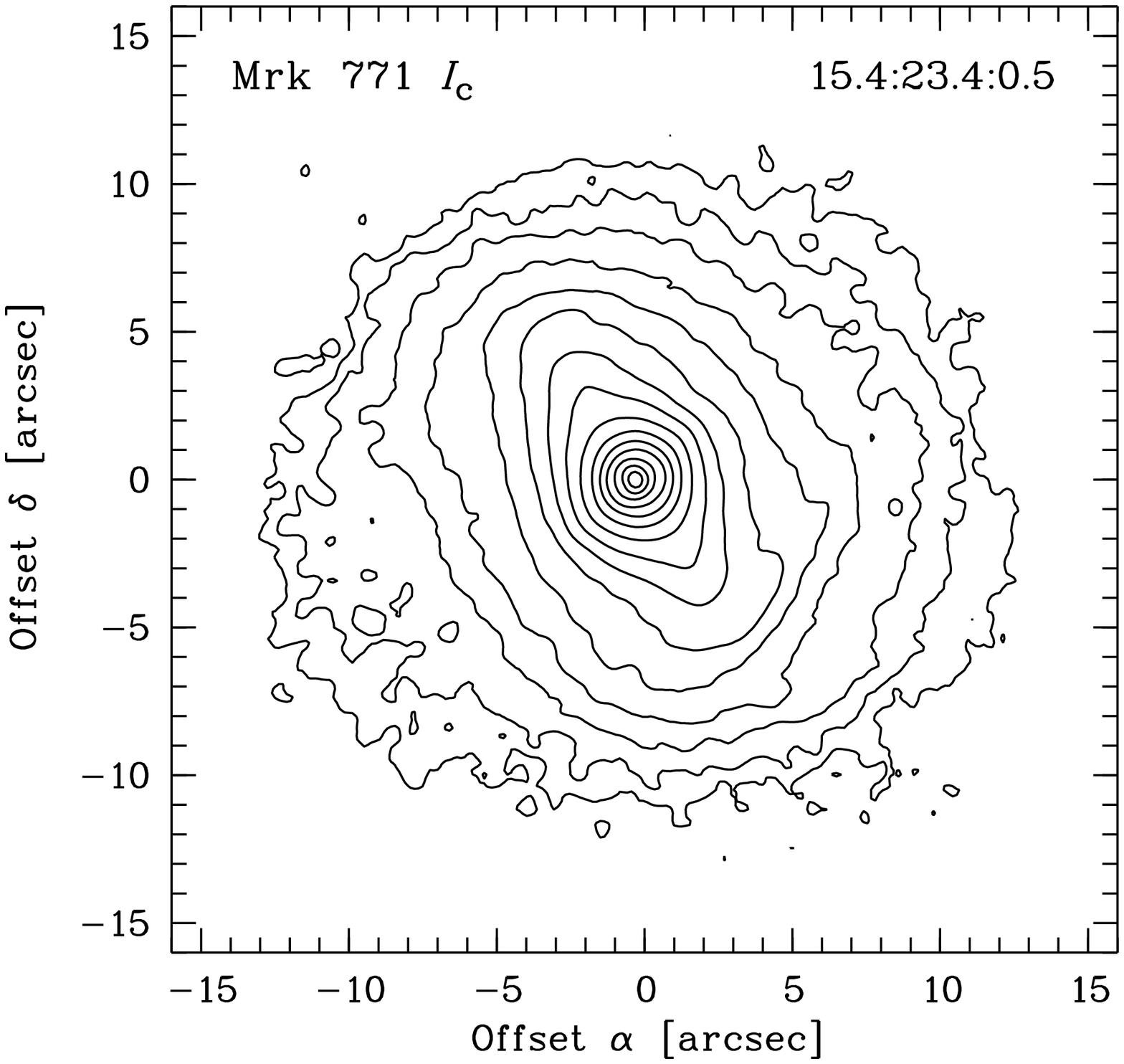}
\hspace{0.5cm}
\includegraphics[width=5.6cm]{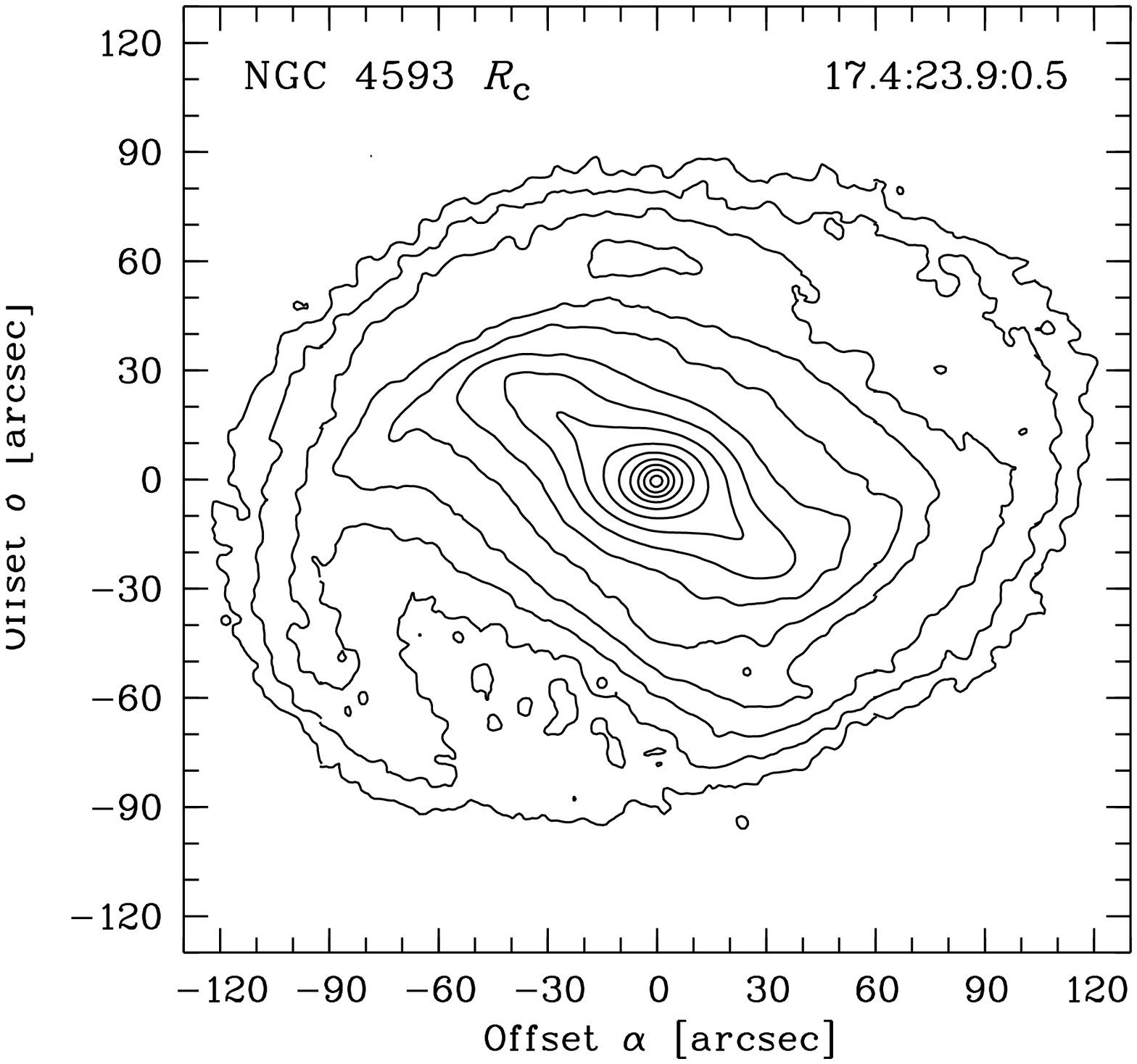}
\hspace{0.5cm}
\includegraphics[width=5.6cm]{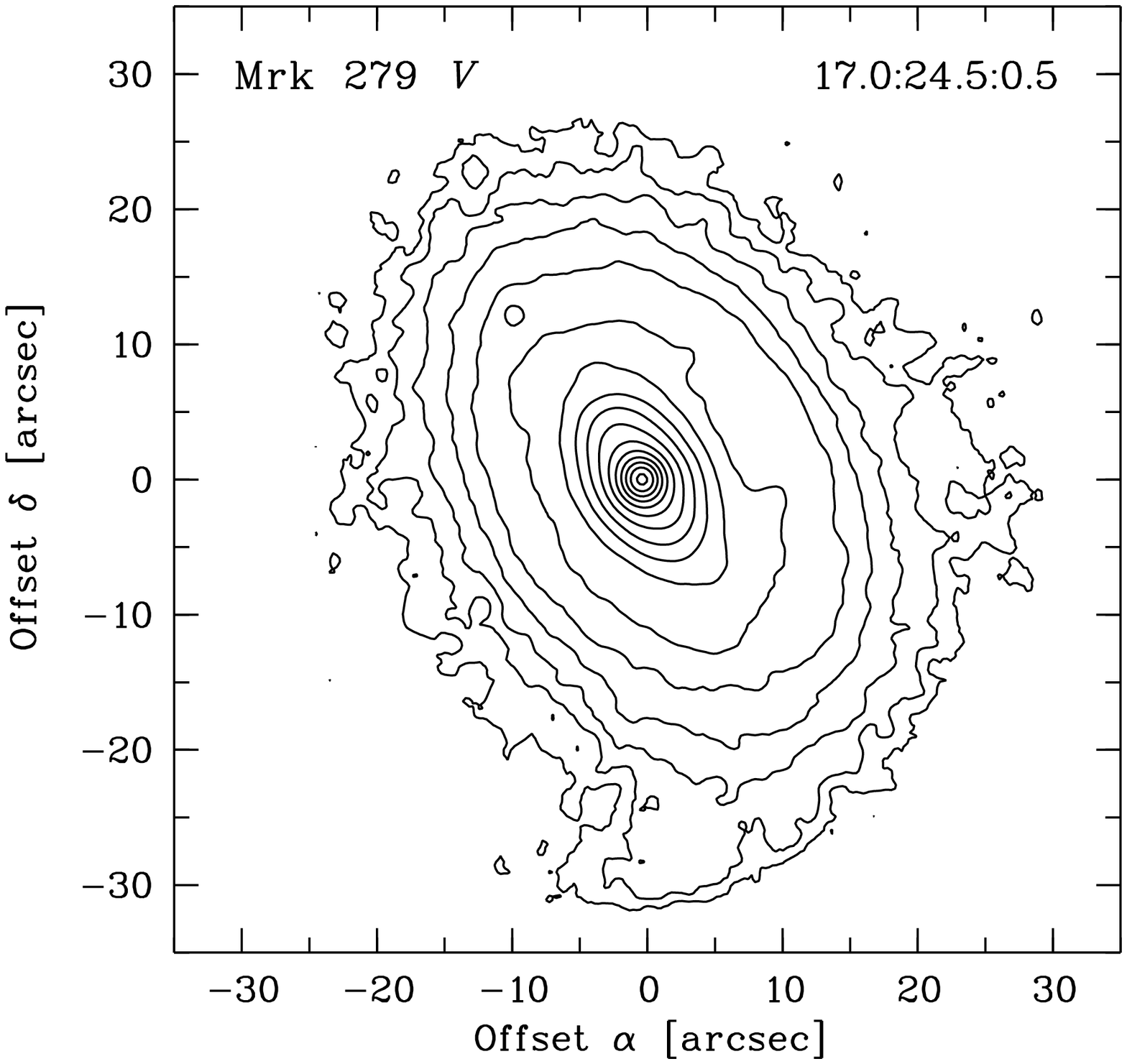}

\vspace{0.3cm}

\includegraphics[width=5.6cm]{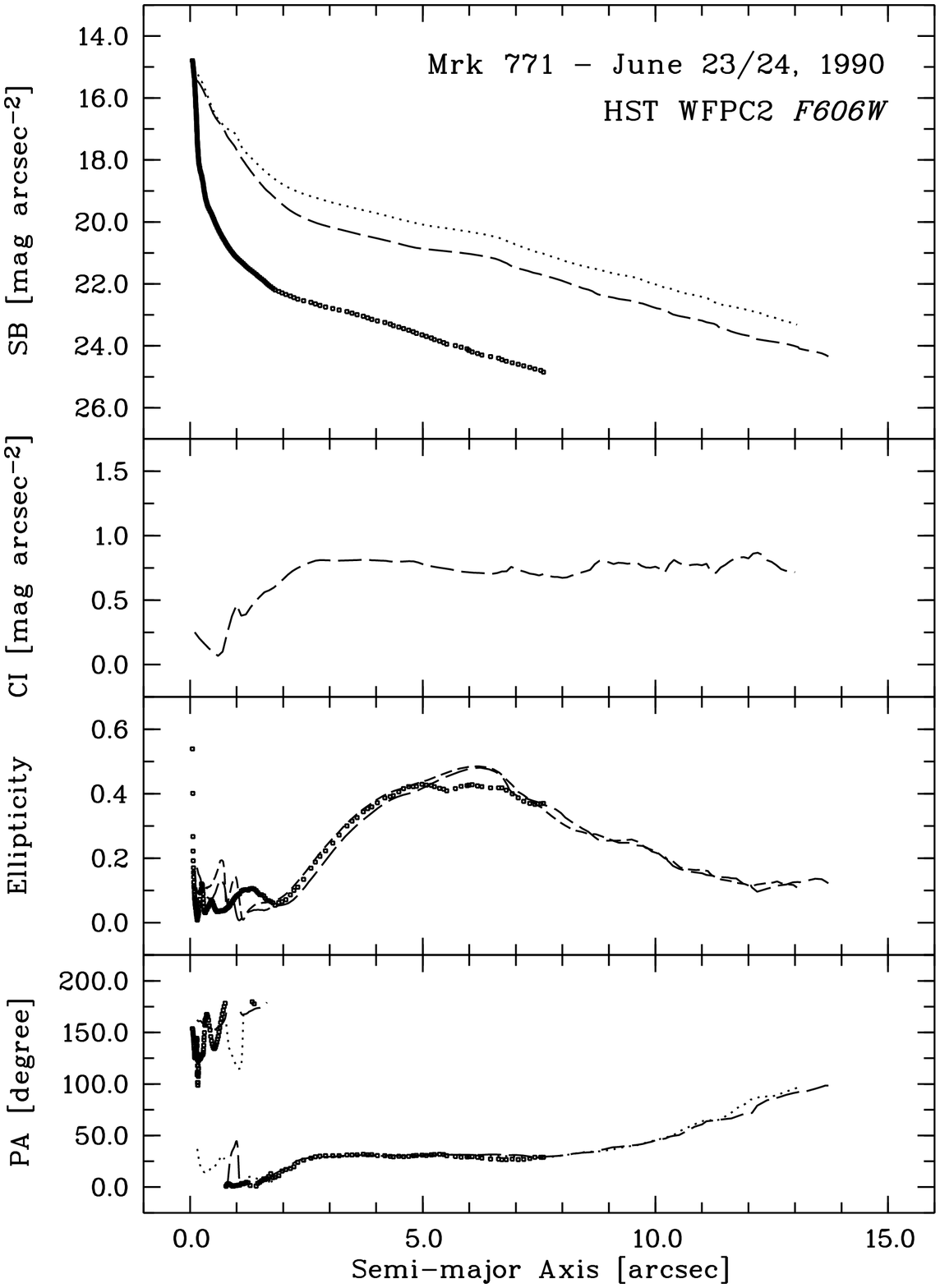}
\hspace{0.5cm}
\includegraphics[width=5.6cm]{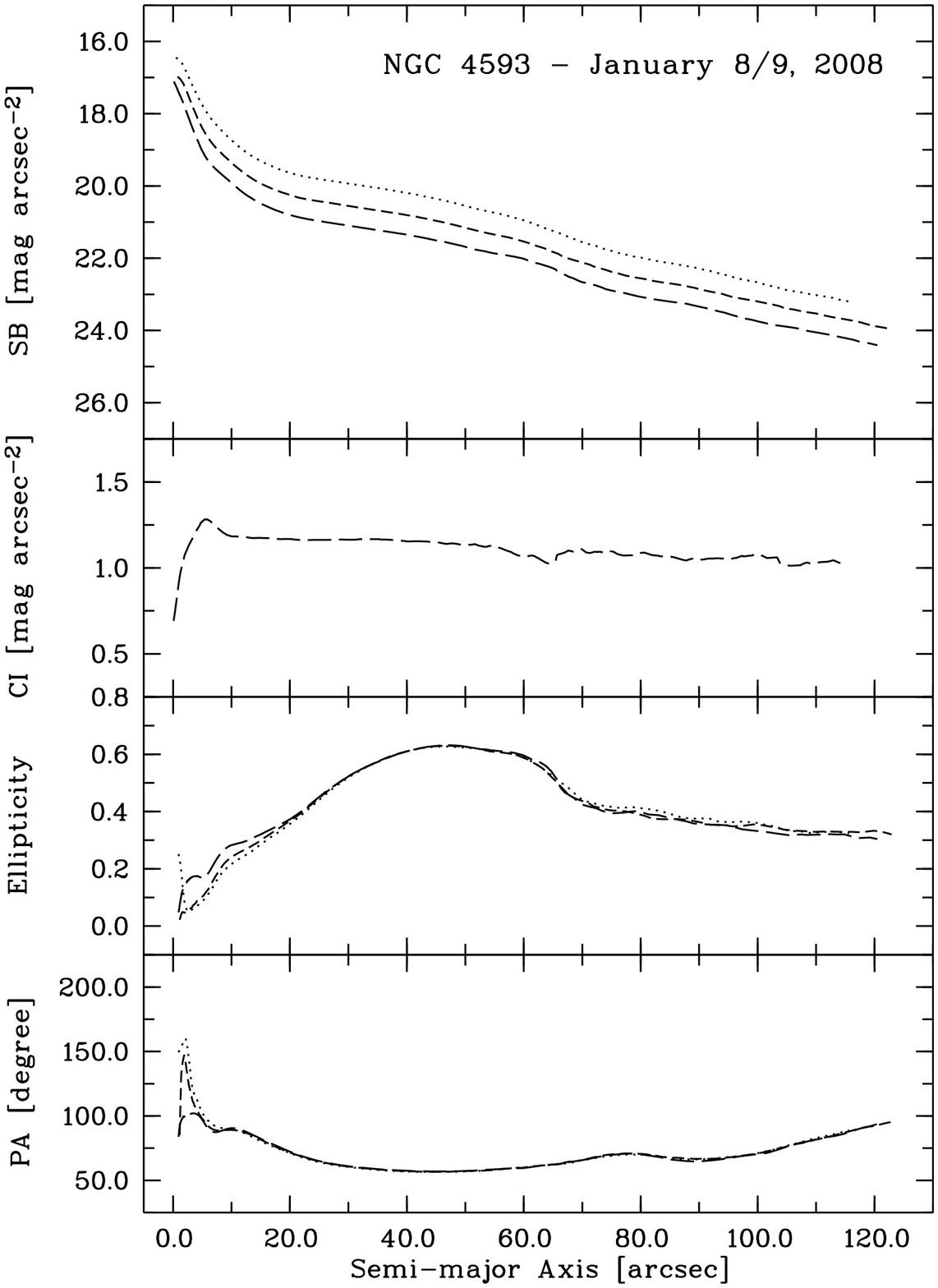}
\hspace{0.5cm}
\includegraphics[width=5.6cm]{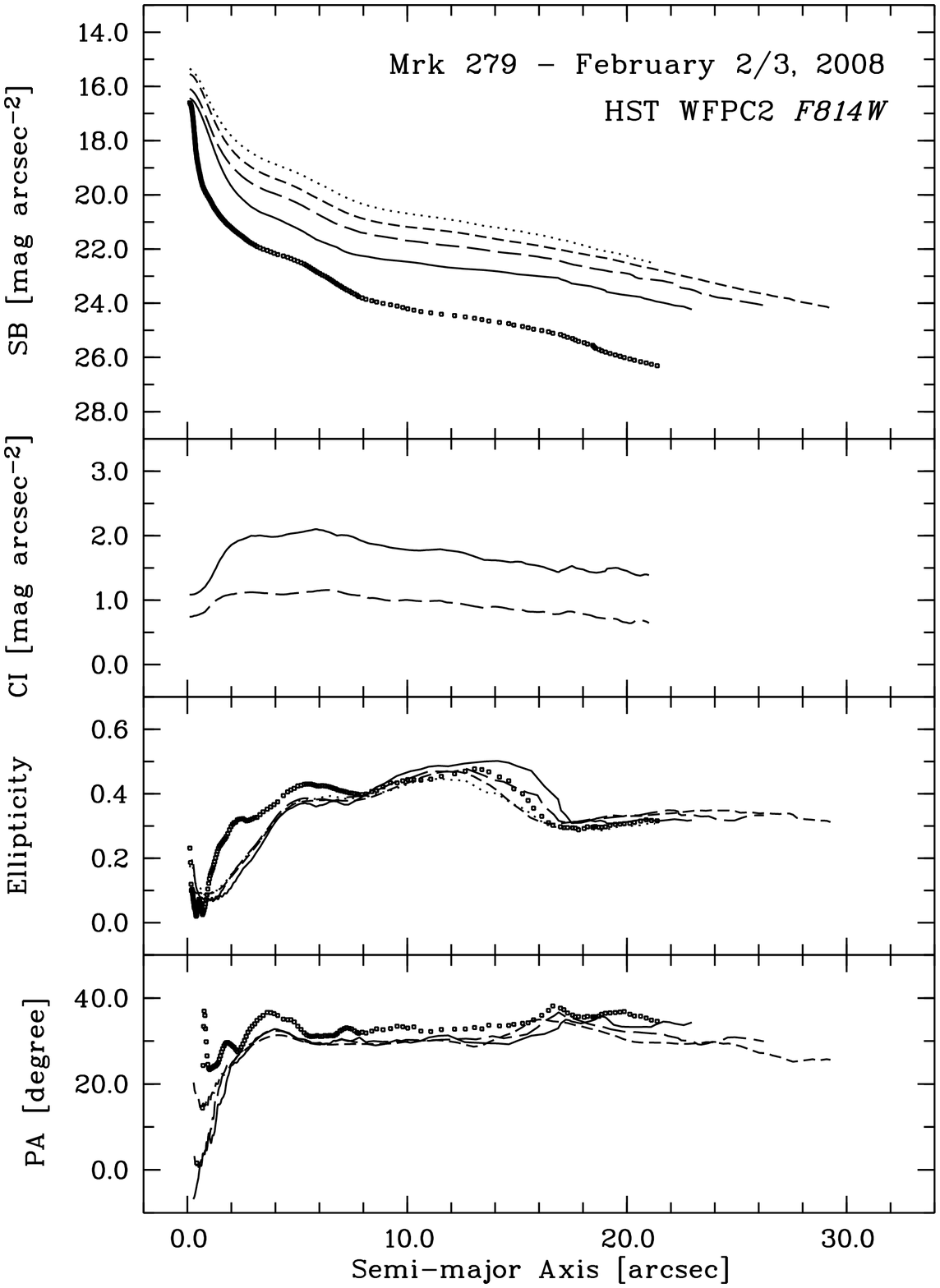}

     \caption{Continued.}
    \end{figure*}
\setcounter{figure}{0}
 
\begin{figure*}[htbp]
\vspace{0.1cm}
   \centering
\includegraphics[width=5.6cm]{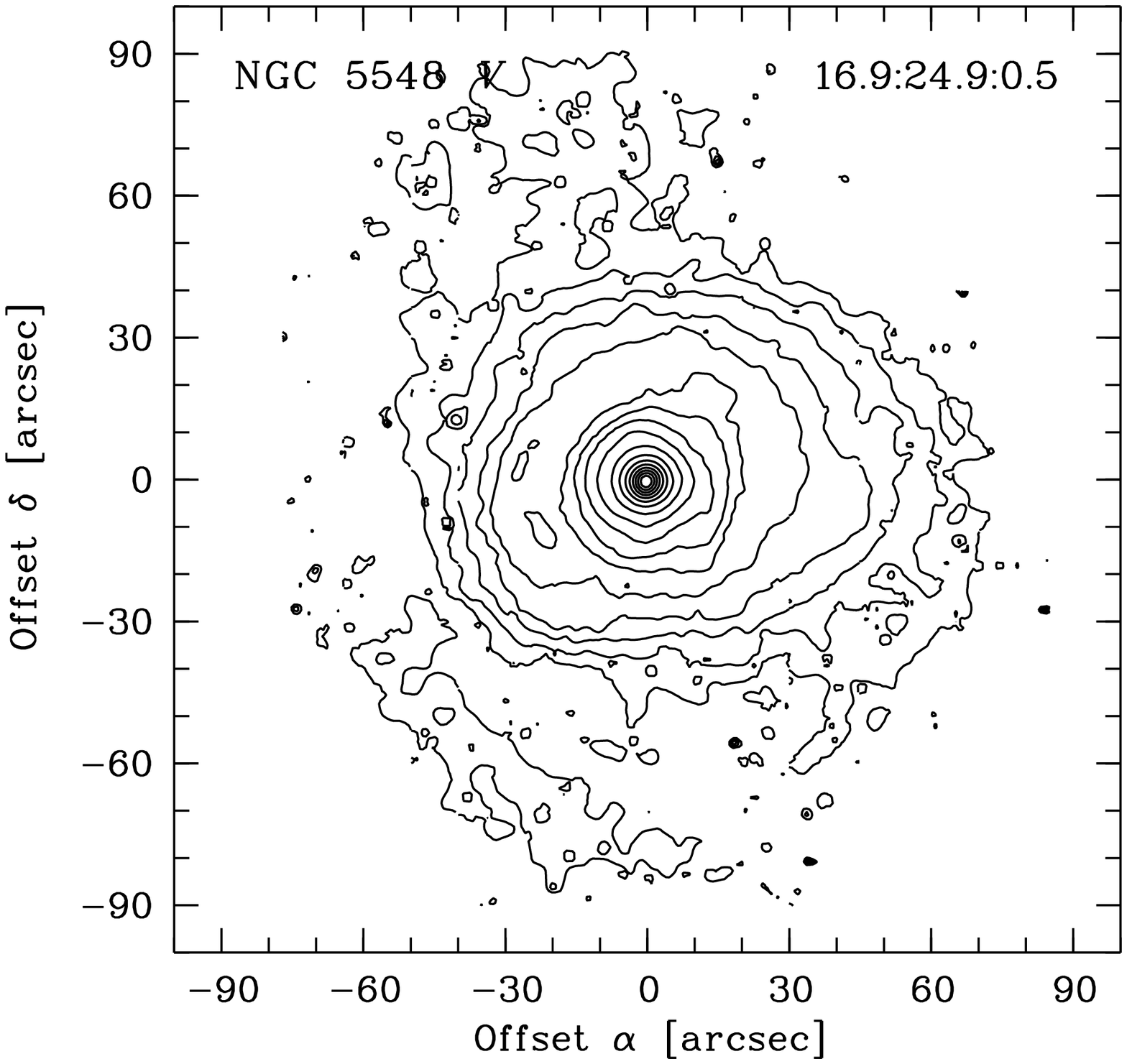}
\hspace{0.5cm}
\includegraphics[width=5.6cm]{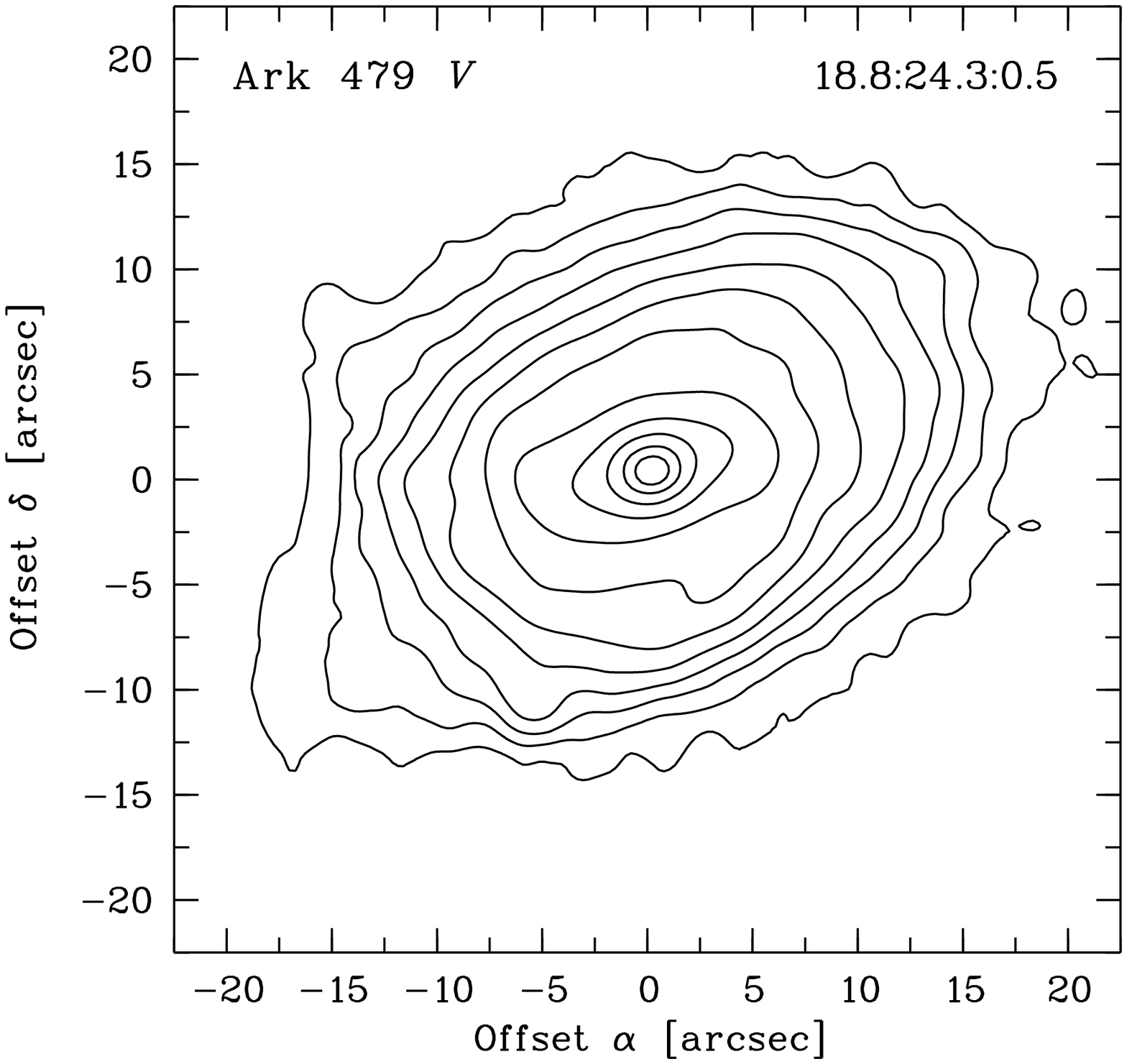}
\hspace{0.5cm}
\includegraphics[width=5.6cm]{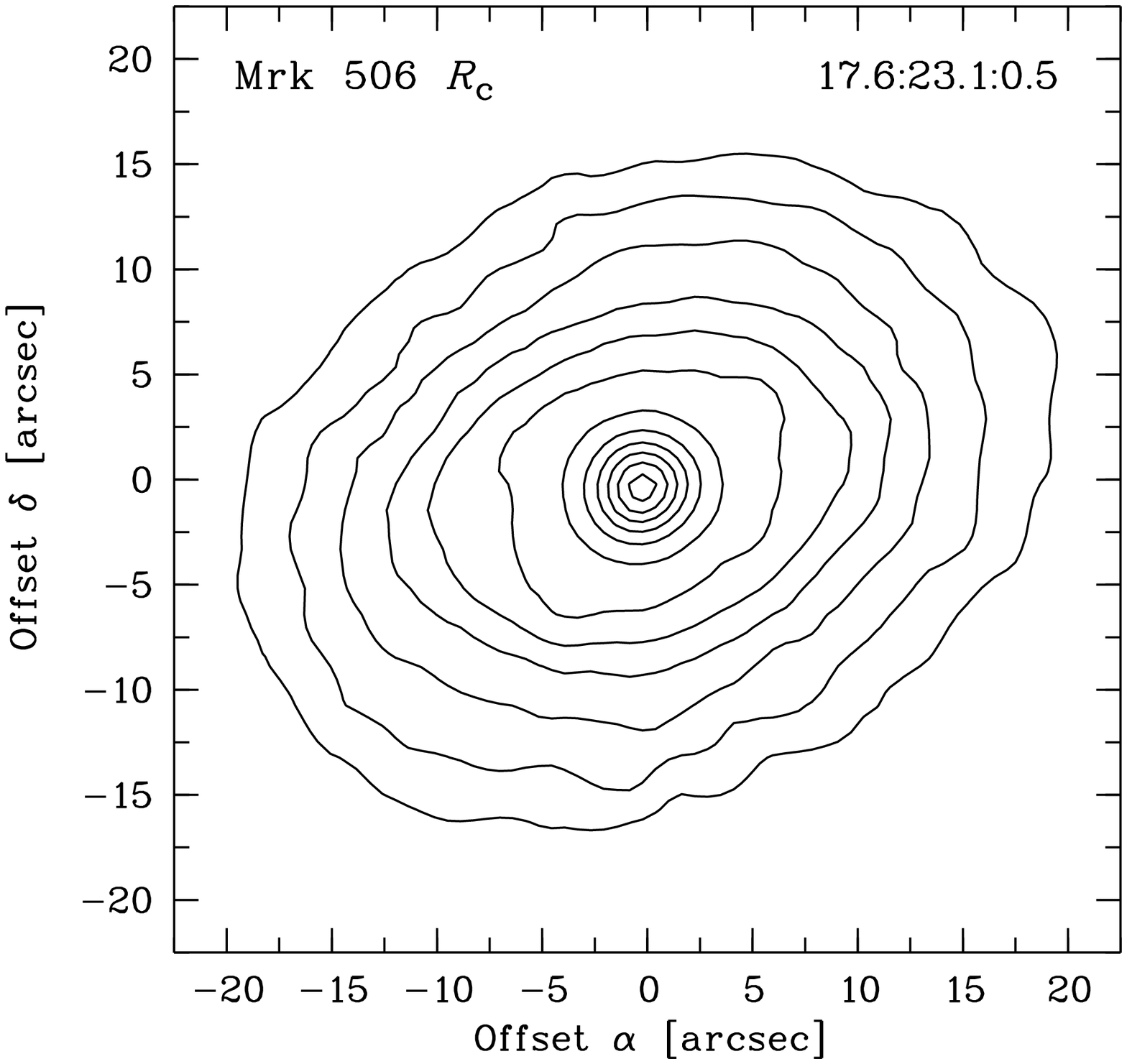}

\vspace{0.3cm}

 \includegraphics[width=5.6cm]{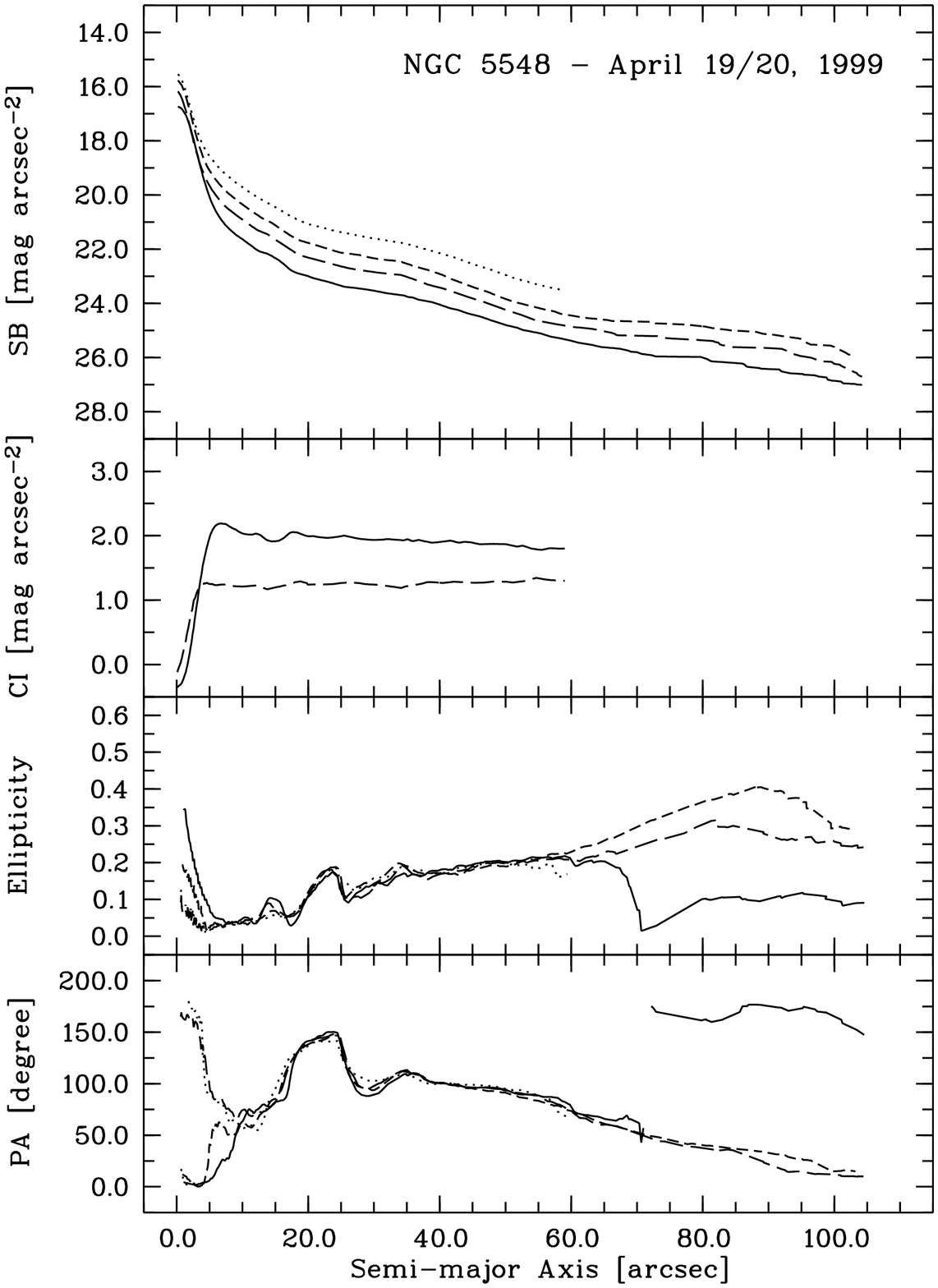}
\hspace{0.5cm}
\includegraphics[width=5.6cm]{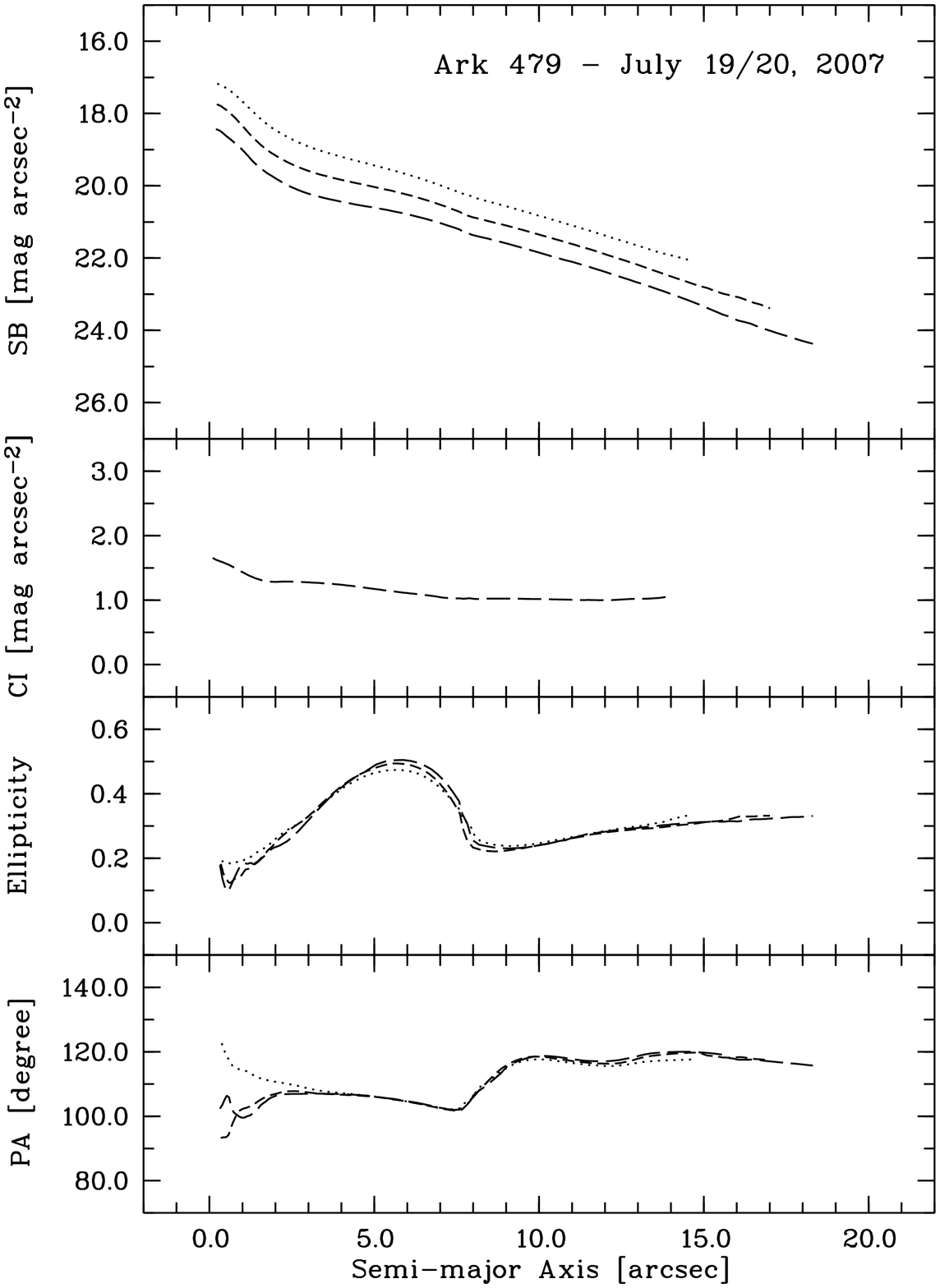}
\hspace{0.5cm}
\includegraphics[width=5.6cm]{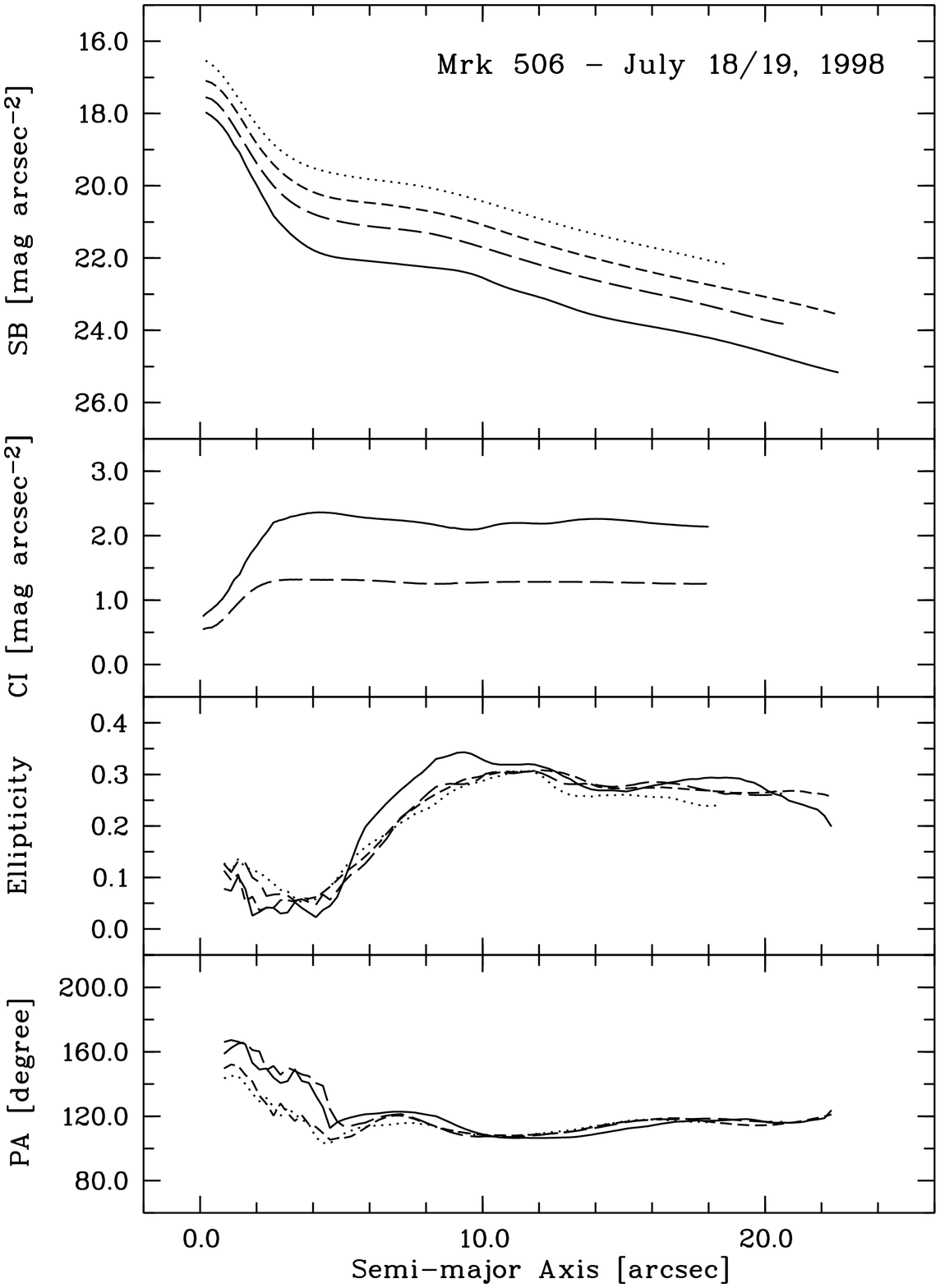}

      \caption{Continued.}
    \end{figure*}
\setcounter{figure}{0}
 
\begin{figure*}[htbp]
\vspace{0.1cm}
   \centering
\includegraphics[width=5.6cm]{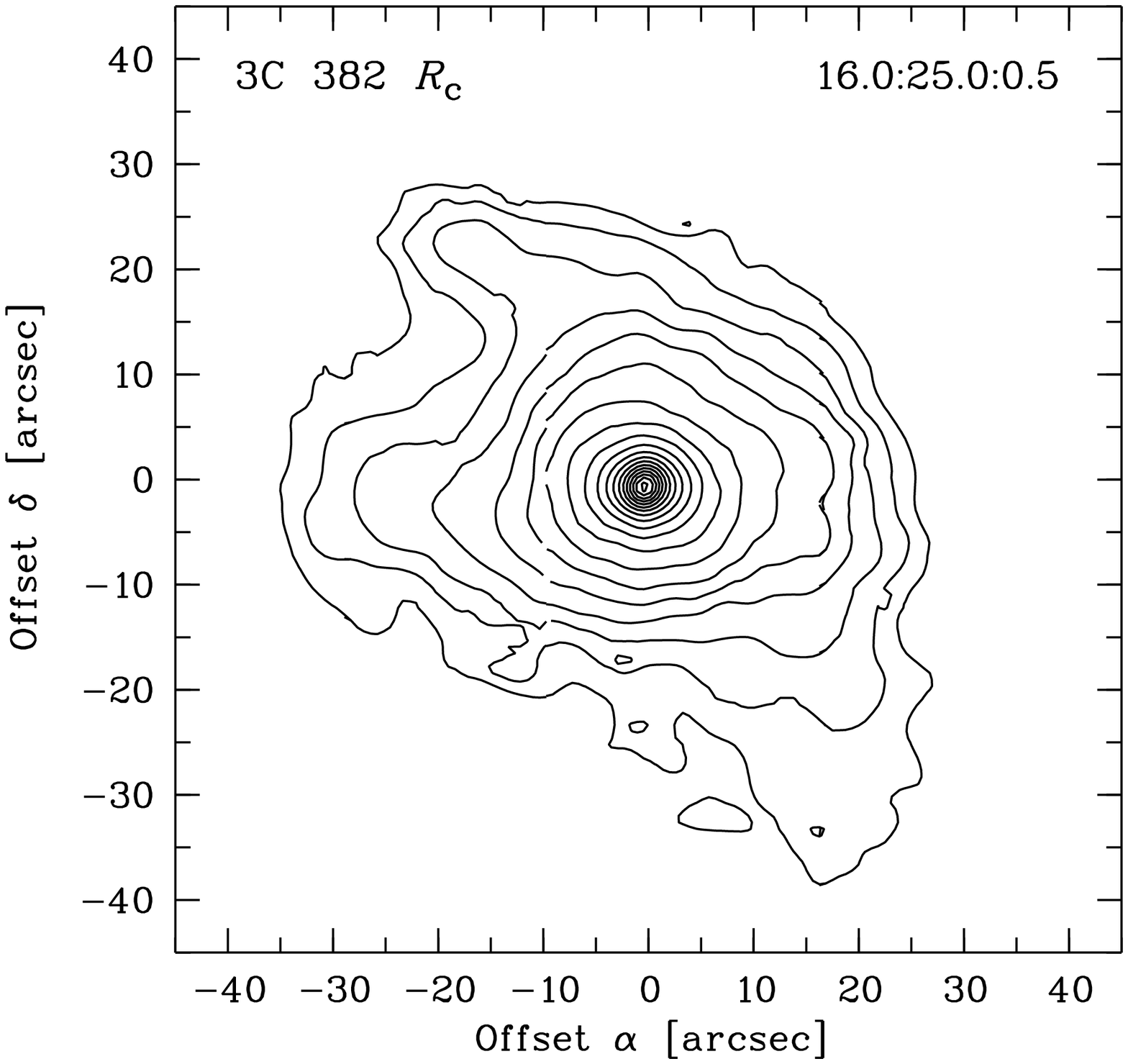}
\hspace{0.5cm}
\includegraphics[width=5.6cm]{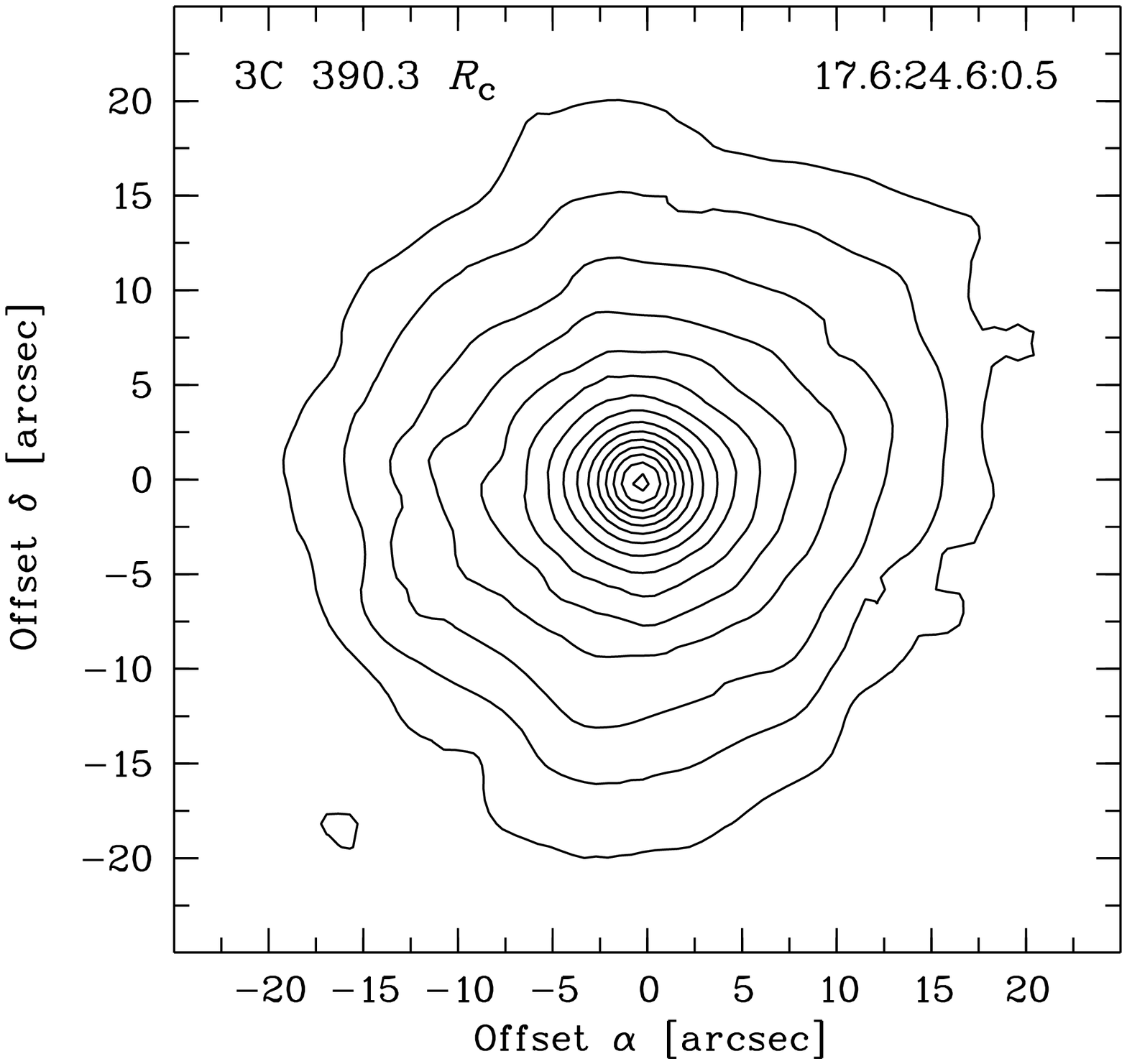}
\hspace{0.5cm}
\includegraphics[width=5.6cm]{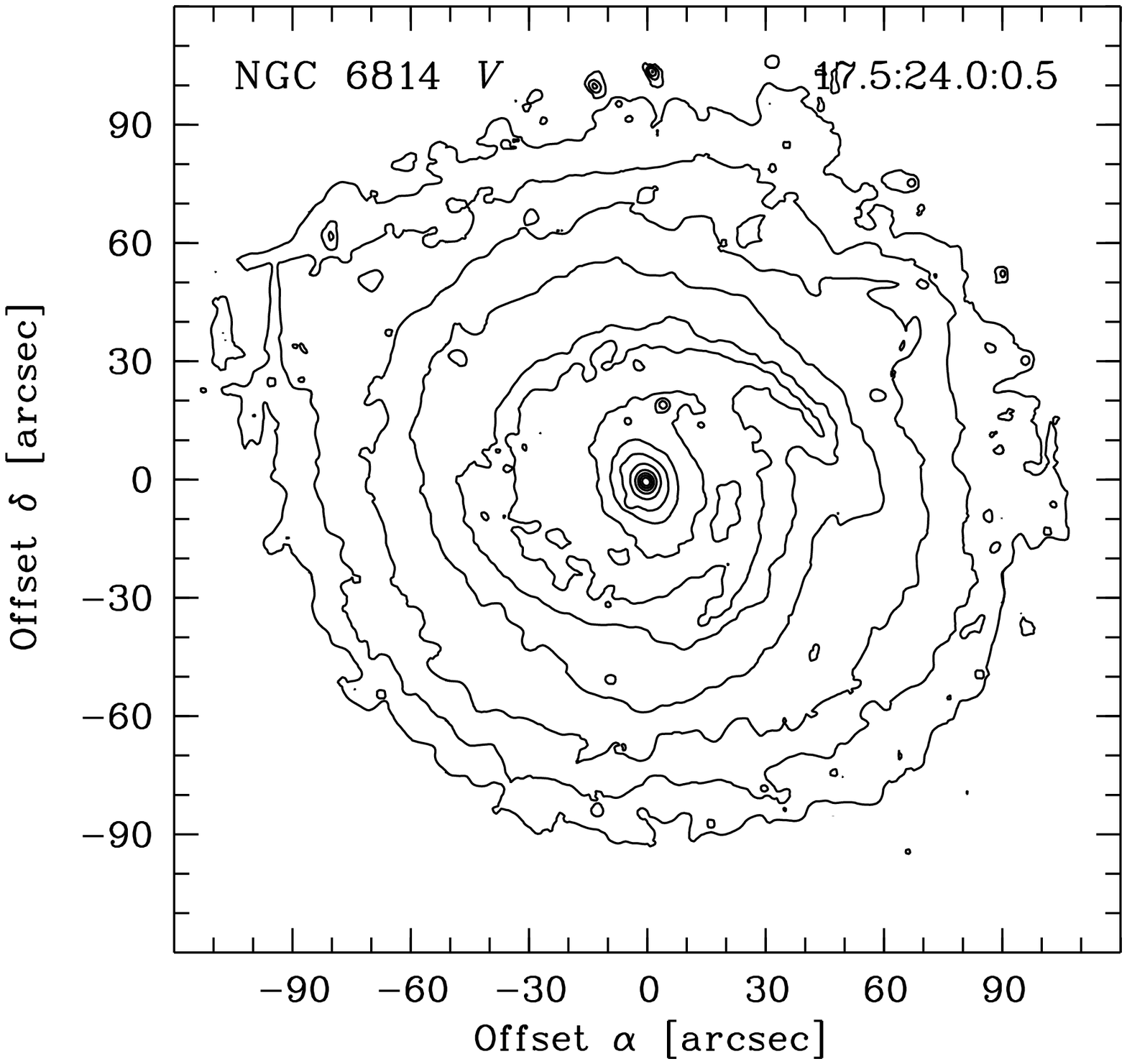}

\vspace{0.3cm}

\includegraphics[width=5.6cm]{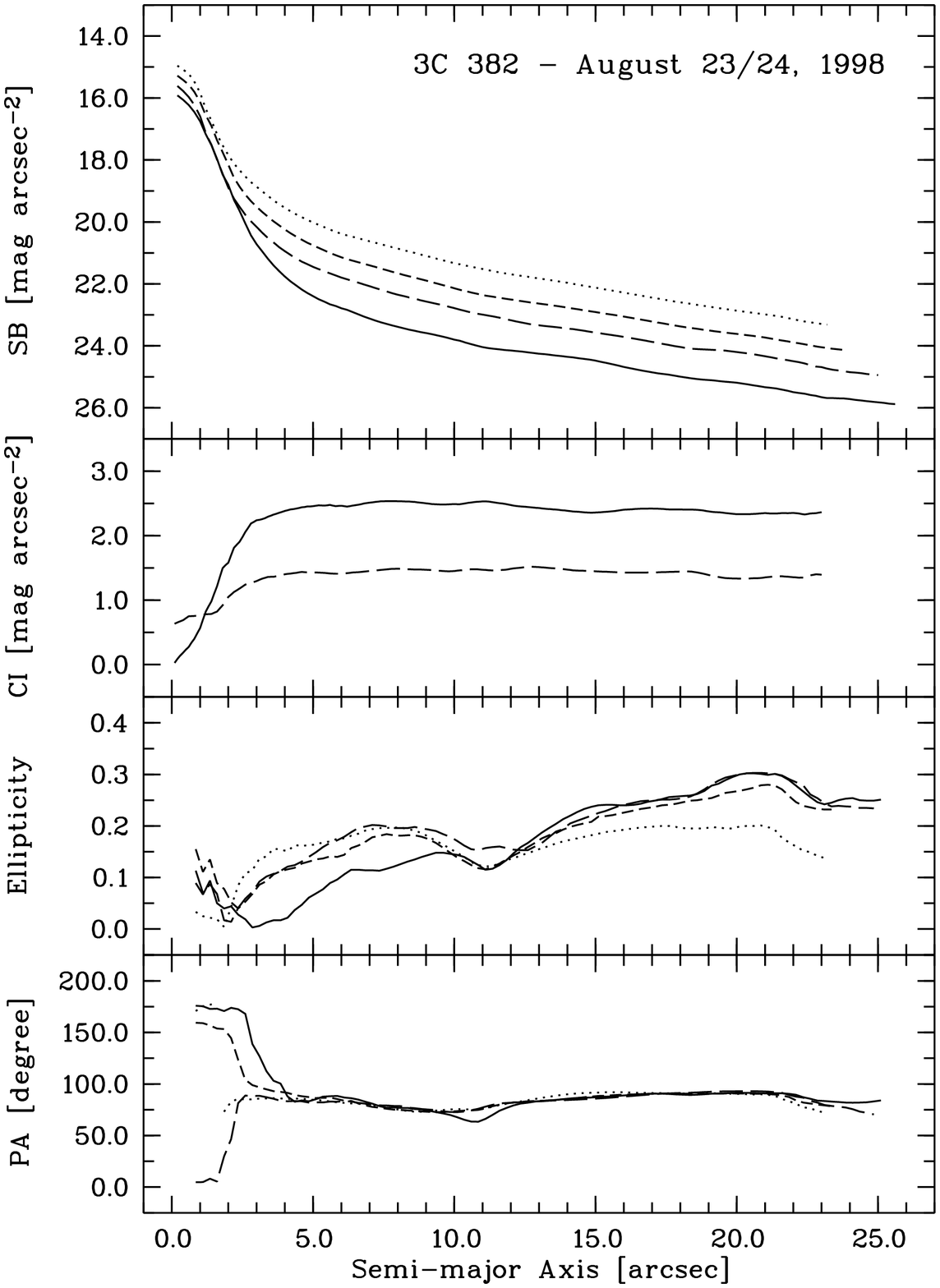}
\hspace{0.5cm}
\includegraphics[width=5.6cm]{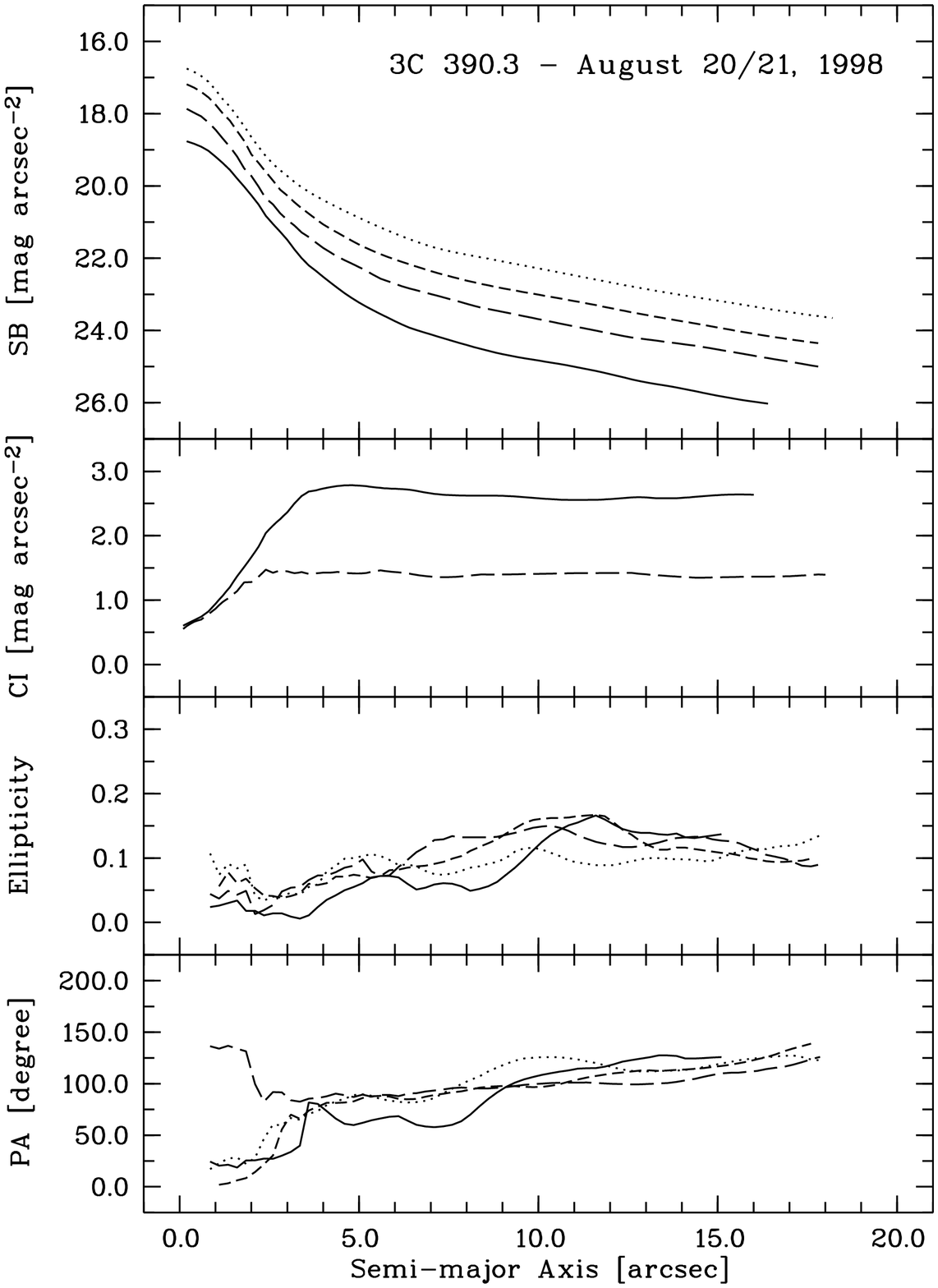}
\hspace{0.5cm}
\includegraphics[width=5.6cm]{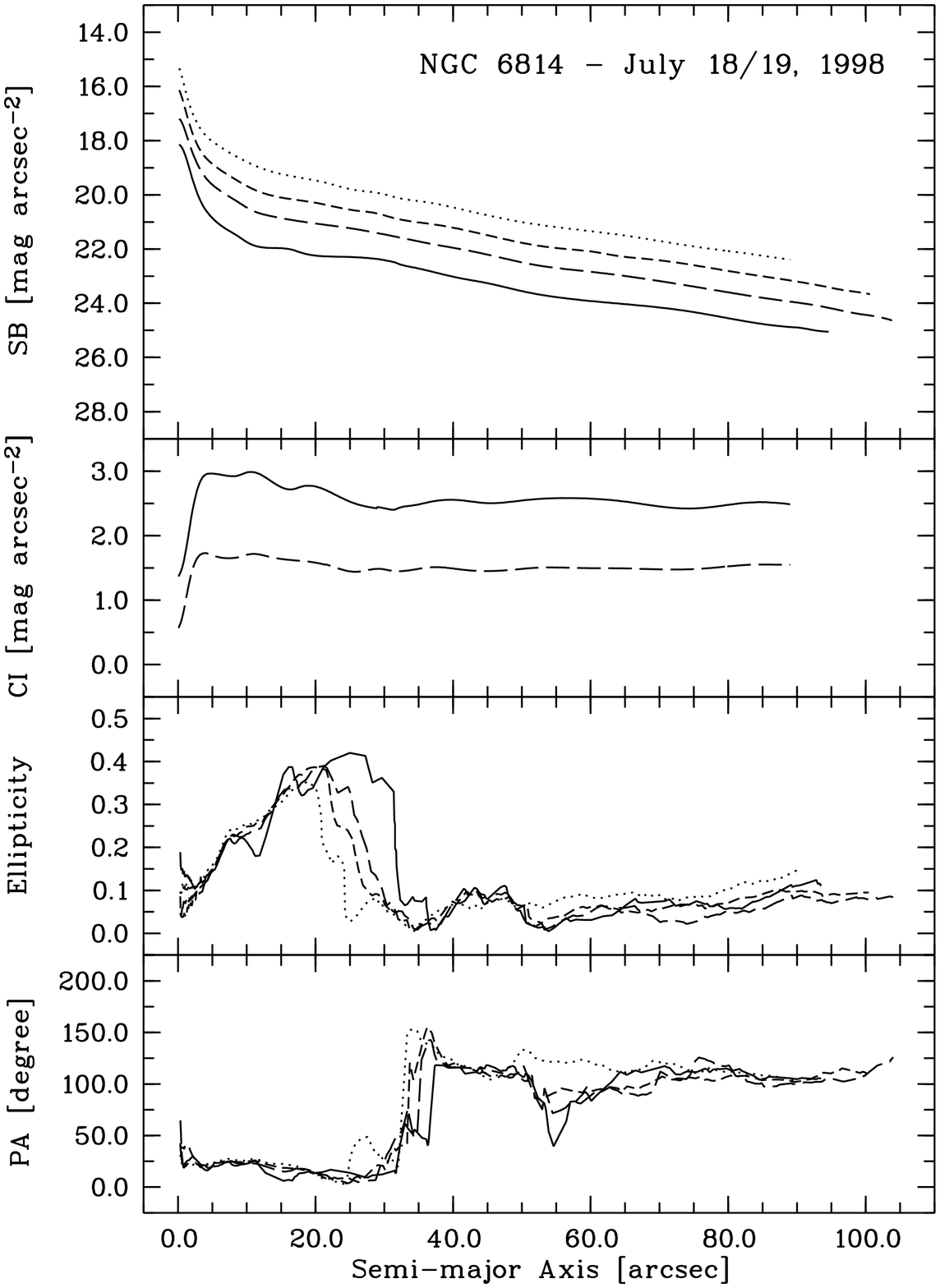}

      \caption{Continued.}
   \end{figure*}

\setcounter{figure}{0}
 
\begin{figure*}[htbp]
\vspace{0.1cm}
   \centering
\includegraphics[width=5.6cm]{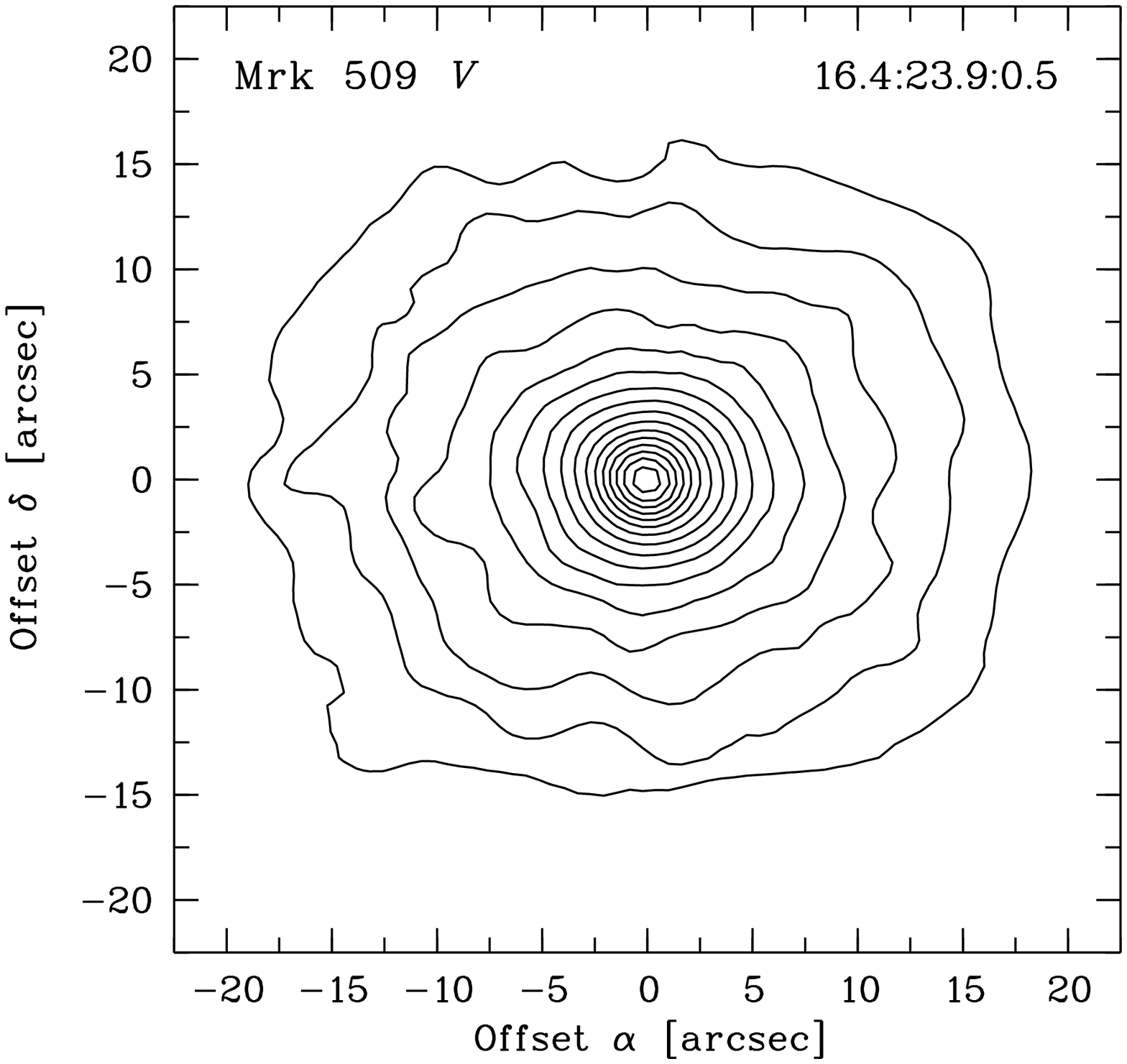}
\hspace{0.5cm}
\includegraphics[width=5.6cm]{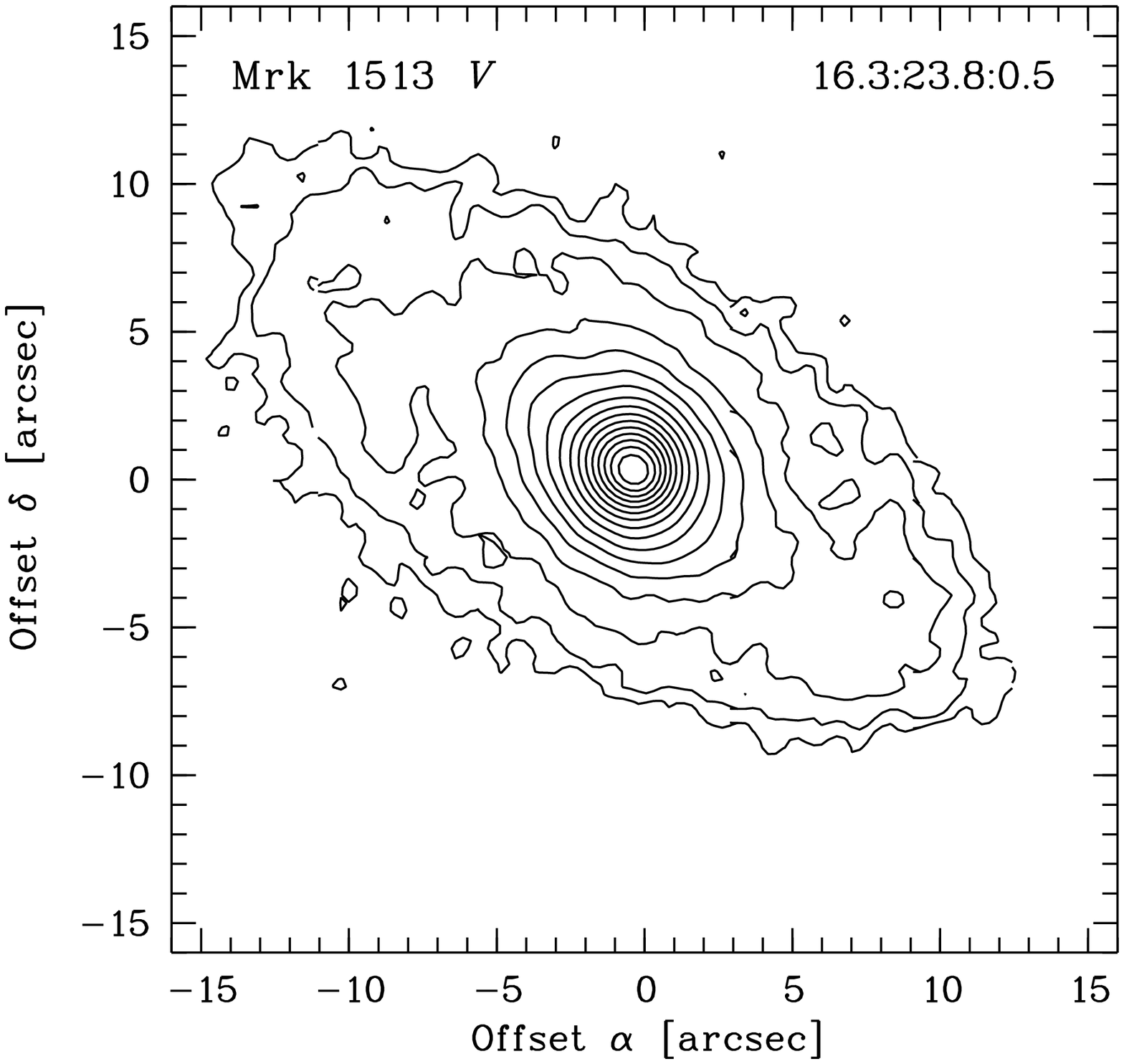}
\hspace{0.5cm}
\includegraphics[width=5.6cm]{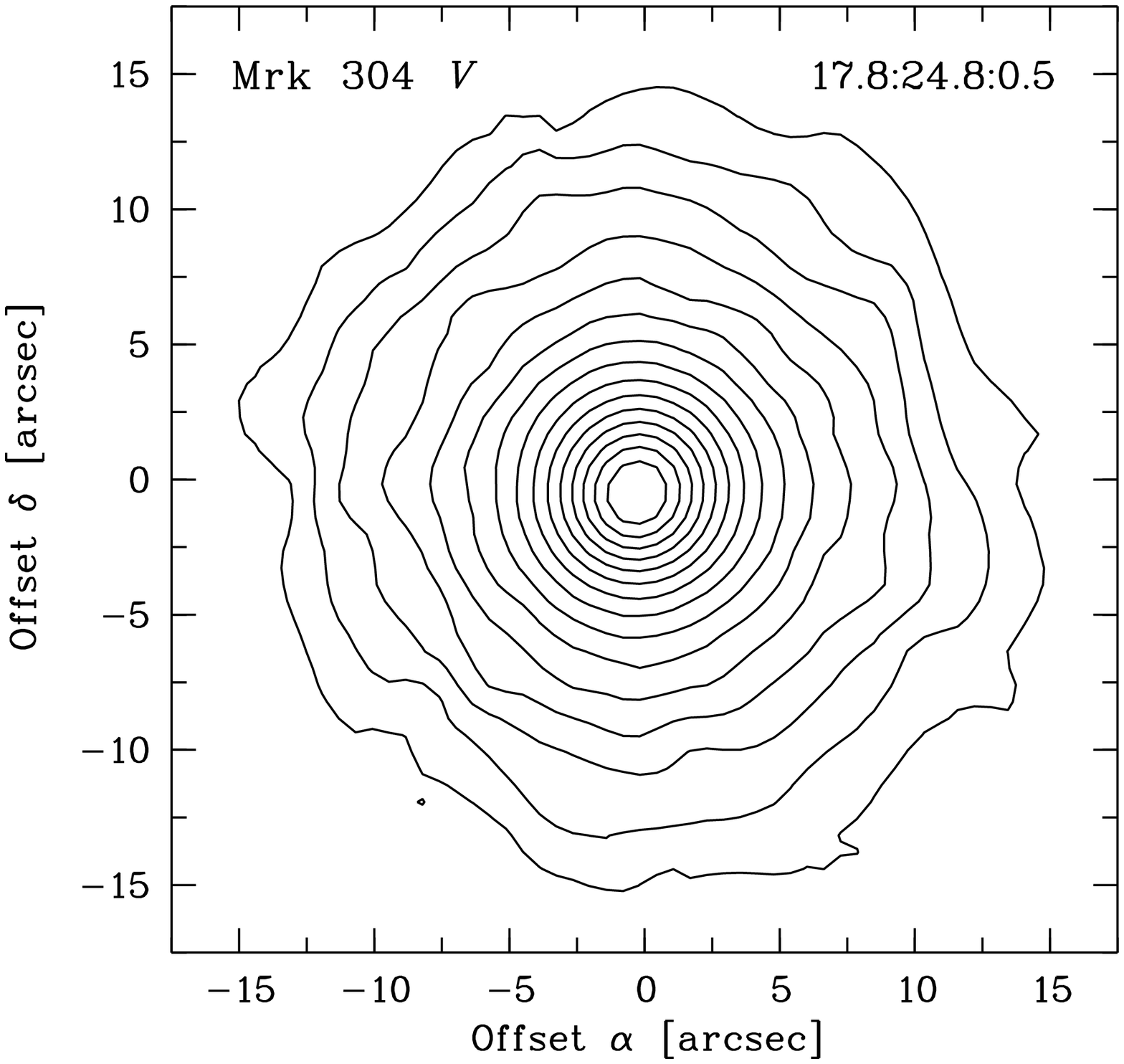}

\vspace{0.3cm}
 
\includegraphics[width=5.6cm]{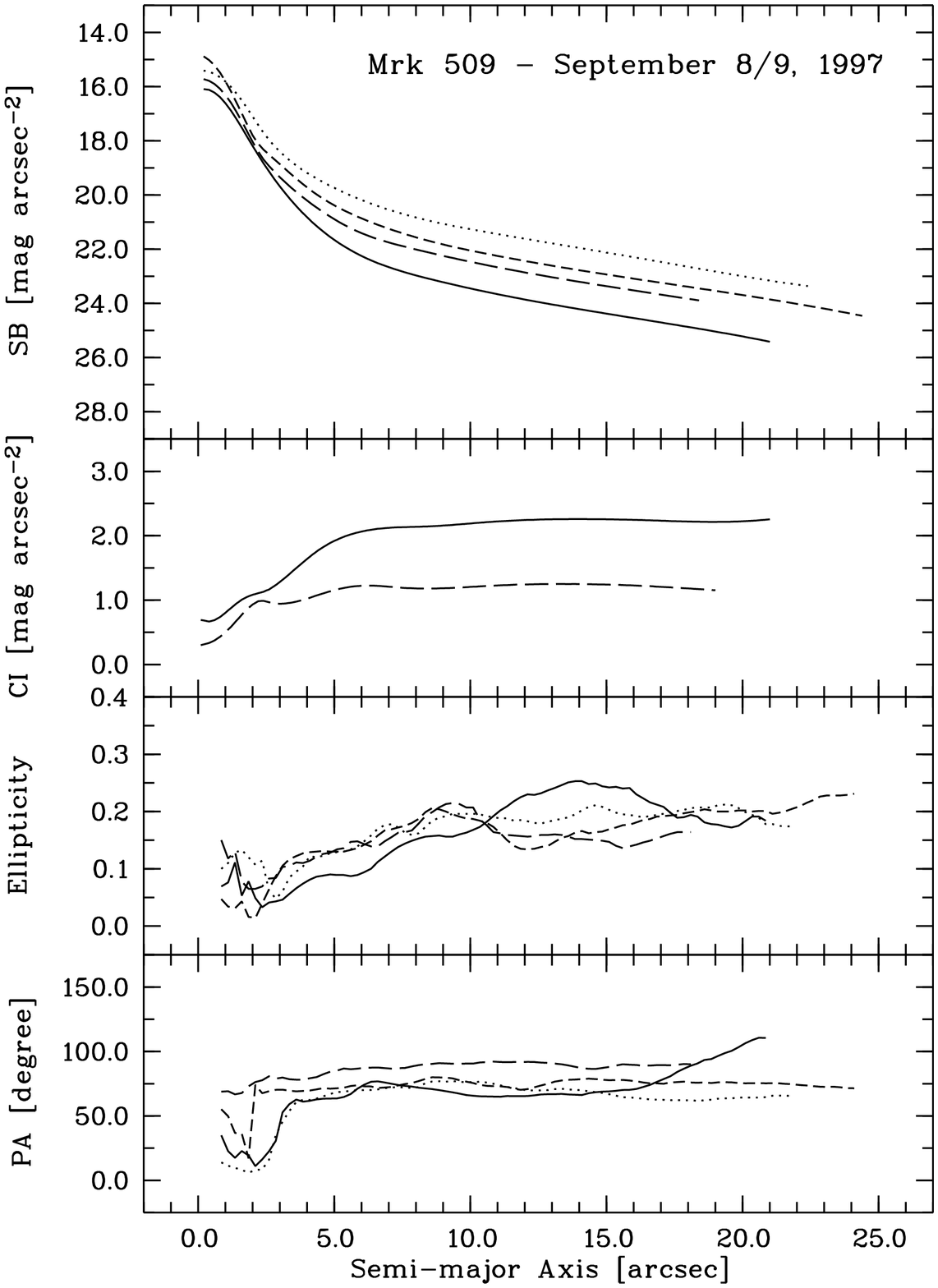}
\hspace{0.5cm}
\includegraphics[width=5.6cm]{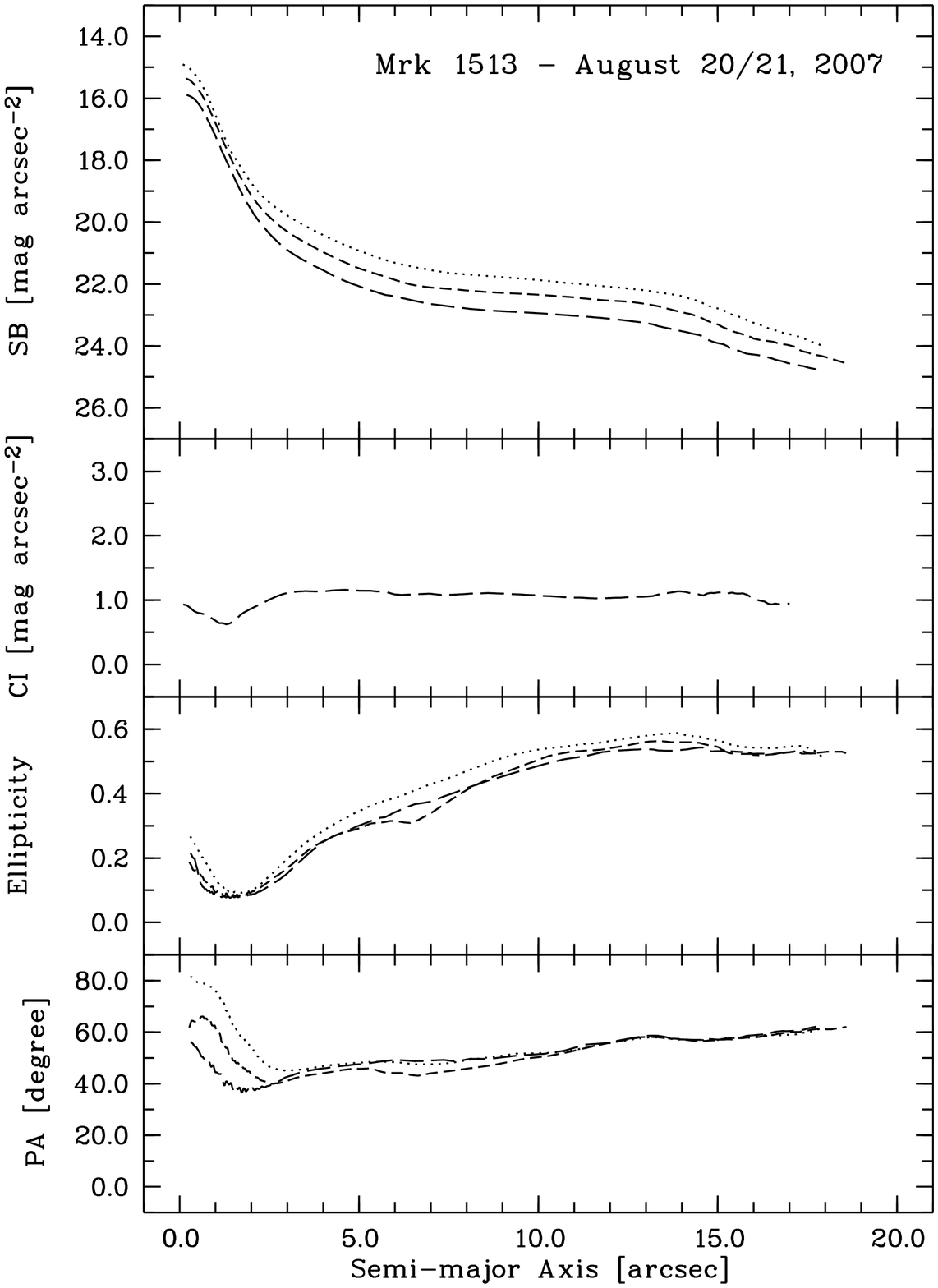}
\hspace{0.5cm}
\includegraphics[width=5.6cm]{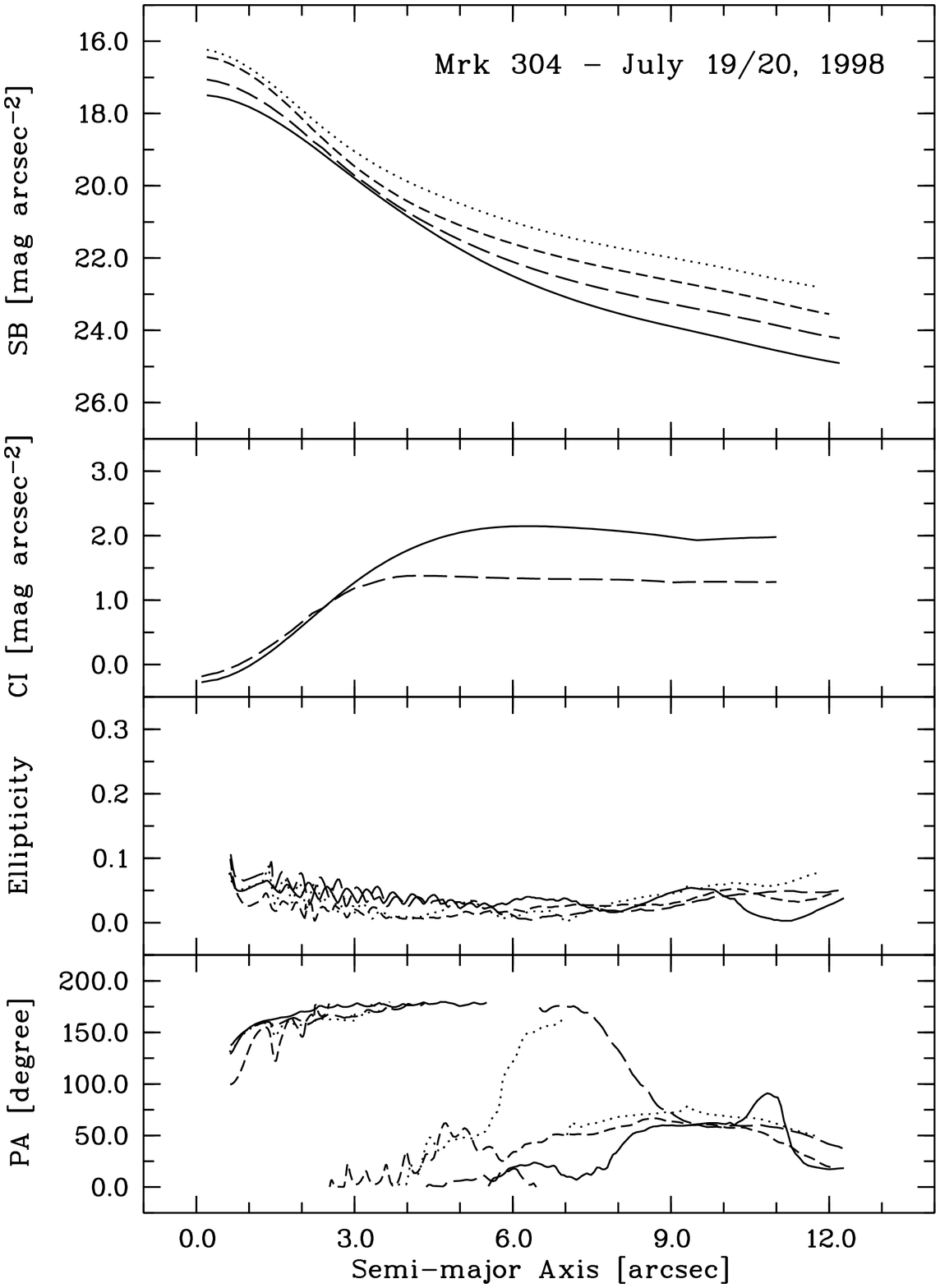}

     \caption{Continued.}
  \end{figure*}
\setcounter{figure}{0}

\begin{figure*}[htbp]
\vspace{0.1cm}
   \centering
\includegraphics[width=5.6cm]{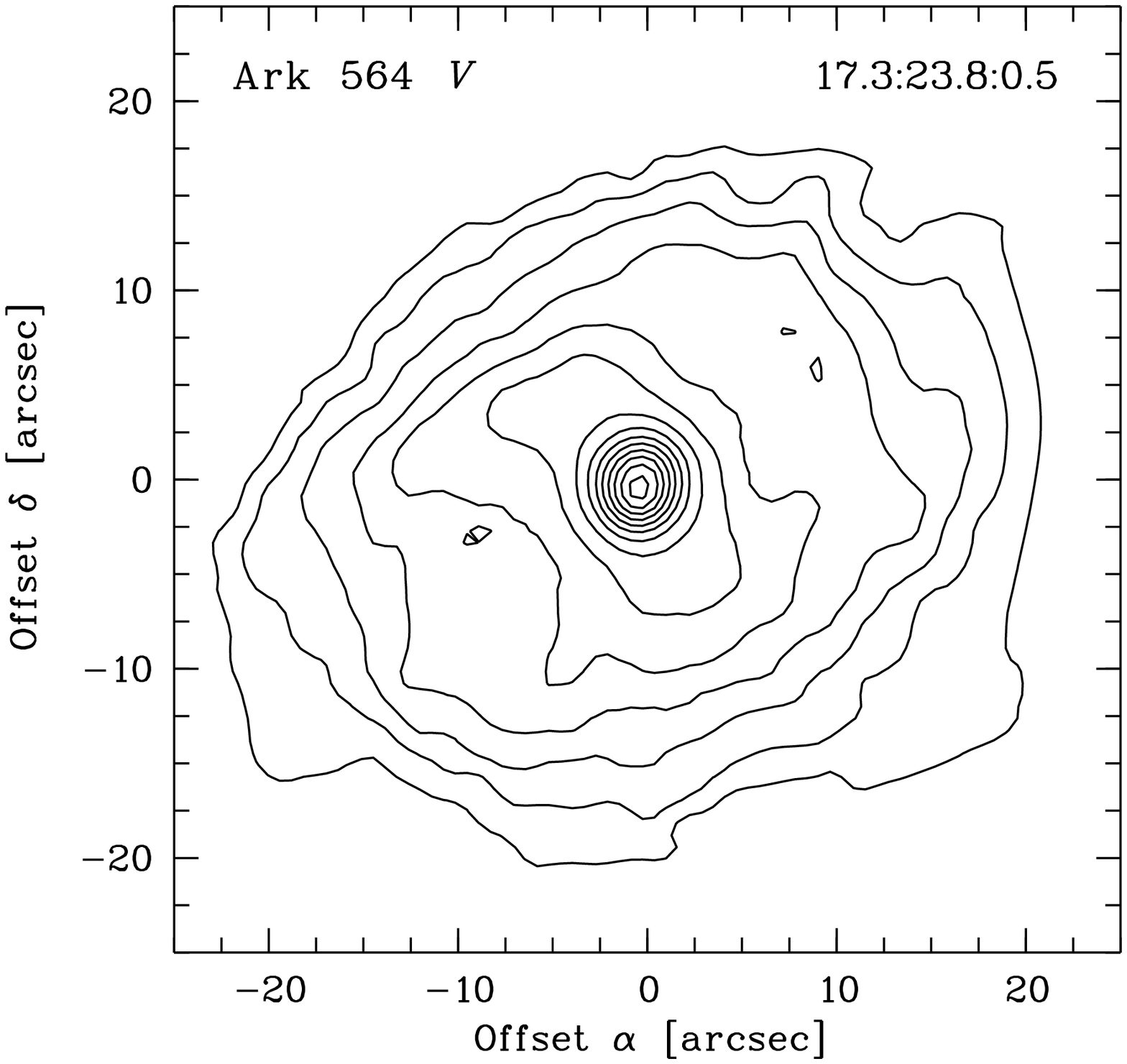}
\hspace{0.5cm}
\includegraphics[width=5.6cm]{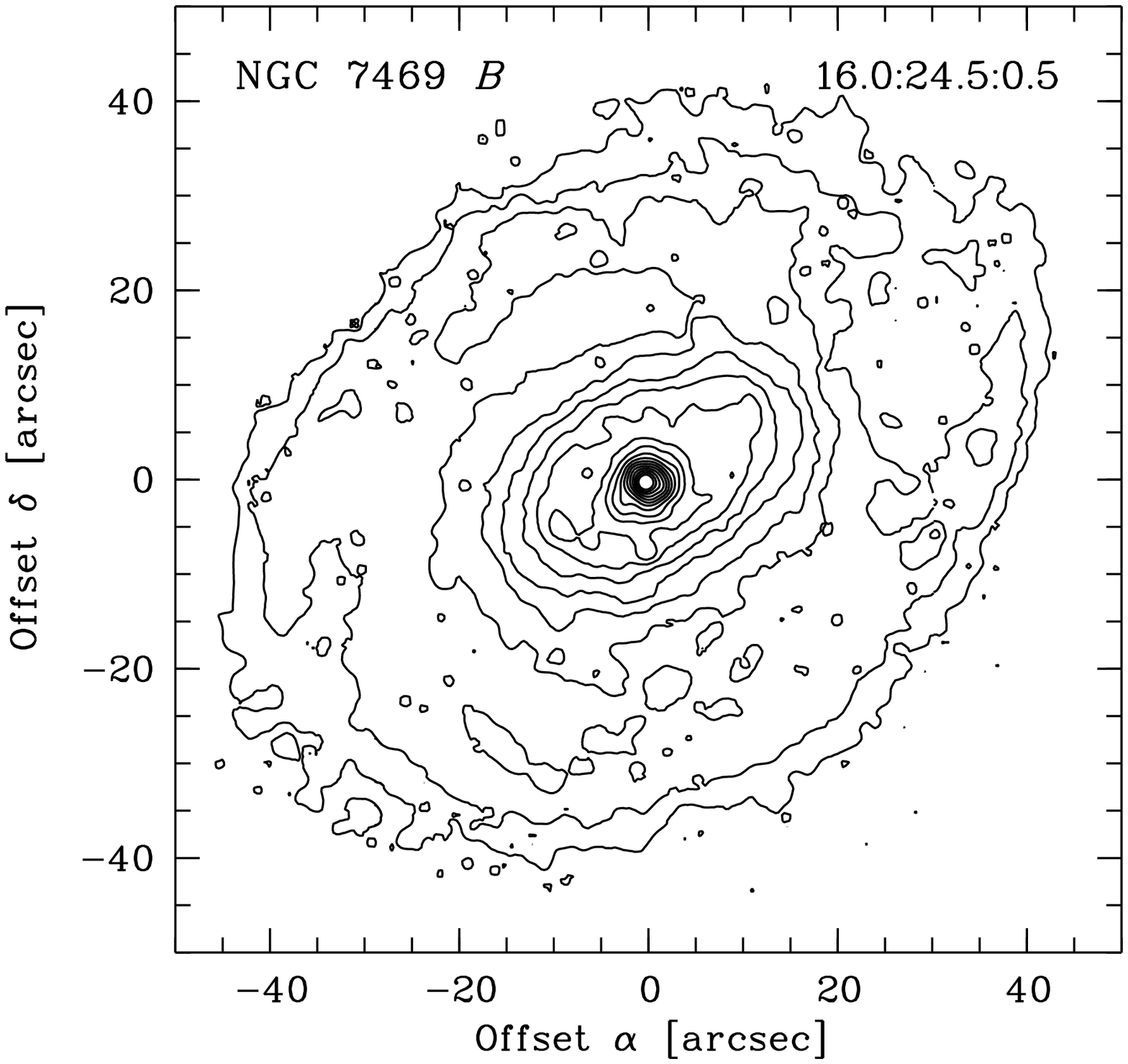}
\hspace{0.5cm}
\includegraphics[width=5.6cm]{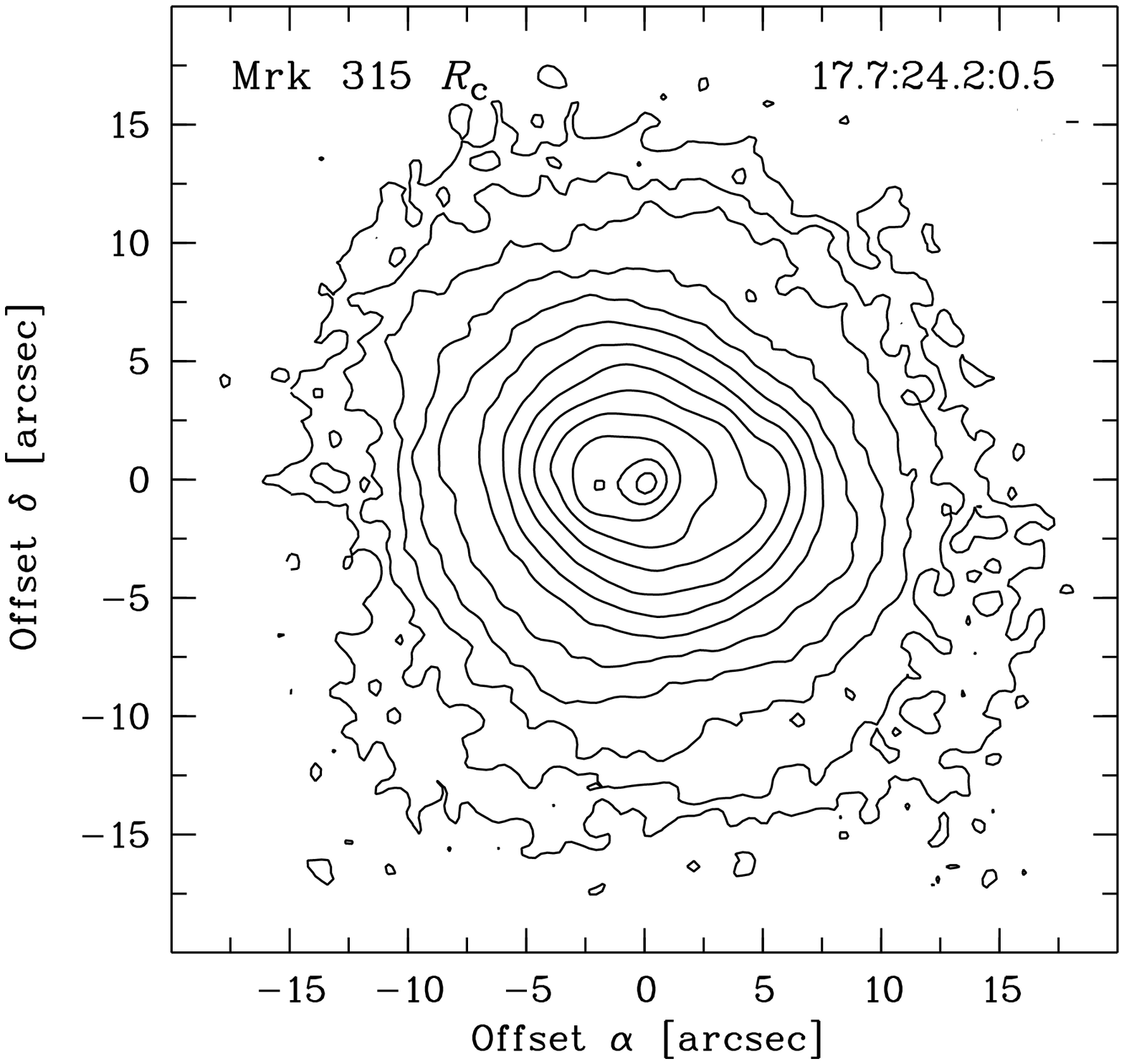}

\vspace{0.3cm}
 
\includegraphics[width=5.6cm]{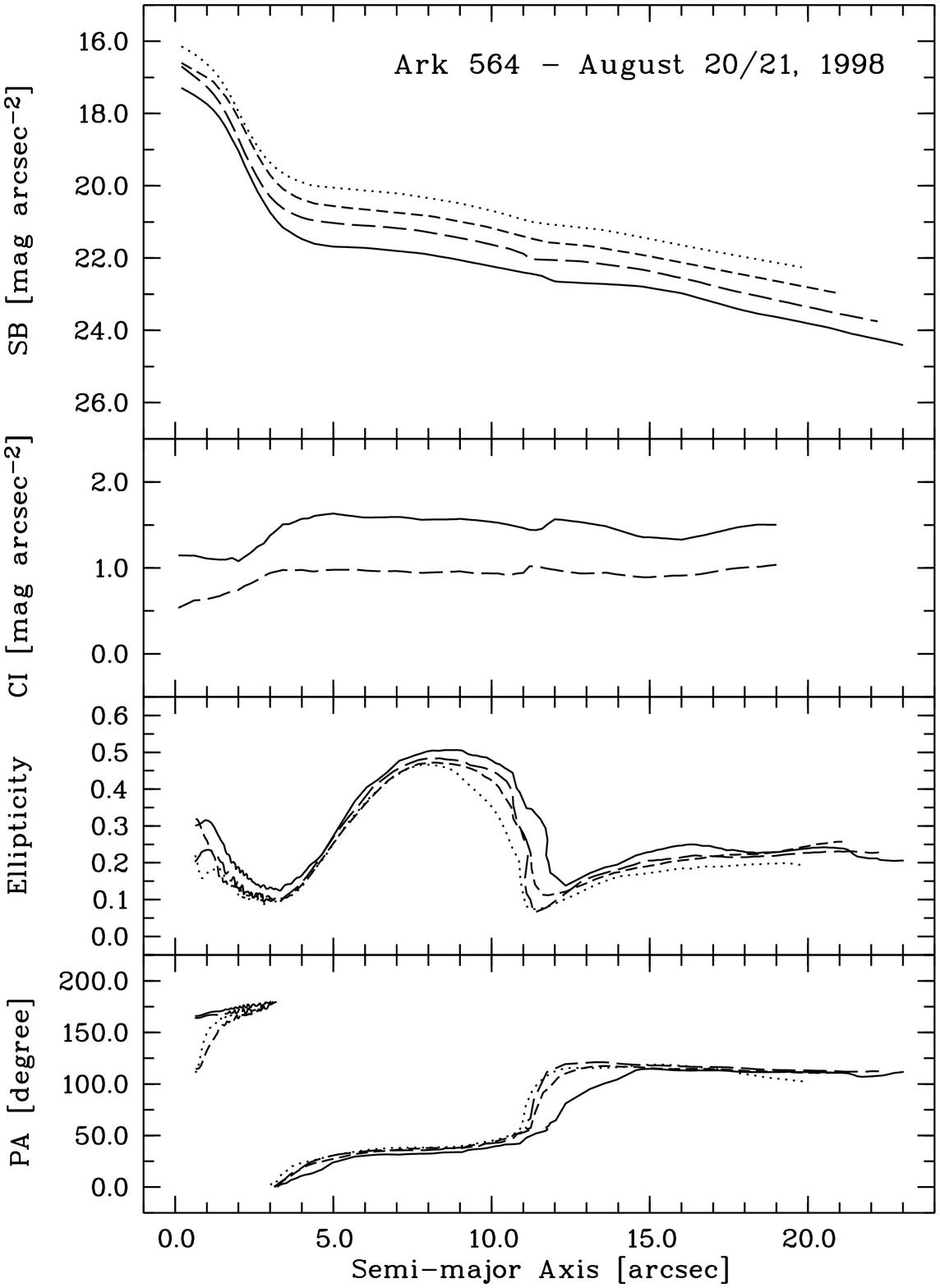}
\hspace{0.5cm}
\includegraphics[width=5.6cm]{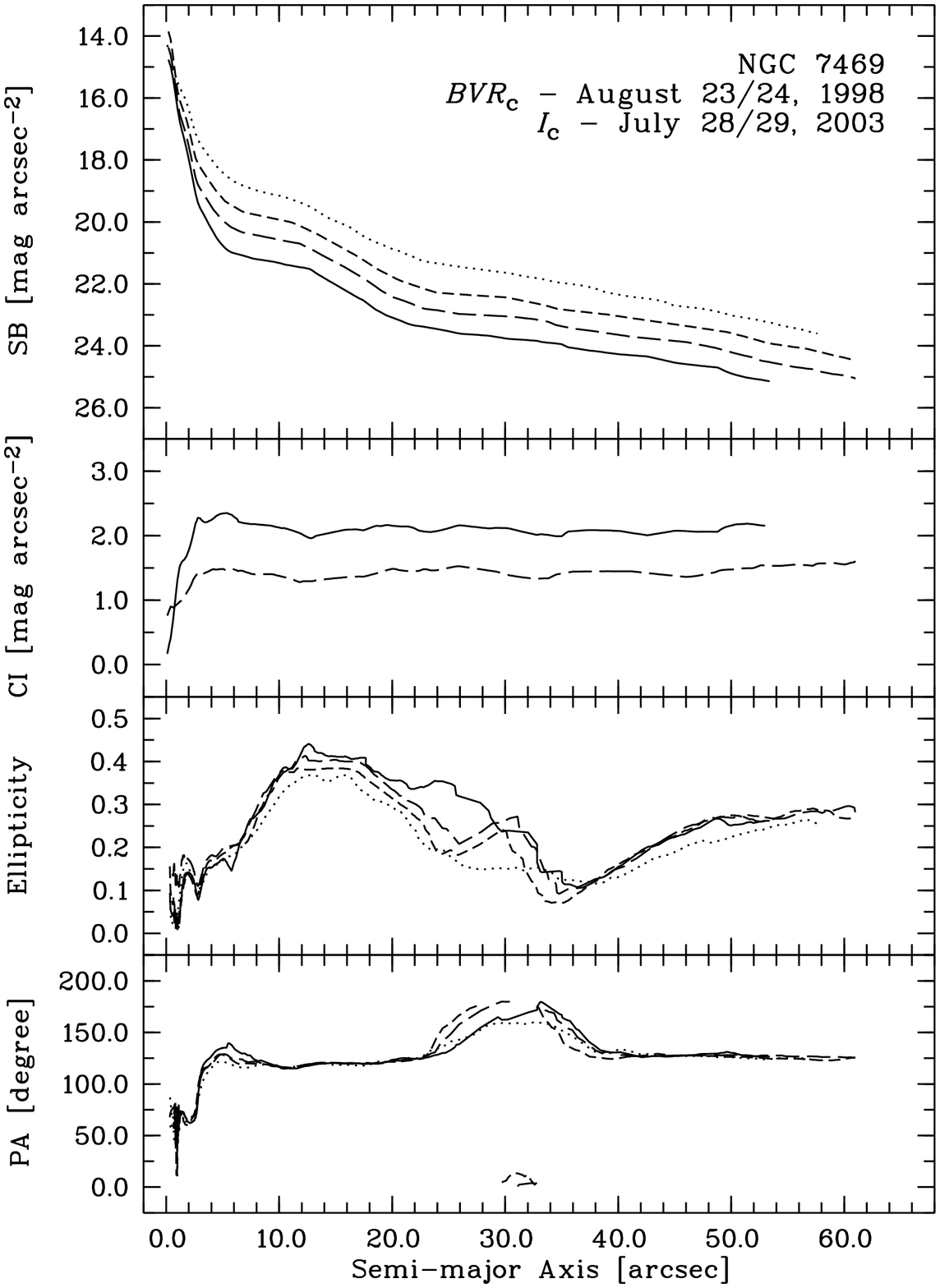}
\hspace{0.5cm}
\includegraphics[width=5.6cm]{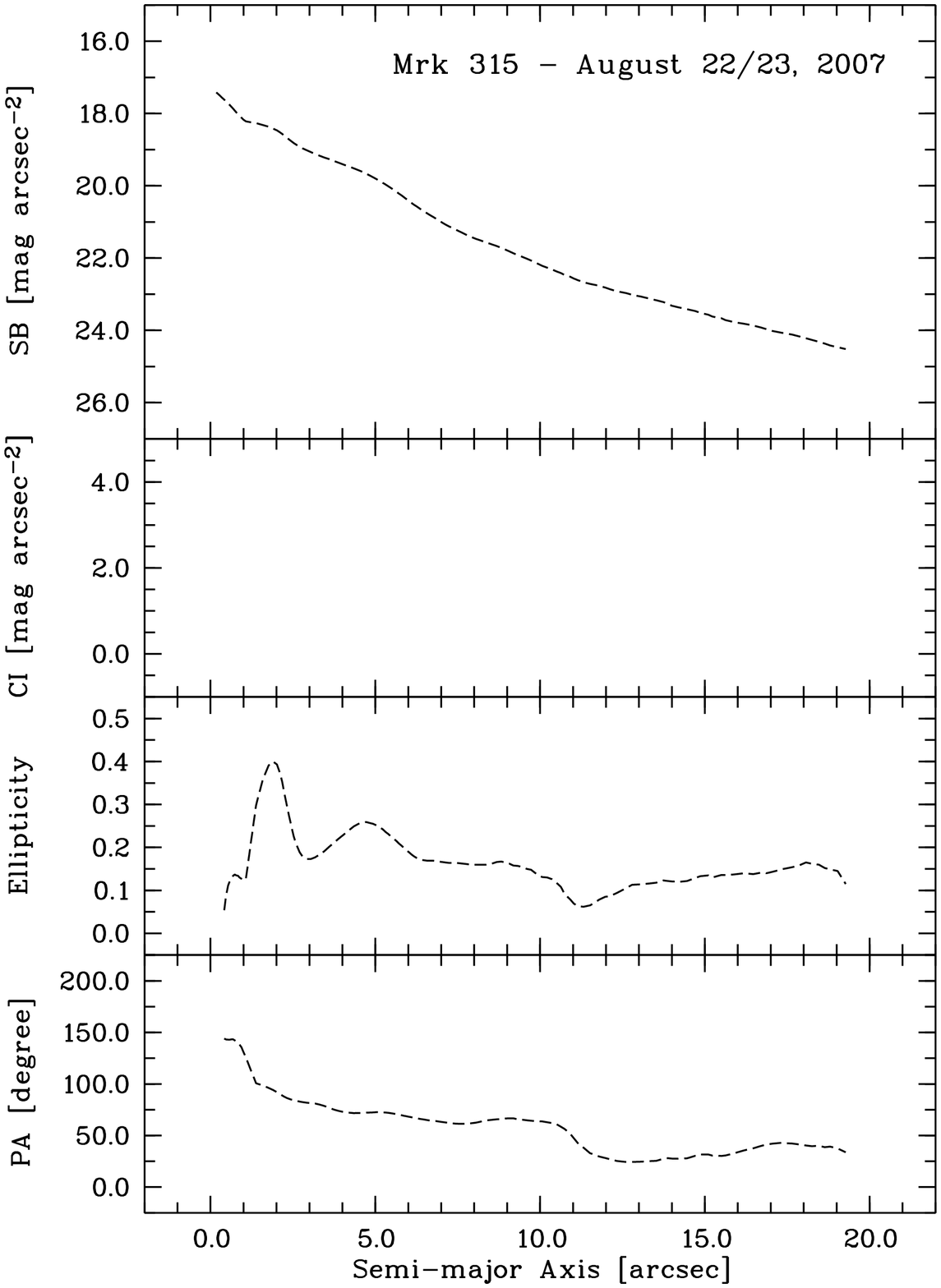}
\caption{Continued.}
   \end{figure*}
\setcounter{figure}{0}

\begin{figure*}[htbp]
\vspace{0.1cm}
   \centering
\includegraphics[width=5.6cm]{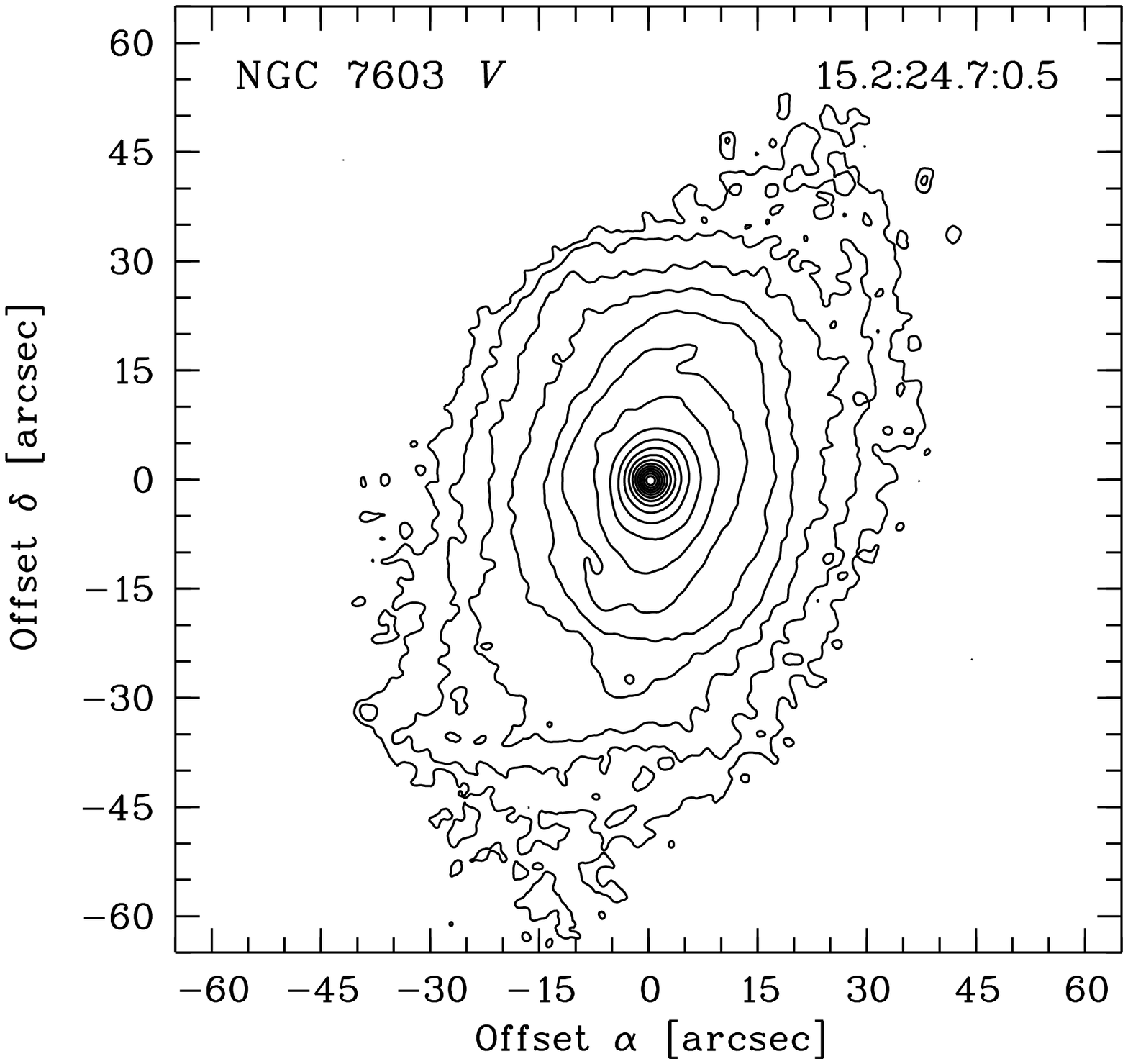}
\hspace{0.5cm}
\includegraphics[width=5.6cm]{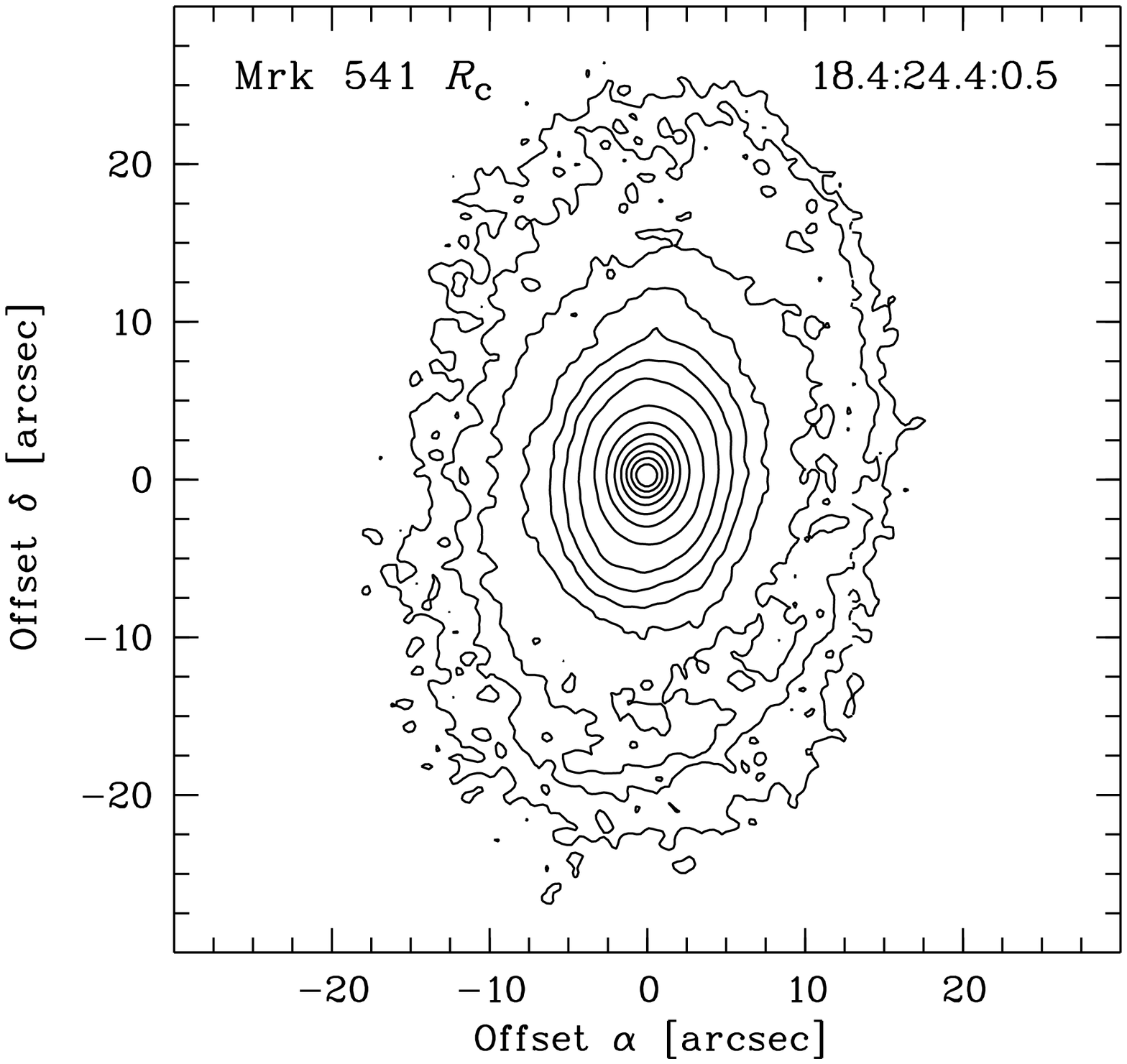}
\vspace{0.3cm}

\includegraphics[width=5.6cm]{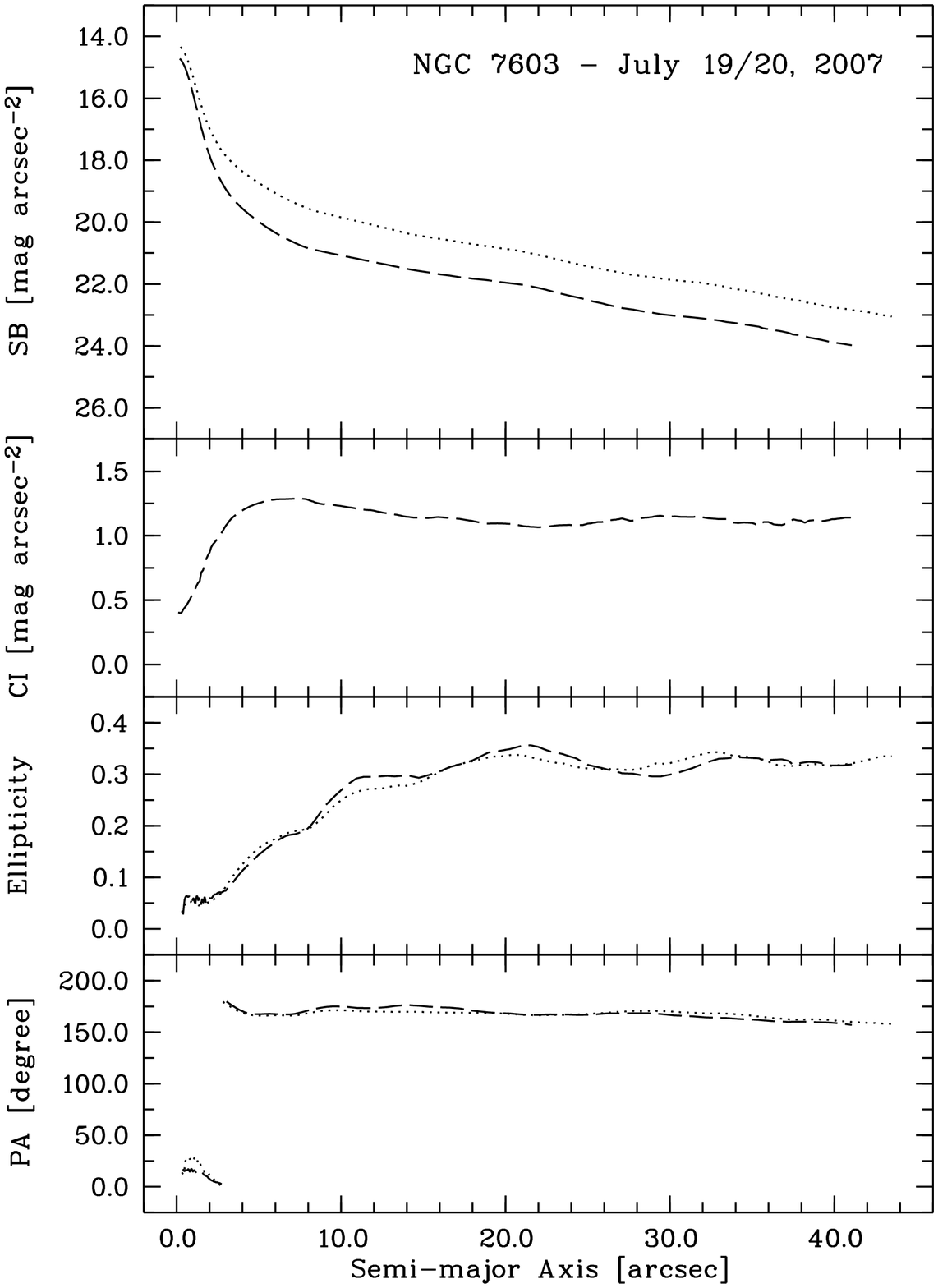}
\hspace{0.5cm}
\includegraphics[width=5.6cm]{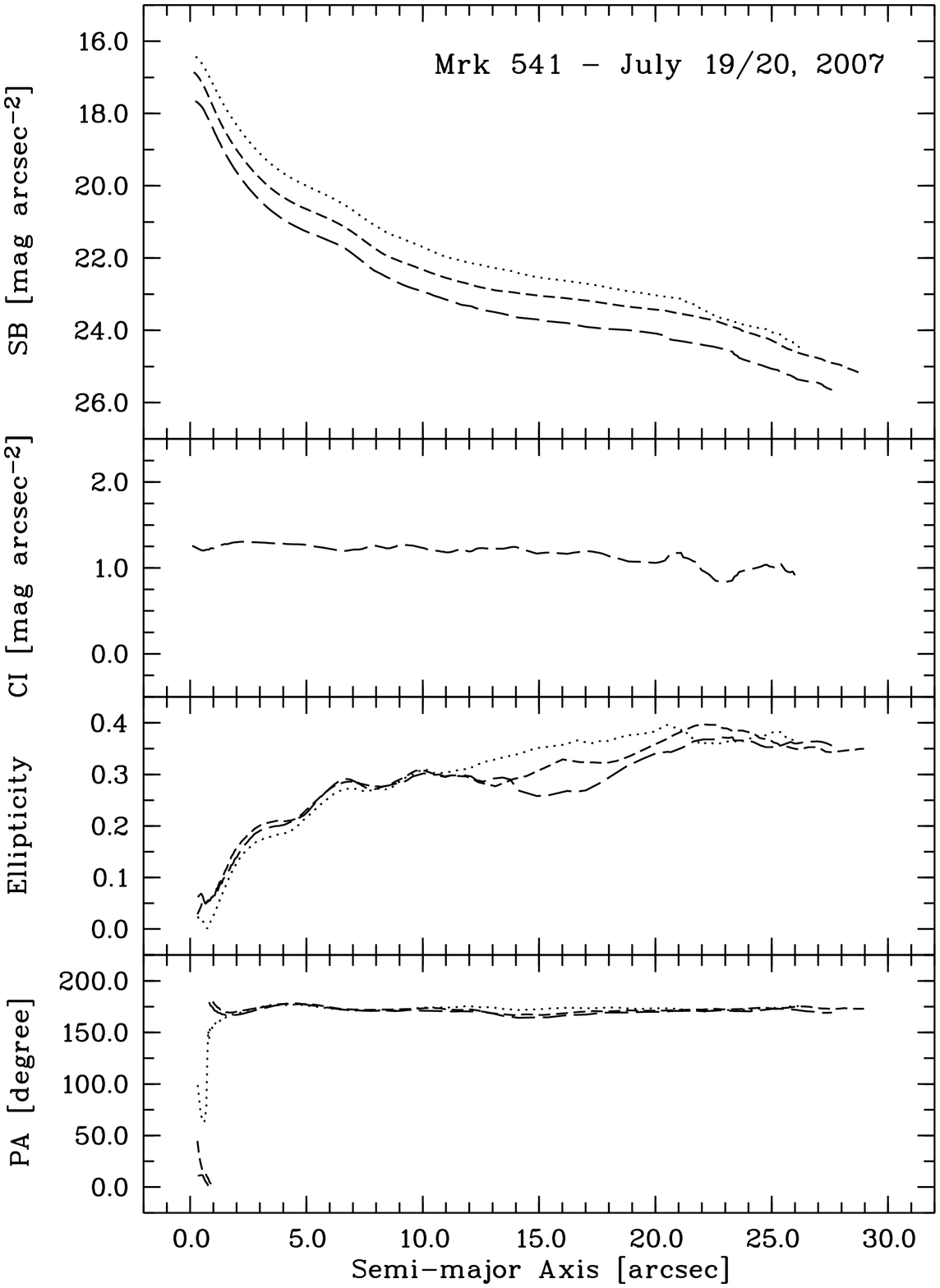}
      \caption{Continued.}
        \end{figure*}

\setcounter{figure}{1}

\section{Comments on Sy sample galaxies}
\label{indiv}

\begin{figure*}[t]
   \centering
\begin{minipage}[t]{5.6cm}   
\includegraphics[width=5.6cm]{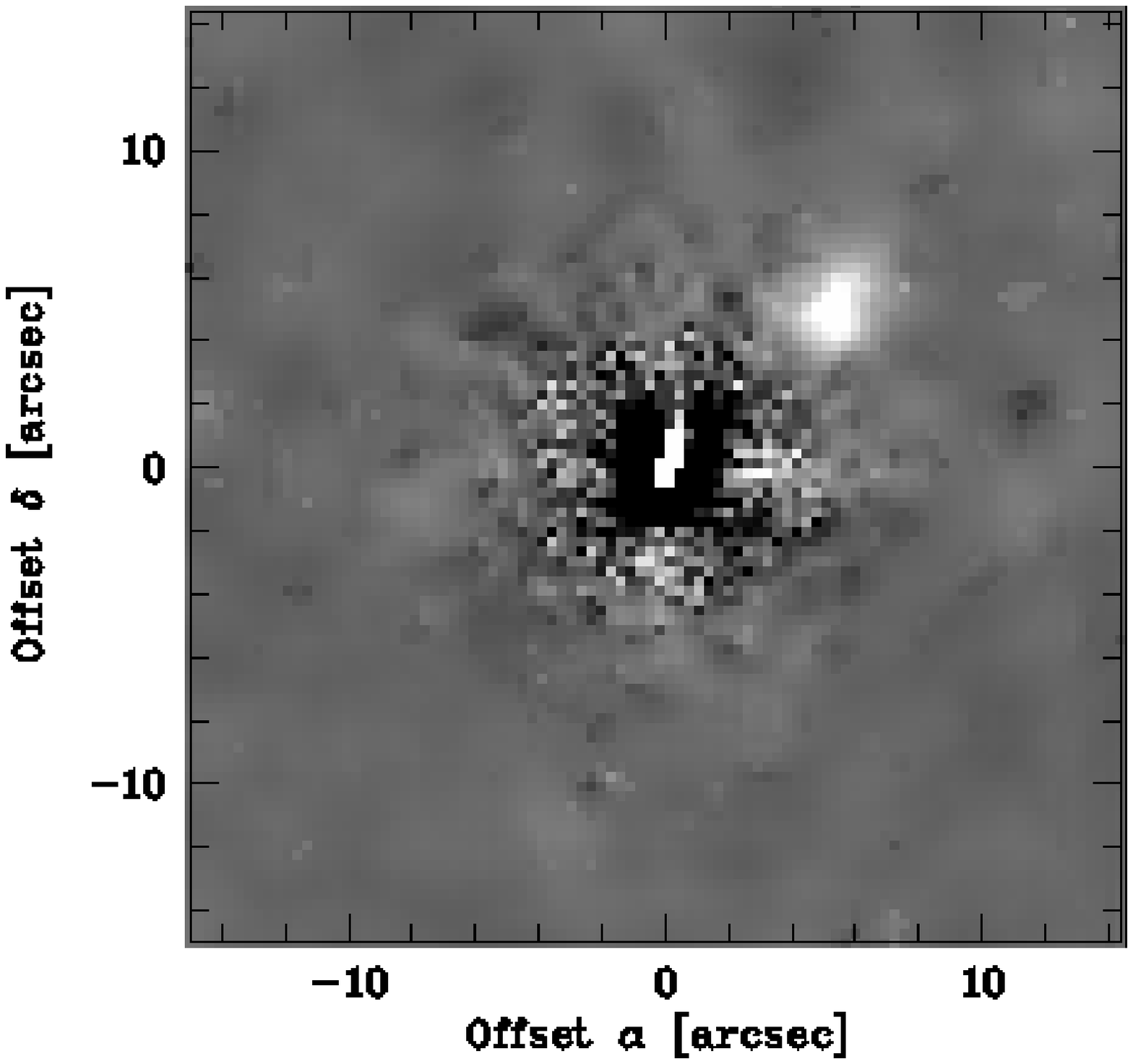}
      \caption{Mrk\,335 $I_{\rm \scriptstyle C}$ model-subtracted residual image. The extended feature is clearly visible.}
\label{Mrk335id}
\end{minipage}
\hspace{0.6cm}
\begin{minipage}[t]{5.6cm}   
\includegraphics[width=5.6cm]{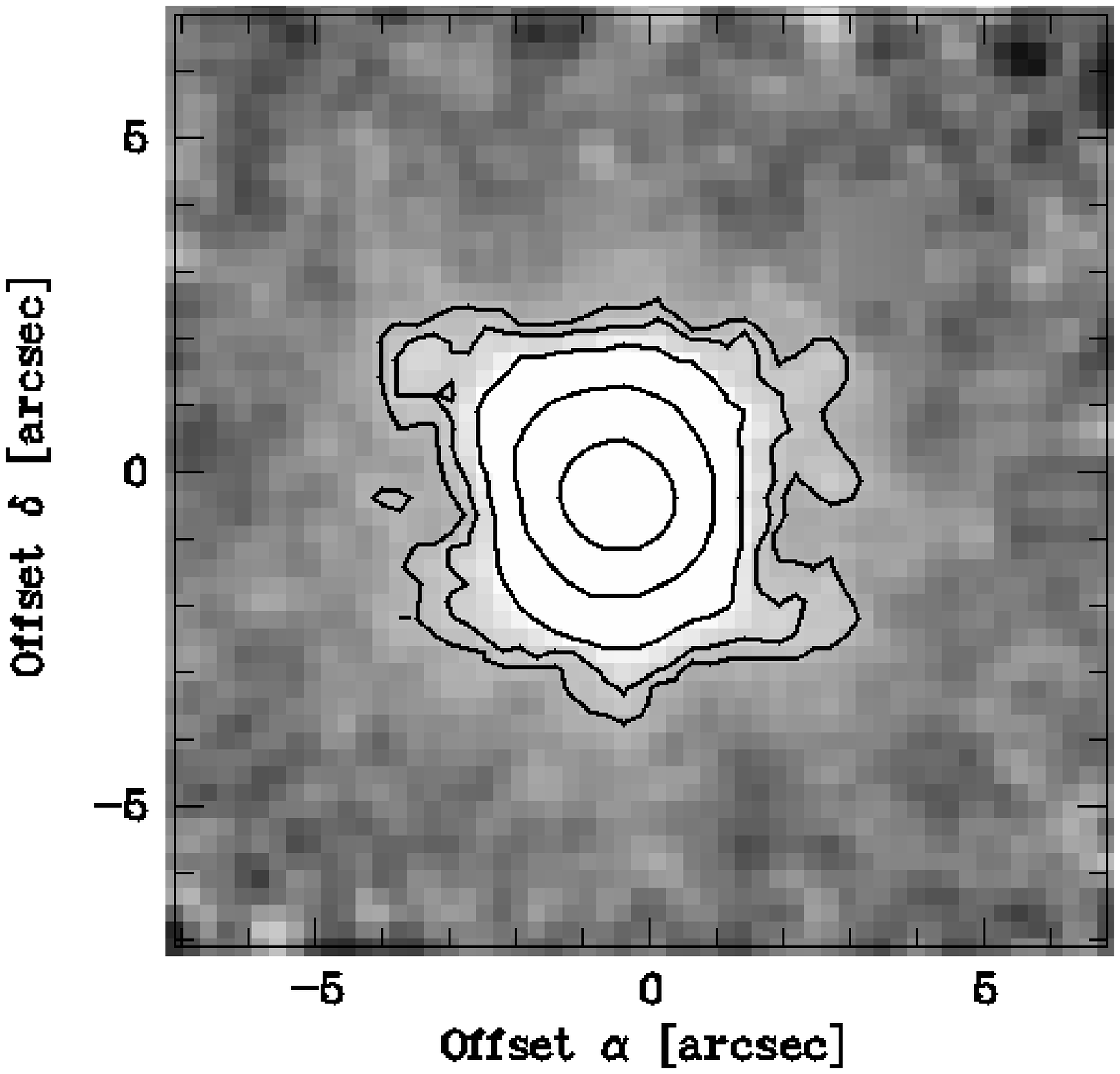}
      \caption{Mrk\,352 $R_{\rm \scriptstyle C}$ unsharp mask-divided residual image. The nuclear ring can be discerned.}
\label{05_Mrk352rud}
\end{minipage}
\hspace{0.6cm}
\begin{minipage}[t]{5.6cm}   
\includegraphics[width=5.6cm]{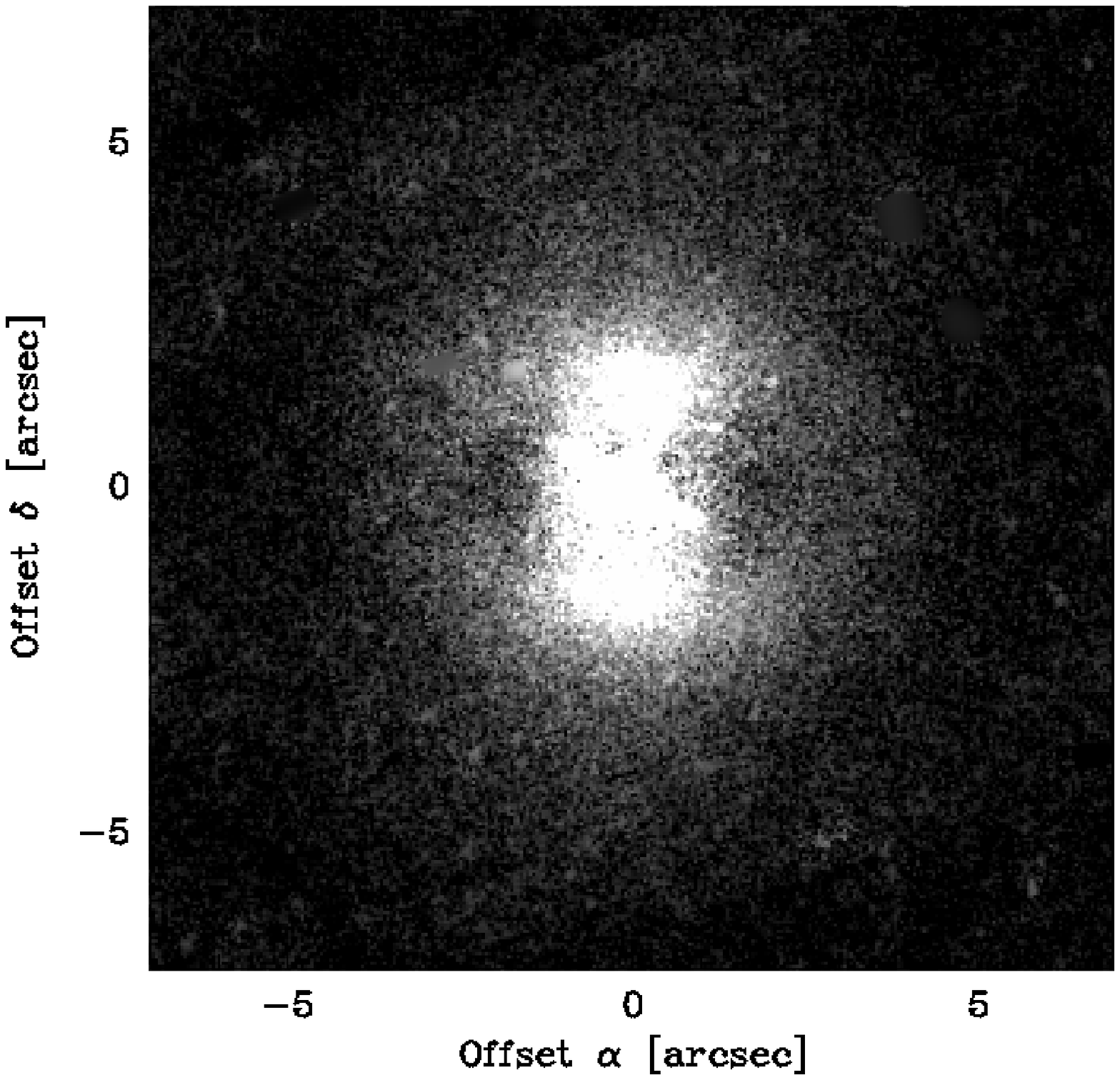}
      \caption{Mrk\,352 HST WFPC2 $F606W$ 2D model-subtracted residual image. The bar encircled by a ring can be traced.}
\label{05_Mrk352hst}
\end{minipage}
\end{figure*}

We discuss the Sy sample galaxies concerning features found or particularized by this study and cases where the influence of some features on the profiles is essential for the true structural parameters estimation via decomposition.

When discussing the contour maps and profiles, reference to Fig.\,\ref{contprof} goes without saying. 
Furthermore, CI and residual images/maps, contour maps, structure maps, and 2D model-subtracted residual images (Figs.\,\ref{Mrk335id}-\ref{38_Mrk541vi}) visualize the individual comments; dashed/dotted lines correspond to ellipticity maxima/minima. In all figures north (N) is up, east (E) to the left; the rest of the directions have been abbreviated as follows: south (S), west (W), northeast (NE), northwest (NW), southeast (SE), southwest (SW). In the CI images black is blue, white is red. Concerning the residual images, white are the excess structures. 

\subsection{Mrk\,335}  
\label{m335}

There is an extended feature at $PA\,$=$\,-43\degr$ with $I_{\rm \scriptstyle C}$ peak SB $7\farcs4$ away from the nucleus (Fig.\,\ref{Mrk335id}). 
The $B\,$--$\,V$, $V\,$--$\,R_{\rm \scriptstyle C}$, and $R_{\rm \scriptstyle C}\,$--$\,I_{\rm \scriptstyle C}$ CIs of this feature, estimated on the model-subtracted residual images by means of aperture photometry, are $1\fm48$, $0\fm75$, and $0\fm89$, respectively \citep[correction for Galactic extinction was applied after][]{SFD_98}. 
The feature is redshifted by about 280 $\rm km\,s^{-1}$ and shows a steep Balmer decrement \citep{FKS_83}. The authors  discussed ejection or merging and found the latter unlikely. \citet{M1_90} proposed an edge-on background galaxy as another interpretation. In our view, given the undisturbed appearance of the galaxy, a companion or a merging satellite at an early stage, seen through Mrk\,335, is a plausible hypothesis. No radio counterpart of the feature has been observed \citep[e.g.,][]{KPB_95}.

\subsection{III\,Zw\,2 (III\,Zw\,2A in NED)}

The contour map and the model-subtracted residual map reveal an elongation to the SE, which is a merging galaxy \citep[separation of about $7\arcsec$;][]{SSE_01}, connected through a tidal bridge with III\,Zw\,2, as can be seen in the residual images of \citet{VKP_06}, and an arm-like extension to the N, which is a tidal counterarm with knots of star formation \citep[see also][]{SSE_01}.
The tidal arm and the merging satellite produce an SB bump, accompanied by blue CI dips and an ellipticity maximum at a continuously changing PA in the region $a\,$$\approx$$\,5\arcsec\,$-$\,10\arcsec$.

\subsection{Mrk\,348} 

The galaxy has a distorted outer spiral structure \citep[e.g.,][]{PE_93} due to interaction with its close companion, which shows a CI gradient \citep{ATP_02}.

 The ellipticity increase and the SB bump beyond  $a\,$$\approx$$\,45\arcsec$ are related to the diffuse stretched outer spiral structure. The inner spiral structure leads to a continuous change of the PA. 
The weak SB bump, visible in $B$, and the blue $B\,$--$\,I_{\rm \scriptstyle C}$ dip at $a\,$$\approx$$\,5\arcsec$ are produced by a blue nuclear ring, more pronounced to the S \citep{ATP_02}.

\subsection{I\,Zw\,1}

The galaxy has asymmetric knotty spiral arms (the NW one being more pronounced) that may be of tidal origin \citep{SS_00,CS_01}. They produce an SB bump, CI dips, and an ellipticity maximum at a continuously increasing PA in the region $a\,$$\approx$$\,7\arcsec\,$-$\,13\arcsec$.

\begin{figure*}[htbp]
   \centering
\begin{minipage}[t]{5.6cm}   
\includegraphics[width=5.6cm]{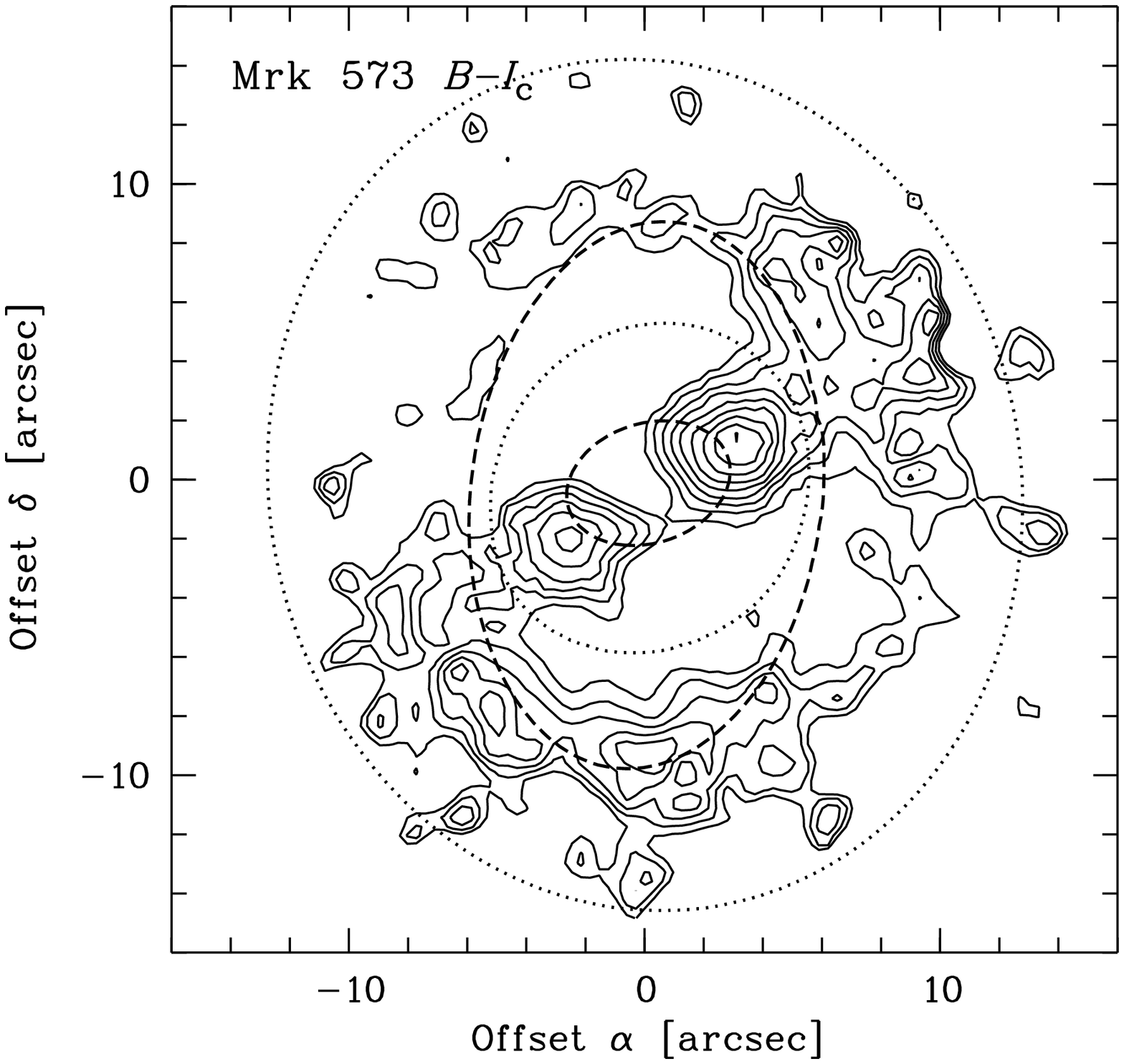}
      \caption{Mrk\,573 $B\,$--$\,I_{\rm \scriptstyle C}$ map. The contours range from 1.2 to 1.95 with a step of $0.05~\rm mag\,\rm arcsec^{-2}$. Overplotted are the $B$ model contours corresponding to the first two ellipticity maxima and the minima following them. The first ellipticity maximum is related to the ionization cones. }
\label{06_m573cbi_Ee}
\end{minipage}
\hspace{0.6cm}
\begin{minipage}[t]{5.6cm}   
\includegraphics[width=5.6cm]{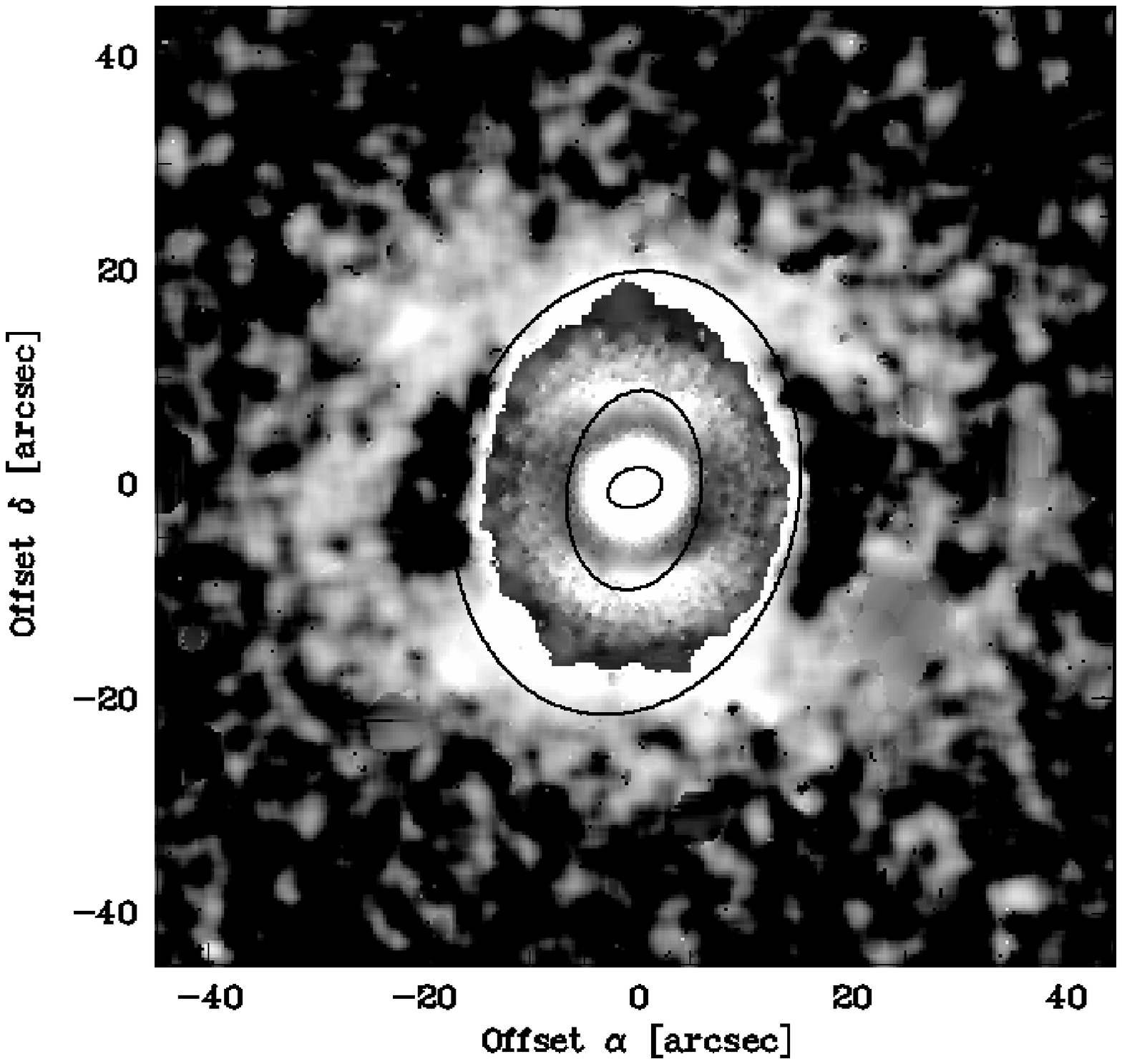}
      \caption{Mrk\,573 $R_{\rm \scriptstyle C}$ residual image, composite of an unsharp mask-divided/-subtracted one, so that there could be traced the arcs and the outer ring. Overplotted are the model contours corresponding to the $V$ ellipticity maxima.}
\label{06_Mrk573rDx2_vE}
\end{minipage}
\hspace{0.6cm}
\begin{minipage}[t]{5.6cm}   
\includegraphics[width=5.6cm]{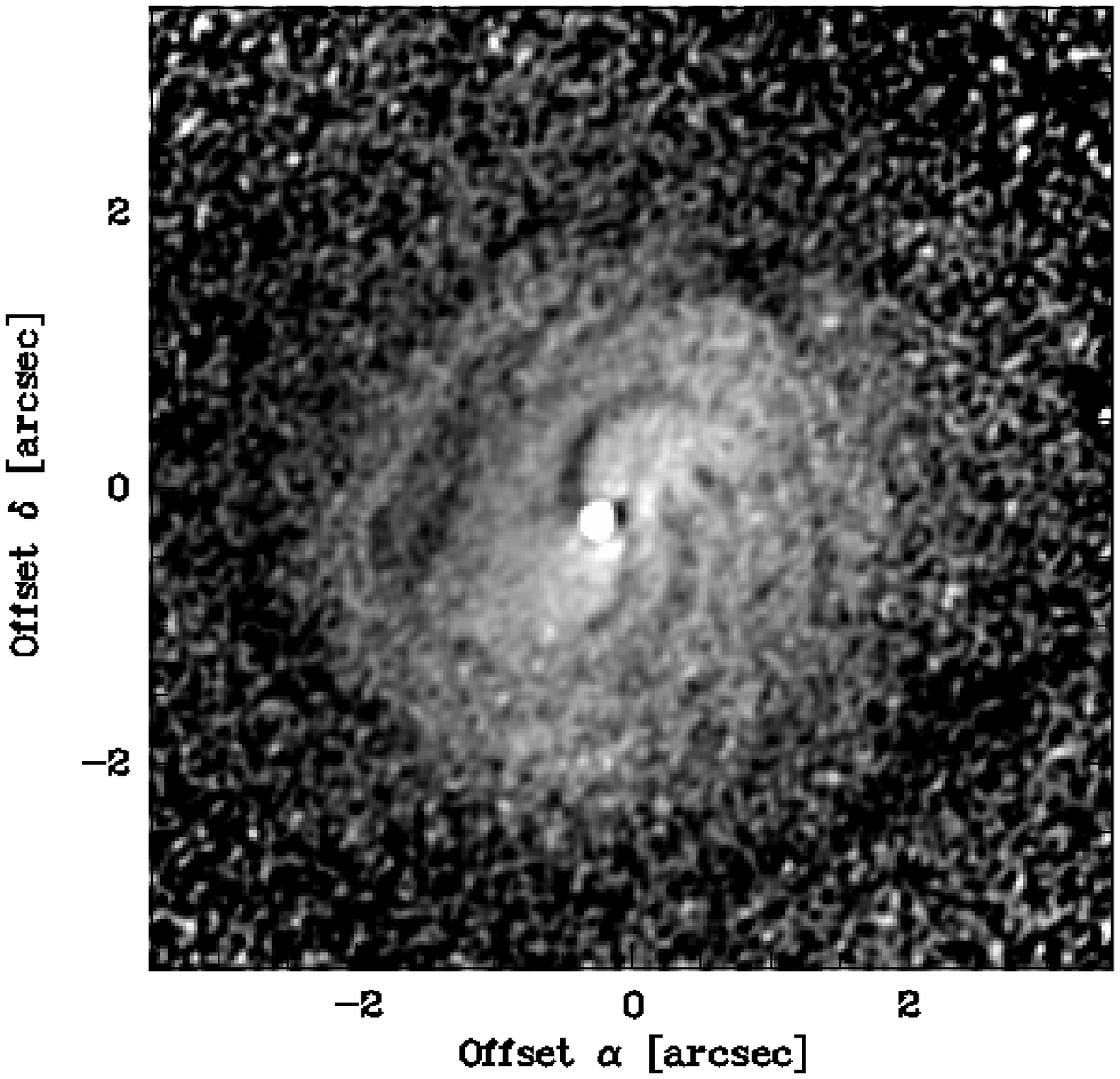}
      \caption{Mrk\,590 HST ACS/HRC $F550M$ structure map. The dust lanes can be traced.}
\label{07_Mrk590_hst2}
\end{minipage}
\end{figure*}

\subsection{Mrk\,352} 

There is evidence of a nuclear bar around $a\,$$=$$\,2\arcsec$: the ellipticity profile shows a peak, accompanied by a plateau on the PA profile; however, there is no obvious SB bump. Moreover, the unsharp masked residual reveals a nuclear ring (Fig.\,\ref{05_Mrk352rud}). To further verify our findings, we processed archival HST WFPC2\footnote{Wide-Field Planetary Camera\,2.} $F606W$ data. Subtraction of a fitted 2D bulge-disk model (the nucleus was masked out) reveals a $\theta$-shaped residual (Fig.\,\ref{05_Mrk352hst}), most probably due to a nuclear bar surrounded by a ring. In the region around $a\,$$=$$\,2\arcsec$ the ellipticity and PA profiles show a similar behaviour like ours and, in addition, the SB profile reveals a weak bump; we used the HST profiles to derive the bar parameters. Furthermore, the SB profile has a weak bump, accompanied by ellipticity peaks at $a\,$$\approx$$\,4\arcsec$, which we associate with the ring. Besides this, there is no significant dust \citep{DCK_06} that could disturb the corresponding isophotes.

\subsection{Mrk\,573} 

The $B\,$--$\,I_{\rm \scriptstyle C}$ map (Fig.\,\ref{06_m573cbi_Ee}) exhibits two blue regions corresponding to the ionization cones \citep{FWF_99}. They result in a blue dip on the CI profiles and $BV$ ellipticity maximum at $a\,$$\approx$$\,3\arcsec$. The $V$ maximum was misinterpreted as evidence of a secondary bar\footnote{The galaxy actually harbours a secondary bar of $a\,$$=$$\,1\farcs2$ \citep{MP_01}.} by \citet{AMS_98}. The second ellipticity maximum (at $a\,$$\approx$$\,9\arcsec$) corresponds to a bar.
The $B\,$--$\,I_{\rm \scriptstyle C}$ map also reveals a couple of arcs \citep[labelled SE3 and NW3 by][]{FWF_99}, which appears as a broken ring in the unsharp masked residual image (Fig.\,\ref{06_Mrk573rDx2_vE}).

 The SB profiles in the region $25\arcsec\,$-$\,40\arcsec$ are affected by an outer ring \citep[][]{PR_95}, clearly outlined in Fig.\,\ref{06_Mrk573rDx2_vE}. The profiles show signs for another bar at $a\,$$\approx$$\,21\arcsec$ as already proposed by \citet{LSK_02}. After a 2D elliptical ring model subtraction, the ellipticity maximum decreases and the weak SB bump vanishes, which keeps us from considering this galaxy triple barred \citep[see also][]{E_04}.

\begin{figure*}[htbp]
   \centering
\begin{minipage}[t]{5.6cm}   
\includegraphics[width=5.6cm]{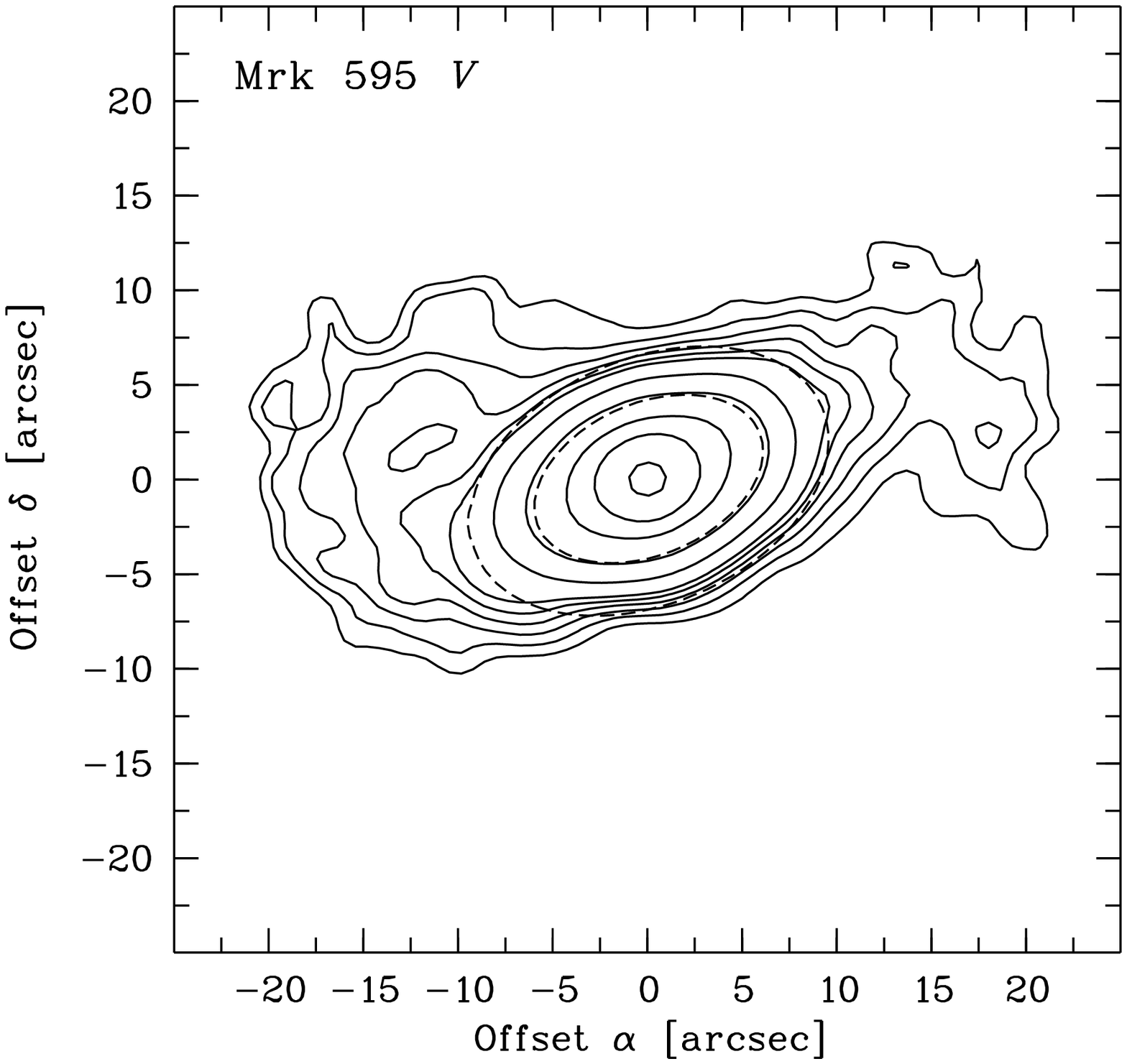}
      \caption{Mrk\,595 $V$ unsharp mask-divided residual map. Overplotted are the $V$ model contours corresponding to the two ellipticity peaks. There can be traced the oval/lens, together with the spiral arm stubs.}
\label{09_m595cvD23}
\end{minipage}
\hspace{0.6cm}
\begin{minipage}[t]{5.6cm}   
\includegraphics[width=5.6cm]{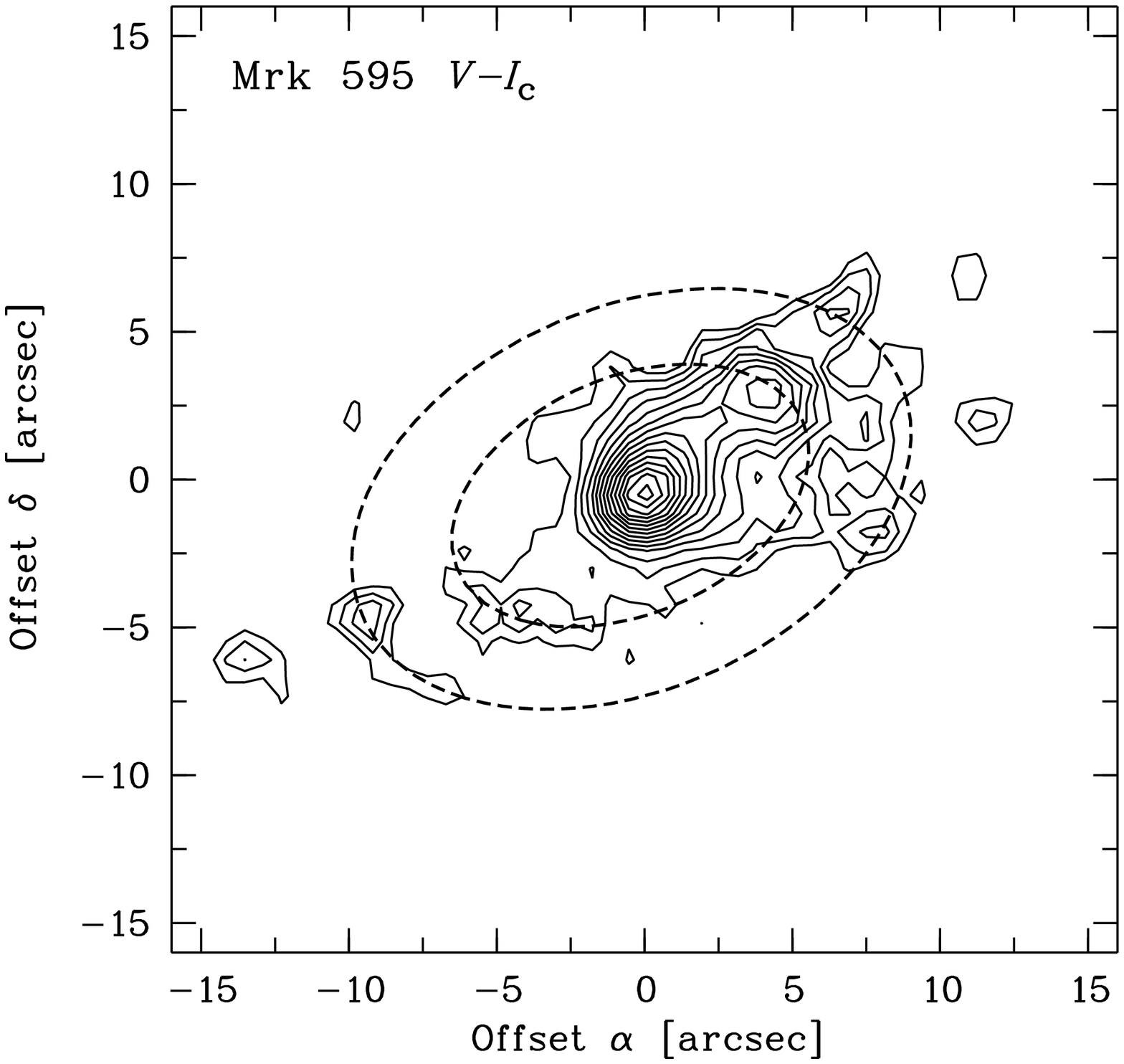}
      \caption{Mrk\,595 $V\,$--$\,I_{\rm \scriptstyle C}$ map. The contour levels range from 1 to 1.4 with a step of $0.025\,\rm mag\,\rm arcsec^{-2}$; overplotted are the $V$ model contours corresponding to the two ellipticity peaks. The mapped region roughly reflects the [\ion{O}{iii}] emission.}
\label{09_m595cvi_Ee}
\end{minipage}
\hspace{0.6cm}
\begin{minipage}[t]{5.6cm}   
\includegraphics[width=5.6cm]{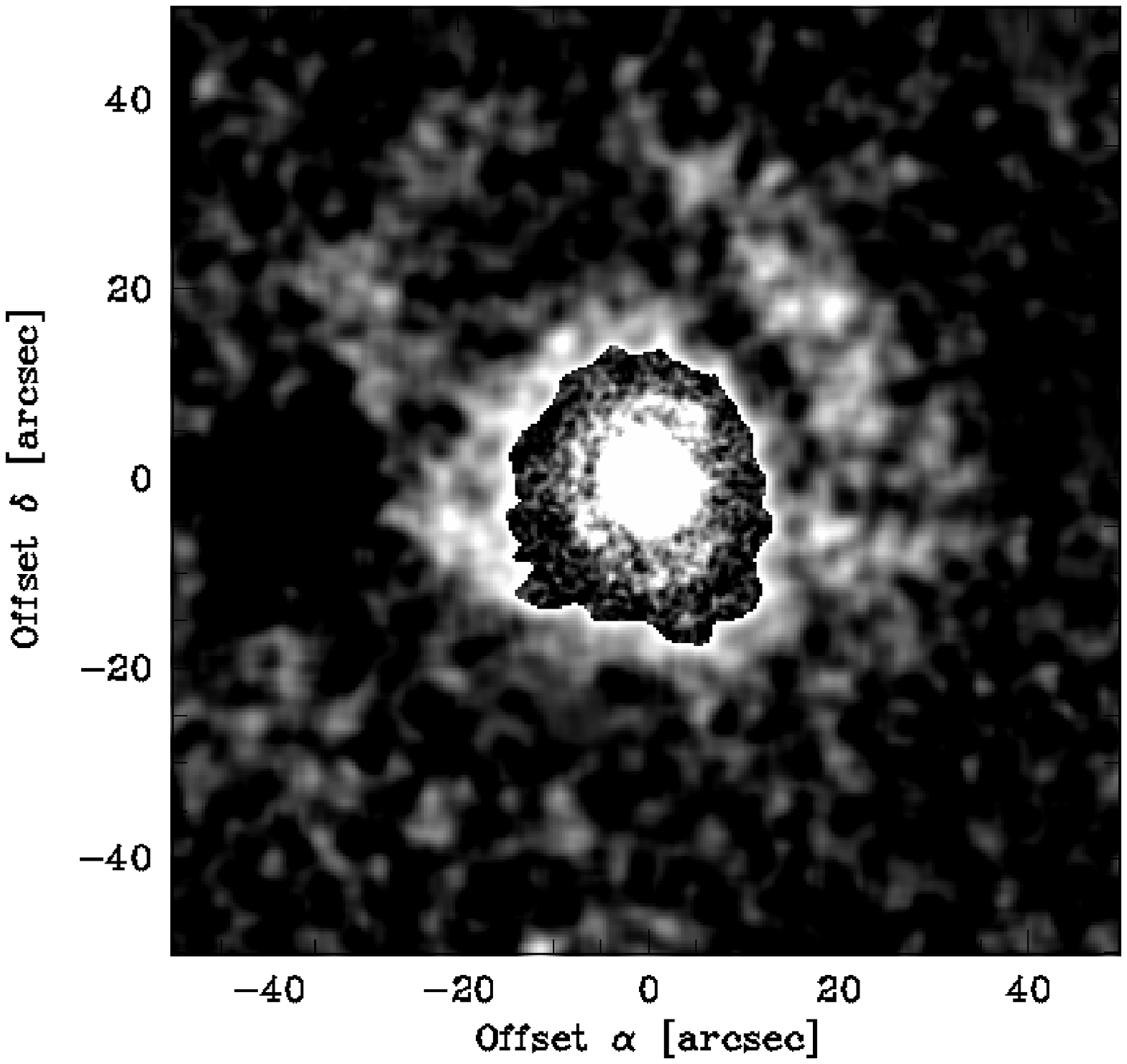}
      \caption{Ark\,120 $B$ residual image, composite of two unsharp mask-subtracted residual images so that there could be traced the inner knotty ring and the two faint arcs running to the N.}
\label{11_Ark120bdm2o}
\end{minipage}
\end{figure*}

\subsection{Mrk\,590} 

\citet{PM_02} reported a nuclear bar, based on HST WFPC2 images. In their structure map, a ring-like structure around the bar could also be seen, best traced to the NE. We processed archival HST ACS/HRC\footnote{Advanced Camera for Surveys/High Resolution Channel.} $F550M$ data and created a structure map, which reveals several dust lanes. A couple of them appear as straight dust lanes along bar leading edges (although the bar signatures are not very evident, Fig.\,\ref{07_Mrk590_hst2}) and could be traced down to about 70\,pc ($0\farcs14$) in radius.

\subsection{Mrk\,595} 

The profiles and contour map suggest a bar-like structure, which is most probably an oval/lens, given its small deprojected ellipticity of 0.13\footnote{Bars and bar-like features detected in the Sy sample will be discussed in Slavcheva-Mihova \& Mihov (in prep.).}. Weak spiral arm stubs can be seen in the unsharp masked residual map (Fig.\,\ref{09_m595cvD23}). 
The $V\,$--$\,I_{\rm \scriptstyle C}$ map (Fig.\,\ref{09_m595cvi_Ee}) reveals a blue region to the NW, which roughly reflects the [\ion{O}{iii}] emission \citep[see also][]{MWT1_96}.

The oval/lens produces an ellipticity peak and an SB bump at an almost constant PA in the region $a\,$$\approx$$\,4\arcsec\,$-$\,8\arcsec$. The $BV$ ellipticity peak, accompanied by a weak $B$ SB bump and blue CI dips around $a\,$$=$$\,11\arcsec$, is caused by the spiral arm stubs. The [\ion{O}{iii}] emission results in enhanced ellipticity in $V$ and in blue $V\,$--$\,I_{\rm \scriptstyle C}$ dips in the regions of the two ellipticity peaks (Fig.\,\ref{09_m595cvi_Ee}).

 \subsection{3C\,120}

The contour map reveals strongly disturbed isophotes due to underlying features \citep{SWB_89,HVS_95}, which affect the SB profiles.

\begin{figure*}[t]
   \centering
\begin{minipage}[t]{5.6cm}   
\includegraphics[width=5.6cm]{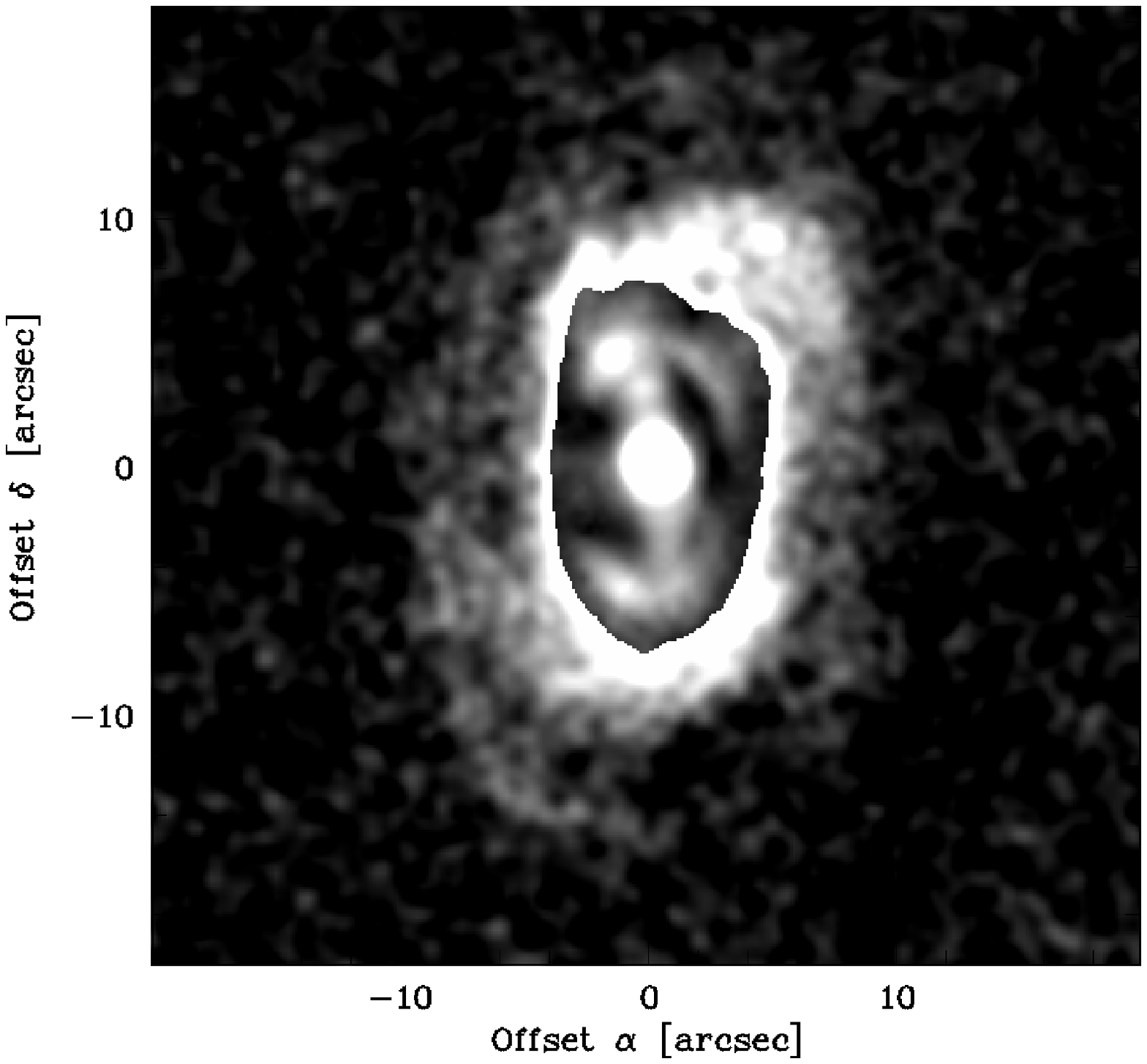}
      \caption{Mrk\,376 $R_{\rm \scriptstyle C}$ residual image, composite of an unsharp mask-divided/-subtracted one. The bent bar, the clumpy ring around it, and the two spiral arms, forming a weak outer pseudo-ring, are clearly outlined.}
\label{12_Mrk376rdmed2n}
\end{minipage}
\hspace{0.6cm}
\begin{minipage}[t]{5.6cm}   
\includegraphics[width=5.6cm]{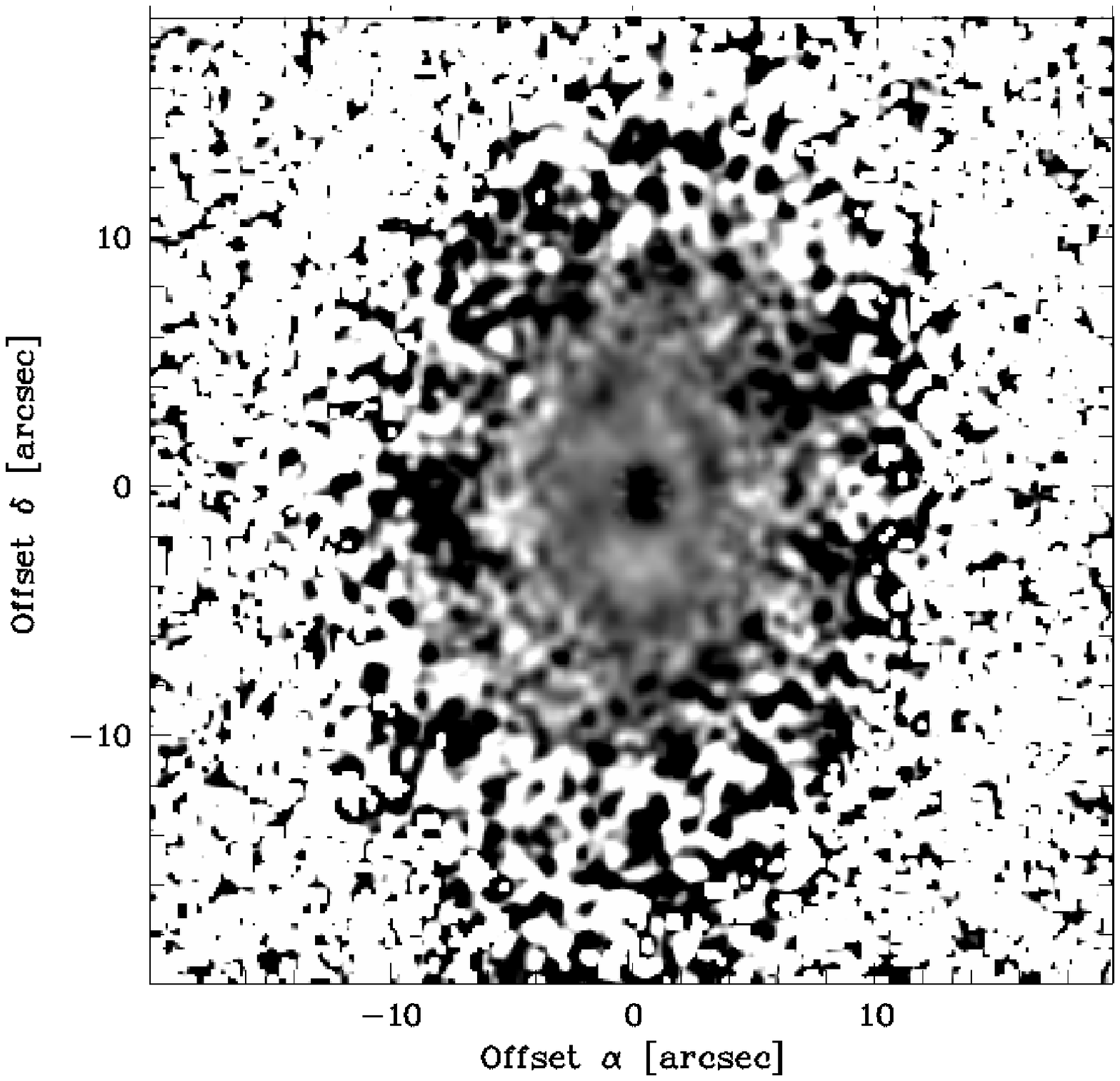}
      \caption{Mrk\,376 $V\,$--$\,I_{\rm \scriptstyle C}$ image; the CI coding ranges from $1.1$ to $1.6\,\rm mag\,\rm arcsec^{-2}$. The clumpy blue ring can be traced.}
\label{12_Mrk376vi}
\end{minipage}
\hspace{0.6cm}
\begin{minipage}[t]{5.6cm}   
\includegraphics[width=5.6cm]{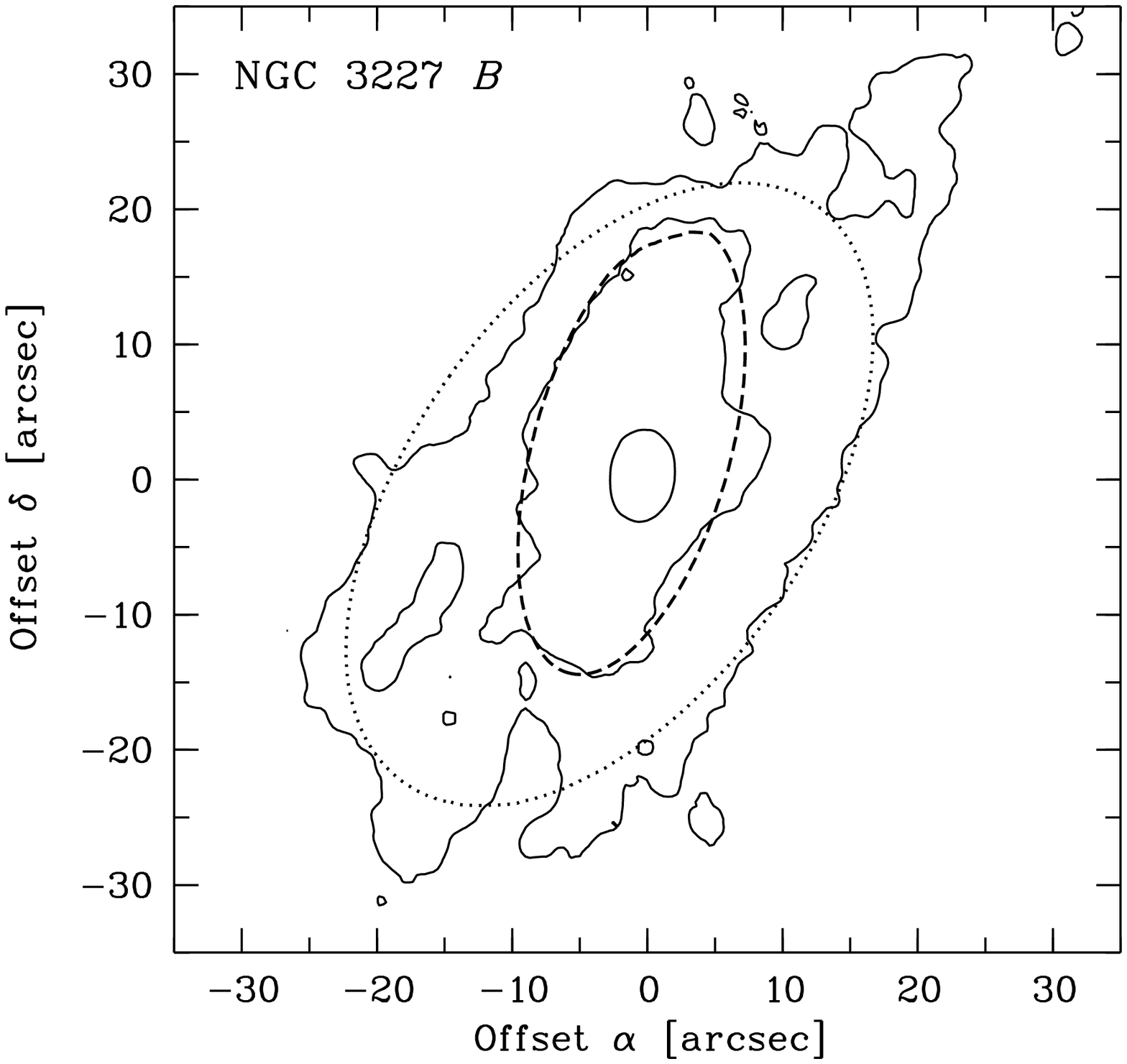}
      \caption{NGC\,3227 $B$ contour map. The model levels, corresponding to the ellipticity maximum and minimum around $a\,$$=$$20\arcsec$, are plotted over the relevant galaxy levels. The dust imparts a disturbed appearance to the isophotes.}
\label{n3227cb_s}
\end{minipage}
\end{figure*}

\subsection{Ark\,120}

The unsharp masked residual image (Fig.\,\ref{11_Ark120bdm2o}) reveals a blue knotty ring ($a\,$$\times$$\,b\,$$\approx$$9\arcsec$$\times$$8\arcsec$) and a couple of arcs extending to the N on either side of the nucleus, the W one being more pronounced. The ring produces an SB bump and a blue dip on the $B\,$--$\,R_{\rm \scriptstyle C}$ profile.

\subsection{Mrk\,376}
 
The unsharp masked residual image (Fig.\,\ref{12_Mrk376rdmed2n}) reveals a bent bar. It is encircled by a ring ($a\,$$\times$$\,b\,$$\approx$$5\arcsec$$\times$$3\arcsec$) with knots of star formation that are outstanding in the $V\,$--$\,I_{\rm \scriptstyle C}$ image (Fig.\,\ref{12_Mrk376vi}). The two spiral arms get noticeably not as bright and form a weak outer pseudo-ring ($a\,$$\times$$\,b\,$$\approx$$14\arcsec$$\times$$9\arcsec$). The bar results in an ellipticity maximum around $a\,$$=$$\,4\arcsec$ (the less pronounced ellipticity maximum in $B$ is due to seeing change), accompanied by an SB bump at an almost constant PA. The inner ring produces an SB bump (best expressed in $B$), which merges with the bar bump, and a CI dip. The spiral structure results in weak SB bumps further out. Note that ``spiral arms or a bar-like structure'' in the NIR were hinted at by \citet{HRD_93}.

\subsection{Mrk\,79} 

 The galaxy is asymmetric: the NW spiral arm is truncated and the NE bar side is diffuse. The outer isophotes have a rectangular shape \citep[see also][]{WW_77}. 
The bar produces an SB bump, together with an ellipticity maximum in the region $a\,$$\approx$$\,6\arcsec\,$-$\,17\arcsec$; the ellipticity is highest in $B$ due to the partial fitting of the spiral arm beginnings. The wavelength-dependent ellipticity bump next to the maximum, accompanied by an SB bump and a blue CI dip, is produced by the inner part of the spiral structure. The behaviour of the profiles\footnote{The profiles are composed of two data sets~-- the last $\approx$\,$20\arcsec$ are extracted from images, which are deeper but with worse seeing (Febr.\,16/17,\,1999).} beyond $a\,$=$\,24\arcsec$, is related to the outer spiral structure.

\subsection{Mrk\,382} 

Mrk\,382 has a bar with a ring ($a\,$$\times$$\,b\,$$\approx$$9\arcsec$$\times$$7\arcsec$) around it. The spiral structure forms an outer pseudo-ring ($a\,$$\times$$\,b\,$$\approx$$17\arcsec$$\times$$12\arcsec$). 
The ellipticity maximum, accompanied by an SB bump around  $a\,$$=$$\,7\arcsec$results from the bar; the second ellipticity maximum and the corresponding SB bump are due to the spiral structure.

\subsection{NGC\,3227}
\label{ngc3227}

The ellipticity profile shows a broad maximum in the region $a\,$$\approx$$\,25\arcsec\,$-$\,85\arcsec$, double-peaked in $B$, accompanied by SB bumps, best expressed in $B$.
The behaviour of the profiles around the inner part of the ellipticity maximum is dominated by the bar, 
and in the region of the outer part of the ellipticity maximum, is due to the spiral arm beginnings, 
which result in blue $B\,$--$\,I_{\rm \scriptstyle C}$ dips and a slight PA shift, best expressed in $B$. 

\begin{figure*}[htbp]
   \centering
\begin{minipage}[t]{5.6cm}   
\includegraphics[width=5.6cm]{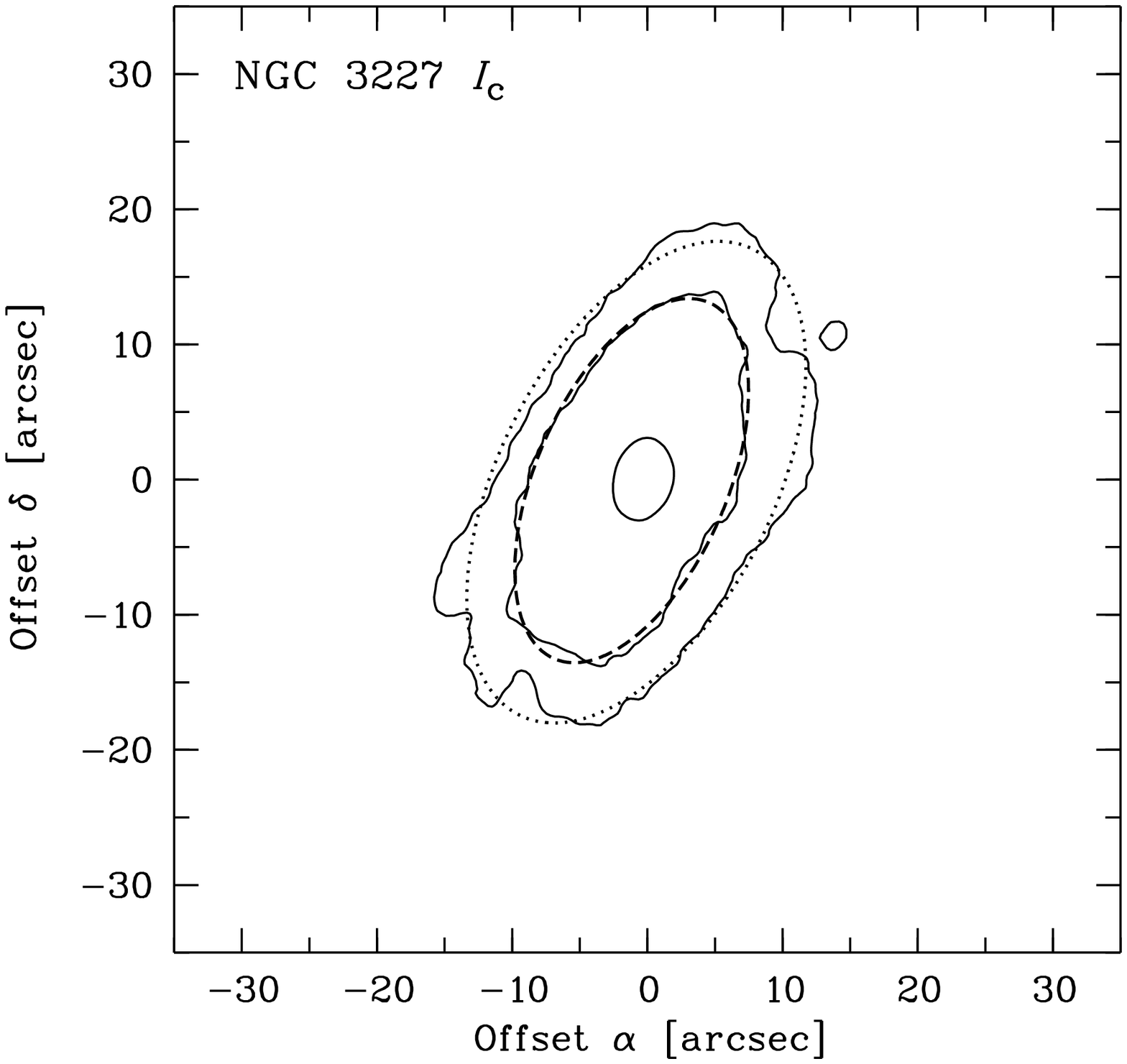}
      \caption{NGC\,3227 $I_{\rm \scriptstyle C}$ contour map. The model levels, corresponding to the ellipticity maximum and minimum around $a\,$$=$$20\arcsec$, are plotted over the relevant galaxy levels. Compare with Fig.\,\ref{n3227cb_s}.}
\label{n3227ci_s}
\end{minipage}
\hspace{0.6cm}
\begin{minipage}[t]{5.6cm}   
\includegraphics[width=5.6cm]{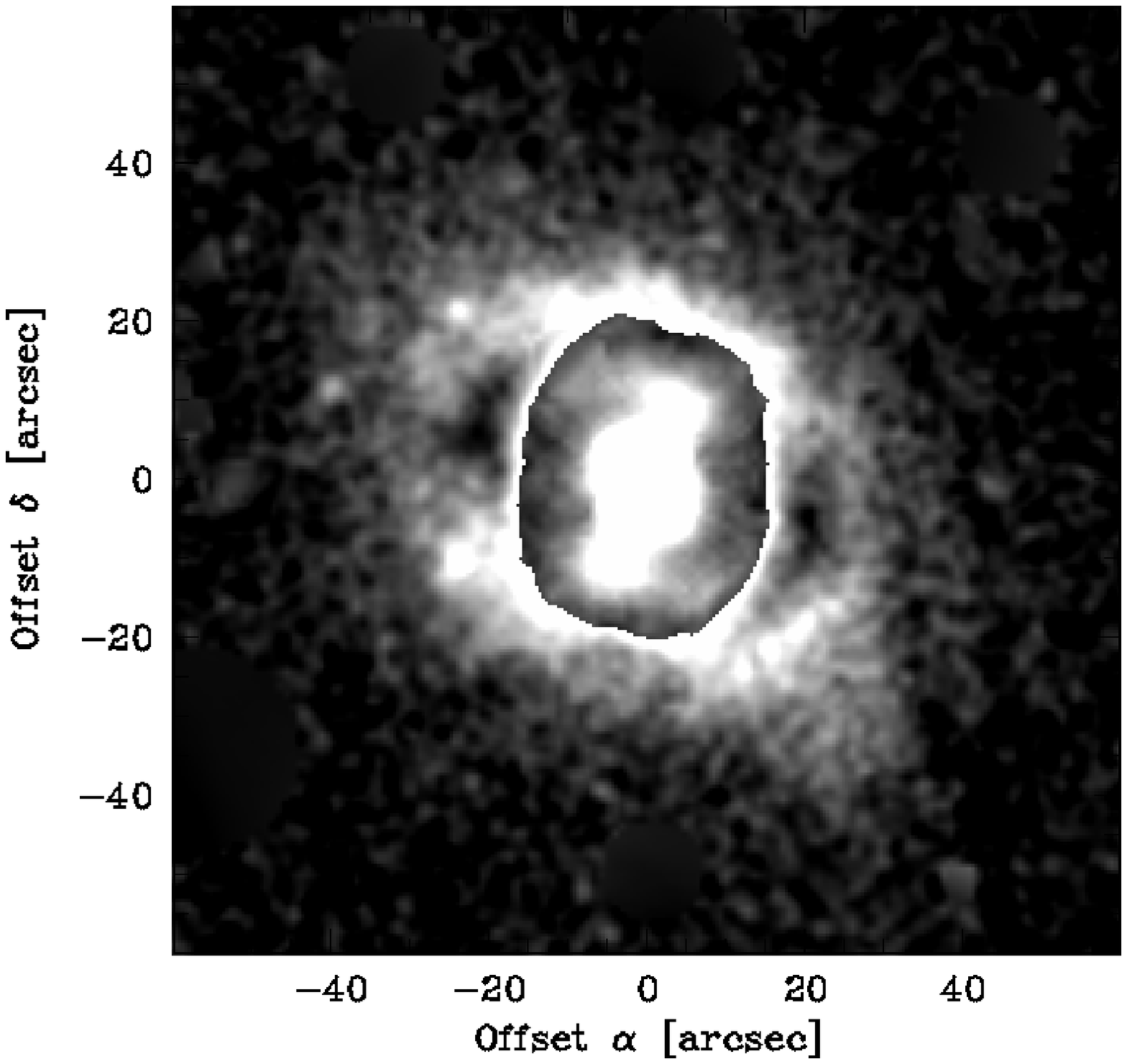}
      \caption{NGC\,3516 $R_{\rm \scriptstyle C}$ residual image, composite of an unsharp mask-divided/-subtracted one, so that the bar and rings can be traced.}
\label{16_NGC3516rsD}
\end{minipage}
\hspace{0.6cm}
\begin{minipage}[t]{5.6cm}   
\includegraphics[width=5.6cm]{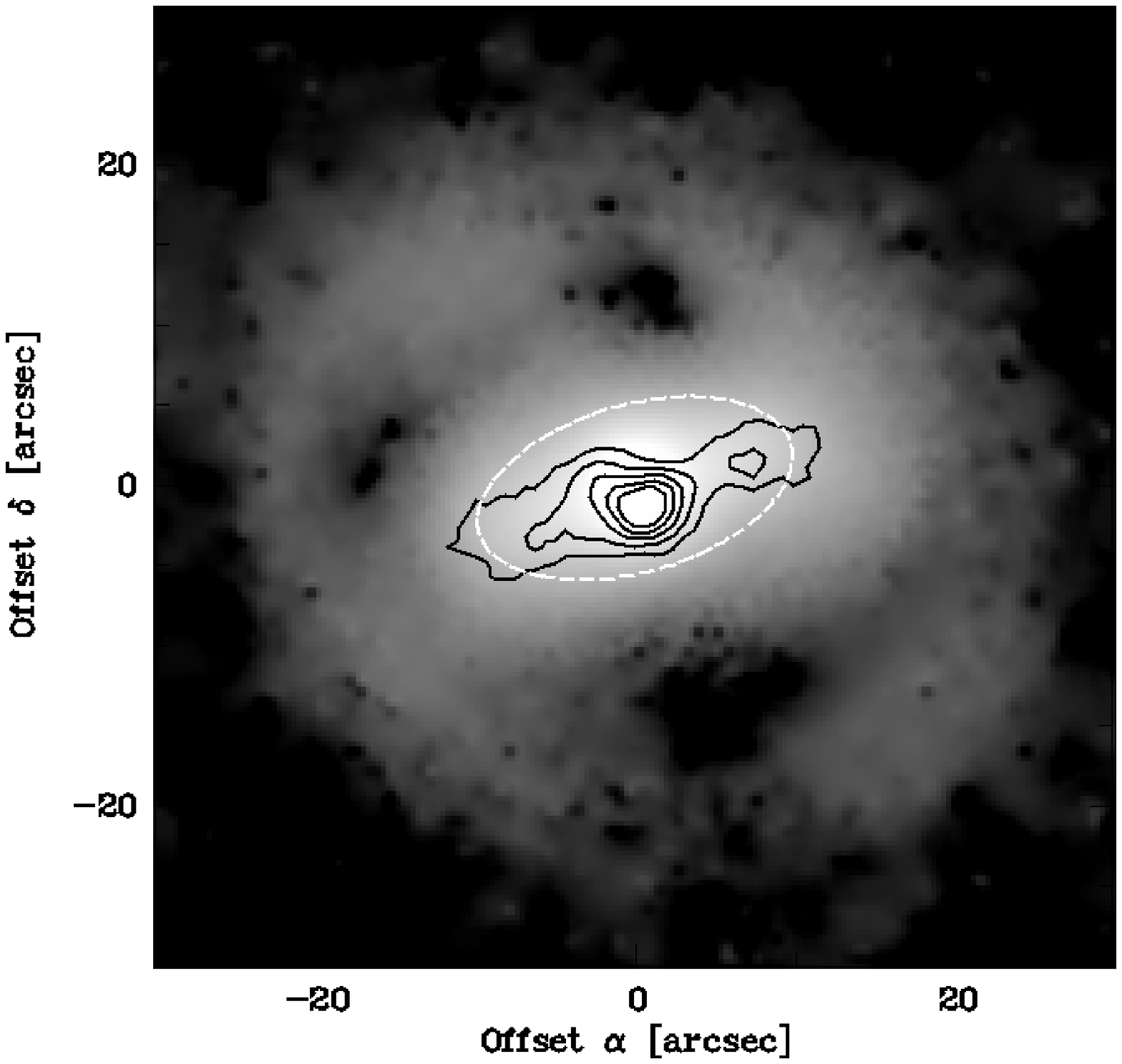}
      \caption{Mrk\,766 $V$ unsharp mask-subtracted residual image. Overplotted is the $V\,$--$\,I_{\rm \scriptstyle C}$ map (the levels range from $0.9$ to $1.15$ with a step of $0.05~\rm mag\,\rm arcsec^{-2}$; solid) and the $V$ ellipticity maximum. Note the outer pseudo-ring, the blue region oriented NW-SE, and the blue protrusion to the NE.}
\label{19_Mrk766vdmedl_viE}
\end{minipage}
\end{figure*}

When analysing a decomposition residual image, \citet{GP_97} argued for an N-S stellar bar of $a\,$$\approx$$\,1.6\,$kpc ($\approx$\,$21\arcsec$, assumed distance to the galaxy 15.6\,Mpc). Furthermore, based on isophotal analysis, \citet{GS_06} reported a bar of $a\,$$=$$1.9\,$kpc ($\approx$\,$22\arcsec$, assumed distance to the galaxy 17.6\,Mpc). In this region~-- more precisely, around $a\,$$=$$\,17\arcsec$~-- the profiles show complex behaviour: the ellipticity profile has a wavelength-dependent maximum, accompanied by a constant (but also wavelength-dependent) PA. At variance with the above authors, we attribute this to dust absorption: the dust location is such that the absorption by it causes an ellipticity increase and a PA shift, relative to the galaxy PA, when moving to shorter wavelengths (compare Figs.\,\ref{n3227cb_s} and \ref{n3227ci_s}). In the NIR the PA does not show any shift relative to the galaxy PA, and the ellipticity profile does not have a maximum in this region \citep[see Fig. 2 of][]{MRK1_97}.

\begin{figure*}[htbp]
   \centering
\begin{minipage}[t]{5.6cm}   
\includegraphics[width=5.6cm]{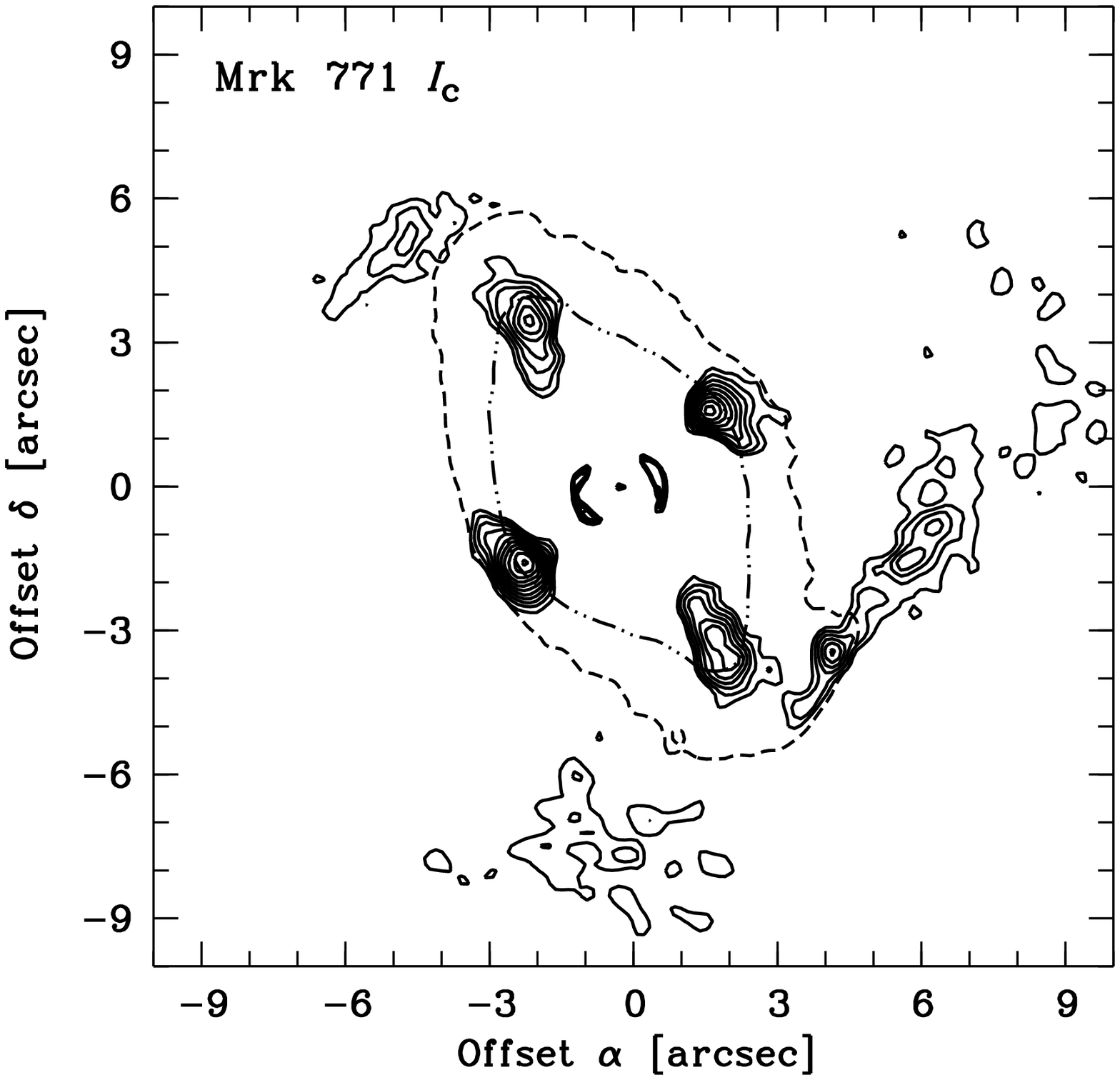}
      \caption{Mrk\,771 $I_{\rm \scriptstyle C}$ model-subtracted residual map. The contours relevant to the $c_4$ coefficient maximum ($a\,$$=$$\,4\arcsec$; dash-double-dotted) and to the ellipticity maximum are overplotted. Note the cross-shaped structure, the disky isophote associated with it, and the features extending on either bar side.}
\label{m771cid_Ei}
\end{minipage}
\hspace{0.6cm}
\begin{minipage}[t]{5.6cm}   
\includegraphics[width=5.6cm]{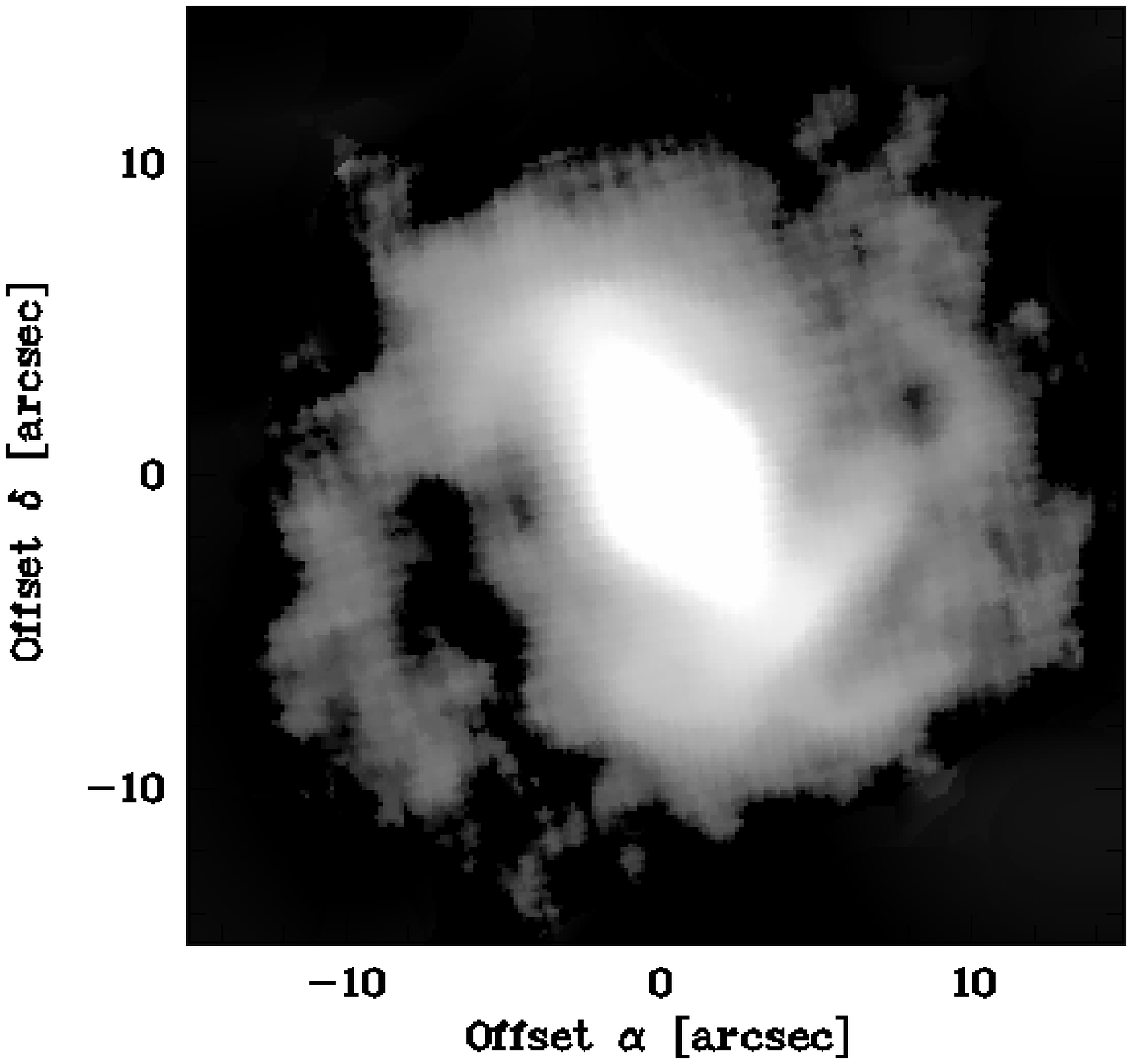}
      \caption{Mrk\,771 $I_{\rm \scriptstyle C}$ unsharp mask-subtracted residual image. The asymmetric spiral structure is 
clearly outlined.}
\label{20_Mrk771ius}
\end{minipage}
\hspace{0.6cm}
\begin{minipage}[t]{5.6cm}   
\includegraphics[width=5.6cm]{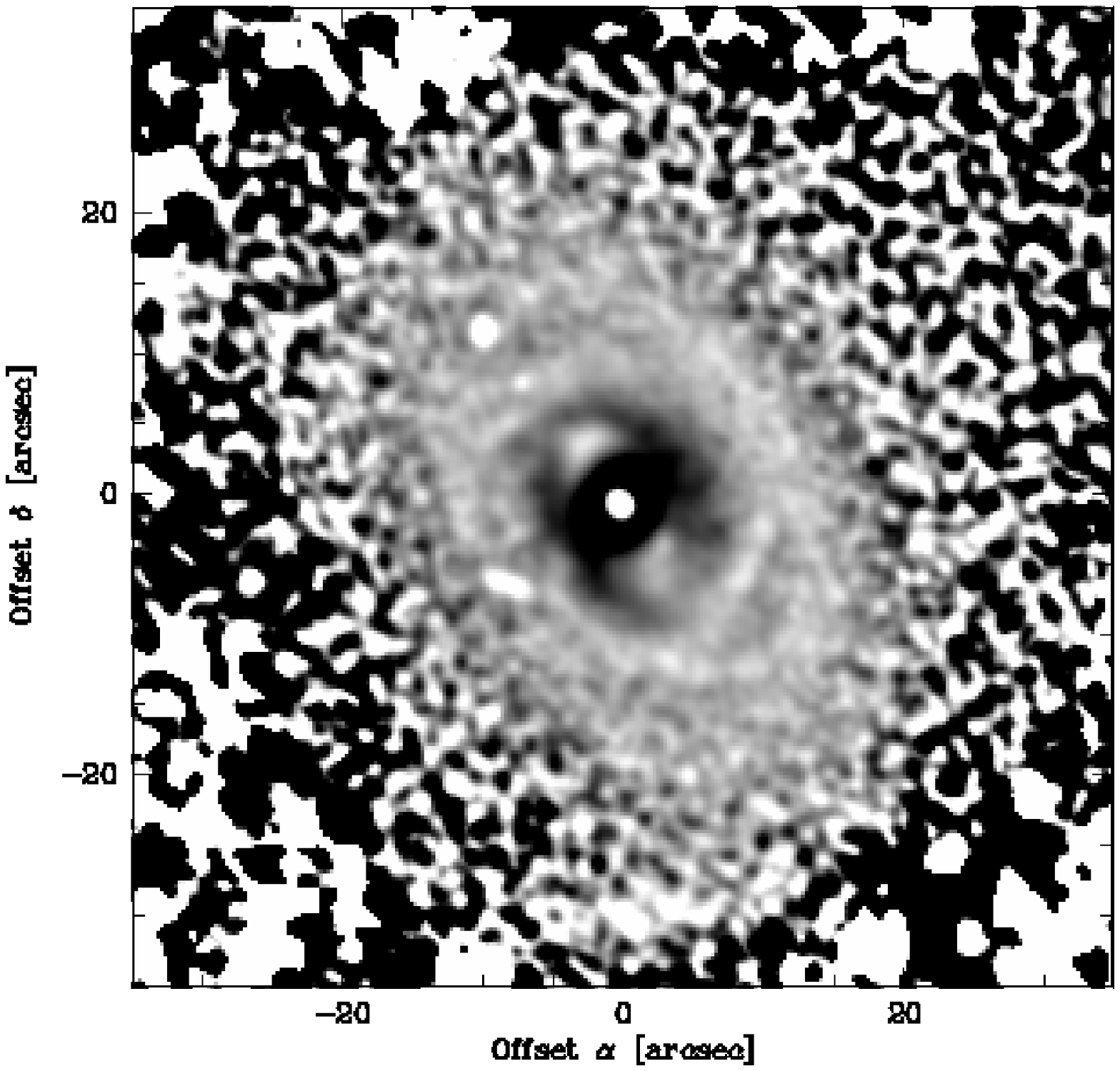}
      \caption{Mrk\,279 $R_{\rm \scriptstyle C}$ structure map. The oval/lens, together with the other features discussed in text could be traced.}
\label{22_Mrk279r_sm0}
\end{minipage}
\end{figure*}

 \begin{figure*}[htbp]
   \centering
\begin{minipage}[t]{5.6cm}   
\includegraphics[width=5.6cm]{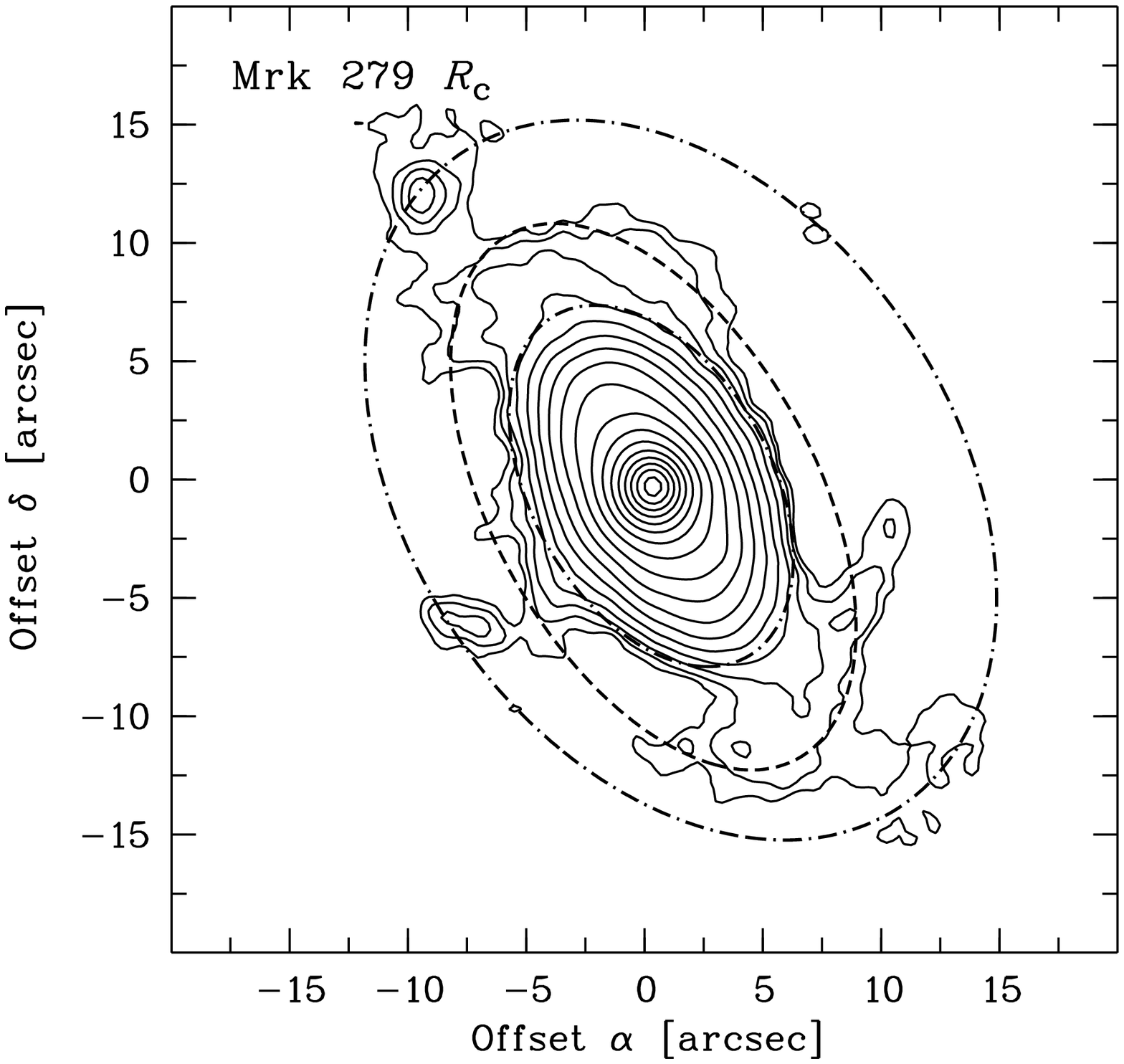}
  \caption{Mrk\,279 $R_{\rm \scriptstyle C}$ unsharp mask-subtracted residual map with the model contour corresponding to the ellipticity maximum and the ones bracketing it ($a\,$$\approx$$\,8\arcsec,\,$$16\arcsec$; dash-dotted) overplotted. The ellipticity maximum is related to the two straight features.}
\label{22_m279crmedl_Ee}
\end{minipage}
\hspace{0.6cm}
\begin{minipage}[t]{5.6cm}   
\includegraphics[width=5.6cm]{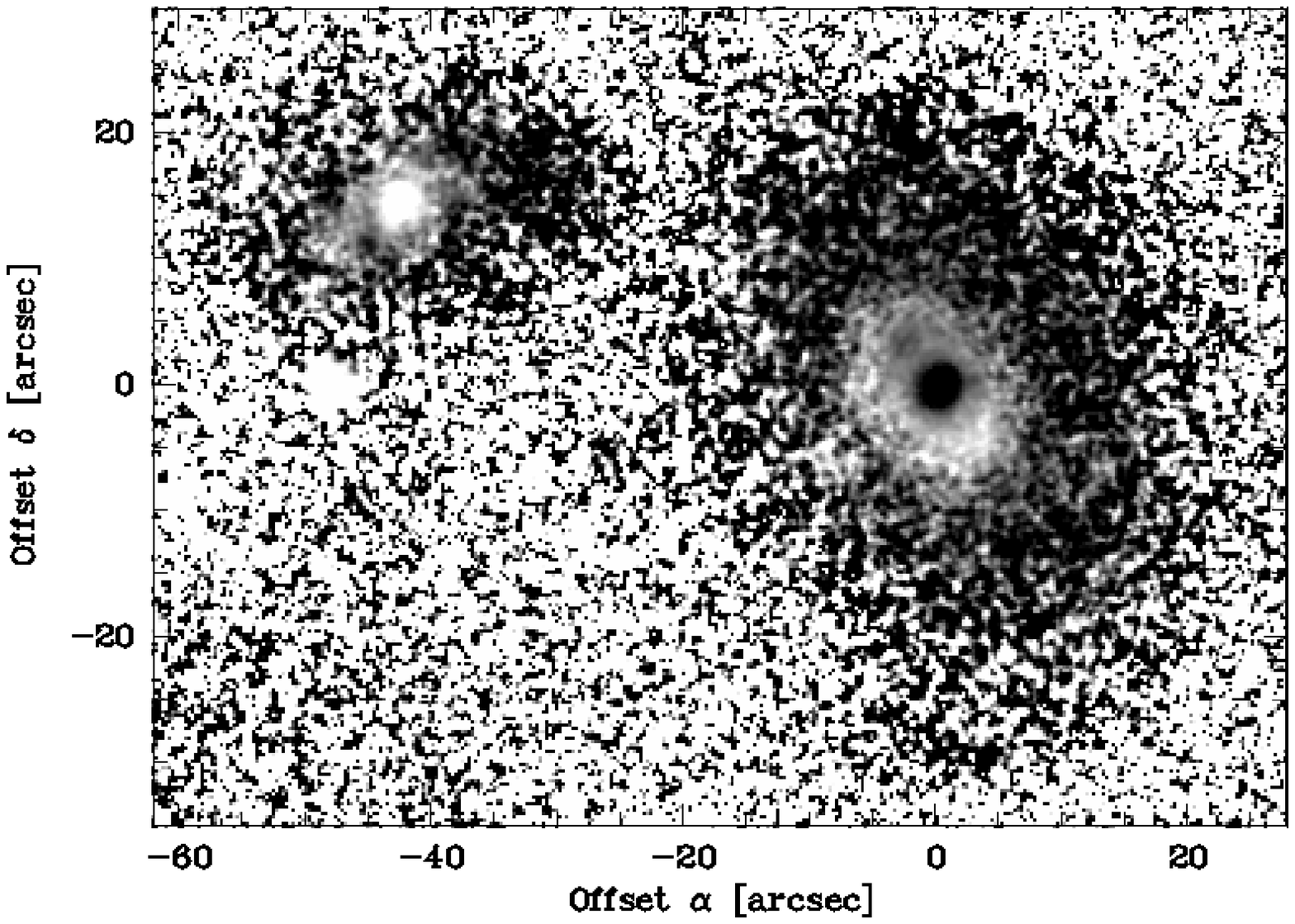}
      \caption{Mrk\,279 $B\,$$-$$\,I_{\rm \scriptstyle C}$ image. The CI coding ranges from $1.7$ to $2.3\,\rm mag\,\rm arcsec^{-2}$. The outer galaxy parts are blue and asymmetric, and the companion gets bluer towards Mrk\,279.}
\label{22_Mrk279bi_b_2}
\end{minipage}
\hspace{0.6cm}
 \begin{minipage}[t]{5.6cm}   
 \includegraphics[width=5.6cm]{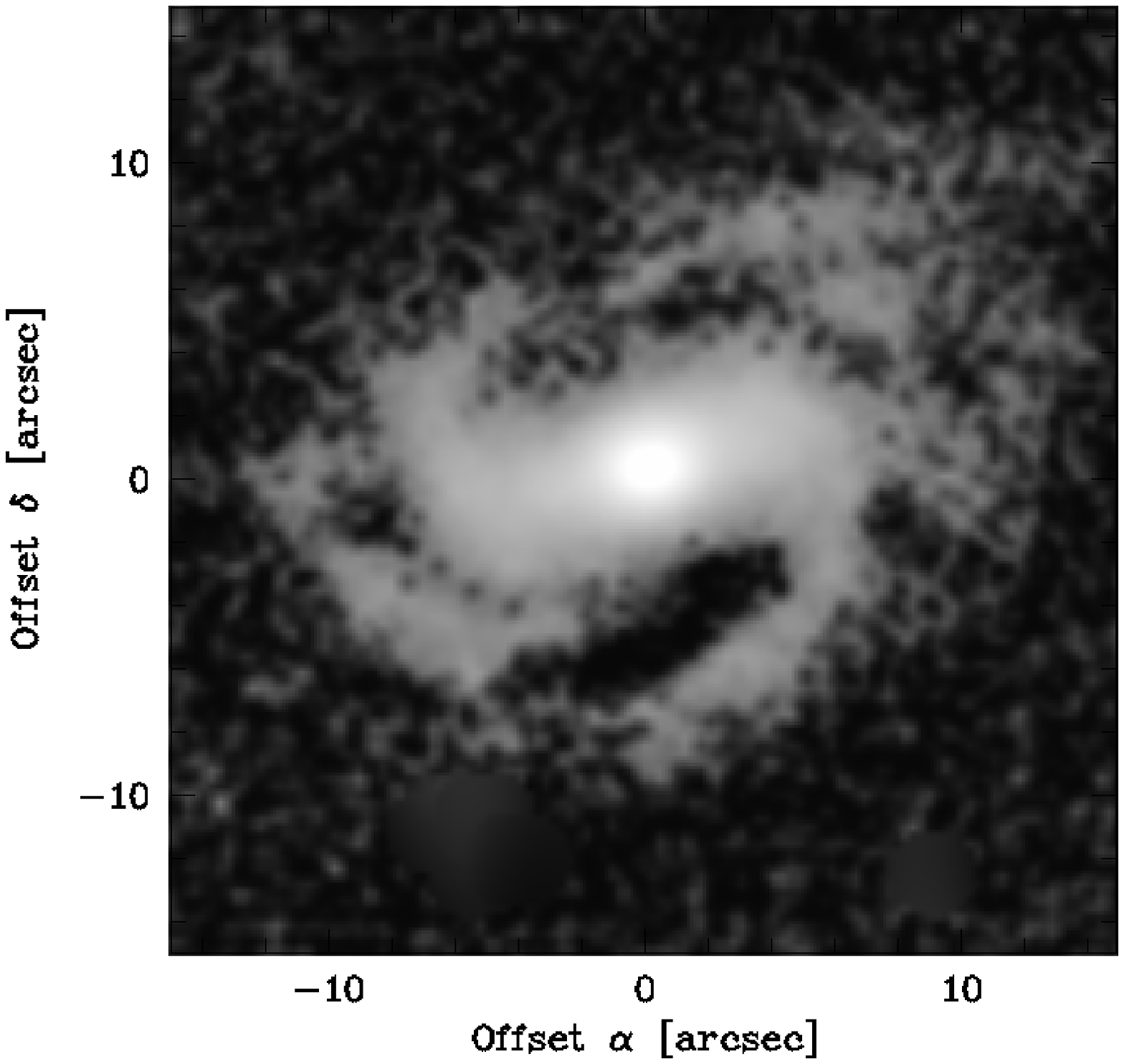}
       \caption{Ark\,479 $V$ unsharp mask-subtracted residual image. The bar and spiral structure can be traced.}
 \label{25_Ark479vsmed19}
\end{minipage}
\end{figure*}

\subsection{NGC\,3516}
 The unsharp masked residual image (Fig.\,\ref{16_NGC3516rsD}) is suggestive of an incomplete inner ring around the bar. The presence of some structures at the bar ends was mentioned by \citet{KPL_02}.

\subsection{NGC\,4051}

The dust lane to the S of the nucleus \citep{DCK_06} causes the $B$ ellipticity peak at $a\,$$\approx$$\,10\arcsec$. The bar and spiral arms together result in a broad ellipticity maximum, accompanied by SB bumps in the region $a\,$$\approx$$\,25\arcsec\,$-$\,110\arcsec$.

\subsection{NGC\,4151}

The behaviour of the profiles in the inner $15\arcsec$ is most probably related to an inner disk on a similar scale \citep[e.g.,][]{S_75,B_81,G_08}, an extended narrow-line region \citep{PGT_89,AMP_98}, and dust arcs \citep[e.g.,][]{U_00}.

\subsection{Mrk\,766}

The higher $V$ ellipticity maximum in the region of the bar is related to the [\ion{O}{iii}] emission, extended along the NW-SE direction \citep[Fig.\,\ref{19_Mrk766vdmedl_viE}, see also][]{MWT1_96}.

\subsection{Mrk\,771}

The cross-shaped structure in the model-subtracted residual map (Fig.\,\ref{m771cid_Ei}), accompanied by disky isophotes \citep[positive fourth-order Fourier cosine coefficient $c_4$; we define $c_4$ after][]{MJ_99} is due to the combined influence of the bulge and bar. 
There is a chain of blue knots near the SW bar end (see Fig.\,\ref{m771cid_Ei}) that has been associated with a small merging companion \citep{HN_92,HHS_94} or with the bar itself, given the similar fainter feature near the opposite bar end \citep{SSE_01}. 
Similar structures have been observed in Mrk\,279. 
The unsharp masked residual image (Fig.\,\ref{20_Mrk771ius}) reveals a weak asymmetric spiral structure forming a nearly complete pseudo-ring.
 
In the region of the bar, the ellipticity profile has a broad maximum at a roughly constant PA; however, the expected SB bump is not present as the outer bar parts remain unfitted (Fig.\,\ref{m771cid_Ei}). The SB bump around $a\,$$=$$\,6\arcsec$, accompanied by a $V\,$--$\,I_{\rm \scriptstyle C}$ dip and a peak, superposed on the ellipticity maximum, is related to the features on either side of the bar. 
To estimate the bar parameters, we extracted archival HST WFPC2 $F606W$ profiles, on which the ellipticity peaks due to the bar and the features discussed are detached.
The innermost variations in the profiles are artifacts from the guide hole of $a\,$$=$$\,0\farcs75$ \citep{HN_92}.

\begin{figure*}[htbp]
   \centering
\begin{minipage}[t]{5.6cm}   
\includegraphics[width=5.6cm]{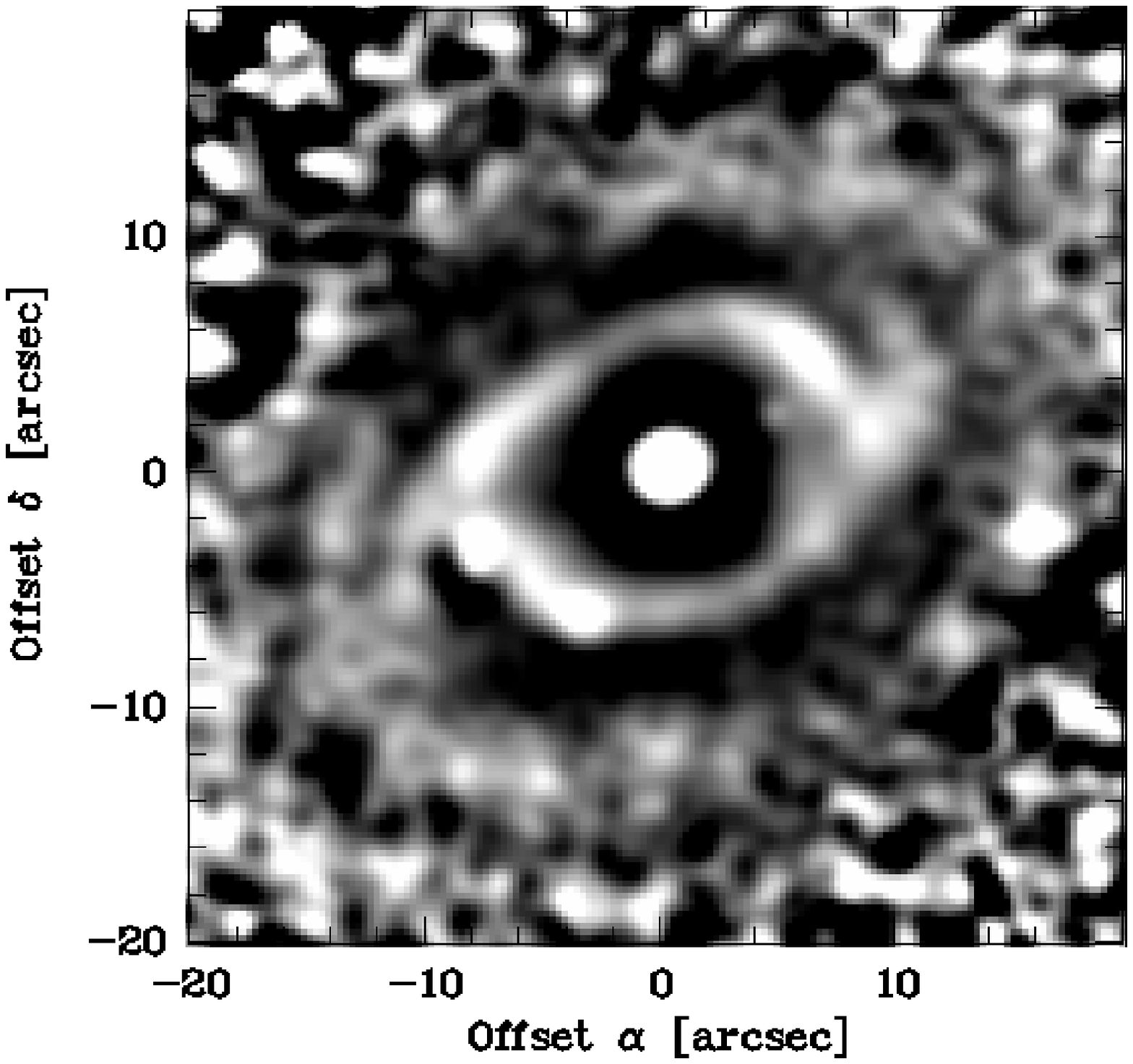}
      \caption{Mrk\,506 $R_{\rm \scriptstyle C}$ structure map. The rings, together with the spiral arms, are clearly outlined.}
\label{26_Mrk506smr}
\end{minipage}
 \hspace{0.6cm}
\begin{minipage}[t]{5.6cm}   
\includegraphics[width=5.6cm]{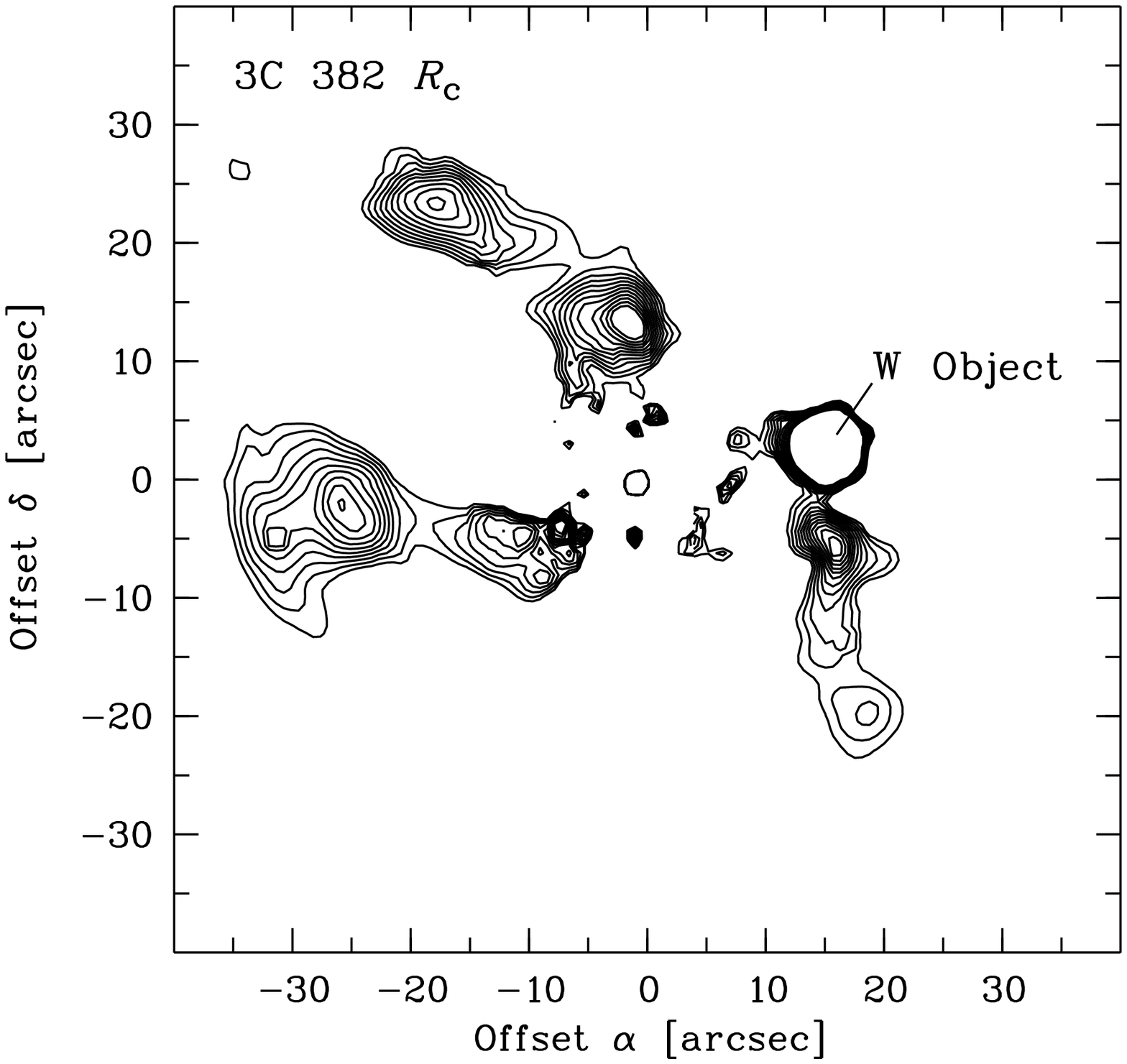}
      \caption{3C\,382 $R_{\rm \scriptstyle C}$ model-subtracted residual map. Three filaments are outstanding. The object 16\,$\arcsec$ to the W, cleaned from the images, is shown here for illustration. }
\label{c382crd_comp2}
\end{minipage}
\hspace{0.6cm}
\begin{minipage}[t]{5.6cm}   
\includegraphics[width=5.6cm]{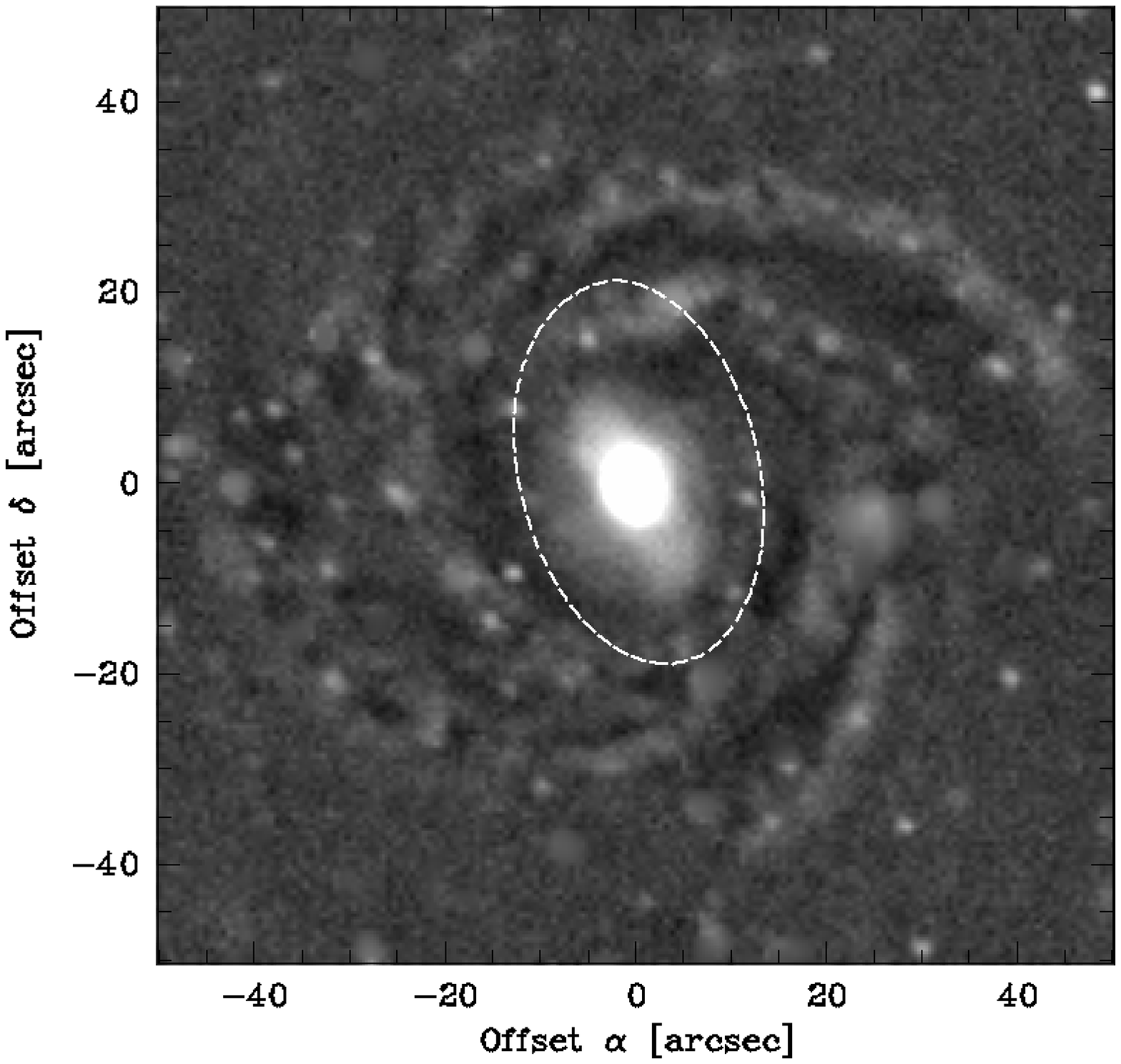}
      \caption{NGC\,6814 $R_{\rm \scriptstyle C}$ unsharp mask-divided residual image. The model contour, corresponding to the ellipticity maximum, is overplotted. The bar and the spiral structure can be traced.}
\label{30_NGC6814rDmed11_rE}
\end{minipage}
\end{figure*}

\subsection{NGC\,4593} 

The bar causes a broad ellipticity maximum around $a\,$$=$$\,47\arcsec$, accompanied by an SB bump.
The blue $V\,$--$\,I_{\rm \scriptstyle C}$ dip, together with an SB bump overlapping with the bar SB bump and an ellipticity peak superposed on the bar ellipticity maximum around $a\,$$=$$\,65\arcsec$, is related to an inner ring.

\subsection{Mrk\,279} 

The direct images reveal an outer ring displaced to the NW compared to the inner galaxy parts \citep[the annular appearance of the outer galaxy parts has already been noticed by][]{A_77} and a tail-like feature to the S. Around $a\,$$=$$\,5\arcsec$ there is an SB bump, accompanied by almost constant ellipticity and PA, and the ellipticity profile, extracted from the archival HST WFPC2 $F814W$ images, has a peak. In the corresponding region, the structure map (Fig.\,\ref{22_Mrk279r_sm0}) reveals a bar-like structure (its parameters were estimated using the HST profiles), which is most probably an oval/lens, given the small deprojected ellipticity of 0.13. The bar hypothesis has already been discussed \citep{KSP2_00,PM_02,SKL_04}. 
Furthermore, there are two straight features on either side of the oval/lens (similar to those in Mrk\,771) and some more compact structures about $15\arcsec$ NE and $9\arcsec$ SE of the nucleus (see Fig.\,\ref{22_m279crmedl_Ee}). The NE structure appears blue and elongated in the HST images, and the SE one, which is actually two objects that are most probably projected, is red.
Furthermore, the straight features \citep[hinted at by][]{A_77} and the tail-like one are blue; as a whole, the outer galaxy parts appear blue and asymmetric (Fig.\,\ref{22_Mrk279bi_b_2}). The disturbed morphology of Mrk\,279 could be a result of interaction with the companion, which is bluer and more extended toward it. The unsharp masked residual image reveals spiral arm stubs in the companion, suggesting SA0/a\,pec morphology.

The wavelength-dependent ellipticity maximum at $a\,$$\approx$$\,13\arcsec$ is related to the straight features. The ring produces an SB bump, accompanied by blue CI dips around $a\,$$=$$\,17\arcsec$. The spiral dust lanes, best traced at $a\,$$\approx$$\,5\arcsec\,$-$\,8\arcsec$ in the HST images \citep[see e.g.,][]{PM_02}, produce the red CI bump around $a\,$$=$$\,6\arcsec$.

\begin{figure*}[htbp]
   \centering
\begin{minipage}[t]{5.6cm}   
\includegraphics[width=5.6cm]{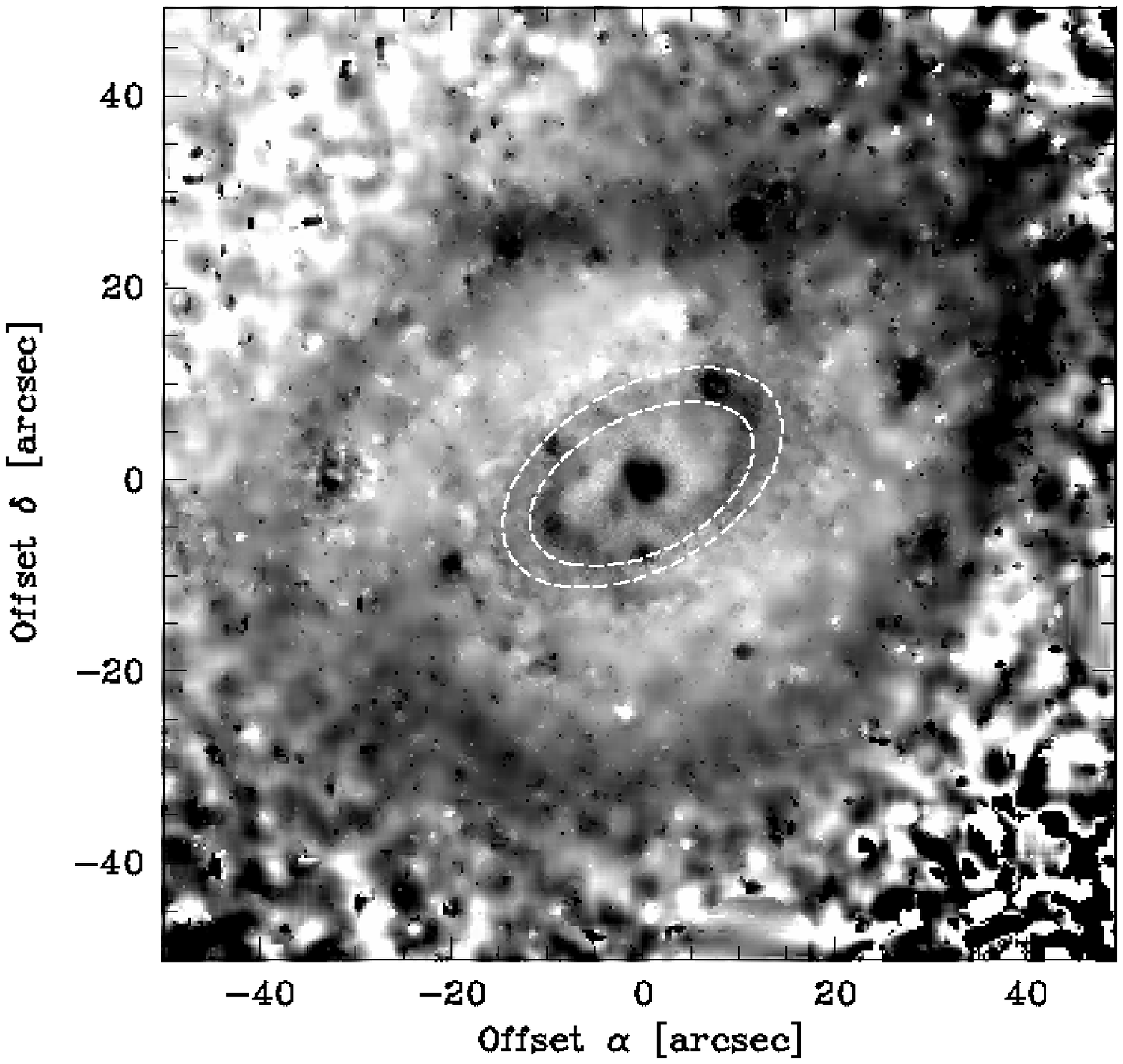}
      \caption{NGC\,7469 $B\,$--$\,I_{\rm \scriptstyle C}$ image. The CI coding ranges from $1.6$ to $2.7\,\rm mag\,\rm arcsec^{-2}$. Overplotted are the model contours corresponding to the two $I_{\rm \scriptstyle C}$ ellipticity peaks. The pseudo-rings and the dust can be traced.}
\label{35_NGC7469bi_Ew}
\end{minipage}
\hspace{0.6cm}
\begin{minipage}[t]{5.6cm}   
\includegraphics[width=5.6cm]{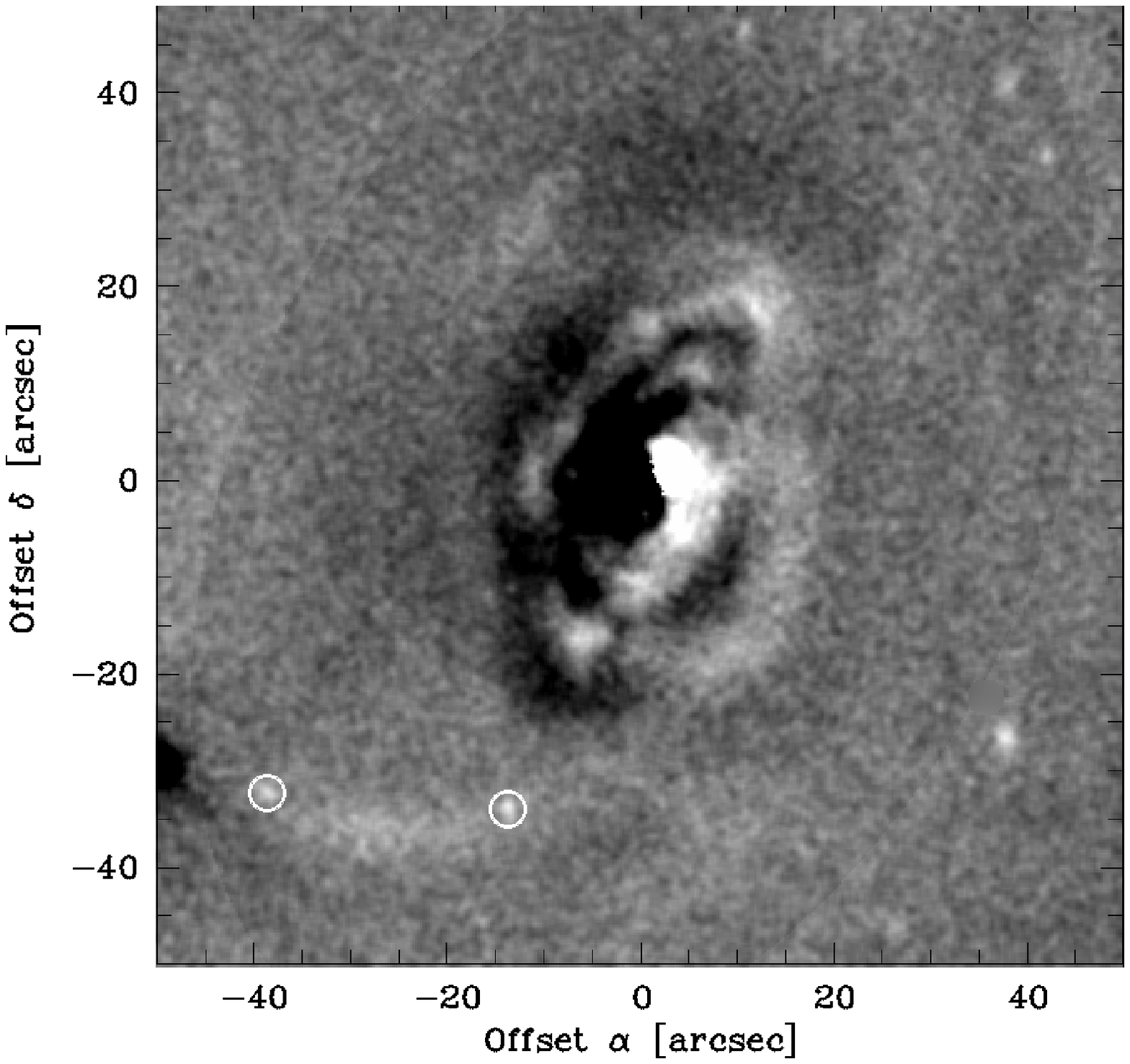}
      \caption{NGC\,7603 $V$ model-subtracted residual image. There can be traced a complex of loop-like features and a filament with two emission-line galaxies overposed (encircled).}
\label{37_Mrk530vd_c}
\end{minipage}
\hspace{0.6cm}
\begin{minipage}[t]{5.6cm}   
\includegraphics[width=5.6cm]{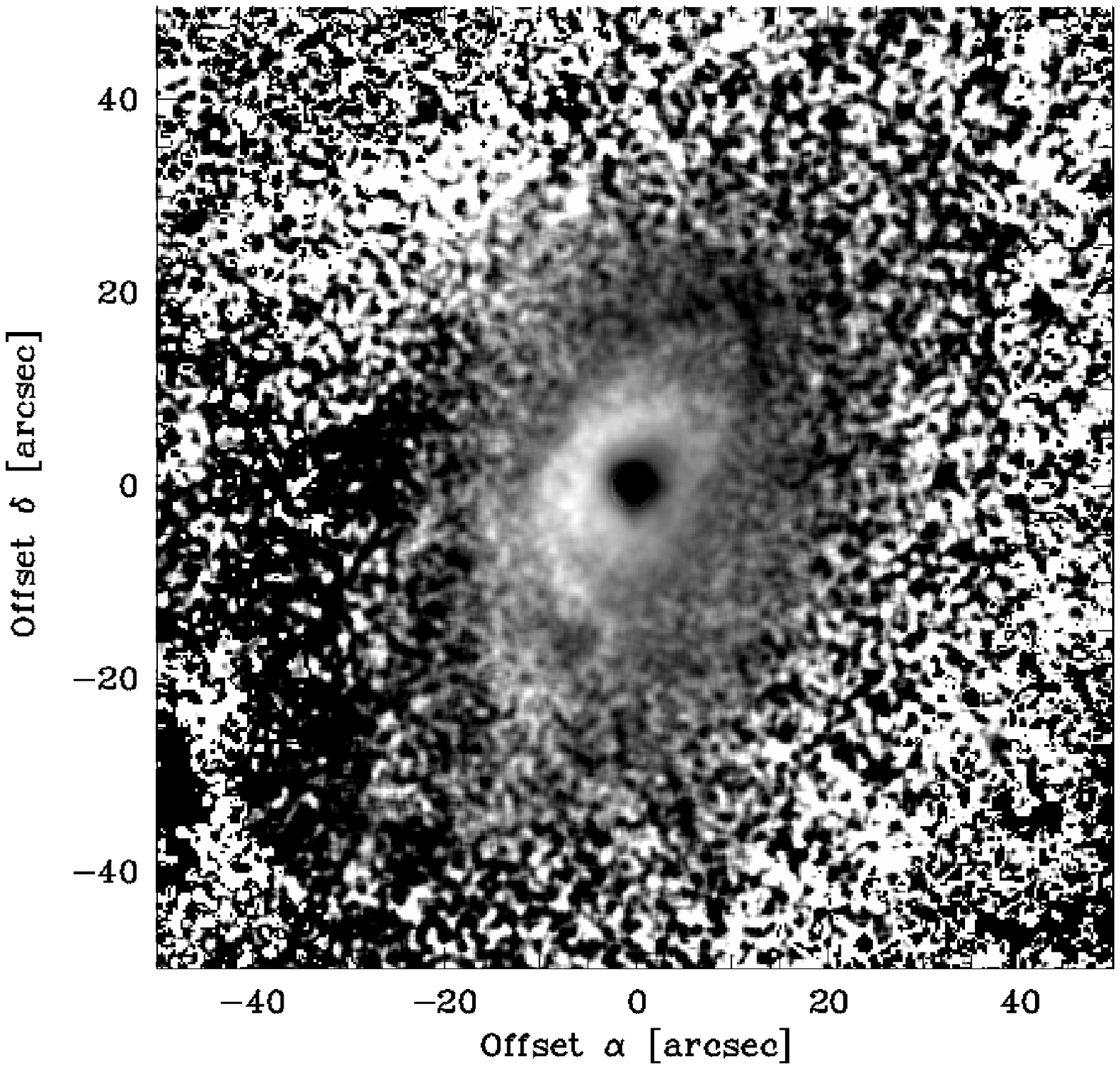}
      \caption{NGC\,7603 $V\,$--$\,I_{\rm \scriptstyle C}$ image. The CI coding ranges from $1.15$ to $1.7\,\rm mag\,\rm arcsec^{-2}$. Note the red dust lanes and the blue loop-like features.}
\label{37_Mrk530vi}
\end{minipage}
\end{figure*}

\subsection{NGC\,5548} 

The galaxy has shells and tidal tails that are suggestive of a merger event \citep[][]{SS_88,NHS_90,TFG_98}. 
The shells and the curved tail result in SB bumps and in ellipticity and PA peaks.

\subsection{Ark\,479}

The ellipticity profile shows a maximum around $a\,$$=$$\,6\arcsec$ accompanied by a weak SB bump and an almost constant PA, which corresponds to a bar. 
The unsharp masked residual image (Fig.\,\ref{25_Ark479vsmed19}) reveals the bar and spiral structure. The latter causes wavelength dependence of the ellipticity maximum and produces a weak SB bump (best expressed in $V$) further out.

\begin{figure*}[htbp]
   \centering
\begin{minipage}[t]{5.6cm}   
\includegraphics[width=5.6cm]{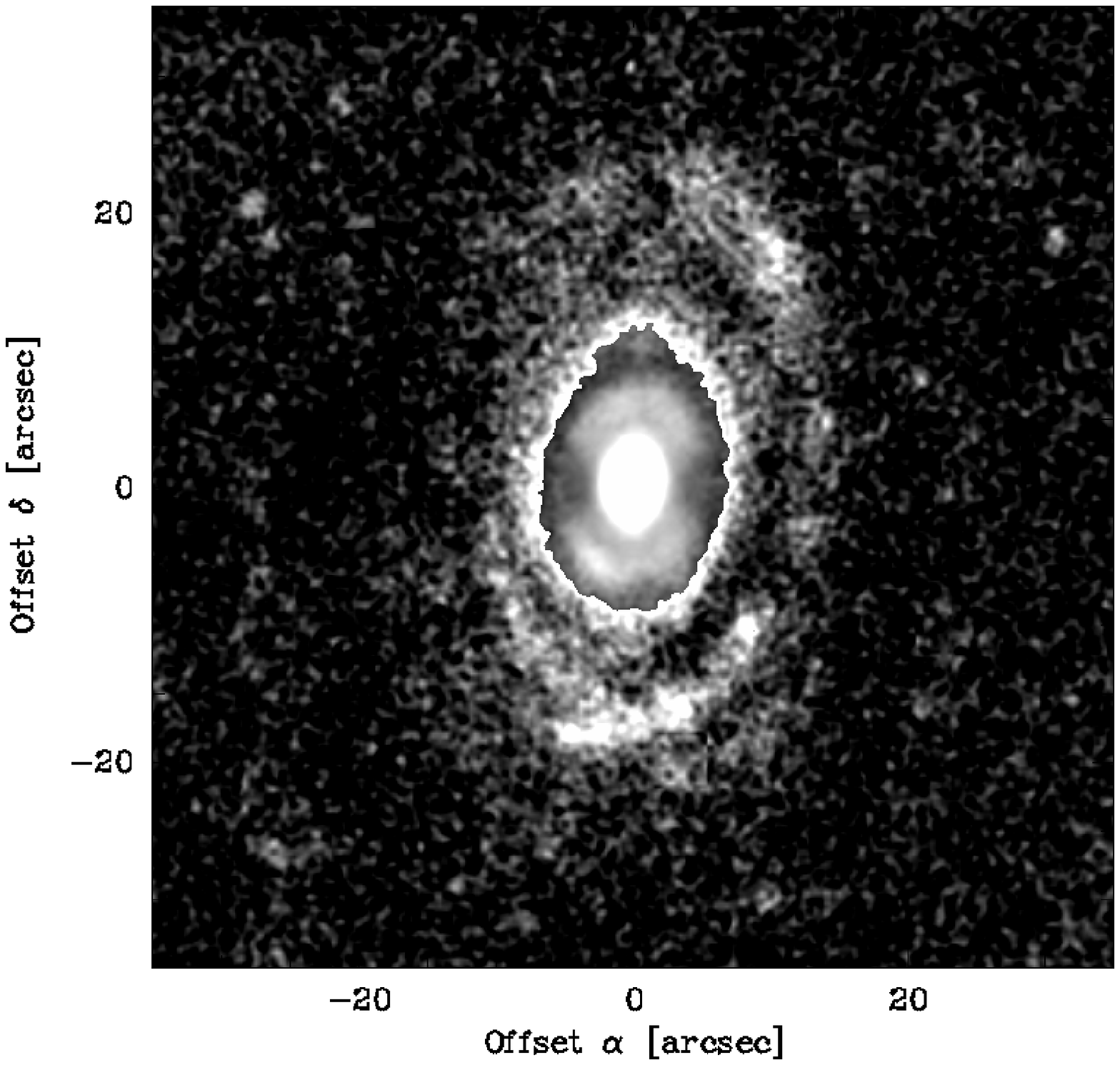}
      \caption{Mrk\,541 $R_{\rm \scriptstyle C}$ residual image, composite of an unsharp mask-divided/-subtracted one. The broken inner ring and the knotty outer ring are clearly outlined.}
\label{38_Mrk541rdDmed2}
\end{minipage}
\hspace{0.6cm}
\begin{minipage}[t]{5.6cm}   
\includegraphics[width=5.6cm]{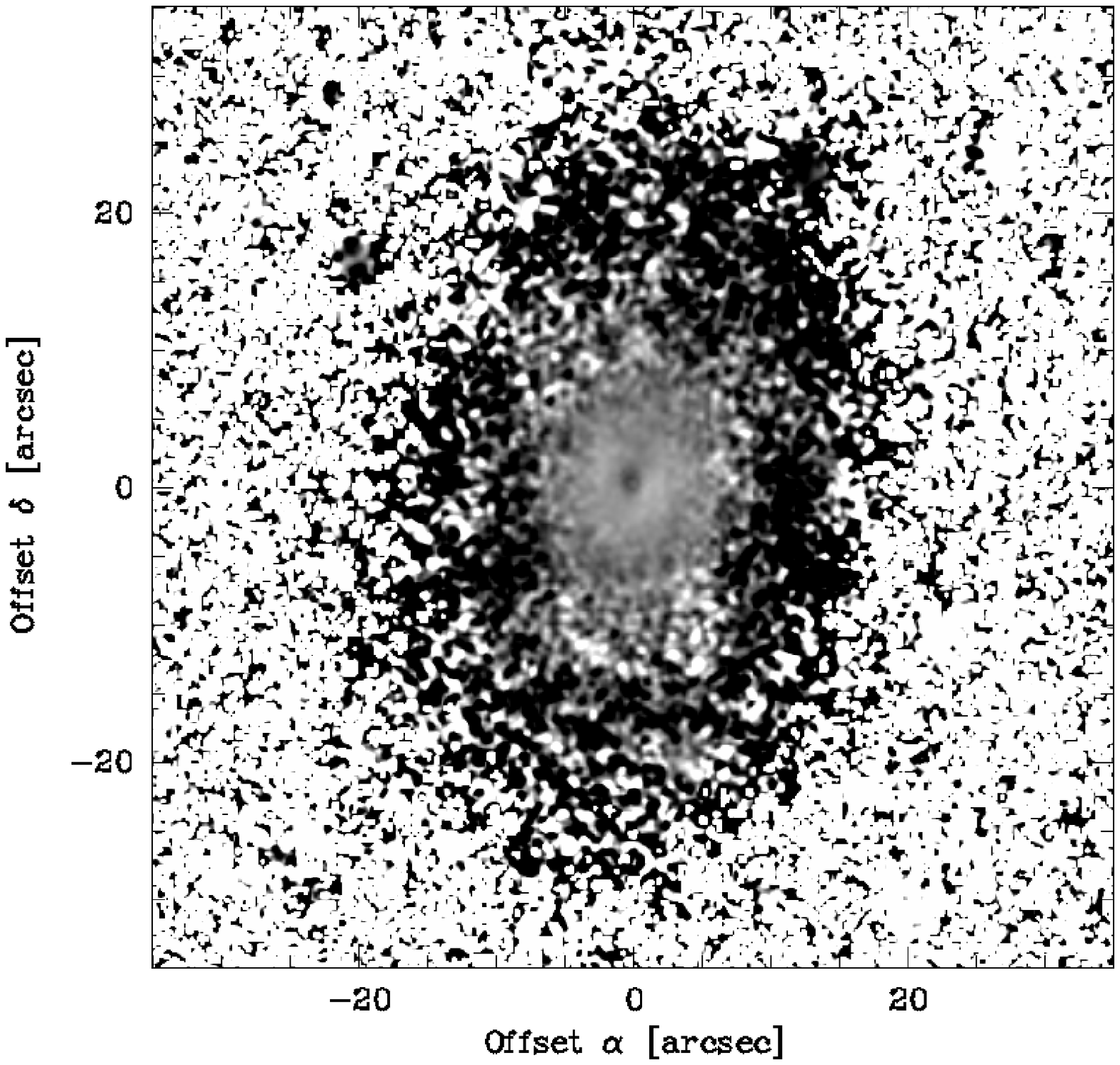}
      \caption{Mrk\,541 $V\,$--$\,I_{\rm \scriptstyle C}$ image. The CI coding ranges from $0.9$ to $1.5\,\rm mag\,\rm arcsec^{-2}$.}
\label{38_Mrk541vi}
\end{minipage}
\end{figure*}

\subsection{Mrk\,506} 

The structure map (Fig.\,\ref{26_Mrk506smr}) reveals a blue inner ring ($a\,$$\times$$\,b\,$$\approx$$9\arcsec$$\times$$6\arcsec$), noticed already by \citet{A_77}, and a couple of faint spiral arms, emerging out of it and reaching a faint outer ring ($a\,$$\times$$\,b\,$$\approx$$16\arcsec$$\times$$13\arcsec$). It is useful to note that \citet{SS_80} included Mrk\,506 in the group of double-ringed galaxies. The inner ring produces a $B\,$--$\,I_{\rm \scriptstyle C}$ dip and an SB bump, and the weak $B$ bump following the latter is due to the spiral arms.

Although Mrk\,506 is classified as weakly barred (RC3), the behaviour of  the profiles does not indicate a bar. Furthermore, after a 2D elliptical ring model subtraction, the SB bump around $a\,$$=$$\,9\arcsec$ practically disappears.

\subsection{3C\,382}

Three filaments can be discerned in the model-subtracted residual map (Fig.\,\ref{c382crd_comp2}). The NE and E filaments are oriented toward a barred spiral galaxy about $1\farcm4$ to the NE. According to \citet{RE_00}, the two galaxies are interacting and the blue object $16\arcsec$ to the W (Fig.\,\ref{c382crd_comp2}) might be a gas-rich starburst dwarf galaxy in the process of merging into 3C\,382. 
In our view, this object is most probably projected. It has a stellar-like appearance in archival HST WFPC2 images; subtraction of Tiny Tim generated PSF \citep{K_95} left no significant residual structure.

The SW filament could also be of tidal origin.
The ellipticity dip around $a\,$$=$$\,11\arcsec$ is associated with the NE filament. The filaments result in weak SB bumps.

\subsection{NGC\,6814}

A weak bar can be traced in the unsharp masked residual image (Fig.\,\ref{30_NGC6814rDmed11_rE}). Its signature on the profiles, however, is masked by the spiral structure, that produces SB bumps, best pronounced in $B$, blue CI dips, and a wavelength-dependent ellipticity maximum in the region $a\,$$\approx$$\,10\arcsec\,$-$\,40\arcsec$. To estimate the bar parameters, we extracted $J$ profiles (using Two Micron All Sky Survey images), on which the ellipticity peaks due to the bar and the spiral arm beginnings are detached.

\subsection{Mrk\,1513}  

The SB bump around $a\,$$=$$\,14\arcsec$ is produced by the outer pseudo-ring.

\subsection{Ark\,564} 

The bar results in an ellipticity maximum, accompanied by an SB bump around $a\,$$=$$\,8\arcsec$. 
The partial fitting of the spiral arm beginnings by the model causes wavelength dependence of the ellipticity maximum, a blue $B\,$--$\,I_{\rm \scriptstyle C}$ dip, and an SB bump, overlapping with the bar bump and largest in $B$. 
The spiral structure forms a blue pseudo-ring ($a\,$$\times$$\,b\,$$\approx$$16\arcsec$$\times$$12\arcsec$), which produces an SB bump and a $B\,$--$\,I_{\rm \scriptstyle C}$ dip.

\subsection{NGC\,7469}

The ellipticity profile shows a maximum around $a\,$$=$$\,14\arcsec$, accompanied by an SB bump, and an almost constant PA. The behaviour of the profiles there is not typical~-- in the inner part it is dominated by an inner pseudo-ring ($a\,$$\times$$\,b\,$$\approx$$13\arcsec$$\times$$7\arcsec$), and in the region of the outer part it is due to a bar-like structure (Fig.\,\ref{35_NGC7469bi_Ew}). This is best illustrated in $I_{\rm \scriptstyle C}$ by a double-peaked ellipticity maximum and a corresponding weak double structure of the SB bump. Given the small deprojected ellipticity of 0.12, the bar-like structure is most probably an oval/lens. \citet{MM_94} suggested the bump is due to a lens but based mainly on the fact that they could not find a reasonable fit by a bar. The wavelength dependence of the ellipticity maximum is caused by the inner pseudo-ring and the spiral structure. The variations in the PA profile are related to the spiral structure.

\subsection{Mrk\,315} 

The inner $7\arcsec$ of the profiles are influenced by underlying features. \citet{CAM_05} associated them with a dwarf galaxy remnant and star formation regions; a faint spiral structure could be traced in their residual images.

\subsection{NGC\,7603}

NGC\,7603 and the galaxy about $1\arcmin$ to the SE are an example of an anomalous redshift association \citep{A_71}.
NGC\,7603 is disturbed and shows evidence of tidal interaction \citep[see][and references therein]{LG_04}. 

There are a number of loop-like features (Fig.\,\ref{37_Mrk530vd_c}), which appear blue in the $V\,$--$\,I_{\rm \scriptstyle C}$ image (Fig.\,\ref{37_Mrk530vi}). They result in an SB bump, accompanied by a blue $V\,$--$\,I_{\rm \scriptstyle C}$ dip and a weak ellipticity peak at $a\,$$\approx$$\,21\arcsec$.

\subsection{Mrk\,541}

The unsharp masked residual image (Fig.\,\ref{38_Mrk541rdDmed2}) unveils an inner ring ($a\,$$\times$$\,b\,$$\approx$$7\arcsec$$\times$$5\arcsec$), broken roughly along the galaxy minor axis and a knotty outer ring ($a\,$$\times$$\,b\,$$\approx$$20\arcsec$$\times$$13\arcsec$), more pronounced to the W and displaced to the N with respect to the nucleus. Both rings appear blue in the $V\,$--$\,I_{\rm \scriptstyle C}$ image (Fig.\,\ref{38_Mrk541vi}). The outer ring was noted already by Adams (1977), and \citet{SS_80} included Mrk\,541 in the group of double-ringed galaxies. The rings produce SB bumps, accompanied by blue dips on the CI profile.

\end{document}